\documentclass[a4paper,12pt]{article}
\usepackage{amsmath}
\usepackage{amssymb}
\usepackage{bm}
\usepackage[dvipdfmx]{graphicx,xcolor}
\usepackage{enumerate}
\usepackage[compat=1.1.0]{tikz-feynhand}
\usepackage{ulem}
\usepackage{wrapfig}
\newcommand{\Slash}[1]{{\ooalign{\hfil/\hfil\crcr$#1$}}} % from Okumura Bibunsho
\newcommand{\ctext}[1]{\raise0.2ex\hbox{\textcircled{\scriptsize{#1}}}}
\numberwithin{equation}{subsection}
\numberwithin{figure}{subsection}
\DeclareMathOperator{\Real}{Re}

\DeclareMathOperator{\Tr}{Tr}
\begin{document}

\thispagestyle{empty}
\vspace*{-15mm}
{\bf OCHA-PP-372}\\
%%%%%%
\vspace{10mm}
\begin{center}
{\Large\bf
Infrared Divergence and Low Energy Theorem\\
 in Non-Abelian Gauge Theory%
}
 
\vspace{7mm}

\baselineskip 18pt

\vspace{2mm}
Akio Sugamoto

\vspace{2mm}

{\it
%Tokyo Bunkyo Study Center, The Open University of Japan (OUJ), \\
%Tokyo 112-0012, Japan \\
Ochanomizu University, 2-1-1 Ohtsuka, Bunkyo-ku, Tokyo 112-8610,
Japan
}

\end{center}

\vspace{10mm}
%%%%%%%%%%%%%%%%%%%%%%%%%%%%%%%%%%%%%%%%
%%%%%                              %%%%%
%%%%%          Abstract            %%%%%
%%%%%                              %%%%%
%%%%%%%%%%%%%%%%%%%%%%%%%%%%%%%%%%%%%%%%
\begin{abstract}

This is the English translation of the Doctor thesis submitted to the
 University of Tokyo in March 1978.  
 Its Japanese version was published in Soryushiron Kenkyu 60-2 (1979-11) pp. 47--117.
 An appendix is newly added, which is the excerption of a relevant mathematical part
 of a report submitted to Prof.\ Mikio Sato in August 1976.

In the thesis, first, the cancellation of infrared divergences is reviewed.  The cancellation occurs between a virtual process with soft photon corrections and the other process with real soft photon emissions. The factorization of infrared divergences is shown by using the eikonal approximation, or by the (renormalization group like) differential equation controlling the infrared divergences.

Next, two examples in QCD, the fermion (quark)-fermion (quark) scattering and the fermion (quark) gauge-boson (gluon) scattering are examined at one loop, from which the importance of Ward-Takahashi identities becomes manifest for the cancellation to occur. (The identities represent the group properties.)

After these preparations, the factorization of infrared divergences in QCD is proved at all orders in the perturbation theory, by full usage of the Ward-Takahashi identities.  The identites are also proved at all orders.  In these proofs, the axial gauge condition is used, which simplifies the derivation of the Ward-Takahashi identities as well as the proof of the unitarity.  The cancellation of infrared divergences in QCD occurs among \underline{``the gauge invariant set of graphs''}, if the quantum numbers of color are averaged and summed over the initial and final states, respectively.  In this way, the low energy theorem by F. E. Low, for emission of one or two soft gauge bosons, are proved at all orders in the perturbation expansion.  

By joining the two emitted soft gauge bosons, the differential equation controlling the infrared divergences in QCD can be derived.  In QCD, the coupling constant renormalization introduces the other infrared divergence, which is governed by the beta function $\beta(g)$, or by the ultraviolet divergences in the pure Yang-Mills theory.

In the Appendix, the cancellation of infrared divergences is examined mathematically in terms of the singular spectrum $\widehat{S.S}$ of Mikio Sato's  microfunction.

In the Epilogue (2022), a motivation for recently translating the thesis in English is briefly stated. 
\end{abstract}

%\tableofcontents

\newpage

% Section 1
\section{Introduction and Summary}

%%%%%%%%%%%%%%%%%%%%%%%%%%%
%IRsdtr1 (Section 1)%
%
The force potential acting between quarks via strong interaction is considered
to rise linearly, when the distance between quarks becomes larger.
To search for its reasoning in quantum field theory, we have to investigate
the long distance behavior [{\it i.e.}, the infrared behavior
in momentum space] of the non-Abelian gauge theory, since it describes
the dynamics of gluon fields connecting quarks.  From the infrared regions
the divergences called infrared divergences appear, due to the masslessness of
the gluon particle.  It is considered from some time ago, that the information
on the infrared behavior of the non-Abelian gauge theory,
or on the confinement of quarks in other words, can be obtained
from the study of infrared divergences in the theory \cite{1}.%

The Quantum Electrodynamics (QED) has infrared divergences similar
to non-Abelian gauge theory (in the following we will use QCD for it,
Quantum Chromodynamics), since photon is massless.  The infrared divergences
in QED have been studied well for a long time \cite{2}--\cite{4}.
We have to refer to the well-known QED and extract the special characteristics
in QCD, when we study the infrared divergences in QCD.

The results obtained by the past several years' investigation
on the infrared region of QCD are summarized roughly as follows:
\begin{enumerate}[A)]
\item Cancellation of infrared divergences:
It is shown by the lower order graphs that,
by emitting non-detectable extra gauge bosons, a physically meaningful
differential cross section has no infrared divergence
\cite{5}--\cite{8}, \cite{12}.
These works are based on the renormalization carried out at off-shell;
the cancellation of infrared divergences is violated
when on-mass-shell renormalization is performed.
The proof of the cancellation in general is an unsolved problem.

\item Problem of summing up infrared divergences:
Renormalization group-like differential equations are predicted
in various processes at the lower orders \cite{9}--\cite{12},
and a proof at all orders is given \cite{13}.
Its proof is complicated compared with the simplicity of its answer.

\item Potential between quarks:
Some progress has been made in the idea of attributing
the origin of the linearly rising potential between quarks
to the infrared divergences in QCD \cite{14}.

\item Low energy theorem:
One of the low energy theorems has been proved at all orders
of the perturbation \cite{15}.
For this proof QCD in the axial gauge has been used.
Generalization of the proof to another low energy theorem has been done and
it is applied to B) Problem of summing up infrared divergences \cite{16}.

\item Others \cite{17}, \cite{18}.
\end{enumerate}

The above characteristics can be summarized in a word as follows:
the infrared behavior of QCD is characterized by
\emph{the coexistence of the behavior identical to QED
and the
QCD-specific divergences appearing in the coefficients of expanding
the on-mass-shell charge by the off-mass-shell charge}.

\vspace{3mm}
This thesis consists of the author's work \cite{7}
[reference paper I, classified as A) in the above classification],
his work done in collaboration with Norio Nakagawa and Hiroaki Yamamoto \cite{15}
[reference paper II, classified as D)],
and its generalization afterwards \cite{15} [classified as D) and B)].

\vspace{3mm}
%\newpage
Let us summarize the contents.
Section 2 is a summary of infrared divergences in QED,
in Section 3 cancellation of infrared divergences at one loop is discussed,
and Section 4 gives the proof of low energy theorem
using the axial gauge condition, and its application.

Sections 2--4 are divided into a number of subsections
the contents of which are explained in the following.

Firstly, taking the electron scattering by an external field as an example,
the way to extract infrared divergence factors at one loop is stated
in Subsection 2.1.  This is generalized to all orders in Subsection 2.2,
by deriving a differential equation with respect to $\lambda$,
where $\lambda$ is introduced as a photon mass,
acting as an infrared divergence cutoff.
This differential equation is easily solved and shows that
the infrared divergence factor at all orders becomes an exponential
with the one-loop result raised to its power.
It is very recently that the differential equation begins to be used
in the study of infrared divergences in QED \cite{19}.
In this subsection we use a method a little different from the usual one.
Our method is the QED version of the low energy theorem proposed by
Cornwall and Tiktopoulos in the study of infrared divergences in QCD.
They assumed the low energy theorem in QCD, without giving the proof of it
at all orders.
[The proof of this low energy theorem in QCD is a main theme of Section 4.]
In QED, however,
this low energy theorem can be easily derived,
by using the so-called eikonal identity.
Therefore, this Subsection 2.2 is the training place of various techniques
to be used later in Section 4.
In the next Subsection 2.3, the so-called Bloch-Nordsieck theory is reviewed;
that is, the infrared divergences studied in Subsection 2.2 can be cancelled
by adding the soft photon emission processes which are not separable by detection.
Here, in comparison with QCD, it is shown by the explicit estimation of $Z_3$
at one loop that
no infrared divergence newly appears in QED via the coupling renormalization.

The next Section 3 is a summary of the reference paper I,
checking the cancellation of infrared divergences in QCD at one loop.
In this section the covariant gauge QCD is used.
First, in Subsection 3.1, by examining one-loop Feynman diagrams
for fermion-fermion and fermion-gauge boson scatterings,
graphs having infrared divergences are picked up by power counting.
As a result, it becomes a good way to sum up a number of graphs
[one graph for the fermion-fermion scattering,
but three graphs for the fermion-gauge boson scattering].
The next Subsection 3.2 is important, giving a proof of the factorization
of infrared divergences (at one loop) in QCD.
It is trivial that the factorization in QCD of a soft gauge boson
coupled to the external fermion is identical to that in QED,
except for a color factor, but it is also proved that the factorization works
also for a soft gauge boson coupled to the external gauge boson,
by summing up a number of Feynman graphs.
This set of Feynman graphs is that for which Ward-Takahashi identities hold,
and is equal to the set of Feynman graphs appearing in Subsection 3.1.
[This recognition will be deepened in the later Section 4 in which
Ward-Takahashi identities play a crucial role for the factorization of
infrared divergences.]
%%%%%%%%%%%%%%%%%%%%%%%%%

In the next Subsection 3.3, it is cheeked that the cancellation of
infrared divergences occurs by adding the extra-emission of soft gauge bosons,
leading to the cancellation of infrared divergences at scattering amplitudes
expanded in the bare coupling.
Here, it is also shown that the cancellation of infrared divergences
does not occur unless the color indices are averaged and summed
over the initial and final states, respectively
[{\it i.e.}, under the condition that color can not be observed].
Furthermore, the classification of the divergences is done in this Subsection.
For the one-loop diagrams, we consider three cases in which
the internal lines are massless or not, and the infrared divergences are shown
to be classified, by comparing the analysis in the Feynman parameter space
and that in the momentum space.  Introducing two cutoffs $\lambda'$ and $\eta$
[$\lambda'$ cuts the lower end of the momentum $k_{\mu}$
as $|k_{\mu}| > \lambda'$ [no other infrared divergences],
and $\eta$ cuts the angle $\theta$ between two gauge bosons on the mass shell,
as $1-\cos \theta > \eta$], the divergences of $\ln \frac{1}{\lambda'}$ cancel
in the fermion-fermion scattering, while the divergences of
$\ln \frac{1}{\lambda'}$ and
$\ln \frac{1}{\lambda'}\ln \frac{1}{\eta}$
[the highest infrared divergence] cancel in the fermion-gauge boson scattering.
The divergence $\ln \frac{1}{\eta}$ is not discussed here,
but is partly addressed in the reference paper I.
[Similar cancellation of infrared divergences are checked independently
by a number of people \cite{5}--\cite{7}, \cite{12}.
A short summary on what was done so far for what process is given
in the Reference of the reference paper I and its Note added.]

The next Subsection 3.4 gives the infrared divergences specific to QCD
appearing in the charge renormalization.
The coefficients of expanding the coupling constant defined on the mass shell $g_R$
in the bare coupling $g_B$, have an infrared divergence specific to QCD.
This infrared divergence is equal to the coupling constant renormalization of the pure Yang-Mills theory (without fermions),  
when its ultraviolet cutoff is replaced by the infrared cutoff.
This is an infrared divergence specific to QCD,
being controlled by $\beta(g)$ of pure Yang-Mills theory.
[This quantity is pointed our by a number of people.%
\cite{7}, \cite{9}--\cite{12}]
Due to this infrared divergence, the cancelation of infrared divergences
is violated up to this divergence when the on-mass-shell renormalization
is performed.
On the other hand, under the off-mass-shell renormalization,
the divergence appears in the on-mass-shell charge.
This provides a suggestion towards the quark confinement.\cite{14}

From Section 4, the properties of infrared divergences are examined,
which hold at all orders in perturbation expansion.
The main theme is the proof of the low energy theorem in QCD at all orders
in perturbation.

%%%%%%%%%%%%%%%%%%%%%%%%%%%%%
% Section 2
\section{Summary of infrared divergence in QED}

%%%%%%%%%%%%%%%%%%%%%%%%
A number of results have been obtained so far
concerning infrared divergences and the low energy theorem in QED%
\cite{2}--\cite{4}, \cite{19}.  Let us summarize some of the results
that are useful
when we investigate such problems as infrared divergence,
low energy theorem, and quark confinement in non-Abelian gauge theories,
also known as quantum chromodynamics (hereinafter abbreviated as QCD).
%
% 2.1
\subsection{Factorization of infrared divergence (one loop)}
\[
\begin{minipage}{12cm}
\centering
\includegraphics[width=3cm, clip]{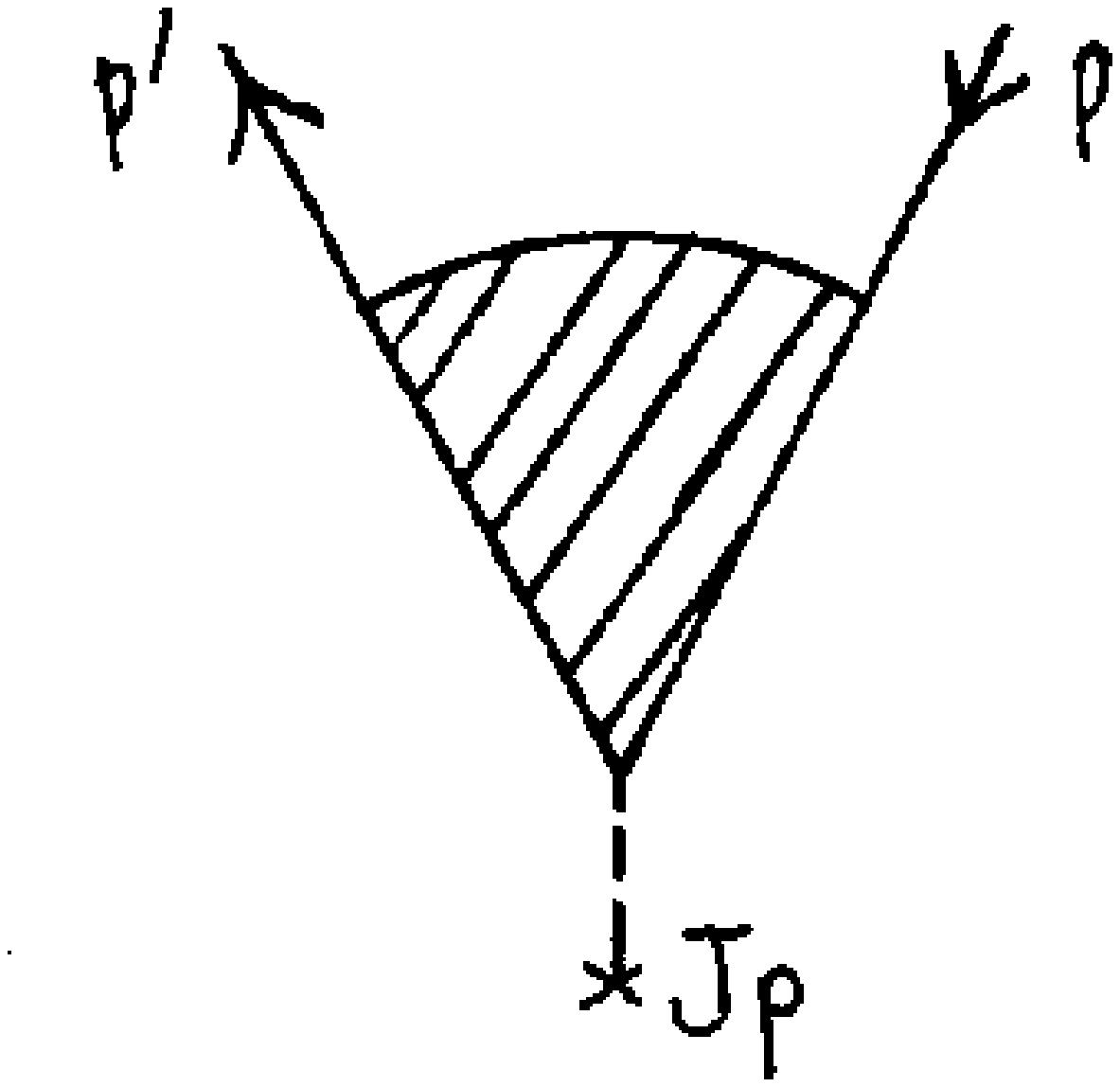}
\text{(Figure 2.1.1)}
\end{minipage}
\]
For simplicity, we consider electron scattering by an external field.
Figure 2.1.1 shows a process in which an electron with momentum $p$
is scattered by an external field $J_\rho$ and flies away with
momentum $p'$.  Let us begin with one-loop graphs.
The following three graphs are of interest (Figure 2.1.2).

\vspace{1cm}

\begin{flushright}
\begin{figure}[h]
\centering
\includegraphics[width=140mm, clip]{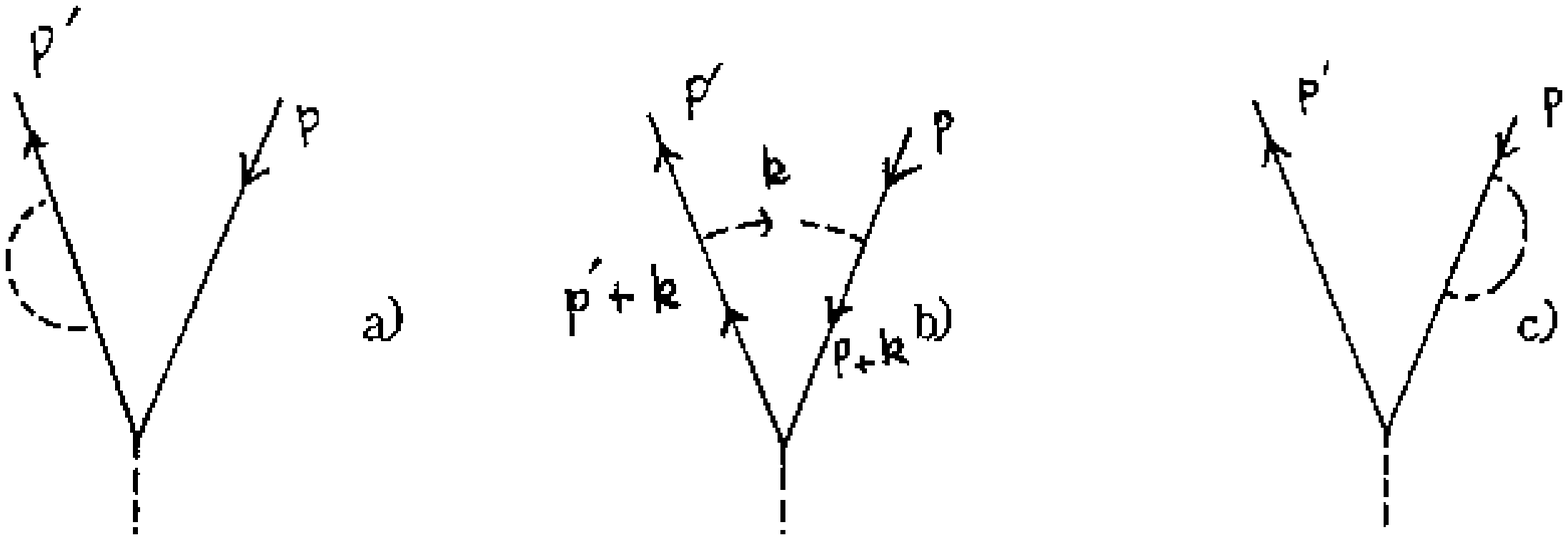}
\caption{}
\end{figure}
\end{flushright}

First, Figure b) is considered.  The scattering amplitude $S_b$
corresponding to this Figure b) is given by
\begin{equation}
S_b = \int\frac{d^N k}{(2\pi)^N}
\bar{u}(p') ie\gamma^\mu
\frac{i(\Slash{p}'+\Slash{k}+m)}{(p'+k)^2 - m^2 + i\epsilon}
\Slash{J}
\frac{i(\Slash{p} +\Slash{k}+m)}{(p +k)^2 - m^2 + i\epsilon}
ie\gamma^\nu u(p) D_{\mu\nu}(k), \label{eq2.1.1}
\end{equation}
where $N$ (being complex in general) is the spacetime dimension and
$D_{\mu\nu}(k)$ represents a photon propagator.
The propagator $D_{\mu\nu}(k)$ has a different form for various
gause conditions.  As an example, the forms for the covariant
and axial gauge are given by
\begin{align}
D_{\mu\nu}(k)
&= \frac{-i}{k^2 + i\epsilon}\left[
g_{\mu\nu} - (1-\alpha)\frac{k_\mu k_\nu}{k^2 + i\epsilon}\right]
\quad\text{(covariant gauge)} \label{eq2.1.2} \\
\intertext{and}
D_{\mu\nu}(k)
&= \frac{-i}{k^2 + i\epsilon}\left[
g_{\mu\nu} - \frac{k_\mu n_\nu + n_\mu k_\nu}{k\cdot n}
+ (\alpha k^2 + n^2)\frac{k_\mu k_\nu}{(k\cdot n)^2}\right].
\quad\text{(axial gauge)} \label{eq2.1.3}
\end{align}
In the case of QED, the effect of changing the gauge condition is
exhibited only in the difference of the gauge boson propagator.
However, in the case of QCD, it also affects the presence or
absence of Faddeev-Popov ghost fields (see Section 4.1).
Now, in Eq.\ (\ref{eq2.1.1}), we estimate
the value of the integral for small $k_\mu$.  First we rewrite
\begin{align}
\bar{u}(p') ie\gamma^\mu
\frac{i(\Slash{p}'+\Slash{k}+m)}{(p'+k)^2 - m^2 + i\epsilon}
\times\dotsb
&\underset{|k_\mu|\ll m}{\approx}
\bar{u}(p')
(-e)
\frac{2p'^\mu + (-\Slash{p}'+m)\gamma^\mu}{2p'\cdot k + k^2 + i\epsilon}
\times\dotsb \notag \\
&= (-e)
\frac{2p'^\mu}{2p'\cdot k + k^2 + i\epsilon}\times \bar{u}(p')
\times\dotsb \label{eq2.1.4}
\end{align}
where the commutation relation of the Dirac matrices
$\{ \gamma^\mu, \gamma^\nu\} = 2g^{\mu\nu}$ is used in the second line.
This transformation takes advantage of $p'$ being on the mass shell.
That is, the propagator of the electron (momentum $p'+k$)
coupled to an on-mass-shell external line and
a soft photon of momentum $k_\mu$ behaves as $O(1/k)$ in the limit of
$|k_\mu|\ll m$.
Similarly,
\begin{equation}
\dotsb\times
\frac{i(\Slash{p} +\Slash{k}+m)}{(p +k)^2 - m^2 + i\epsilon}
ie\gamma^\nu u(p)
\underset{|k_\mu|\ll m}{\approx}
\dotsb\times
u(p)(-e)\frac{2p^\nu}{2p\cdot k + k^2 + i\epsilon} \label{eq2.1.5}
\end{equation}
holds.  Using (\ref{eq2.1.4}) and (\ref{eq2.1.5}),
the integral (\ref{eq2.1.1}) around $|k_\mu|\ll m$ gives
\begin{equation}
S_b \approx e^2\int\frac{d^N k}{(2\pi)^N}
\frac{2p'^\mu}{2p'\cdot k + k^2 + i\epsilon}
\frac{2p^\nu}{2p\cdot k + k^2 + i\epsilon}
D_{\mu\nu}(k) S^{(0)}, \label{eq2.1.6}
\end{equation}
where $S^{(0)}=\bar{u}(p')\Slash{J}u(p)$ represents the amplitude
at the lowest order.

Setting $N=4$, we find that Eq.\ (\ref{eq2.1.6}) produces
a logarithmic divergence from the integral in the region $|k_\mu|\ll m$
(its specific calculation will be described later).
This logarithmic divergence is called \emph{infrared divergence}.
Now, the scattering amplitudes for Figures a) and b) are both given by
\begin{equation}
S_a = S_b = \frac{1}{2}Z_2^{(1)}\times S^{(0)}, \label{eq2.1.7}
\end{equation}
where $Z_2^{(1)}$ is the value of $Z_2$ factor at one loop.
Since $Z_1=Z_2$ holds in QED, this $Z_2^{(1)}$ can be determined
from (\ref{eq2.1.6}) (that is, from $Z_1^{(1)}$) as
\begin{align}
Z_2^{(1)} &= Z_1^{(1)} = -\frac{S_b(p=p')}{S^{(0)}} \notag \\
&= -e^2\int\frac{d^N k}{(2\pi)^N}
\frac{2p^\mu}{2p\cdot k + k^2 + i\epsilon}
\frac{2p^\nu}{2p\cdot k + k^2 + i\epsilon}
D_{\mu\nu}(k). \label{eq2.1.8}
\end{align}
Thus, the infrared divergence part included in the one-loop graphs
of Figures a)--c) is completely extracted as a function as
\begin{align}
S_a + S_b + S_c
&\approx -\frac{1}{2}e^2\int\frac{d^N k}{(2\pi)^N}
\left( \frac{2p'^\mu}{2p'\cdot k + k^2 + i\epsilon}
- \frac{2p^\mu}{2p\cdot k + k^2 + i\epsilon} \right) \notag \\
&\quad\times
\left( \frac{2p'^\nu}{2p'\cdot k + k^2 + i\epsilon}
- \frac{2p^\nu}{2p\cdot k + k^2 + i\epsilon} \right)
D_{\mu\nu}(k) S^{(0)}. \label{eq2.1.9}
\end{align}
When the propagators (\ref{eq2.1.2}), (\ref{eq2.1.3})
for various gauge conditions are substituted into this formula,
the term proportional to $k_\mu$, $k_\nu$ is dropped in both cases,
and the formula
yields the following value independent of the gauge conditions (since
$k^2$ in the denominator in (\ref{eq2.1.9}) is negligible):
\begin{align}
S^{(1)} &\equiv S_a + S_b + S_c \notag \\
&\approx -\frac{1}{2}e^2\int\frac{d^N k}{(2\pi)^N}
\left( \frac{2p'^\mu}{2p'\cdot k + k^2 + i\epsilon}
- \frac{2p^\mu}{2p\cdot k + k^2 + i\epsilon} \right)^2
\frac{-i}{k^2 + i\epsilon} S^{(0)}. \label{eq2.1.10}
\end{align}
We have seen above how the infrared divergence factor
(the coefficient of $S^{(0)}$ in Eq.\ (\ref{eq2.1.10})) for one loop
is decomposed.

Next, we specifically calculate the value of
this infrared divergence factor.
Various methods can be conceived
as the regularization of infrared divergence.
\begin{enumerate}[1)]
\item $N$-dimensional method:
The dimensionality of space is analytically continued to
a complex number $N$ with $\Real N > 4$.
The divergent quantity is extracted as a pole $\frac{1}{N-4}$
in the limit $N\to 4$.
\item Give the photon a virtual mass $\lambda$:
The infrared divergence is extracted as $\ln\frac{1}{\lambda}$
in the limit $\lambda\to 0$.
\item Keep the external line off the mass shell, {\it i.e.},
$p^2 = p'^2 = m^2 - \delta^2$:
The infrared divergence appears as $\ln\frac{1}{\delta}$
in the limit $\delta\to 0$.
\item Cut the lower end of the momentum integral, {\it i.e.},
set the domain of integration to $|k_\mu| > \lambda'$:
The infrared divergence is extracted as $\ln\frac{1}{\lambda'}$
in the limit $\lambda'\to 0$.
\end{enumerate}
In general, infrared divergence arises when the condition of
the external line being on the mass shell is satisfied simultaneously with
the condition of the photon being massless (refer to one-loop examples).
Therefore, the infrared divergence can be removed by violating
one of these two conditions.  The regularization 2) is designed to
violate the latter, while the regularization 3) the former.

Using the regularization 1), the factor corresponding to $S_b$ yields
\[
I_b \equiv e^2\int\frac{d^N k}{(2\pi)^N}
\frac{2p'^\mu}{2p'\cdot k + k^2 + i\epsilon}
\frac{2p^\nu}{2p\cdot k + k^2 + i\epsilon}
\frac{-ig_{\mu\nu}}{k^2 + i\epsilon},
\]
and introducing Feynman parameters $\alpha$, $\beta$, $\gamma$
into respective factors of the integrand,
\begin{align}
I_b
&= -e^2\,4(p\cdot p')\,2! \int\frac{d^N k}{(2\pi)^N}
\int_0^1 \frac{d\alpha\,d\beta\,d\gamma\,\delta(1-\alpha-\beta-\gamma)}
{[k^2 + 2(\alpha p' + \beta p)\cdot k + i\epsilon]^3} \notag \\
&= -e^2\,4(p\cdot p')\,2! \int\frac{d^N k}{(2\pi)^N}
\int_0^1 \frac{d\alpha\,d\beta\,d\gamma\,\delta(1-\alpha-\beta-\gamma)}
{[k^2 - (\alpha p' + \beta p)^2 + i\epsilon]^3}
\quad\text{($k$-integral shifted)} \notag \\
&= -\frac{e^2}{(4\pi)^2}\,4(p\cdot p')\,
\Gamma\left(3-\frac{N}{2}\right)
\int_0^1 \frac{d\alpha\,d\beta\,d\gamma\,\delta(1-\alpha-\beta-\gamma)}
{[(\alpha p' + \beta p)^2 - i\epsilon]^{3-\frac{N}{2}}}.
\quad\text{($k$-integral performed)} \label{eq2.1.11}
\end{align}
The above calculation is based on the Feynman parameter formula
and the formula
\begin{equation}
\int\frac{d^N k}{(2\pi)^N}\frac{1}{[k^2 - V(\alpha) + i\epsilon]^m}
= i\frac{(-1)^m}{(4\pi)^{N/2}}\frac{\Gamma(m-\frac{N}{2})}{\Gamma(m)}
\frac{1}{[V(\alpha) - i\epsilon]^{m-\frac{N}{2}}}. \label{eq2.1.12}
\end{equation}
In (\ref{eq2.1.11}),
$\Gamma(3-\frac{N}{2}) \xrightarrow[N\to 4]{} 1$, which
does not produce a pole of $\frac{1}{N-4}$.
This shows that this integral $I_b$ has no ultraviolet (UV) divergence
\cite{30}.
Infrared (IR) divergence arises from the integral of the Feynman parameters,
since in (\ref{eq2.1.11}), the denominator of the integral becomes zero
for $\alpha=\beta=0$ and contributes to divergence \cite{31}.
Now we change the variables as follows:
\begin{equation}
  \begin{split}
    \alpha &= xy, \\
    \beta &= (1-x)y, \\
    \gamma &= 1-y.
  \end{split} \label{eq2.1.13}
\end{equation}
Then $\alpha=\beta=0$ corresponds to $y=0$ (the Jacobian being $y$), and
\[
I_b = -\frac{e^2}{(4\pi)^2}\,4(p\cdot p')\,
\Gamma\left(3-\frac{N}{2}\right)
\int_0^1 dx \int_0^1 y\,dy\,(y^2)^{\frac{N}{2}-3}
\frac{1}{[(xp'+(1-x)p)^2 - i\epsilon]^{3-\frac{N}{2}}}.
\]
The $y$-integral produces a divergence from around $y\approx 0$:
\[
\int_0^1 y^{N-5}\,dy = \frac{1}{N-4}\left[y^{N-4}\right]_0^1 = \frac{1}{N-4}.
\quad\text{(where $\Real N > 4$)}
\]
This is a pole at $N=4$, corresponding to infrared divergence.  Thus
\begin{align}
I_b &= -\frac{e^2}{(4\pi)^2}\frac{1}{N-4}
\int_0^1 dx\,\frac{4(p\cdot p')}{(xp'+(1-x)p)^2 - i\epsilon} \notag \\
&= +\frac{e^2}{(4\pi)^2}\frac{1}{N-4}
\frac{4 - 8m^2/t}{\sqrt{1 - 4m^2/t}}
\ln\frac{\sqrt{1 - 4m^2/t} - 1}{\sqrt{1 - 4m^2/t} + 1} \label{eq2.1.14}
\end{align}
Using this equation and (\ref{eq2.1.8}),
the infrared divergence factor for Figures a) and c) yields
\begin{equation}
I_a = I_c = -\frac{1}{2}I_b(p=p', \text{\it i.e.}, t=0)
= 2\frac{e^2}{(4\pi)^2}\frac{1}{N-4}. \label{eq2.1.15}
\end{equation}
Thus the infrared divergence parts at one loop are grouped into the
following form:
\begin{equation}
[S^{(1)}]_{\text{IR}} = e^2\frac{1}{N-4}
B\left(\frac{1}{m^2}\right)^{\text{QED}}\times S^{(0)}, \label{eq2.1.16}
\end{equation}
where
\begin{equation}
B\left(\frac{1}{m^2}\right)^{\text{QED}}
= \frac{1}{(4\pi)^2}\left( 4 + \frac{4 - 8m^2/t}{\sqrt{1 - 4m^2/t}}
\ln\frac{\sqrt{1 - 4m^2/t} - 1}{\sqrt{1 - 4m^2/t} + 1} \right).
\label{eq2.1.17}
\end{equation}
The replacement rule for other regularizations of infrared divergence
is given by
\begin{equation}
\ln\frac{1}{\lambda},\
2\ln\frac{1}{\delta},\
\ln\frac{1}{\lambda'}
\leftrightarrow \frac{1}{N-4} \quad (\Real N > 4). \label{eq2.1.18}
\end{equation}
%
% 2.2
\subsection{Factorization of infrared divergence in terms of
a differential equation (all order)}
In this subsection, the result for one loop in the previous subsection
is generalized to all orders.  That is, we prove that the strongest
infrared divergences included in the scattering amplitude of the process
of Figure 2.1.1 are grouped into
\begin{equation}
[S]_{\text{IR}} = e^{[I^{(1)}]_{\text{IR}}}\times S^{(0)}
\label{eq2.2.1}
\end{equation}
(in the form of an exponential
with the one-loop value $[I^{(1)}]_{\text{IR}}$ raised to its power)
at all orders.  In this case the exponent in (\ref{eq2.2.1}) is the
one-loop IR factor
\begin{align}
[I^{(1)}]_{\text{IR}}
&= -\frac{1}{2}e_R^2\int\frac{d^N k}{(2\pi)^N}
\left( \frac{p'^\mu}{p'\cdot k} - \frac{p^\mu}{p\cdot k} \right)
\left( \frac{p'^\nu}{p'\cdot k} - \frac{p^\nu}{p\cdot k} \right)
\frac{-ig_{\mu\nu}}{k^2 - \lambda^2 + i\epsilon} \notag \\
&= e_R^2\ln\frac{m}{\lambda}B\left(\frac{1}{m^2}\right)^{\text{QED}},
\label{eq2.2.2}
\end{align}
where the renormalized charge $e_R$ ($e_R = eZ_3^{1/2}$) appears due to
photon radiative correction.
Instead of showing (\ref{eq2.2.1}), what happens if we derive a differential
equation that it satisfies?
As is easily understood, the equation satisfied by $[S]_{\text{IR}}$
is given by
\begin{equation}
\lambda\frac{\partial}{\partial\lambda}[S]_{\text{IR}}
= -e_R^2 B\left(\frac{1}{m^2}\right)^{\text{QED}}[S]_{\text{IR}}.
\label{eq2.2.3}
\end{equation}
We derive this equation below.
The action of $\lambda\frac{\partial}{\partial\lambda}$ induces
mass insertion in the photon propagator $D_{\mu\nu}(k)$
(in the following the Feynman gauge is taken):
\begin{equation}
\begin{minipage}{12cm}
\centering
\includegraphics[width=8cm, clip]{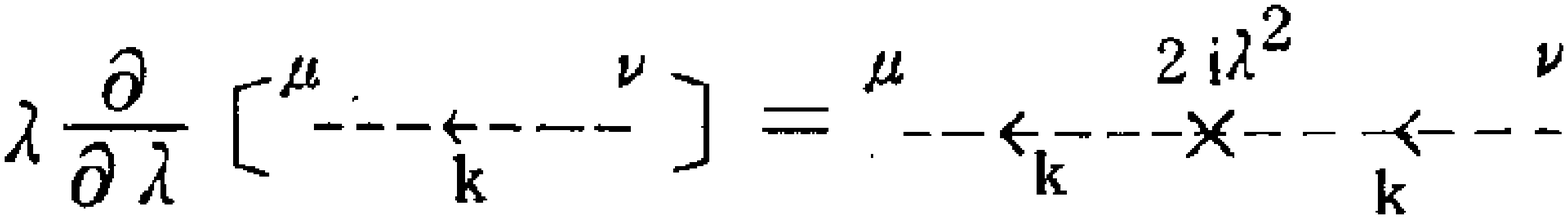}
\end{minipage}
\label{eq2.2.4}
\end{equation}
This represents the identity
\begin{equation}
\lambda\frac{\partial}{\partial\lambda}
\frac{-ig_{\mu\nu}}{k^2 - \lambda^2 + i\epsilon}
= \frac{-ig_{\mu\rho}}{k^2 - \lambda^2 + i\epsilon}2i\lambda^2
\frac{-ig_{\rho\nu}}{k^2 - \lambda^2 + i\epsilon}
\label{eq2.2.5}
\end{equation}
in a graphical form.

(In this paper,
\begin{tikzpicture}
\begin{feynhand}
\vertex (i) at (-1,0);
\vertex (f) at (1,0);
\propag [scalar] (i) to (f);
\end{feynhand}
\end{tikzpicture}
is used for the photon propagator,
\begin{tikzpicture}
\begin{feynhand}
\vertex (i) at (-1,0);
\vertex (f) at (1,0);
\propag [photon] (i) to (f);
\end{feynhand}
\end{tikzpicture}
for the propagator of gauge bosons, and
\begin{tikzpicture}
\begin{feynhand}
\vertex (i) at (-1,0);
\vertex (f) at (1,0);
\propag [anti charged ghost] (i) to (f);
\end{feynhand}
\end{tikzpicture}
for the propagator of Faddeev-Popov ghost fields.)

The bare propagator becomes a dressed propagator
$\tilde{D}_{\mu\nu}(k)$ when all the graphs are summed.
Thus the $\lambda$-dependence is exhibited as the $\lambda$-dependence
of $\tilde{D}_{\mu\nu}(k)$.

Application of $\lambda\frac{\partial}{\partial\lambda}$
to the scattering amplitude of Figure 2.1.1 results in
differentiation of $\tilde{D}$ at every site and yields
\begin{equation}
\lambda\frac{\partial}{\partial\lambda}S
= \frac{1}{2}\int\frac{d^N k}{(2\pi)^N}K^{\mu\nu}(p, p', k, -k)
\lambda\frac{\partial}{\partial\lambda}\tilde{D}_{\mu\nu}(k; \lambda),
\label{eq2.2.6}
\end{equation}
\[
\begin{minipage}{12cm}
\centering
\includegraphics[width=3cm, clip]{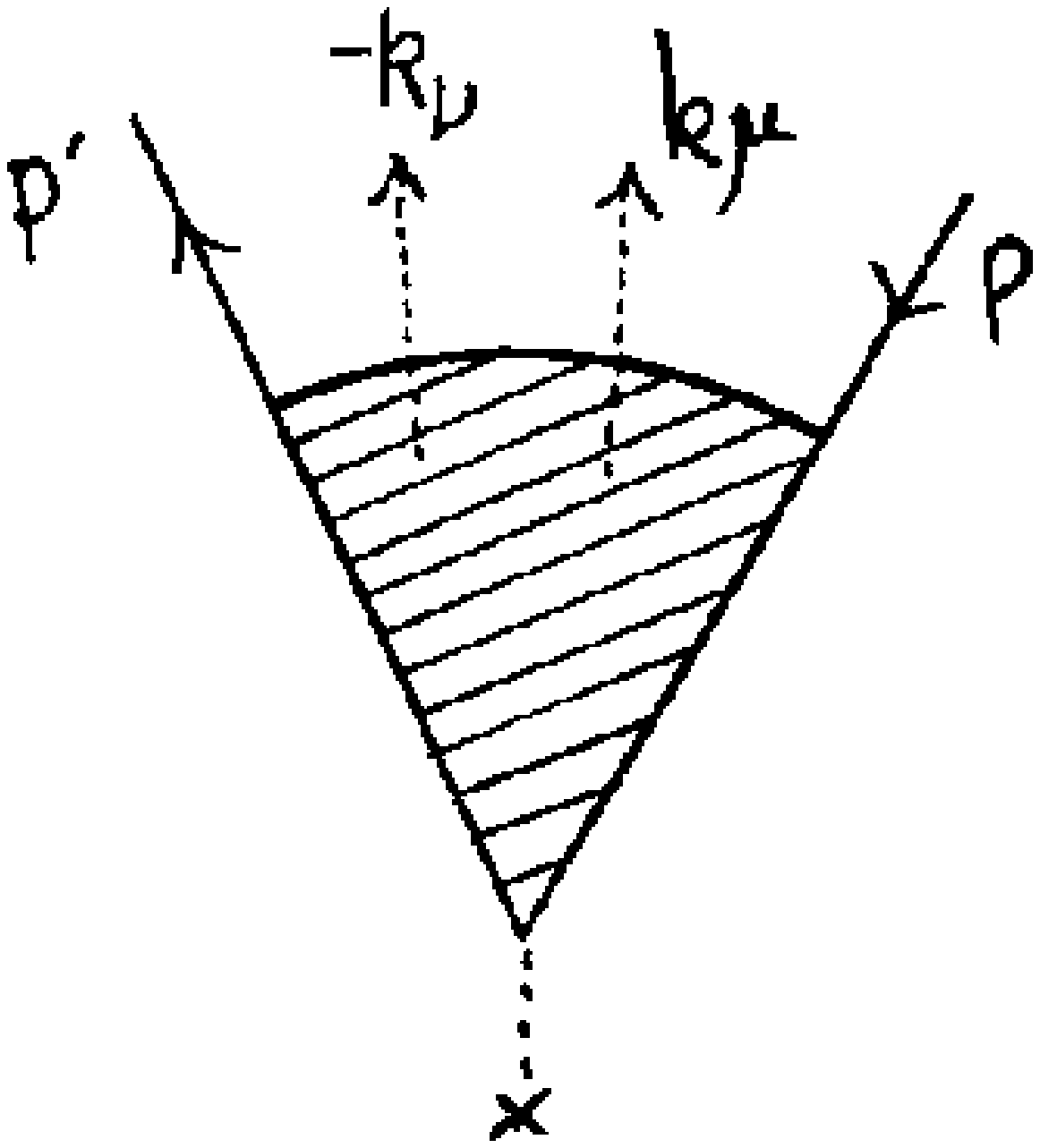}
\text{(Figure 2.2.1)}
\end{minipage}
\]
where $K^{\mu\nu}$ is shown by the following diagram (Figure 2.2.1),
representing
all the graphs in which photons with momenta $k_\mu$, $-k_\nu$
(which are off the mass shell) are emitted from any site in Figure 2.1.1.
Here the right hand side of (\ref{eq2.2.6}) includes $\frac{1}{2}$
because of symmetry with respect to the exchange of
($\mu\leftrightarrow\nu$), ($k\leftrightarrow -k$).
Now IR divergence can be determined by examining the behavior of
$K^{\mu\nu}(p, p', k, -k)$ around $k_\mu\to 0$ and the behavior of
$\lambda\frac{\partial}{\partial\lambda}
\tilde{D}_{\mu\nu}(k; \lambda)$.  First, with regard to $K^{\mu\nu}$,
we prove
\begin{equation}
K^{\mu\nu}(p, p', k, l) \underset{k_\mu, l_\nu \to 0}{\approx}
+e^2
\left( \frac{p'^\mu}{p'\cdot k} - \frac{p^\mu}{p\cdot k} \right)
\left( \frac{p'^\nu}{p'\cdot l} - \frac{p^\nu}{p\cdot l} \right)
\times S.
\label{eq2.2.7}
\end{equation}
This is the low energy theorem of F. E. Low type \cite{2}, indicating that
the effect of soft photons $k$, $l$ can be factored out completely.
An example of one loop can be easily seen in Eq.\ (\ref{eq2.1.9}).
(A substantially identical formula can also be derived in QCD,
but the derivation will be performed in Subsection 4.5.)
First we show that photons $k_\mu$ and $l_\nu$ coupled to other than
the incoming and outgoing electron paths, do not contribute to
(\ref{eq2.2.7}).  These photons occur from the internal fermion loop,
as shown in Figure 2.2.2.
\[
\begin{minipage}{12cm}
\centering
\includegraphics[width=3cm, clip]{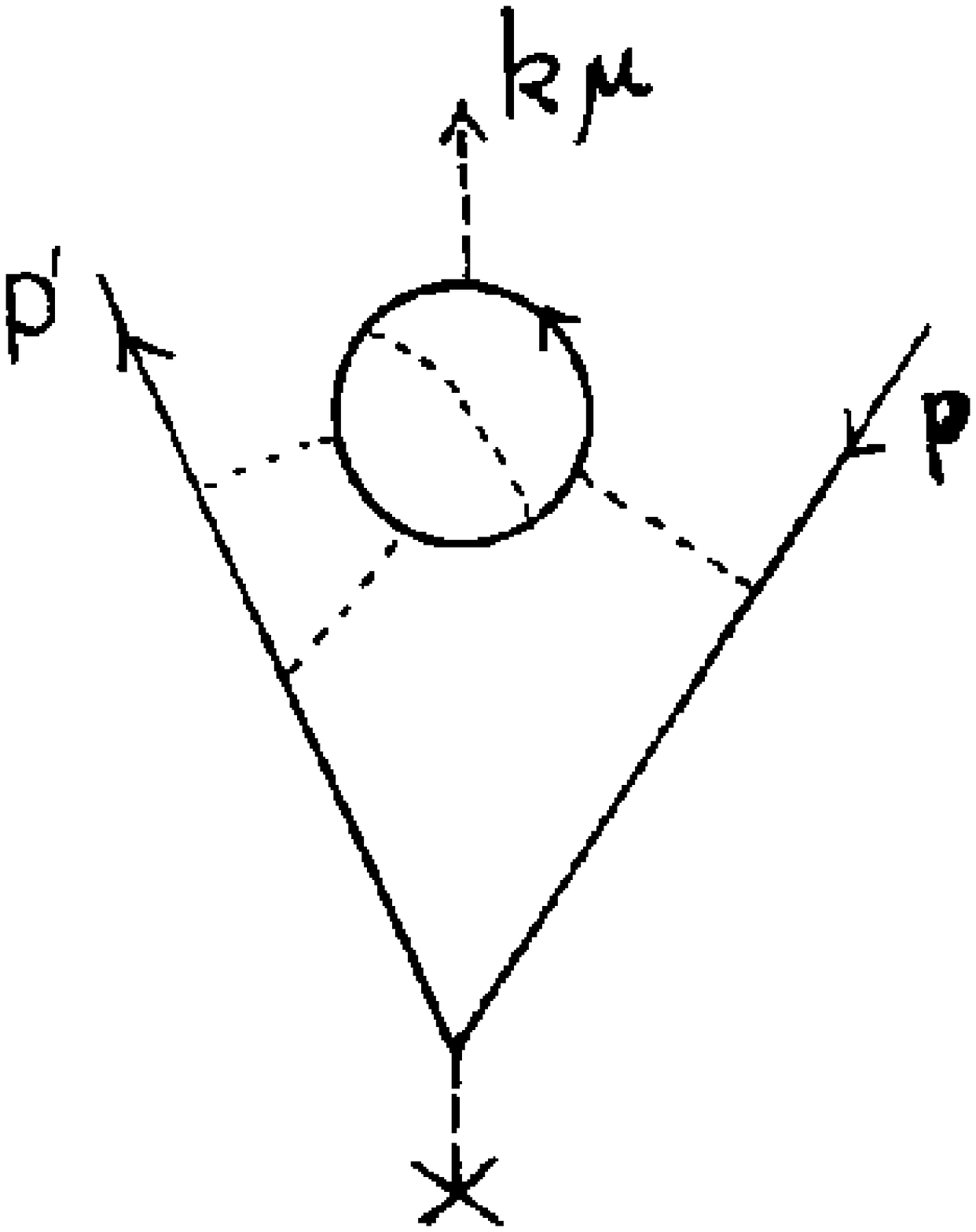}
\text{(Figure 2.2.2)}
\end{minipage}
\]
The reason that the photons occurring from the internal fermion loop
do not contribute to (\ref{eq2.2.7}) is shown using the Ward identity
(WI).

That is, emission of $k_\mu$ from any site of the internal fermion yields
\begin{equation}
\begin{minipage}{12cm}
\centering
\includegraphics[width=10cm, clip]{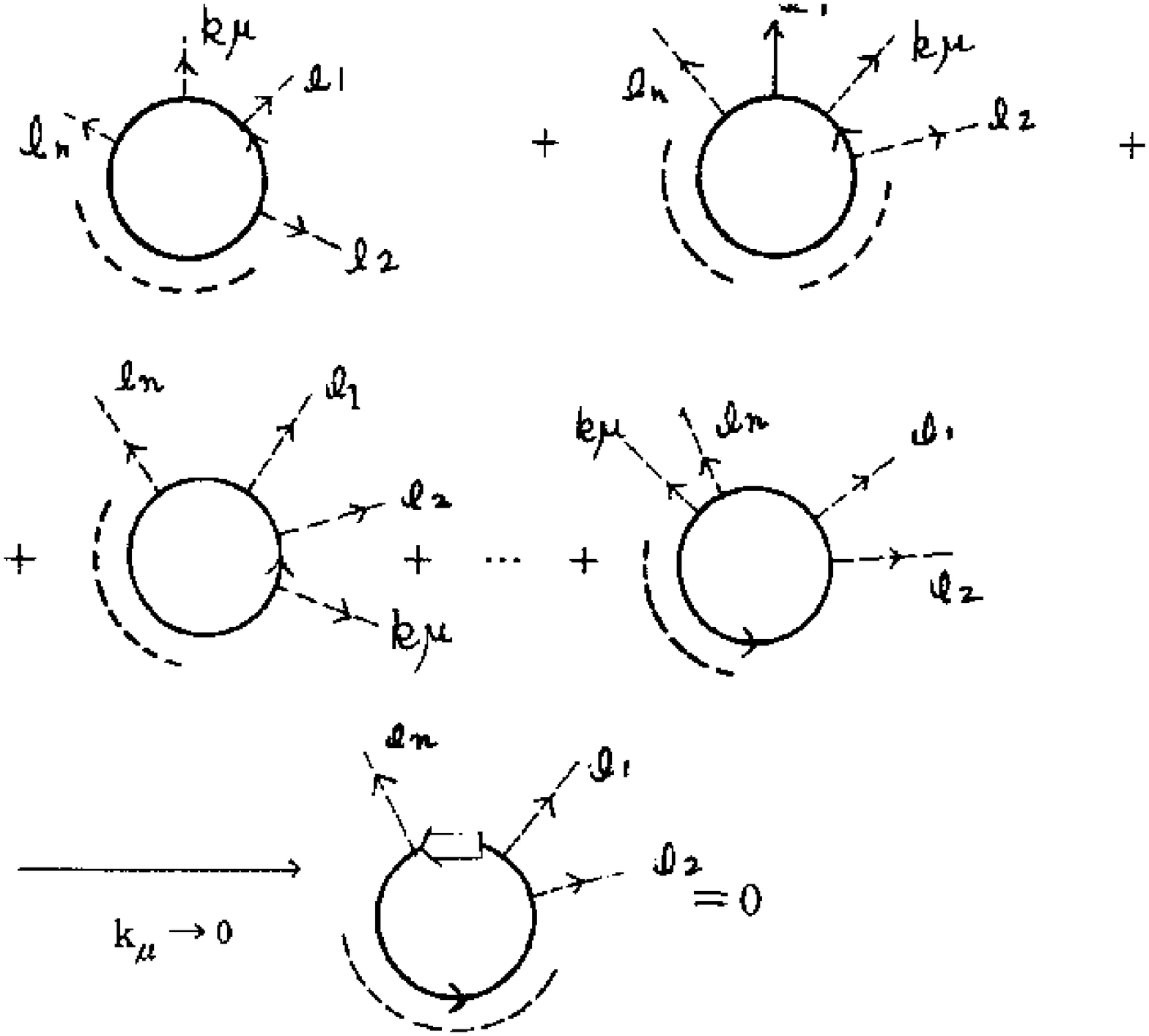}
\end{minipage}
\label{eq2.2.8}
\end{equation}
where the arrow $\Leftarrow$ represents the following differential
operation \cite{32}.
\begin{equation}
\begin{minipage}{12cm}
\centering
\includegraphics[width=6cm, clip]{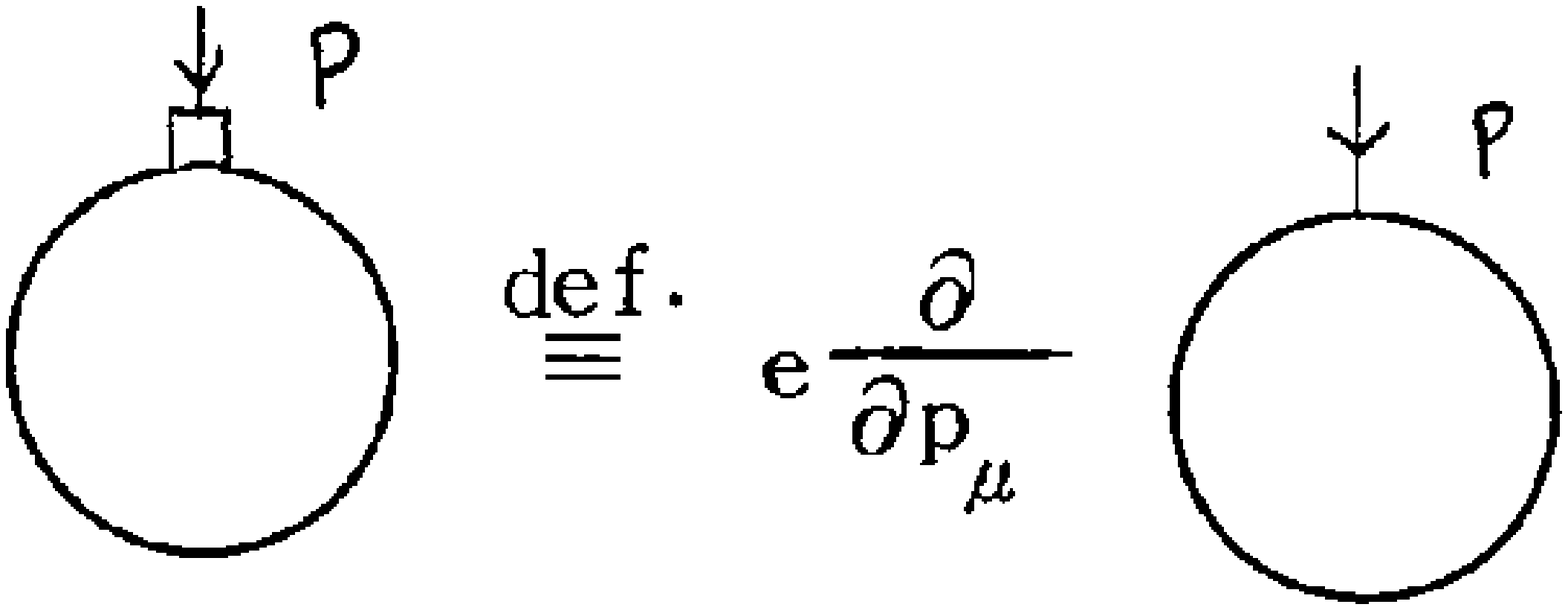}
\end{minipage}
\label{eq2.2.9}
\end{equation}
In (\ref{eq2.2.8}) this differentiation is a differentiation
for the loop momentum $l_\mu$ and must be performed
before loop integration, because in the transformation of (\ref{eq2.2.8})
the Ward identity, represented in our notation as
\begin{subequations}
\begin{align}
&\begin{minipage}{12cm}
\centering
\includegraphics[width=8cm, clip]{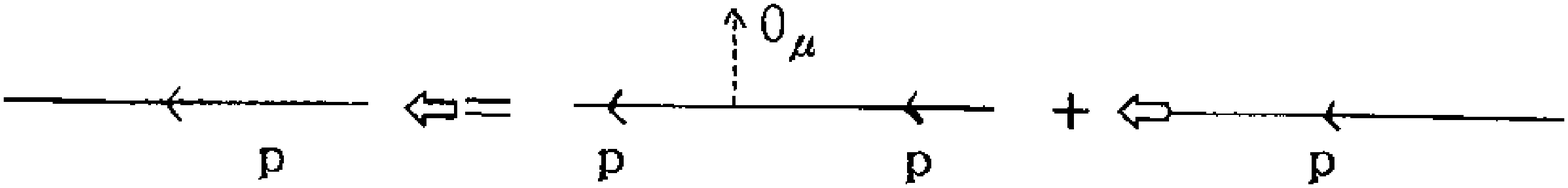}
\end{minipage}
\label{eq2.2.10a} \\
\intertext{or in mathematical expression as}
&\dotsb\frac{i}{\Slash{p}-m}e\overleftarrow{\frac{\partial}{\partial p_\mu}}
\dotsb
= \dotsb\frac{i}{\Slash{p}-m}ie\gamma_\mu\frac{i}{\Slash{p}-m}\dotsb
+ \dotsb e\overleftarrow{\frac{\partial}{\partial p_\mu}}
\frac{i}{\Slash{p}-m}\dotsb, \label{eq2.2.10b}
\end{align}
\end{subequations}
is used in the integrand.  The arrow $\Leftarrow$ points to
the part to be differentiated.  The reason that (\ref{eq2.2.8}) vanishes
is that it yields a surface term of the integral.  Now we specifically
calculate this electron loop.  All the electron propagators are grouped
by the Feynman parameter formula as
\begin{align}
&\sum A^{\nu_1\dots\nu_m}\int\frac{d^N l}{(2\pi)^N}
\frac{\partial}{\partial l_\mu}
\frac{l_{\nu_1}\dotsb l_{\nu_m}}{(l^2 - c + i\epsilon)^n} \notag \\
&\quad= \sum A^{\nu_1\dots\nu_m}\Bigl\{
\int\frac{d^N l}{(2\pi)^N}\sum_{i=1}^m
\frac{l_{\nu_1}\dotsb l_{\nu_{i-1}}g_{\mu\nu_i}
l_{\nu_{i+1}}\dotsb l_{\nu_m}}{(l^2 - c + i\epsilon)^n} \notag \\
&\qquad + \int\frac{d^N l}{(2\pi)^N}(-2)n
\frac{l_\mu\times l_{\nu_1}l_{\nu_2}\dotsb l_{\nu_m}}
{(l^2 - c + i\epsilon)^{n+1}}\Bigr\}, \label{eq2.2.11}
\end{align}
where the domain of integration is shifted beforehand.
It can be seen that the first and second terms of the right hand side
cancel out each other upon symmetric integration.
The formula necessary for it is
\begin{align}
\int d^N l\,l_{\nu_1}\dotsb l_{\nu_m} f(l^2)
&= c_m\times\{\text{sum of independent tensors of
($g_{\nu_1\nu_2}\dotsb g_{\nu_{m-1}\nu_m}$) type}\} \notag \\
&\quad \times\int d^N l\,(l^2)^{\frac{m}{2}} f(l^2), \label{eq2.2.12}
\end{align}
where
\[
c_m = \begin{cases}
0, & \text{($m$: odd)} \\
1/N(N+2)\dotsb(N+m-2). & \text{($m$: even)}
\end{cases}
\]
Another necessary formula is
\begin{equation}
\int\frac{d^N l}{(2\pi)^N}\frac{(l^2)^m}{(l^2 - c + i\epsilon)^n}
= i\frac{(-1)^{n+m}}{(4\pi)^{N/2}}
\frac{\Gamma\left(n-m-\frac{N}{2}\right)
\Gamma\left(\frac{N}{2}+m\right)}
{\Gamma(n)\Gamma\left(\frac{N}{2}\right)}
\frac{1}{(c-i\epsilon)^{n-m-\frac{N}{2}}}. \label{eq2.2.13}
\end{equation}
Thus we have confirmed above that emission of soft photons occurring from
the internal fermion loop is negligible.

Therefore we may limit the following discussion to only the case where
$k_\mu$, $l_\nu$ are emitted from the incoming and outgoing electron paths.
This case can be addressed using the following formula,
called the eikonal identity \cite{2}.
\begin{align}
&\begin{minipage}{14cm}
\centering
\includegraphics[width=14cm, clip]{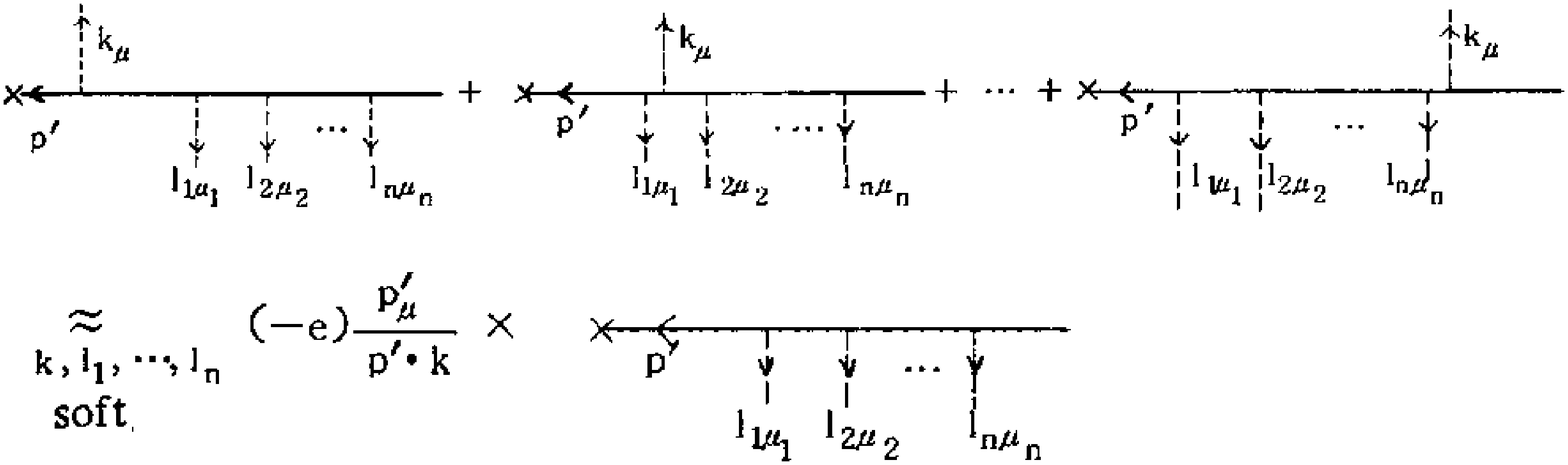}
\end{minipage}
\label{eq2.2.14} \\
&\begin{minipage}{14cm}
\centering
\includegraphics[width=14cm, clip]{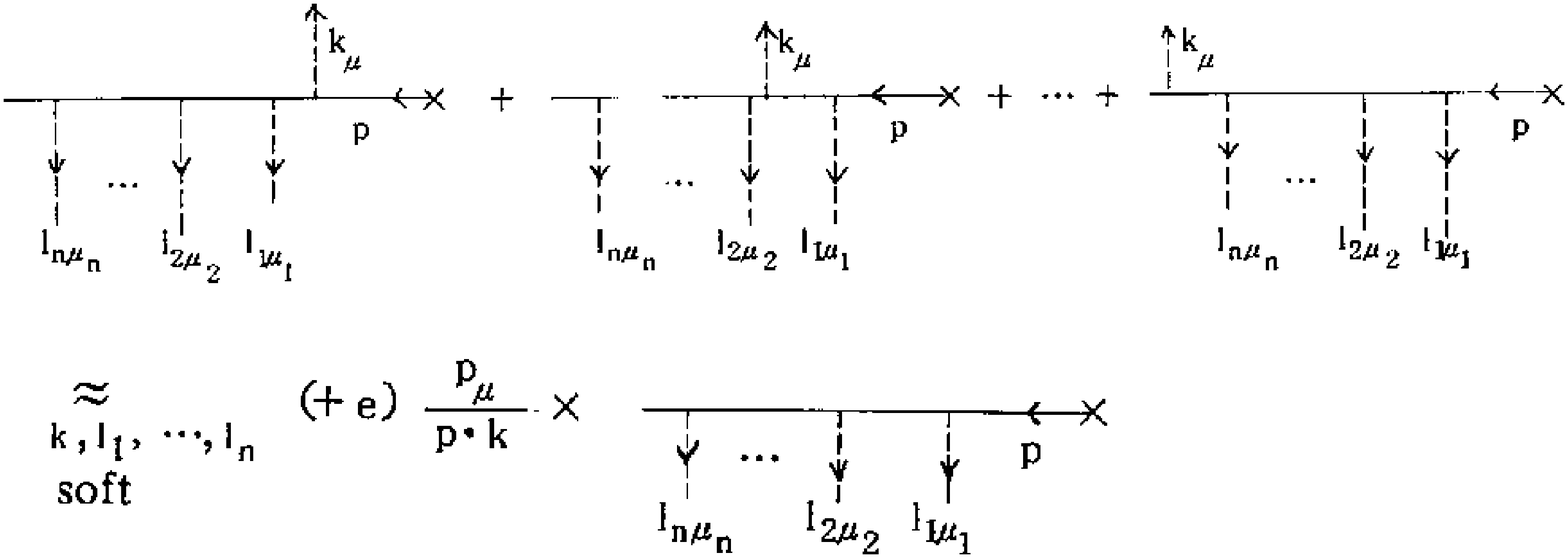}
\end{minipage}
\label{eq2.2.15}
\end{align}
As described above, the symbol $\times$ here indicates that
$p'$ and $p$ are on the mass shell.
As to the proof of the identity, it may be sufficient
to prove (\ref{eq2.2.14}) only.
Since $k, l_1, \dots, l_n$ are soft, the approximation used in
(\ref{eq2.2.4}) can be used, and the left hand side of
(\ref{eq2.2.14}) yields the following expression except for
$(-e)^{n+1} p'_\mu p'_{\mu_1} \dotsb p'_{\mu_n} \bar{u}(p')$:
\begin{align}
&\uwave{\frac{1}{p'\cdot k}}
\frac{1}{p'\cdot (k + l_1)}
\frac{1}{p'\cdot (k + l_1 + l_2)}\dotsb
\frac{1}{p'\cdot (k + l_1 + l_2 +\dotsb+ l_n)} \notag \\
&+ \frac{1}{p'\cdot l_1}
\frac{1}{p'\cdot (k + l_1)}
\frac{1}{p'\cdot (k + l_1 + l_2)}\dotsb
\frac{1}{p'\cdot (k + l_1 + l_2 +\dotsb+ l_n)} \notag \\
&+ \frac{1}{p'\cdot l_1}
\frac{1}{p'\cdot (l_1 + l_2)}
\frac{1}{p'\cdot (k + l_1 + l_2)}\dotsb
\frac{1}{p'\cdot (k + l_1 + l_2 +\dotsb+ l_n)} + \dotsb \notag \\
&+ \frac{1}{p'\cdot l_1}
\frac{1}{p'\cdot (l_1 + l_2)}
\frac{1}{p'\cdot (l_1 + l_2 + l_3)}\dotsb
\frac{1}{p'\cdot (l_1 + l_2 +\dotsb+ l_n)}
\frac{1}{p'\cdot (k + l_1 + l_2 +\dotsb+ l_n)}. \label{eq2.2.16}
\end{align}
First, the first and second terms of Eq.\ (\ref{eq2.2.16}) are
summed and reduced to a common denominator:
\[
\frac{1}{p'\cdot l_1}
\uwave{\frac{1}{p'\cdot k}}
\frac{1}{p'\cdot (k + l_1 + l_2)}\dotsb
\frac{1}{p'\cdot (k + l_1 + l_2 +\dotsb+ l_n)}.
\]
This is added to the third term of Eq.\ (\ref{eq2.2.16})
and reduced to a common denominator:
\[
\frac{1}{p'\cdot l_1}
\frac{1}{p'\cdot (l_1 + l_2)}
\uwave{\frac{1}{p'\cdot k}}
\dotsb
\frac{1}{p'\cdot (k + l_1 + l_2 +\dotsb+ l_n)}.
\]
This is added to the fourth term of Eq.\ (\ref{eq2.2.16}),
\dots and the same operation is repeated, finally yielding
the following factor:
\[
\frac{1}{p'\cdot l_1}
\frac{1}{p'\cdot (l_1 + l_2)}
\dotsb
\frac{1}{p'\cdot (l_1 + l_2 +\dotsb+ l_n)}
\uwave{\frac{1}{p'\cdot k}}.
\]
This factor corresponds to the right hand side of Eq.\ (\ref{eq2.2.14}).
Thus (\ref{eq2.2.14}) has been proved.

(In this above proof, it should be noted how the factor
$\frac{1}{p'\cdot k}$ moves.)  (\ref{eq2.2.15}) is proved similarly.
This eikonal identity can be used to derive the following:
\begin{equation}
\begin{minipage}{14cm}
\centering
\includegraphics[width=14cm, clip]{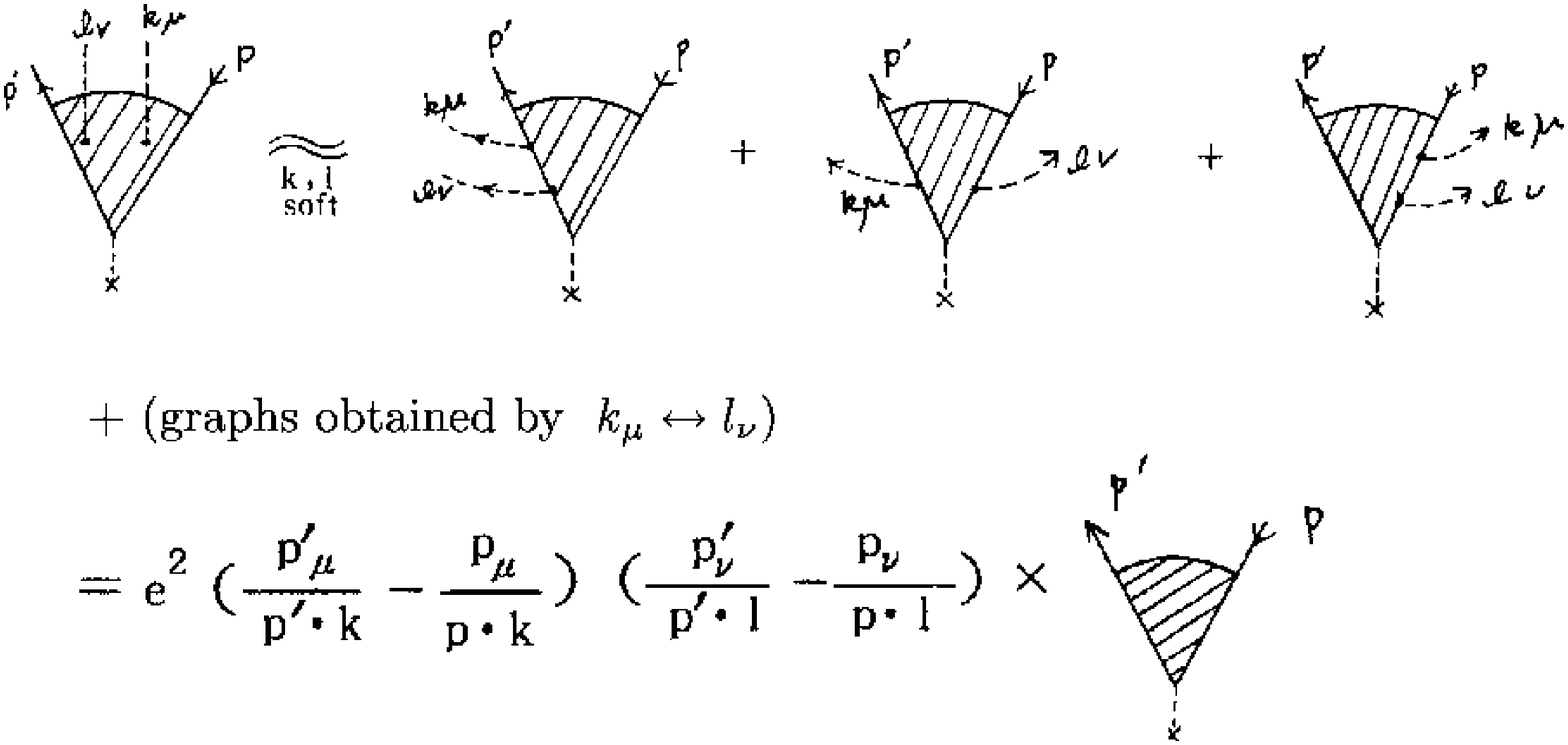}
\end{minipage}
\label{eq2.2.17}
\end{equation}
(For instance, the term of
$-e^2\frac{p'_\mu}{p'\cdot k}\frac{p_\nu}{p\cdot l}$ corresponds to
all the graphs in which a soft photon of momentum $l$ is emitted
from the right electron path and
a soft photon of momentum $k$ is emitted from the left electron path.)
Thus Eq.\ (\ref{eq2.2.7}) has been proved.

Next,
$\lambda\frac{\partial}{\partial\lambda}\tilde{D}_{\mu\nu}(k, \lambda)$
is simply evaluated.  The dressed propagator $\tilde{D}_{\mu\nu}$
is related to the self-energy  $\Pi_{\mu\nu}$ of a proper photon
as follows:
\begin{align}
\Pi_{\mu\nu}(k) &= (k^2 g_{\mu\nu} - k_\mu k_\nu)\Pi(k^2),
\label{eq2.2.18} \\
\tilde{D}_{\mu\nu}(k) &= \frac{-ig_{\mu\nu}}{k^2 - \lambda^2 + i\epsilon}
\times \frac{1}{1 + \Pi(k^2)}
+ \text{a term proportional to $k_\mu k_\nu$}. \label{eq2.2.19}
\end{align}
When
$\lambda\frac{\partial}{\partial\lambda}\tilde{D}_{\mu\nu}(k, \lambda)$
is substituted into the original equation (\ref{eq2.2.6}),
contribution to IR divergence arises around $k_\mu\to 0$.
It is known that in QED, in general, $\Pi(0)$ has no IR divergence,
that is, there is no infrared divergence in
$Z_3=\frac{1}{1+\Pi(0)}$ \cite{2}, \cite{19}.
Now we specifically calculate $\Pi(k^2)$ at one loop to show
that $\Pi(0)$ has no IR divergence:
\begin{align}
&\Pi^{(1)}(k^2)(g_{\mu\nu}k^2 - k_\mu k_\nu) \notag \\
&\quad =\frac{1}{i}\Tr\int\frac{d^N l}{(2\pi)^N}ie\gamma_\mu
\frac{i(\Slash{l} + m)}{l^2 - m^2 + i\epsilon}ie\gamma_\nu
\frac{i(\Slash{l} + \Slash{k} + m)}{(l+k)^2 - m^2 + i\epsilon} \notag \\
&\quad =(g_{\mu\nu}k^2 - k_\mu k_\nu)\times
\frac{e^2}{(4\pi)^{N/2}}
\Gamma\left(2-\frac{N}{2}\right)2N
\int_0^1 d\alpha
\frac{\alpha(1-\alpha)}
{[m^2 - \alpha(1-\alpha)k^2 - i\epsilon]^{2-\frac{N}{2}}}. \label{eq2.2.20}
\end{align}
Here the pole of $N=4$ existing in $\Gamma\left(2-\frac{N}{2}\right)$
is an UV pole (since it is a pole existing irrespective of whether
the external line $p$ is on the mass shell or off the mass shell).
The limit $k_\mu\to 0$ can be taken in the remaining parameter integral to
give
\begin{align}
\Pi^{(1)}(0) &= \frac{e^2}{(4\pi)^{N/2}}
\Gamma\left(2-\frac{N}{2}\right)2N\frac{1}{(m^2)^{2-\frac{N}{2}}}\frac{1}{6}
\notag \\
&\quad \underset{N\to 4}{\sim}\frac{e^2}{(4\pi)^2}\frac{8}{3}\frac{1}{4-N}
(\text{pole of UV divergence}). \label{eq2.2.21}
\end{align}
Thus we have specifically confirmed that
$\Pi^{(1)}(0)$ has no IR divergence.
In the following we treat $\Pi^{(1)}(0)$, and hence $Z_3$, as a finite
quantity with regard to infrared divergence.

Returning to Eq.\ (\ref{eq2.2.6}), first, (\ref{eq2.2.7}) is used at
$k_\mu\to 0$, and
$\lambda\frac{\partial}{\partial\lambda}\tilde{D}_{\mu\nu}(k)$ is
replaced by
$\lambda\frac{\partial}{\partial\lambda}
\frac{-ig_{\mu\nu}}{k^2 - \lambda^2 + i\epsilon}Z_3$.
Then the strongest infrared divergence of $S$ yields
\begin{align}
\lambda\frac{\partial}{\partial\lambda}[S]_{\text{IR}}
&= -\frac{1}{2}\int\frac{d^N k}{(2\pi)^N}e^2
\left( \frac{p'^\mu}{p'\cdot k} - \frac{p^\mu}{p\cdot k} \right)
\left( \frac{p'^\nu}{p'\cdot l} - \frac{p^\nu}{p\cdot l} \right)
[S]_{\text{IR}}
\times\lambda\frac{\partial}{\partial\lambda}
\frac{-ig_{\mu\nu}}{k^2 - \lambda^2 + i\epsilon}Z_3 \notag \\
&= e^2 Z_3\lambda\frac{\partial}{\partial\lambda}
\ln\frac{m}{\lambda}B\left(\frac{1}{m^2}\right)^{\text{QED}}
\times [S]_{\text{IR}} \notag \\
&= -e^2 Z_3 B\left(\frac{1}{m^2}\right)^{\text{QED}}
[S]_{\text{IR}} \label{eq2.2.22}
\end{align}
where the one-loop result is used here.  The differential equation
thus determined is
\begin{equation}
\lambda\frac{\partial}{\partial\lambda}[S]_{\text{IR}}
= -\underbrace{e^2 Z_3}_{e_R^2} B\left(\frac{1}{m^2}\right)^{\text{QED}}.
\label{eq2.2.23}
\end{equation}
This is easily solved to give
\begin{equation}
[S]_{\text{IR}} = \exp\left(\underbrace{e^2 Z_3}_{e_R^2}
\ln\frac{m}{\lambda}
B\left(\frac{1}{m^2}\right)^{\text{QED}}\right)S^{(0)} \label{eq2.2.24}
\end{equation}
where the initial condition is taken as
\[
[S]_{\text{IR}}(\lambda=m) = S^{(0)}.
\]
$e^2 Z_3$ is a renormalized charge $e_R^2$.  In the case of QED,
however, $Z_3$ has no IR divergence.  Thus, as far as infrared
divergence is concerned, $e$ (bare charge) and $e_R$ (observed charge)
are interchangeable.  This concludes the proof of the factorization of
infrared divergence at all orders.
%
% 2.3
\subsection{Cancellation of infrared divergences}
The scattering amplitude $S$ considered up to the previous subsection
includes infrared divergence as seen in (\ref{eq2.2.24}).  If we
calculate the scattering cross section while keeping the infrared divergence
as it is, we would have $\infty$ or zero at $\lambda\to 0$, which is
a meaningless result.  However, our actual observation necessarily
involves energy resolution $\Delta E (\neq 0)$.  Thus, there may be
several soft photons that carry away energy within the
unobservable range.  Therefore it can be said that
a scattering cross section calculated
including the process of emitting several
unobserved soft photons is physically meaningful.

First, using the eikonal identity ((\ref{eq2.2.14}) and
(\ref{eq2.2.15})) derived in the previous subsection, we derive
an amplitude $S_R(n)$ (the subscript $R$ refers to real emission)
for emission of soft photons with small momenta $k_1, k_2, \dots, k_n$.
The amplitude for no emission of soft photons ($S$ up to the previous
subsection) is denoted $S_V$ ($V$ refers to virtual photon)
from this subsection.  Then we have
\begin{equation}
\begin{minipage}{16cm}
\centering
\includegraphics[width=16cm, clip]{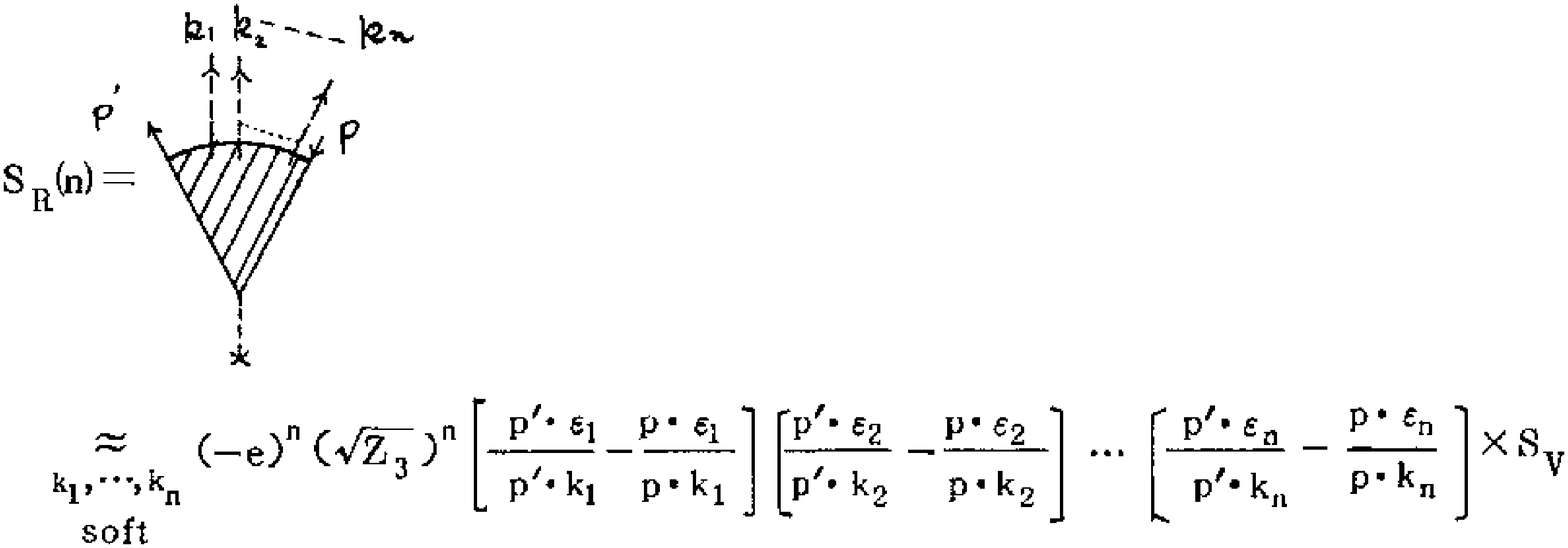}
\end{minipage}
\label{eq2.3.1}
\end{equation}
where $\epsilon_1, \dots, \epsilon_n$ are polarization vectors for $n$
soft photons, and $\sqrt{Z_3}$ indicates the effect of radiative
correction for soft photons of external lines.
It is obvious from (\ref{eq2.2.8}) of the previous subsection that
soft photons do not contribute to (\ref{eq2.3.1})
in the case of emission from the internal fermion loop.
The scattering cross section can be made from (\ref{eq2.3.1}) as
\begin{align}
d\sigma &\approx \sum_n \frac{1}{n!}\sum_{\text{pol.\ sum}}
\int_0^\delta\frac{d^{N-1}k_1}{(2\pi)^{N-1}2\sqrt{|\bm{k}_1|^2 + \lambda^2}}
\int_0^\delta\frac{d^{N-1}k_2}{(2\pi)^{N-1}2\sqrt{|\bm{k}_2|^2 + \lambda^2}}
\notag \\
&\quad\dotsb
\int_0^\delta\frac{d^{N-1}k_n}{(2\pi)^{N-1}2\sqrt{|\bm{k}_n|^2 + \lambda^2}}
\times |S_R(n)|^2, \label{eq2.3.2}
\end{align}
where $1/n!$ is the Bose factor resulting from $n$ particles being identical,
and the $k_1, \dots, k_n$-integrations are performed in the domain of
$|\bm{k}_i|<\delta$ ($i=1,\dots, n$) for sufficiently small $\delta$.
Using (\ref{eq2.3.1}) in (\ref{eq2.3.2}), we have
\begin{equation}
d\sigma \approx \exp\left(
+e^2 Z_3
\int_0^\delta\frac{d^{N-1}k}{(2\pi)^{N-1}2\sqrt{|\bm{k}|^2 + \lambda^2}}
\sum_{\text{pol.\ sum}}\left|
\frac{p'\cdot\epsilon}{p'\cdot k} - \frac{p\cdot\epsilon}{p\cdot k}
\right|^2 \right)\times|S_V|^2. \label{eq2.3.3}
\end{equation}
Substituting further the result (\ref{eq2.2.2}) of the previous
subsection into $S_V$, we have
\begin{align}
d\sigma &\approx \exp\left(
+e^2 Z_3
\int_0^\delta\frac{d^{N-1}k}{(2\pi)^{N-1}2\sqrt{|\bm{k}|^2 + \lambda^2}}
\sum_{\text{pol.\ sum}}\left|
\frac{p'\cdot\epsilon}{p'\cdot k} - \frac{p\cdot\epsilon}{p\cdot k}
\right|^2 \right) \notag \\
&\quad\times\exp\left(
-\frac{1}{2}e^2 Z_3\int\frac{d^N k}{(2\pi)^N}
\left( \frac{p'^\mu}{p'\cdot k} - \frac{p^\mu}{p\cdot k} \right)
\left( \frac{p'^\nu}{p'\cdot k} - \frac{p^\nu}{p\cdot k} \right)
\frac{-ig_{\mu\nu}}{k^2 - \lambda^2 + i\epsilon} + \text{c.c.}
\right)\times d\sigma^{(0)}, \label{eq2.3.4}
\end{align}
where $d\sigma^{(0)} = |S^{(0)}|^2$ is the differential cross section
at the lowest order.

Reviewing the meaning of (\ref{eq2.3.4}), the first exponential factor
corresponds to the effect of soft photon emission,
and the second exponential factor corresponds to the contribution of
infrared divergence from virtual photons.
We will show below that the factor associated with soft photon emission
also includes infrared divergence, which just cancels out the
infrared divergence associated with virtual photons, and that
the physical cross section $d\sigma$ is a finite quantity without
infrared divergence.  To this end it is sufficient to show
\begin{align}
&\int_0^\delta\frac{d^{N-1}k}{(2\pi)^{N-1}2\sqrt{|\bm{k}|^2 + \lambda^2}}
\sum_{\text{pol.\ sum}}\left|
\frac{p'\cdot\epsilon}{p'\cdot k} - \frac{p\cdot\epsilon}{p\cdot k}
\right|^2 \notag \\
&\quad - \left\{
\frac{1}{2}\int\frac{d^N k}{(2\pi)^N}
\left( \frac{p'^\mu}{p'\cdot k} - \frac{p^\mu}{p\cdot k} \right)
\left( \frac{p'^\nu}{p'\cdot k} - \frac{p^\nu}{p\cdot k} \right)
\frac{-ig_{\mu\nu}}{k^2 - \lambda^2 + i\epsilon} + \text{c.c.}
\right\} \notag \\
&= \text{No IR divergence}. \label{eq2.3.5}
\end{align}
This Eq.\ (\ref{eq2.3.5}) states only the cancellation of infrared
divergence in one-loop approximation.  That is, in the case of QED,
the cancellation of infrared divergence at all orders is
ultimately reduced
to the cancellation of infrared divergence at one loop.

Now we show (\ref{eq2.3.5}).  First, with regard to the sum in
polarization vectors, we note that the following holds:
\begin{equation}
\sum_{\text{pol.}}\epsilon_\mu\epsilon_\nu^\ast
= -g_{\mu\nu} + k_\mu \bar{k}_\nu + \bar{k}_\mu k_\nu,
\label{eq2.3.6}
\end{equation}
where
\[
\bar{k}_\mu \equiv \frac{1}{2|\bm{k}|^2}(k_0, -\bm{k}).
\]
Then the first term of (\ref{eq2.3.5}) yields:
\begin{equation}
\int_0^\delta\frac{d^{N-1}k}{(2\pi)^{N-1}2\sqrt{|\bm{k}|^2 + \lambda^2}}
\left( \frac{p'^\mu}{p'\cdot k} - \frac{p^\mu}{p\cdot k} \right)
\left( \frac{p'^\nu}{p'\cdot k} - \frac{p^\nu}{p\cdot k} \right)
(-ig_{\mu\nu}).  \label{eq2.3.7}
\end{equation}
On the other hand, $k^0$-integration is performed in the second
term of (\ref{eq2.3.5}).  When the $k^0$-integration is performed,
there is concern at the positions of poles of
$\left( \frac{p'^\mu}{p'\cdot k}
- \frac{p^\mu}{p\cdot k} \right)^2$.  Thus returning to the beginning
and recovering $i\epsilon$ (see (\ref{eq2.1.10})), the second term
of (\ref{eq2.3.5}) yields:
\begin{equation}
-\frac{1}{2}\int\frac{d^N k}{(2\pi)^N}
\left( \frac{p'^\mu}{p'\cdot k + i\epsilon}
- \frac{p^\mu}{p\cdot k + i\epsilon} \right)
\left( \frac{p'^\nu}{p'\cdot k + i\epsilon}
- \frac{p^\nu}{p\cdot k + i\epsilon} \right)
\frac{-ig_{\mu\nu}}{k^2 - \lambda^2 + i\epsilon}. \label{eq2.3.8}
\end{equation}
The positions of poles are depicted as follows.
\begin{figure}[h]
\includegraphics[width=140mm, clip]{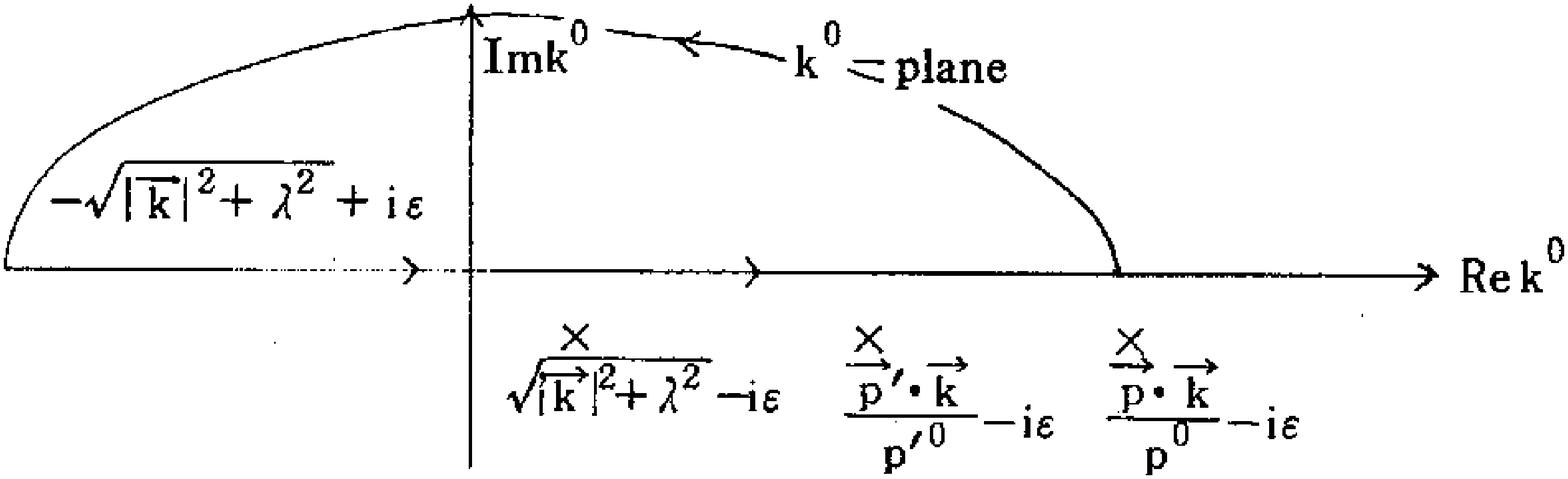}
\caption{Positions of poles}
\end{figure}
The poles from $\frac{1}{p'\cdot k + i\epsilon}$,
$\frac{1}{p\cdot k + i\epsilon}$ are both in the lower half plane.
Taking the $k^0$-integration path
so as to enclose the upper half plane, we have
\begin{align}
(\ref{eq2.3.8})
&= -\frac{1}{2}
\int\frac{d^{N-1}k}{(2\pi)^{N-1}(-2)\sqrt{|\bm{k}|^2 + \lambda^2}}
\left.\left( \frac{p'^\mu}{p'\cdot k} - \frac{p^\mu}{p\cdot k} \right)
\left( \frac{p'^\nu}{p'\cdot k} - \frac{p^\nu}{p\cdot k} \right)
g_{\mu\nu} \right|_{k^0=-\sqrt{|\bm{k}|^2 + \lambda^2}} \notag \\
&= \frac{1}{2}
\int\frac{d^{N-1}k}{(2\pi)^{N-1} 2\sqrt{|\bm{k}|^2 + \lambda^2}}
\left.\left( \frac{p'^\mu}{p'\cdot k} - \frac{p^\mu}{p\cdot k} \right)
\left( \frac{p'^\nu}{p'\cdot k} - \frac{p^\nu}{p\cdot k} \right)
g_{\mu\nu} \right|_{k^0=\sqrt{|\bm{k}|^2 + \lambda^2}},
\label{eq2.3.9}
\end{align}
where the transformation $\bm{k}\to -\bm{k}$ is performed in the
last step of Eq.\ (\ref{eq2.3.9}).  Since the infrared divergence of
(\ref{eq2.3.8}) arises from the sufficiently soft part
$|\bm{k}|<\delta$, the following equation finally holds:
\begin{equation}
(\ref{eq2.3.7}) + (\ref{eq2.3.9}) + \text{c.c. of (\ref{eq2.3.9})}
= \text{No IR divergence}. \label{eq2.3.10}
\end{equation}
That is, we have shown (\ref{eq2.3.5}).  Therefore,
from (\ref{eq2.3.4}), we have shown
\begin{equation}
[d\sigma]_{\text{IR}} = 0. \label{eq2.3.11}
\end{equation}
The foregoing is the proof of cancellation of infrared divergences
at all orders in QED \cite{33}.
Without this cancellation of infrared divergences, it cannot be said
that a particle is physically observable in field theory.  Therefore
infrared divergence remains (previous subsection) in the state of
the electron not accompanied by soft photons.  This is a state
that is not physically observable (confined state).  However,
cancellation of infrared divergences holds in the state of
the electron accompanied by soft photons within the allowable range
of energy resolution (the state of the electron dressed with soft
photons).  Thus it can be said that this is a physically observable
state.

Therefore, when we show quark confinement in QCD, we must first examine
the problem of infrared divergence cancellation.  This is the theme of
the next Section 3.

%%%%%%%%%%%%%%%%%%%%%%%%%%%%%
% Section 3
\section{Cancellation of infrared divergences in QCD (one loop)}

%%%%%%%%%%%%%%%%%%%%%%%%%%%

In the previous section (Section 2) we reviewed the theory of infrared
divergence in QED.  From this section we examine
how the theory is generalized and what difference appears in the case of
QCD with a focus on the author's work.  First in this section
we discuss fermion-fermion scattering (quark-quark scattering) and
fermion-gauge boson scattering (quark-gluon scattering) in one-loop
approximation.  (Quark scattering at one loop by an external field
as seen in Subsection 2.1 is too simple to grasp the characteristics
of QCD.  Investigation of this reaction requires calculation at two loops
at the minimum \cite{5}.)  In this section we use the covariant gauge QCD.
The main theme is the problem of cancellation of infrared divergences,
summarizing Reference paper I.
%
% 3.1
\subsection{Classification of Feynman graphs (one loop)}
The covariant gauge QCD is expressed by an effective Lagrangian
density $\mathcal{L}_{\text{eff}}$ including ghosts:
\begin{align}
\mathcal{L}_{\text{eff}}
&= \bar{\psi}(i\Slash{\nabla} - m)\psi
- \frac{1}{4}F^a_{\mu\nu}F^{a,\mu\nu}
- \frac{1}{2\alpha}(\partial_\mu A^{a\mu})^2
+ g\bar{\psi}\gamma_\mu t^a \psi A^{a\mu} \notag \\
&\quad -\partial^\mu c^{a\dag}(\partial_\mu c^a + gf^{abc}A_\mu^b c^c).
\label{eq3.1.1}
\end{align}
For the group $SU(N)$, the symbols used in (\ref{eq3.1.1}) are given as
follows.  $F^a_{\mu\nu}$ is defined as
\begin{equation}
F^a_{\mu\nu}
= \partial_\mu A_\nu^a - \partial_\nu A_\mu^a - gf^{abc}A_\mu^b A_\nu^c.
\label{eq3.1.2}
\end{equation}
$t^a$ is the representation matrix for fermions, usually an $N$-dimensional
representation for the fundamental representation of $SU(N)$.
The commutation relation is given by
\begin{equation}
[t^a, t^b] = if^{abc}t^c, \label{eq3.1.3}
\end{equation}
where $f^{abc}$ is called the structure constant.
$c^a$ is the Faddeev-Popov ghost field, and $\alpha$ is the gauge parameter.

In the case of fermion-fermion scattering, relevant Feynman graphs are
as follows.
\begin{figure}[h]
\centering
\includegraphics[width=120mm, clip]{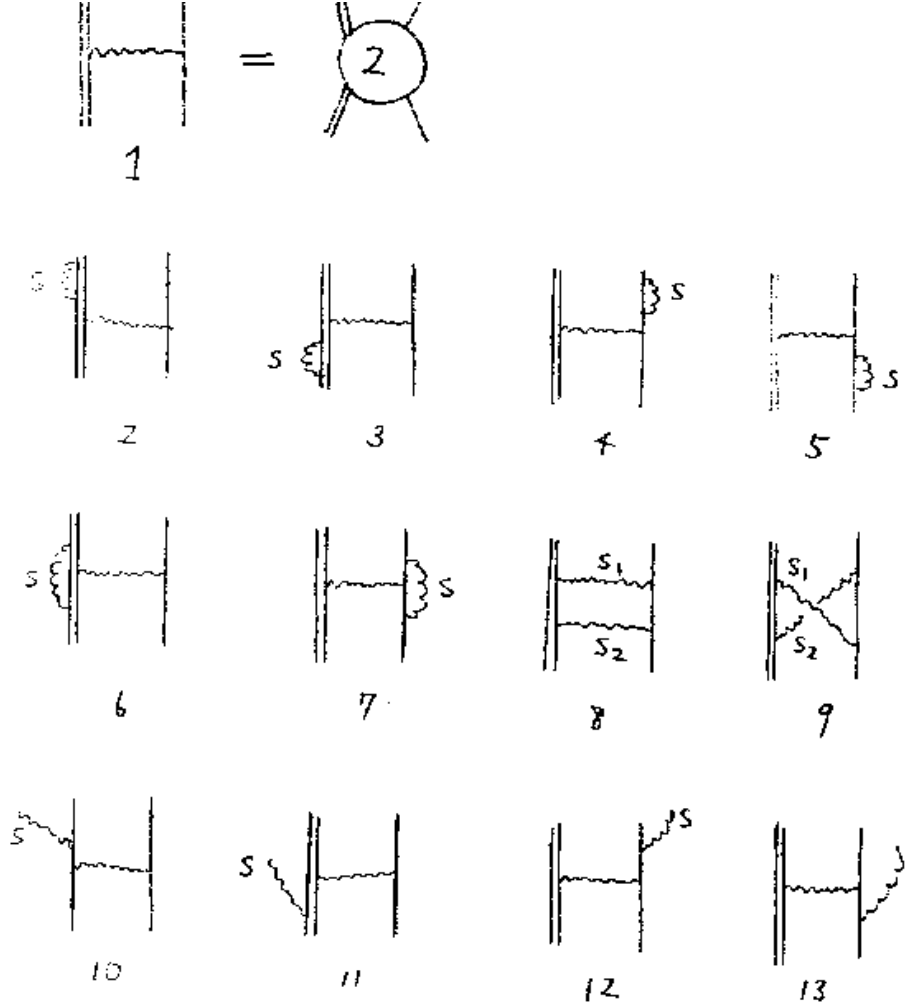}
\caption{fermion-fermion scattering}
\end{figure}
Here the two fermions are different fermions, and the kinematics is taken
as shown in Figure 3.1.2:
\[
\begin{minipage}{12cm}
\centering
\includegraphics[width=3cm, clip]{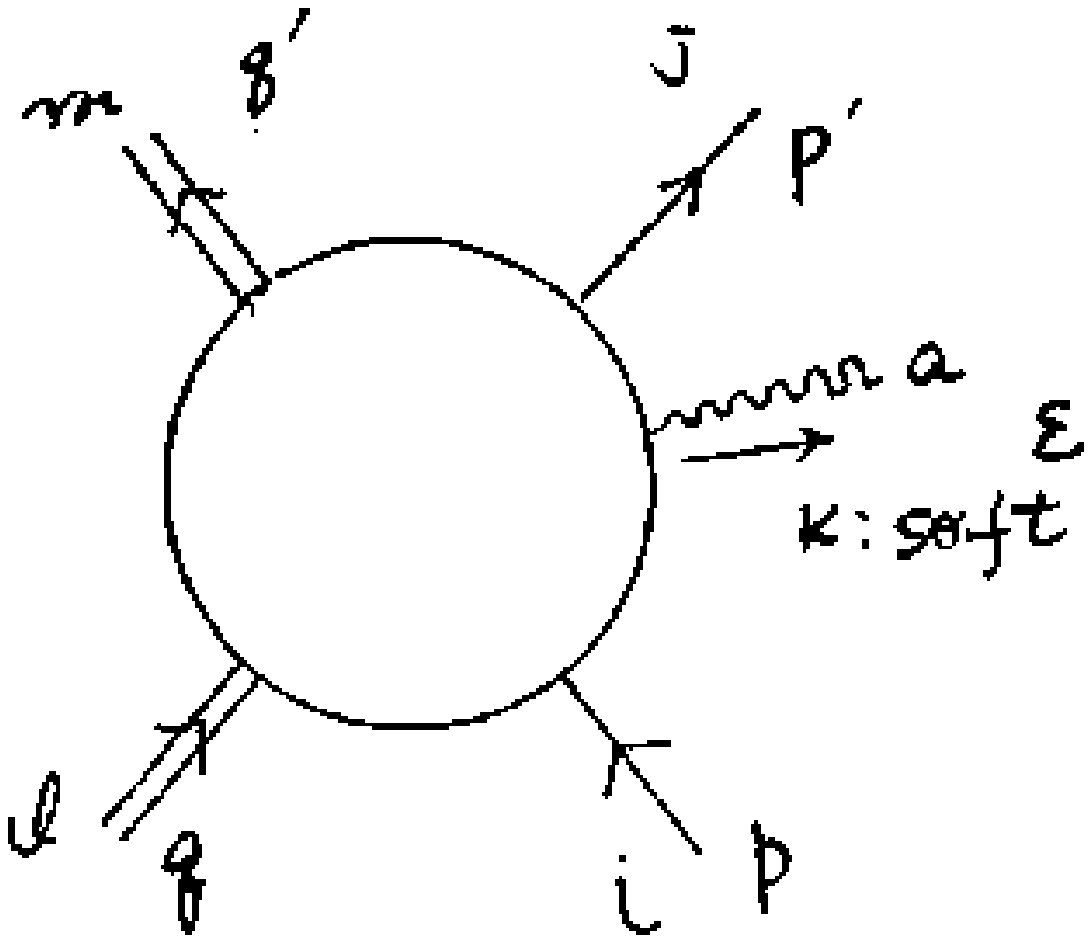}
\text{(Figure 3.1.2)}
\end{minipage}
\]
where $\{p, p', q, q'\}$ represent four-momenta of fermions,
$\{i, j, l, m\}$ represent color indices of fermions, and
$k, a, \epsilon$ represent four-momentum, color index, and polarization
vector for a soft gauge boson (gluon) emitted additionally.

We will limit the following analysis to \emph{non-forward scattering}
($p_\mu \neq p'_\mu$) \cite{34}.  In this case it turns out that
the following three graphs do not contribute to the analysis of
infrared divergence.
\begin{figure}[h]
\centering
\includegraphics[width=120mm, clip]{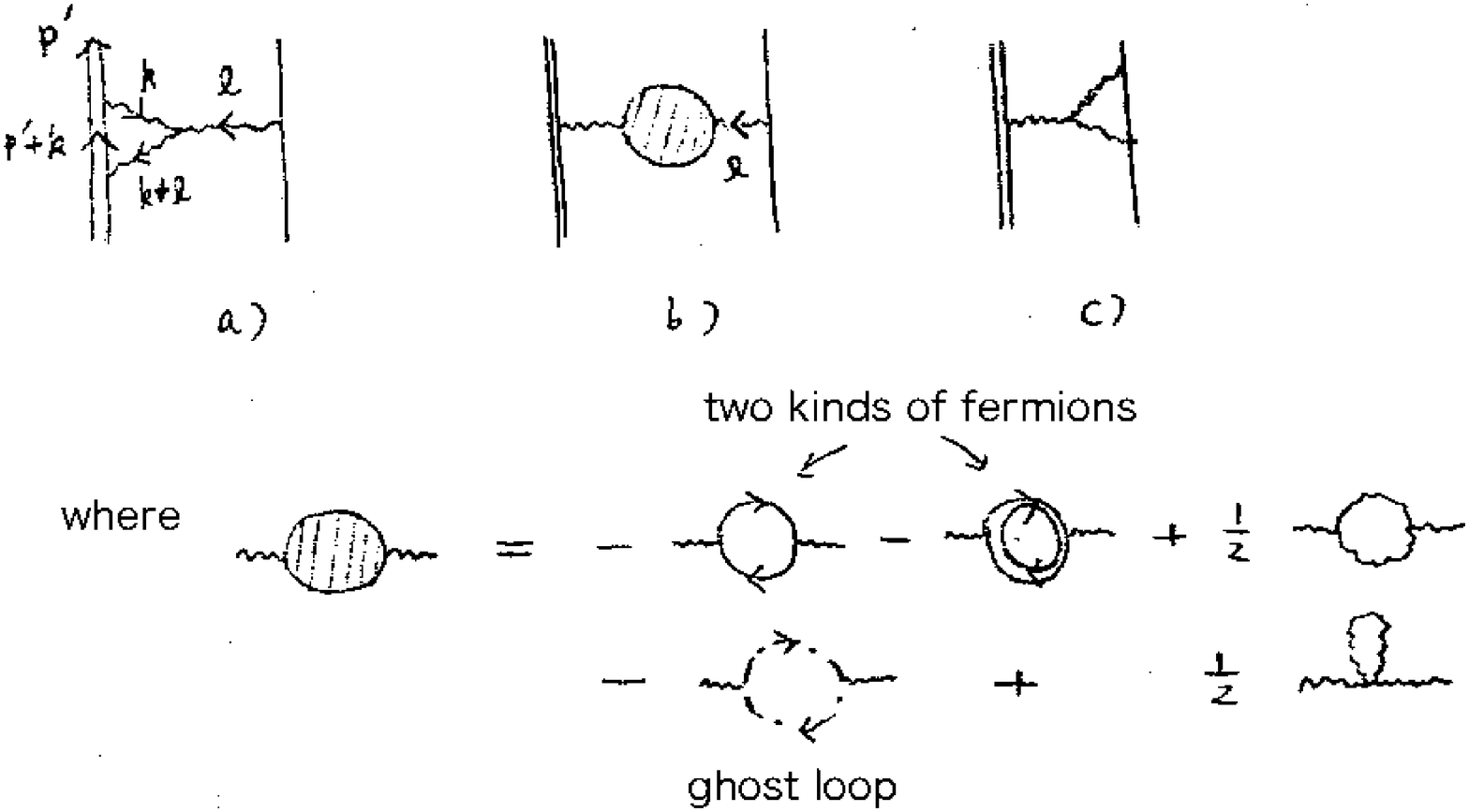}
\caption{}
\end{figure}
In Figure a), we examine the integral around $k_\mu\approx 0$.
The term in which the numerator is $O(1)$ has the strongest
contribution to infrared divergence.  This term has no
infrared divergence, since it yields,
in view of $p'^2 = m^2$, $l_\mu\neq 0$:
\begin{equation}
\int d^4 k\,\frac{1}{(p'+k)^2 - m^2}\frac{1}{k^2}\frac{1}{(k+l)^2}
\sim\int d^4 k\,\frac{1}{k^3}.
\label{eq3.1.4}
\end{equation}
Figure c) has no infrared divergence for the same reason as Figure a).
Also, Figure b) cannot have infrared divergence for $l_\mu\neq 0$
\cite{35}.

However, the infrared divergence included in the $Z_1$ and $Z_3$ factors,
which will be calculated later (Subsection 3.4), correspond to the
infrared divergence included in Figures a)--c) in the case of forward
scattering.
Since two gauge bosons in graphs 8 and 9 of Figure 3.1.1
cannot simultaneously be soft (being simultaneously soft correspond to
forward scattering), infrared divergence occurs when one of them
($S_1$ or $S_2$) is soft.

Next, Feynman graphs related to fermion-gauge boson scattering are
as follows.
\begin{figure}[h]
\centering
\includegraphics[width=120mm, clip]{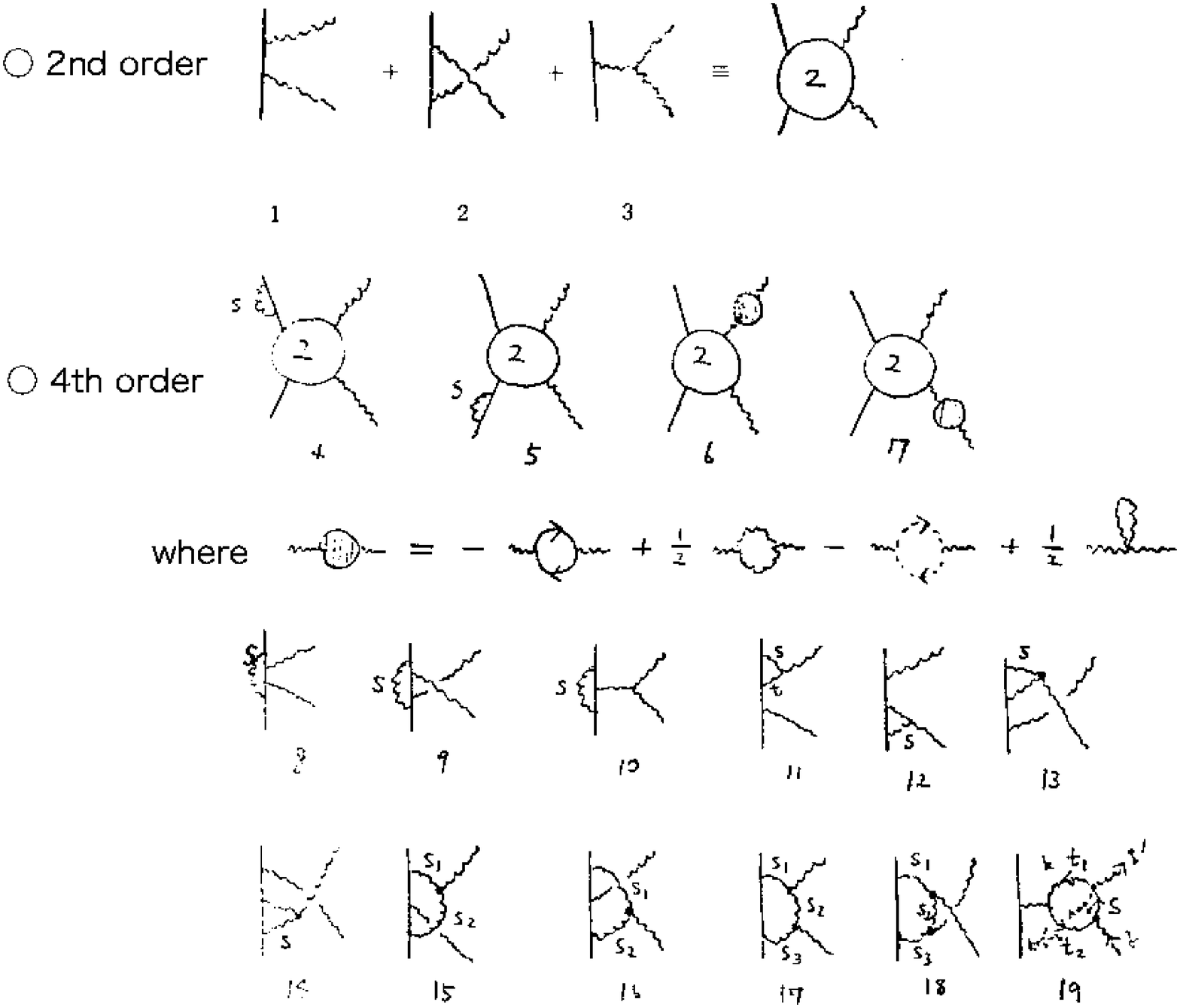} \\
\includegraphics[width=120mm, clip]{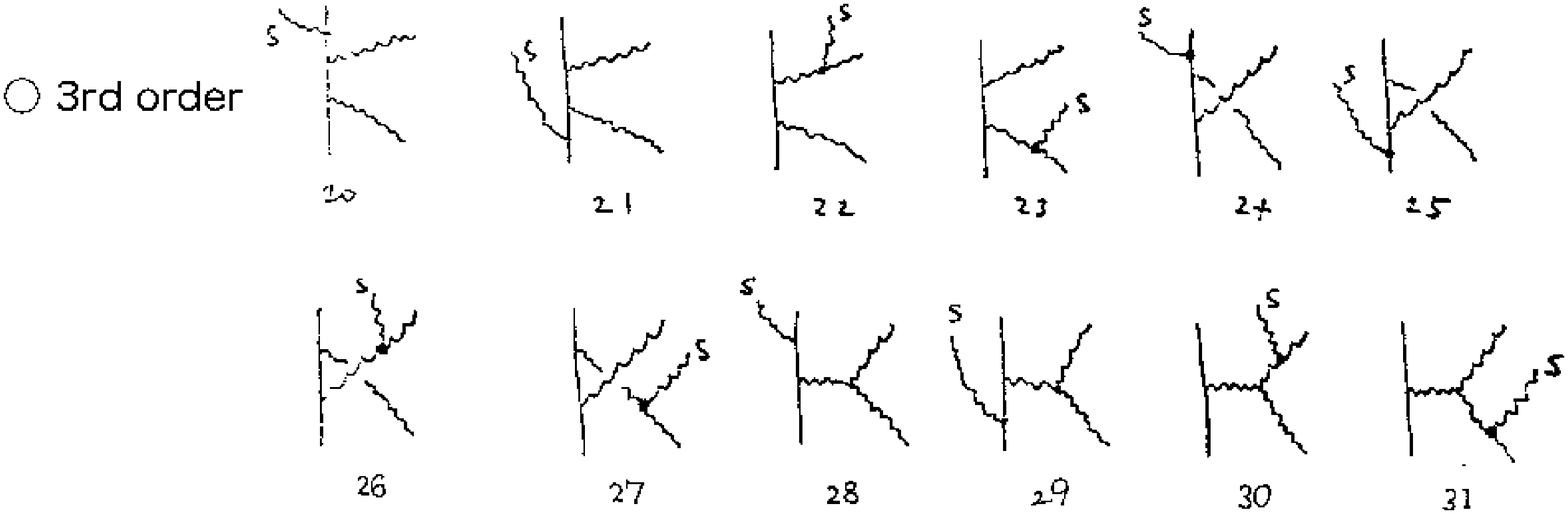}
\caption{fermion-gauge boson scattering}
\end{figure}
\[
\begin{minipage}{12cm}
\centering
\includegraphics[width=3cm, clip]{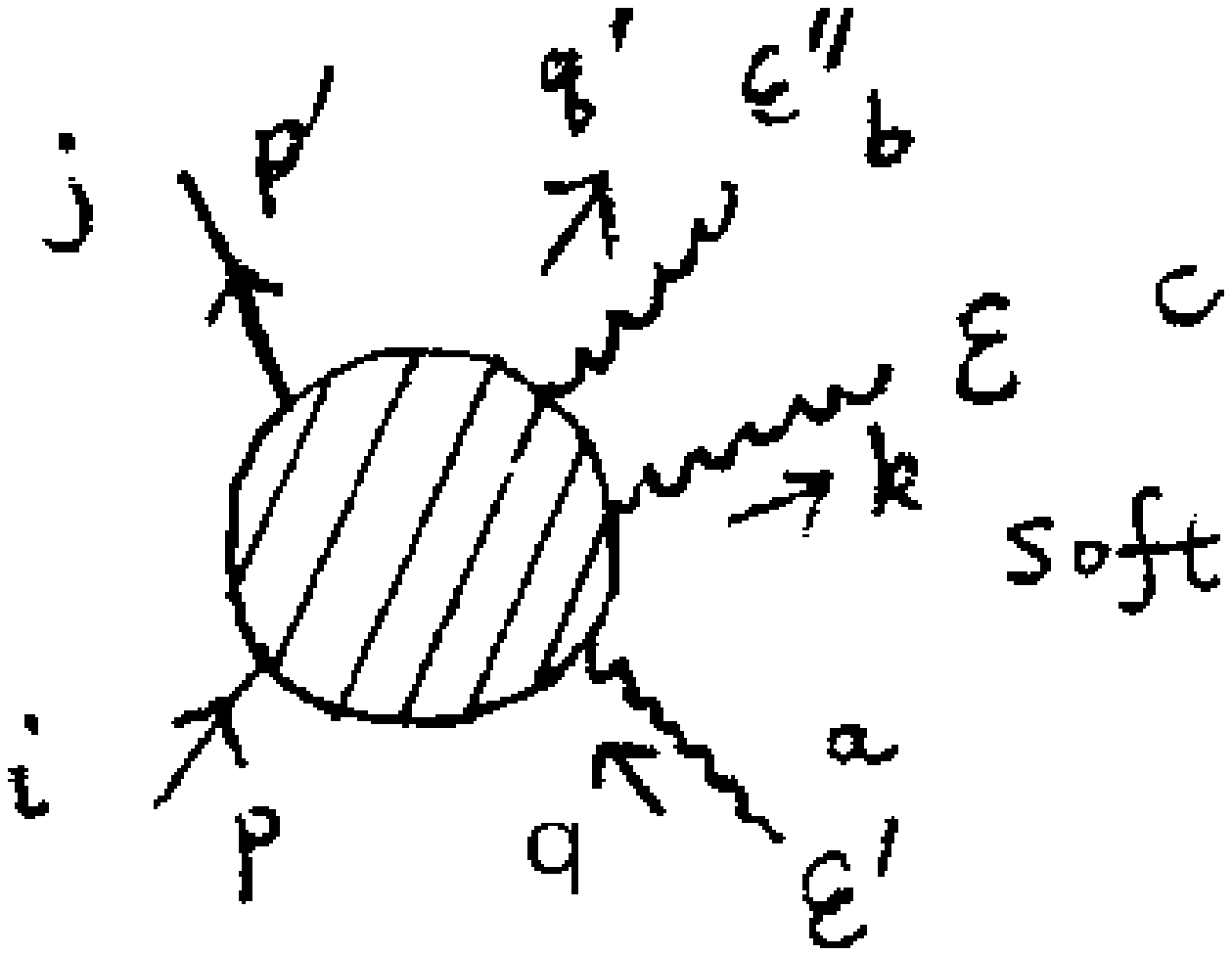}
\text{(Figure 3.1.5)}
\end{minipage}
\]
The kinematics is taken as shown in Figure 3.1.5, 
where $\{q, q'\}$, $\{a, b\}$, $\{\epsilon', \epsilon''\}$
represent four-momenta, color indices, and polarization
vectors of hard gauge bosons, and $k, c, \epsilon$ represent
those of a soft gauge boson.

Here, \emph{non-forward scattering} is considered.
Furthermore, the initial state and the final state must include
at least one hard gauge boson.  Thus, it turns out that in the graphs
of Figure 3.1.4, although there are many gauge bosons that can be
soft, \emph{two or more gauge bosons cannot be simultaneously soft
in one graph}.  For instance, in graph 17 of Figure 3.1.4,
\[
\begin{cases}
\text{Gauge bosons $S_1$ and $S_2$ are soft}
&\Rightarrow\text{External line (momentum $q'$) is soft,} \\
\text{Gauge bosons $S_2$ and $S_3$ are soft}
&\Rightarrow\text{External line (momentum $q$) is soft,} \\
\text{Gauge bosons $S_1$ and $S_3$ are soft}
&\Rightarrow\text{Forward scattering.}
\end{cases}
\]
Thus, these three cases are excluded, and only one of $S_1$, $S_2$
and $S_3$ can be soft.  Considering similarly for other graphs,
it turns out that each graph can include only one soft gauge boson.

Next, some gauge bosons can be soft but do not contribute
to infrared divergence, such as
gauge bosons denoted $t$ in graph 11 and denoted $t_1$, $t_2$
in graph 19.  For instance, if $t_1$ is soft,
the term in which the numerator is $O(1)$ has the strongest
divergence in the loop integral of graph 19.  However, the integral
yields
\begin{align}
&\int d^4 k\,\frac{1}{k^2}\frac{1}{(k+q')^2}\frac{1}{(k+q'-q)^2}
\notag \\
&\underset{k:\text{soft}, q'^2=0}{\sim}
\int d^4 k\,\frac{1}{k^2}\frac{1}{2k\cdot q}\frac{1}{(q'-q)^2}
\sim\int d^4 k\,\frac{1}{k^3}, \label{eq3.1.5}
\end{align}
and does not contribute to infrared divergence.
The same also applies to the other examples.

It should be remarked a little about the graphs omitted from Figure 3.1.4.
For instance, the following graph exists.
\begin{figure}[h]
\centering
\includegraphics[width=110mm, clip]{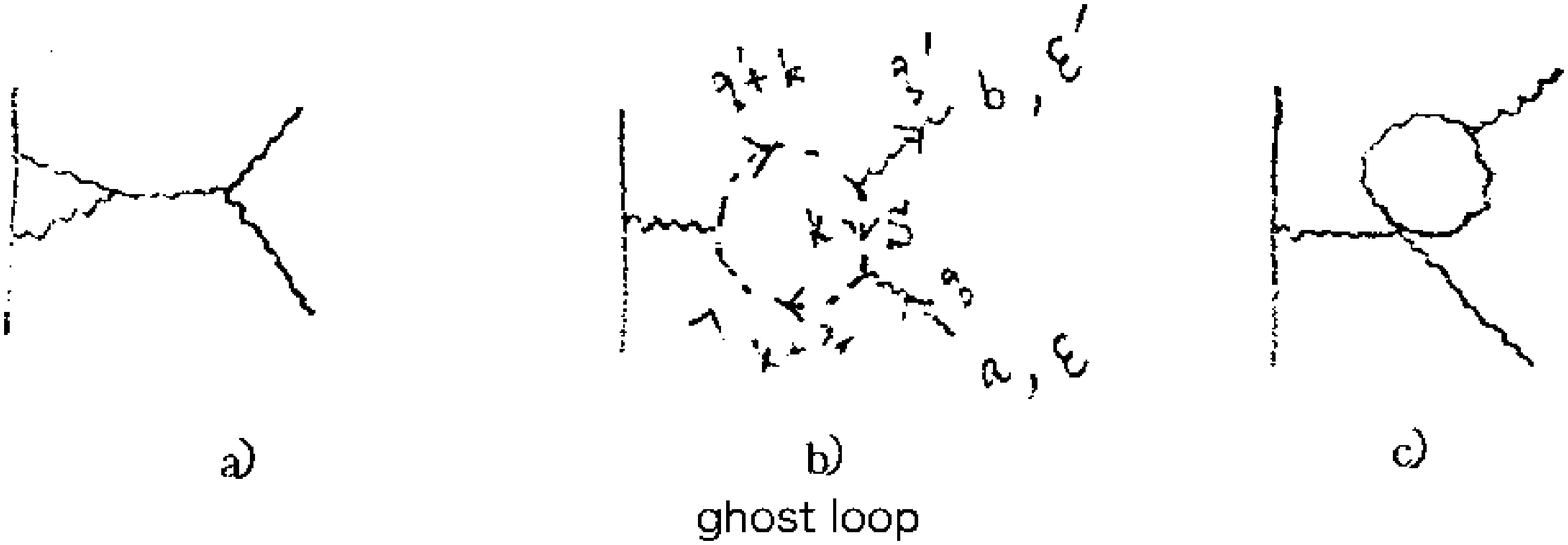}
\caption{}
\end{figure}
Graph a) has no contribution for the same reason as a) of Figure 3.1.3.
Also, graph c) immediately proves to have no contribution by power
counting.  For graph b), calculating the numerator as $O(1)$,
it has the same structure as graph 19 of Figure 3.1.5.  Thus
it may be considered that infrared divergence occurs when
the gauge boson denoted $S$ is soft.  However, the numerator is
actually not $O(1)$, and hence the graph does not contribute to
infrared divergence.  Let the momentum of the ghost denoted $S$ be
$k_\mu$.  Then the relevant part yields
\begin{align}
\text{Graph b)} &\approx \int d^4 k\,
\frac{(k\cdot\epsilon')(k+q)\cdot\epsilon}{k^2 (k+q)^2 (k+q')^2}
\label{eq3.1.6} \\
&\approx \int d^4 k\,\frac{(k\cdot\epsilon')}{k^2 (2kq) (2kq')}.
\quad\text{: no infrared divergence}
\end{align}
That is, the numerator being $O(k)$ is effective \cite{36}.

In view of the above reasoning, below we consider the
infrared divergence in the case where
the gauge boson denoted $S$ in Figure 3.1.4 is soft.
However, we consider all the self-energy parts
of the gauge boson in graphs 16 and 17, irrespective of soft
or hard (for details, see Reference paper I).

Now the three graphs at the lowest order are grouped and represented by
\begin{tikzpicture}[baseline=(b.base), node distance = 5 mm]
\begin{feynhand}
\vertex [ringblob] (b) at (1,0) {2};
\vertex [particle] (a1) [above left = of b] {};
\vertex [particle] (a2) [below left = of b] {};
\vertex [particle] (a3) [above right = of b] {};
\vertex [particle] (a4) [below right = of b] {};
\propag [plain] (a1) to (b);
\propag [plain] (a2) to (b);
\propag [photon] (a3) to (b);
\propag [photon] (a4) to (b);
\end{feynhand}
\end{tikzpicture}
as a single graph.
Then the above graphs are classified into the following.
\begin{figure}[h]
\centering
\includegraphics[width=120mm, clip]{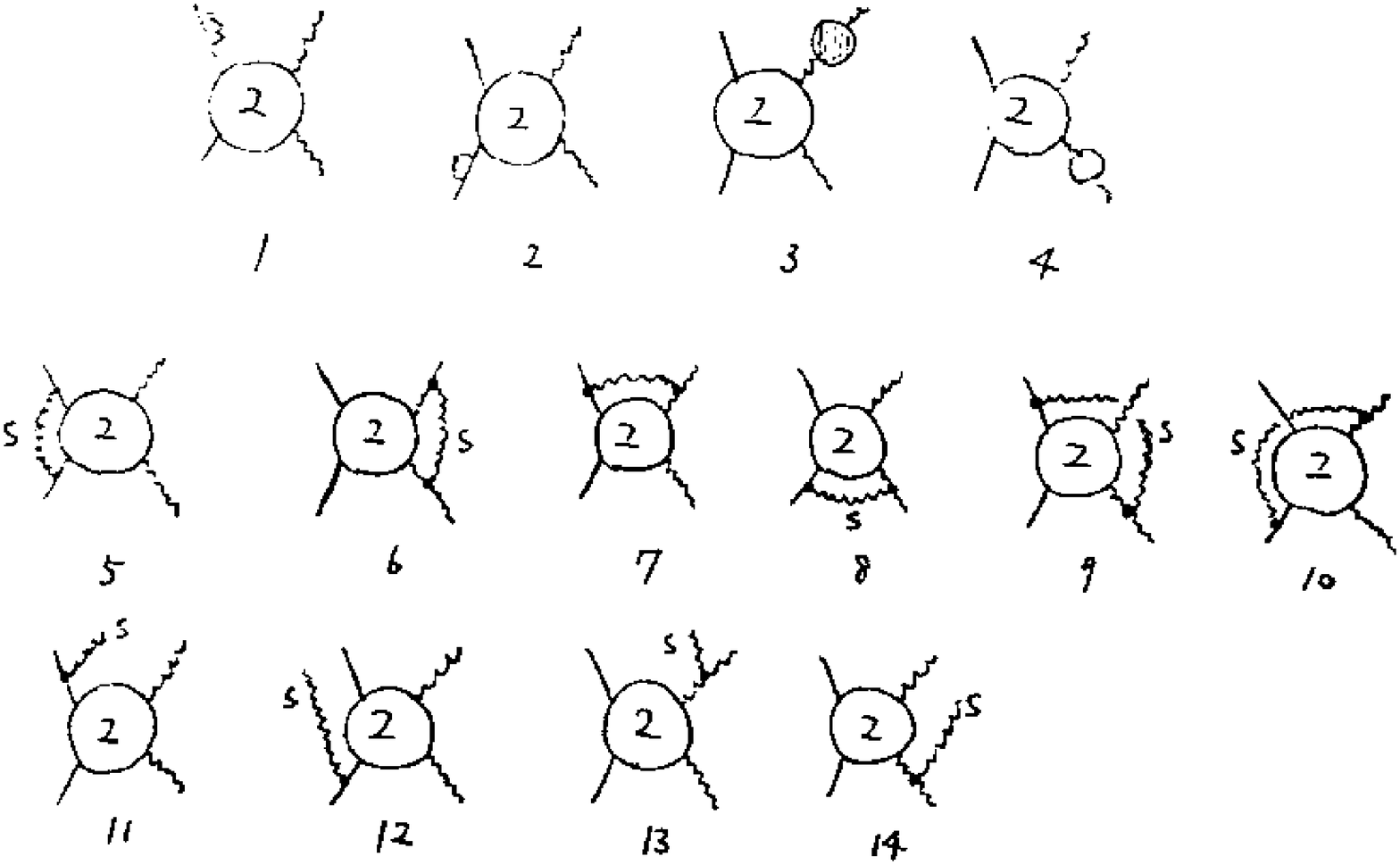}
\caption{}
\end{figure}
That is, they are classified into graphs 1--4 for correction
of an external line, the graphs in which different external lines
are connected by a soft gauge boson, and the graphs in which
a soft gauge boson is emitted from an external line.
For instance, taking graph 18 of Figure 3.1.4, we illustrate below
how the graphs in which $S_1$--$S_3$ are soft are grouped into
a particular graph in Figure 3.1.7.
\begin{figure}[h]
\centering
\includegraphics[width=110mm, clip]{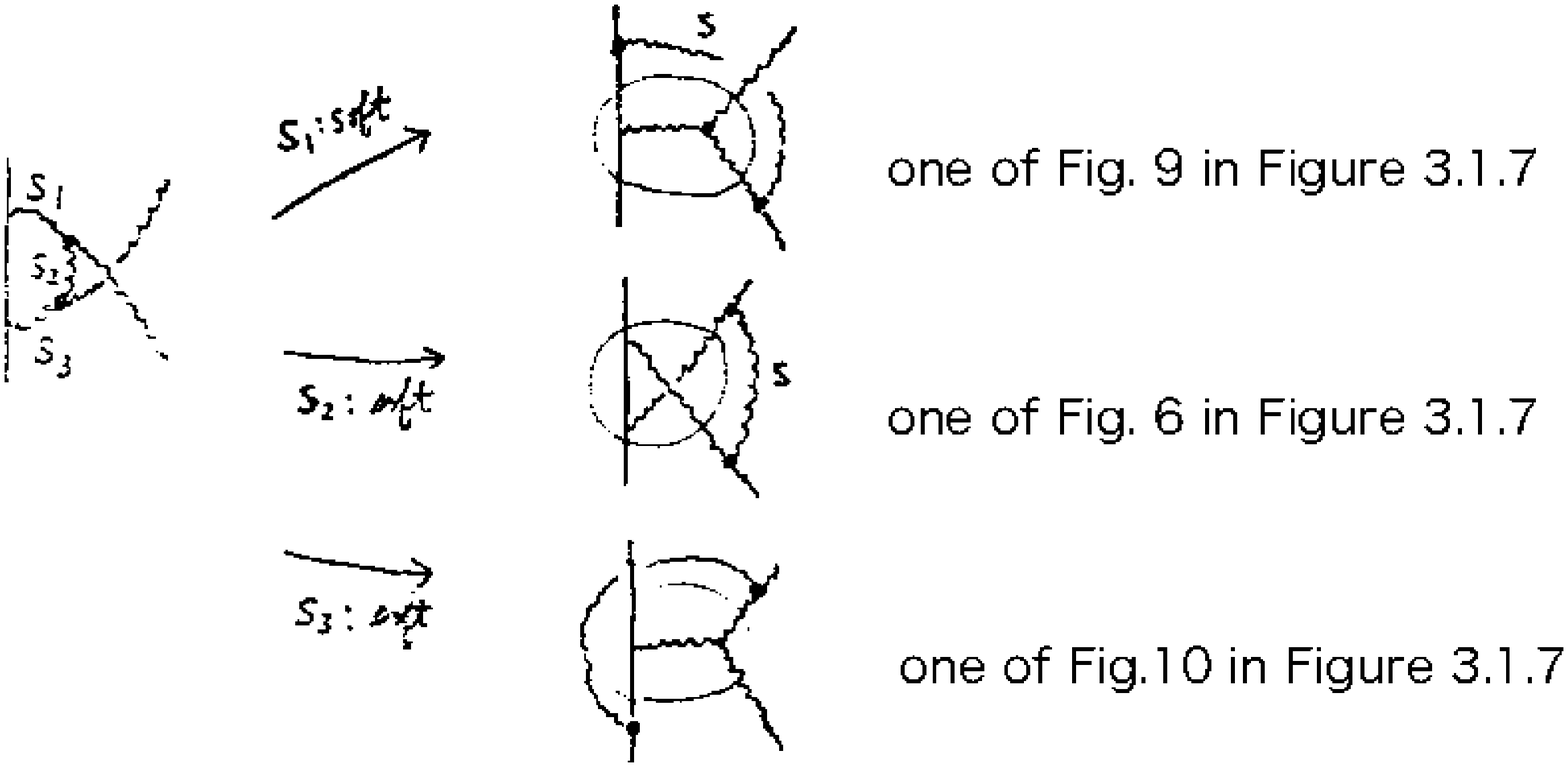}
\caption{illustration}
\end{figure}
Fermion-fermion scattering can also be classified similarly.
In this case, there is only one graph at the lowest order.
Thus the above classification diagram Figure 3.17 is trivial.
In fermion-gauge boson scattering, \emph{grouping the three graphs
at the lowest order} gives a clear view.
%
% 3.2
\subsection{Factorization of infrared divergences in QCD (one loop)}
On the basis of the graph classification in the previous subsection,
we prove in this subsection the following factorization rule.
\begin{align}
&\begin{minipage}{12cm}
\centering
\includegraphics[width=10cm, clip]{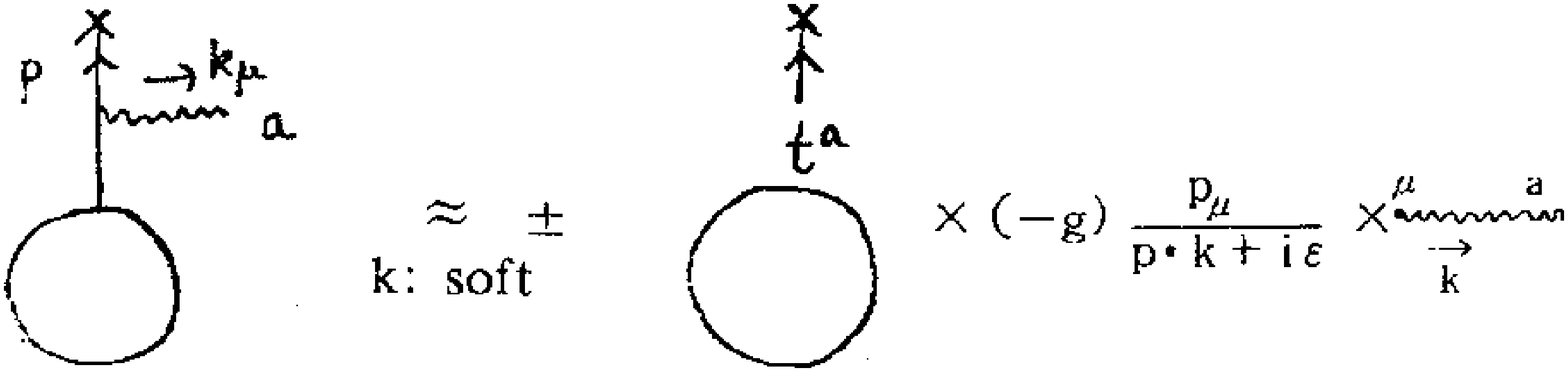}
\end{minipage}
\label{eq3.2.1} \\
&\begin{minipage}{12cm}
\centering
\includegraphics[width=10cm, clip]{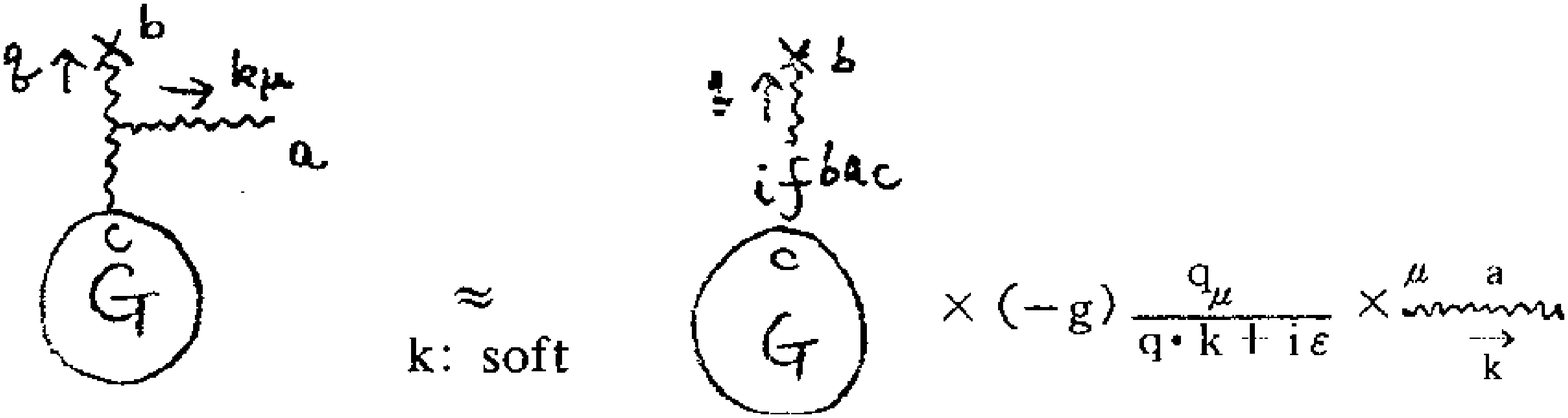}
\end{minipage}
\label{eq3.2.2}
\end{align}
In the above equation, the symbol $\times$ attached to an external line 
represents on-mass-shell (and additionally transverse for gauge bosons).
The $\pm$ sign in (\ref{eq3.2.1}) assumes $+$ for outgoing fermions and 
$-$ for incoming fermions.
The symbol \ctext{G} in Eq.\ (\ref{eq3.2.2}) represents the sum of 
all the graphs at the same order (the set of gauge-invariant graphs).
Thus (\ref{eq3.2.2}) indicates that factorization holds only after 
several graphs are summed up.
It is for the purpose of using (\ref{eq3.2.2}) that the graphs at the 
lowest order are grouped in the classification of graphs in the 
last part of the previous subsection.
There is no need to prove (\ref{eq3.2.1}) because it agrees with Eqs.\ 
(\ref{eq2.1.4}) and (\ref{eq2.1.5}) in \S 2 except for the color factor 
$t^a$.  
[In the subsequent section 4, this eq.\ (\ref{eq3.2.1}) is extended to 
all orders.]
It seems necessary, however, to prove (\ref{eq3.2.2}) since it appears 
in QCD for the first time.  
Let us rewrite the left hand side of (\ref{eq3.2.2}) \cite{36}.
\begin{equation}
\begin{minipage}{12cm}
\centering
\includegraphics[width=12cm, clip]{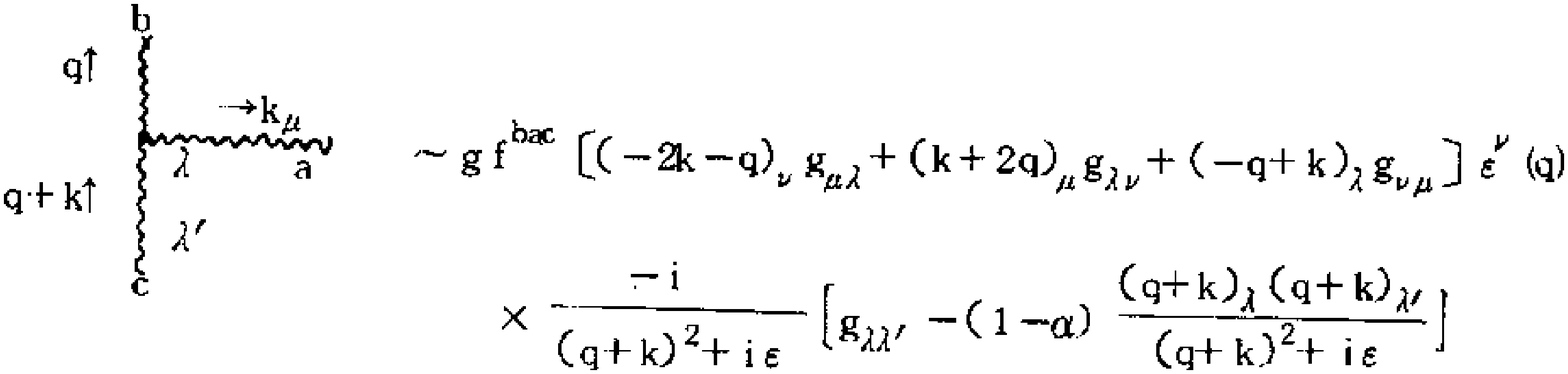}
\end{minipage}
\label{eq3.2.3} 
\end{equation}
In the term proportional to $(1-\alpha)$ in this equation, 
multiplying the three-point vertex by $(q+k)_\lambda$ 
using the general formula
\begin{equation}
\begin{minipage}{12cm}
\centering
\includegraphics[width=12cm, clip]{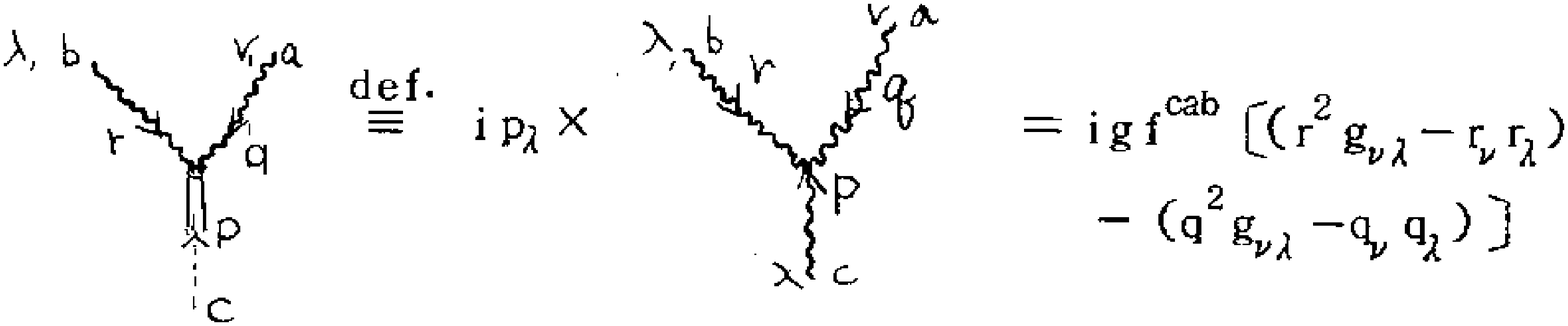}
\end{minipage}
\label{eq3.2.4} 
\end{equation}
(this is a \emph{very important formula}, which is a starting point 
of the general Ward identity) yields the following.
\begin{equation}
\begin{minipage}{12cm}
\centering
\includegraphics[width=12cm, clip]{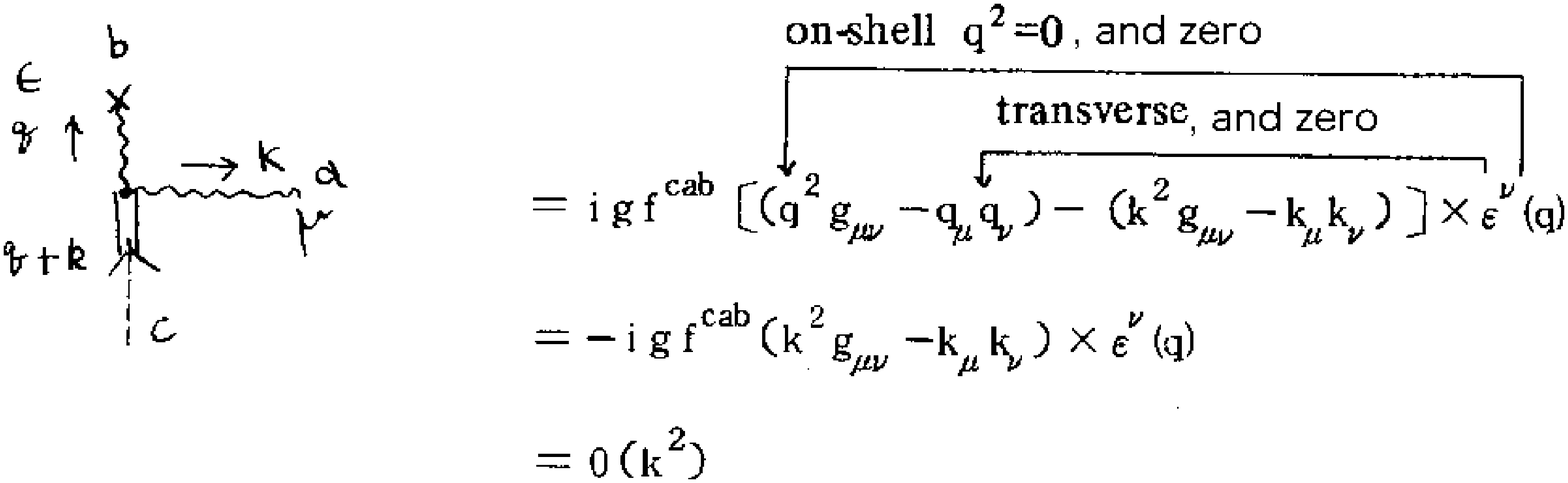}
\end{minipage}
\label{eq3.2.5} 
\end{equation}
In combination with the denominator 
$(q+k)^2 \approx 2q\cdot k$, it is found that 
the term proportional to $(1-\alpha)$ behaves as $O(k)$ relative to 
the $g_{\lambda\lambda'}$ term and can be omitted in the 
soft $k$ limit.  
Then the term proportional to $g_{\lambda\lambda'}$ is rewritten, 
and the most contributing factor under soft $k$ is given by
\begin{equation}
    gf^{bac}[2q_\mu \epsilon_{\lambda'}(q) - 
    q_{\lambda'}\epsilon_{\mu}(q)]
    \frac{-i}{2q\cdot k + i\epsilon}. \label{eq3.2.6}
\end{equation}
If the second term of this expression vanishes, only the first term remains 
and agrees with Eq.\ (\ref{eq3.2.2}).  
The second term of (\ref{eq3.2.6}) vanishes because substituting 
the sum of all the graphs at the same order into \ctext{G} yields 
the following Ward-Takahashi identity.
\begin{align}
& \begin{minipage}{7cm}
\centering
\includegraphics[width=3cm, clip]{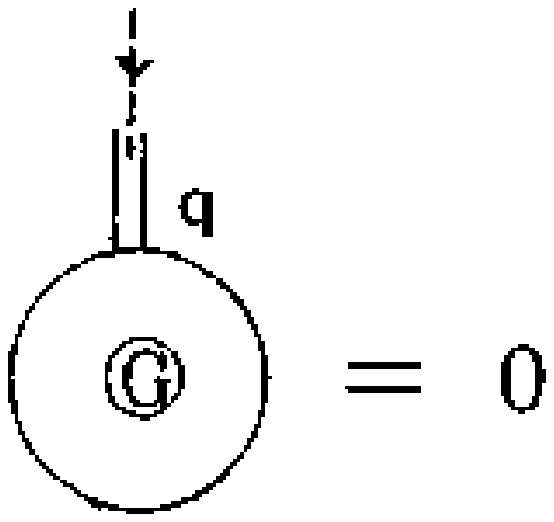}
\end{minipage}
\label{eq3.2.7},  \\
&\begin{minipage}{7cm}
\centering
\includegraphics[width=37mm, clip]{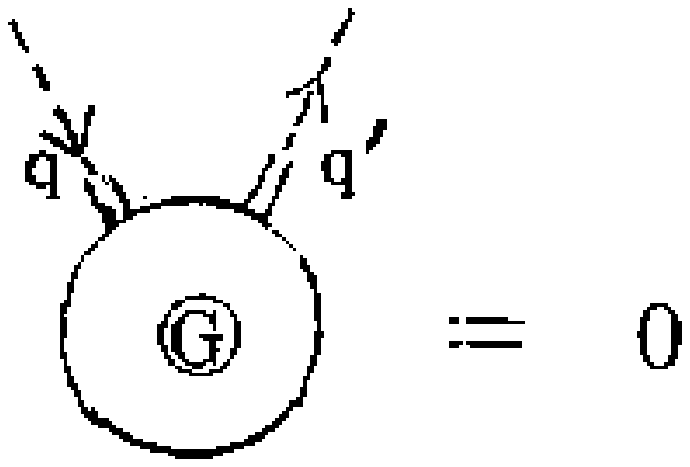}
\end{minipage}
\label{eq3.2.8}
\end{align}
Here it is assumed that the remaining external lines are all on the 
mass shell (and transverse for gauge bosons).  
In the present consideration, the general formula is not needed, 
but it is only necessary to apply a special case of (\ref{eq3.2.7}) 
with $\text{\ctext{G}} = 
\begin{tikzpicture}[baseline=(b.base), node distance = 5 mm]
    \begin{feynhand}
    \vertex [ringblob] (b) at (1,0) {2};
    \vertex [particle] (a1) [above left = of b] {};
    \vertex [particle] (a2) [below left = of b] {};
    \vertex [particle] (a3) [above right = of b] {};
    \vertex [particle] (a4) [below right = of b] {};
    \propag [plain] (a1) to (b);
    \propag [plain] (a2) to (b);
    \propag [photon] (a3) to (b);
    \propag [photon] (a4) to (b);
    \end{feynhand}
\end{tikzpicture}
$ (the lowest order of fermion-gauge boson scattering) given as follows.
\begin{align}
& \begin{minipage}{14cm}
\centering
\includegraphics[width=12cm, clip]{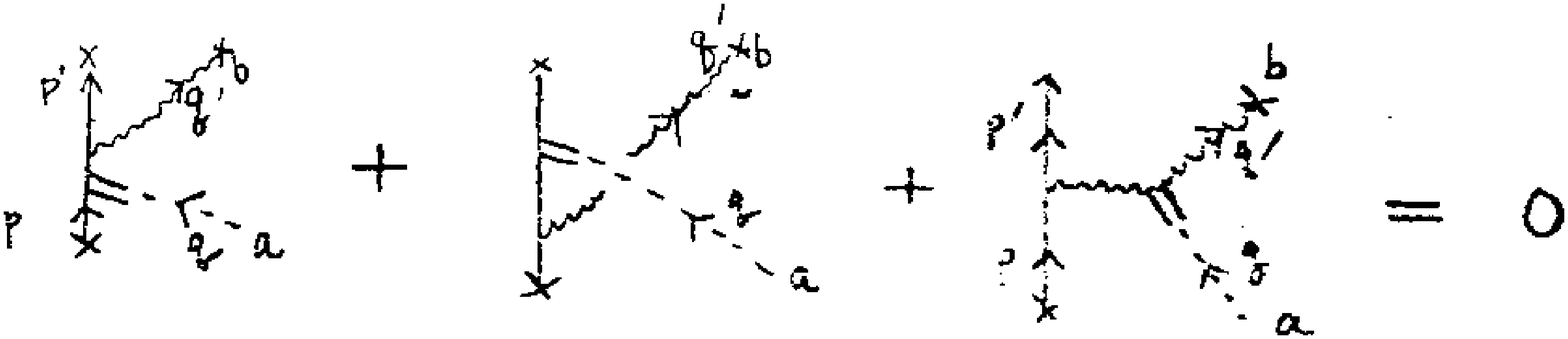}
\end{minipage}
\label{eq3.2.9},  \\
&\begin{minipage}{14cm}
\centering
\includegraphics[width=12cm, clip]{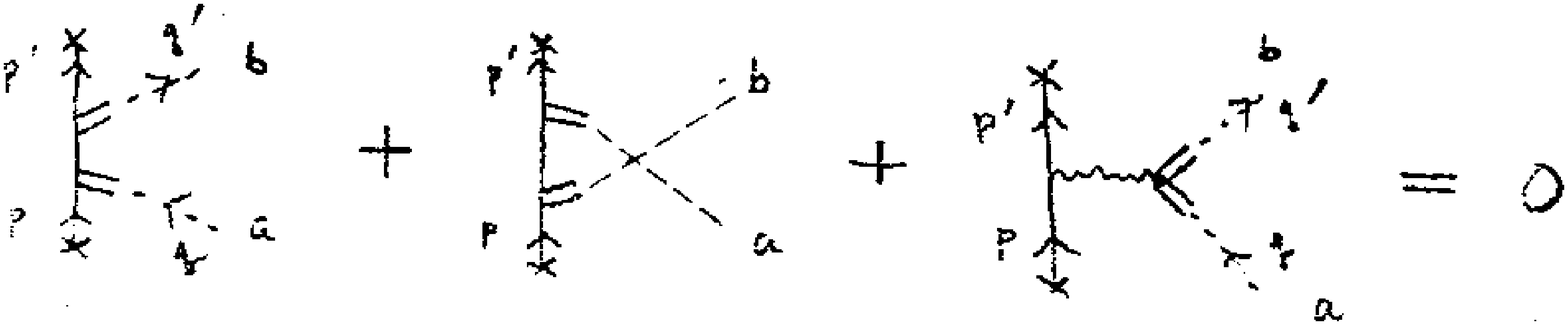}
\end{minipage}
\label{eq3.2.10}
\end{align}
If these hold, the second term of (\ref{eq3.2.6}) is dropped, 
completing the proof of the factorization rule (\ref{eq3.2.2}).
Here (\ref{eq3.2.9}) and (\ref{eq3.2.10}) can be directly checked as
\begin{align}
\text{LHS of (\ref{eq3.2.9})}
    &= (t^b t^a - t^a t^b + it^c f^{abc})
        \times ig^2[\bar{u}(p')\Slash{\epsilon}u(p)] \notag \\
    &= 0, \label{eq3.2.11} \\
\text{LHS of (\ref{eq3.2.10})}
    &= (t^b t^a - t^a t^b + it^c f^{abc})
        \times ig^2[\bar{u}(p')\Slash{q}u(p)] \notag \\
    &= 0. \label{eq3.2.12}
\end{align}
The three graphs correspond to the respective terms of the 
group commutator
\begin{equation}
    t^b t^a - t^a t^b + it^c f^{abc} = 0. \label{eq3.2.13}
\end{equation}
%
% 3.3
\subsection{Cancellation of infrared divergences (in terms of bare 
coupling constant)}
Cancellation of infrared divergences will now be shown at one loop 
using the factorization rule (\ref{eq3.2.1}) and (\ref{eq3.2.2}) 
given in the previous subsection.
However, in the case of QCD, it is necessary to clarify which 
coupling constant is used for expansion when the cancellation occurs.
The coupling constant in this subsection is the \emph{bare 
coupling constant} (denoted by $g_B$).  
First, the scattering amplitude $S_V$ without extra emission of 
soft gauge bosons is expanded in terms of \emph{bare 
coupling constant} as
\begin{equation}
    S_V = S_2(g_B)^2 + S_4(g_B)^4 + \dotsb. \label{eq3.3.1}
\end{equation}
The scattering amplitude $S_B$ (in the thesis, $S_B$ or $S_R$ is used for the amplitude; B means Bremmsstrahlung and R means Real emission) with extra emission of 
soft gauge bosons is similarly expanded in terms of bare 
coupling constant as
\begin{equation}
    S_B = S_3(g_B)^3 + \dotsb. \label{eq3.3.2}
\end{equation}
The scattering cross section formed from $S_V$ is given by
\begin{equation}
    d\sigma_V = |S_2|^2(g_B)^4 + (S_2^\ast S_4 + \text{c.c.})(g_B)^6 + \dotsb,
    \label{eq3.3.3}
\end{equation}
and the scattering cross section formed from $S_B$ is given by
\begin{equation}
    d\sigma_B = |S_3|^2(g_B)^6 + \dotsb. \label{eq3.3.4}
\end{equation}
The sum of them
\begin{equation}
    d\sigma_V + d\sigma_B \label{eq3.3.5}
\end{equation}
is a physically meaningful cross section (see \S 2.3).

Since $|S_2|^2$ includes no infrared divergence, 
the $(g_B)^6$ term 
\begin{equation}
    (S_2^\ast S_4 + \text{c.c.})_{\text{IR}} + |S_3|^2_{\text{IR}}
\end{equation}
is examined below.
Let us consider the following example.
\begin{equation}
\begin{minipage}{14cm}
\centering
\includegraphics[width=12cm, clip]{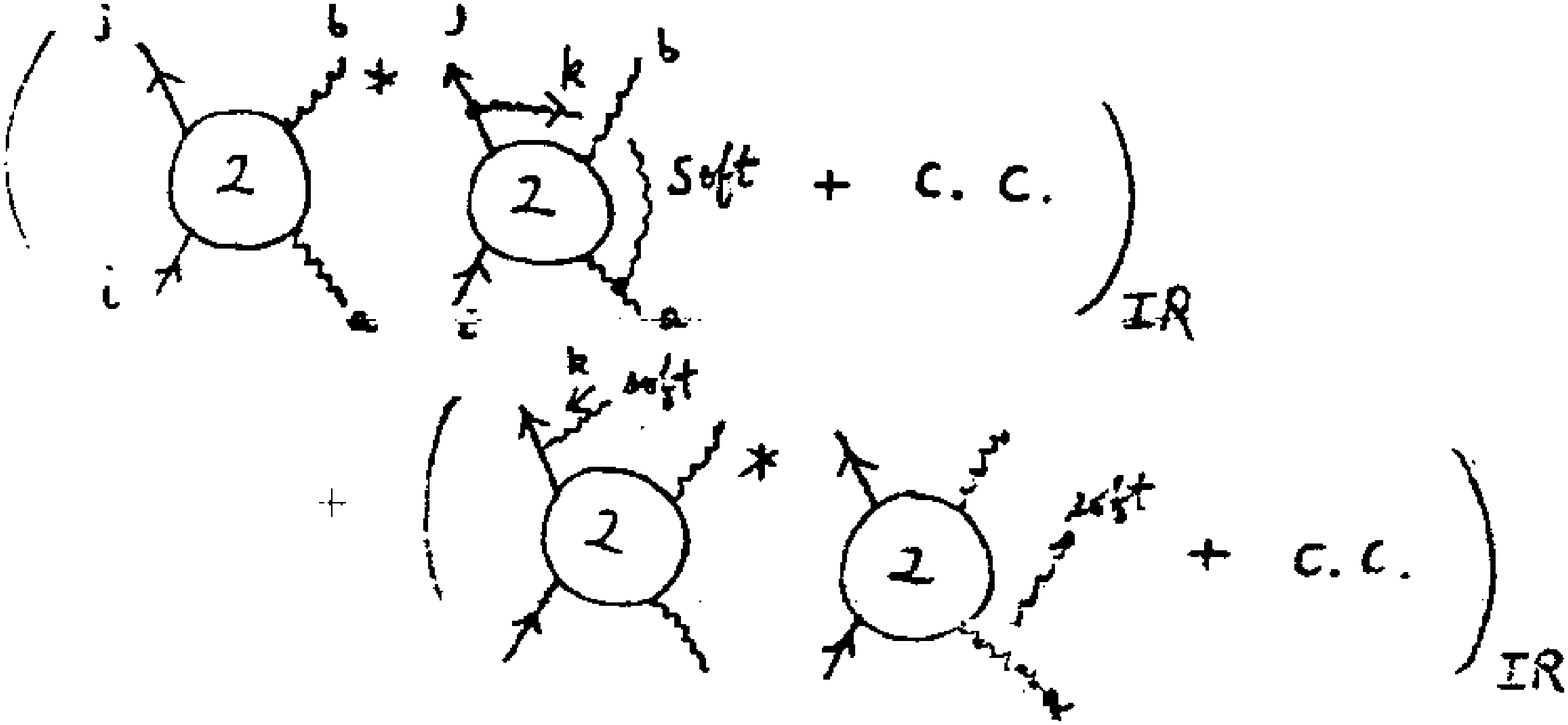}
\end{minipage}
\label{eq3.3.7}
\end{equation}
Here the Feynman gauge is used.

The factorization rule (\ref{eq3.2.1}) and (\ref{eq3.2.2}) can be used 
to rewrite the first term of (\ref{eq3.3.7}) as follows.
\begin{equation}
\begin{minipage}{14cm}
\centering
\includegraphics[width=12cm, clip]{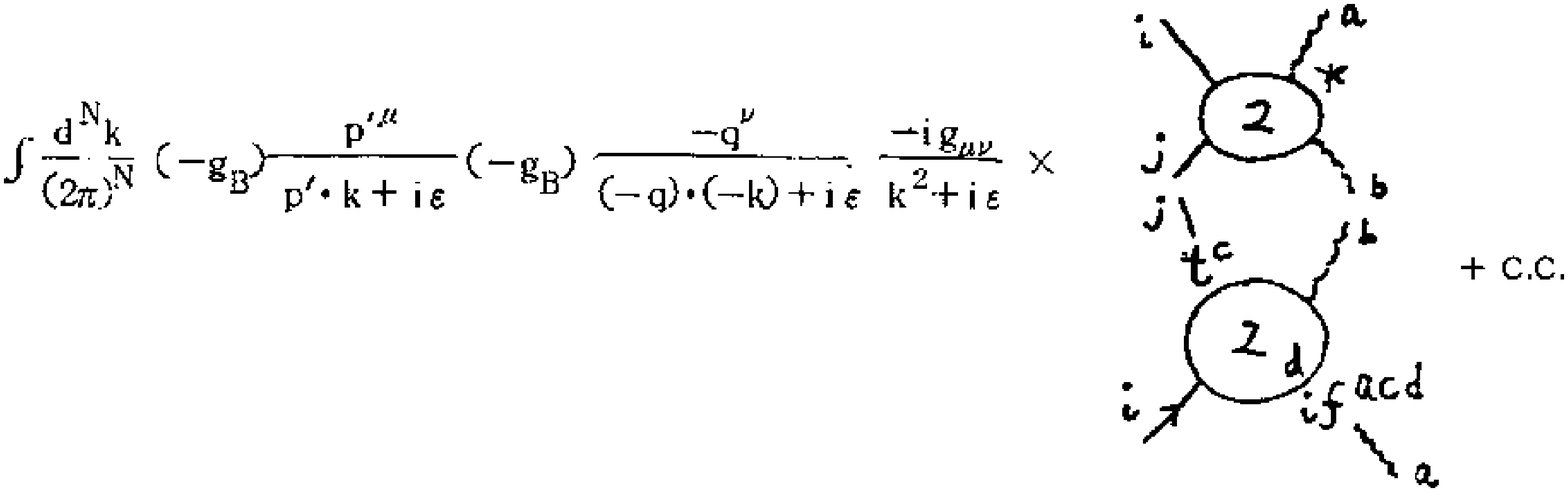}
\end{minipage}
\label{eq3.3.8}
\end{equation}
On the other hand, using the same rule, 
the second term of (\ref{eq3.3.7}) yields the following.
\begin{equation}
\begin{minipage}{14cm}
\centering
\includegraphics[width=12cm, clip]{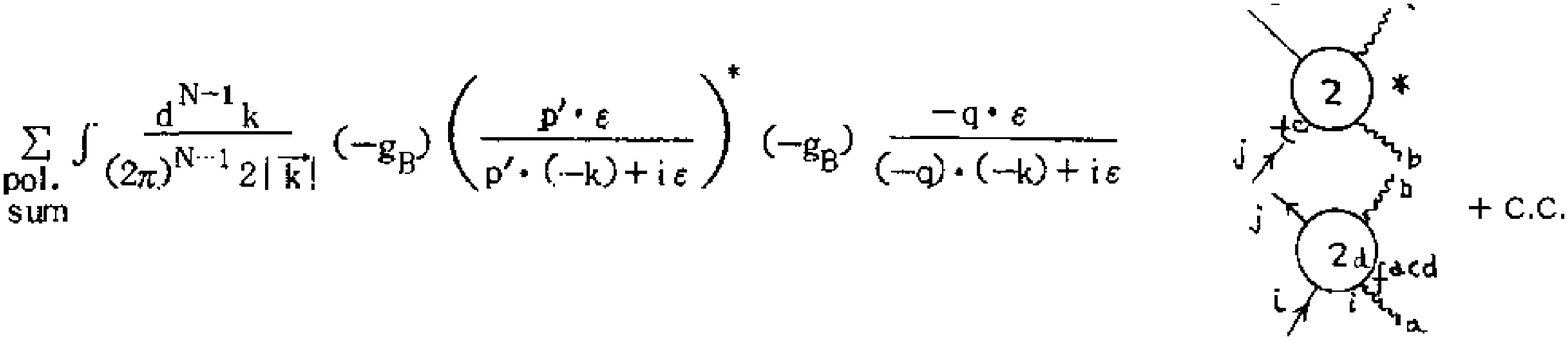}
\end{minipage}
\label{eq3.3.9}
\end{equation}
Here the $N$-dimensional method is used for the regularization of 
infrared divergences (the other regularization methods can 
also be applied in just the same way).  
The sum of polarization vectors can be based on 
Eq.\ (\ref{eq2.3.6}), similarly to QED.  
It is shown later that the $k_\mu \bar{k}_\nu + \bar{k}_\mu k_\nu$ term 
does not contribute [(\ref{eq3.3.12})], which allows the replacement
$\displaystyle\sum_{\text{pol}}\epsilon_\mu \epsilon^\ast_\nu\to -g_{\mu\nu}$. 
Performing $k^0$ integration in (\ref{eq3.3.8}) in the same way as QED 
yields 
\begin{gather}
\int\frac{d^N k}{(2\pi)^N}\,(-g_B)\frac{p'^\mu}{p'\cdot k + i\epsilon}
(-g_B)\frac{-q^\nu}{(-q)\cdot (-k) + i\epsilon}
\frac{-ig_{\mu\nu}}{k^2 + i\epsilon} \notag \\
= \int\frac{d^{N-1} k}{(2\pi)^{N-1}(-2)|\bm{k}|}\,
(-g_B)\frac{p'^\mu}{p'\cdot k + i\epsilon}
(-g_B)\frac{-q_\mu}{(-q)\cdot (-k) + i\epsilon}, \label{eq3.3.10}
\end{gather}
which has the sign opposite to the $k$ integral of (\ref{eq3.3.9}).
Therefore cancellation of (\ref{eq3.3.8}) and (\ref{eq3.3.9}) requires 
equality of the following color factors.
\begin{equation}
\begin{minipage}{14cm}
\centering
\includegraphics[width=12cm, clip]{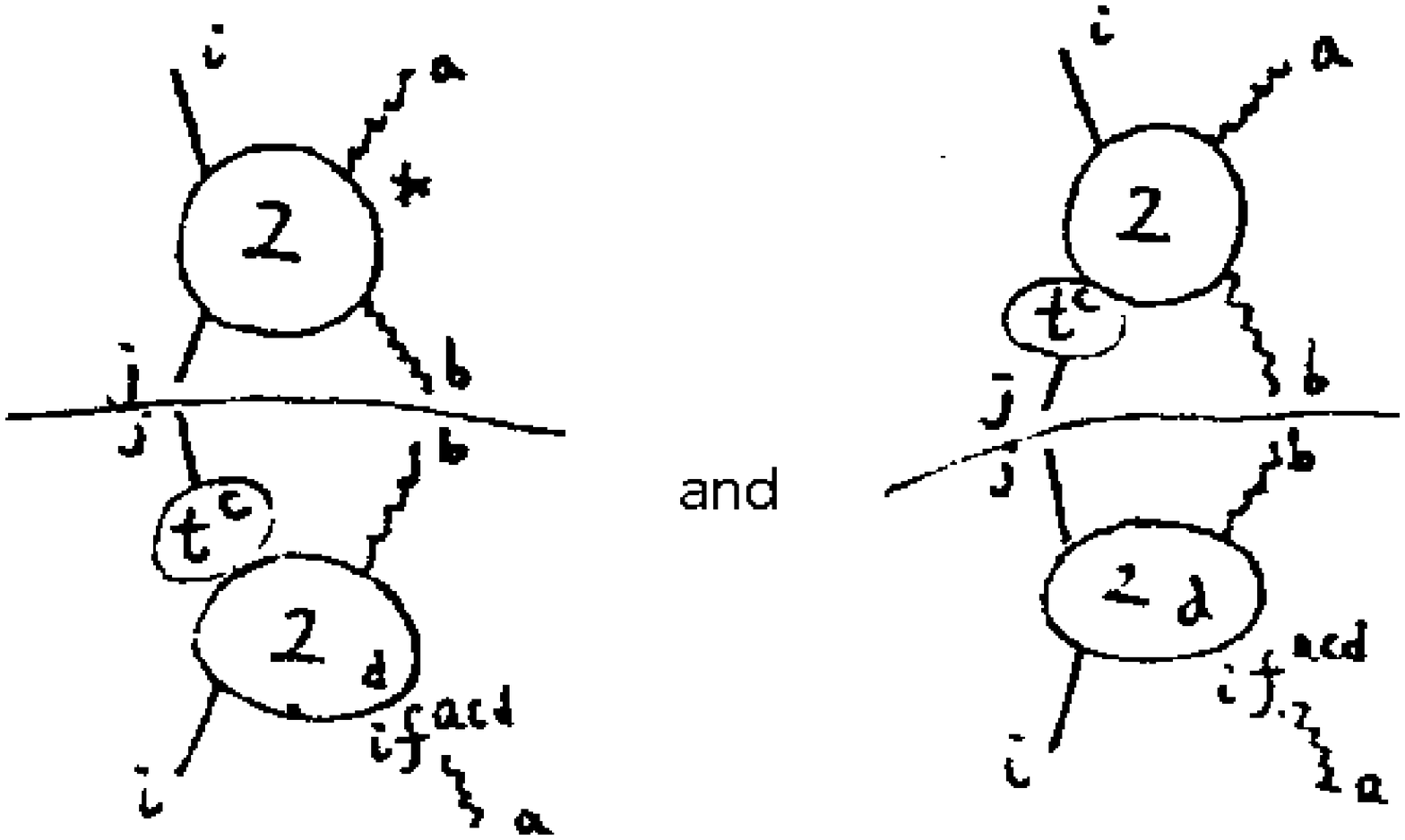}
\end{minipage}
\label{eq3.3.11}
\end{equation}
Equality of these two factors is achieved by \emph{summation over the 
color index $j$ of the fermion in the final state}.  
Similar cancellation of infrared divergences holds for graphs 5--10 
in Figure 3.1.7 where different external lines are connected by 
soft gauge bosons.  This cancellation requires summation over 
color indices $i$, $j$, $a$, $b$ in the initial and final states.

That is, cancellation of infrared divergences occurs for the 
cross section obtained by \emph{averaging 
the colors in the initial state and 
summing over the colors in the final state}, under the assumption that 
the color index is not subjected to our observation.
This is characteristic to QCD.

Let us briefly comment on that the sum of polarization vectors can be replaced by $-g_{\mu\nu}$.  This is manifect from the Ward-Takahashi identity which implies that the sum of external lines of all the possible gauge bosons' emission, being multiplied by $k_{\mu}$, vanishes.

That is, 

\begin{equation}
\begin{minipage}{14cm}
\centering
\includegraphics[width=12cm, clip]{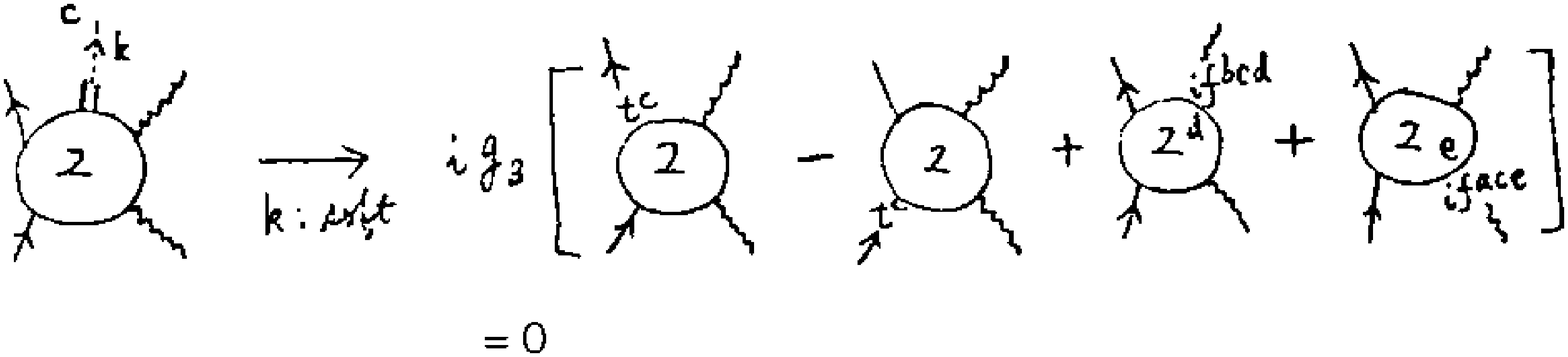}
\end{minipage}
\label{eq3.3.12}
\end{equation}

This is a Ward-(Takahashi) identity in case of soft momentum emission. [Here, factorization rules (\ref{eq3.2.1}) and (\ref{eq3.2.2}) have been used.]  It is easy to check explicitly the vanishing of the right hand side of (\ref{eq3.3.12}) as follows:

\begin{align}
&\begin{minipage}{14cm}
\centering
\includegraphics[width=12cm, clip]{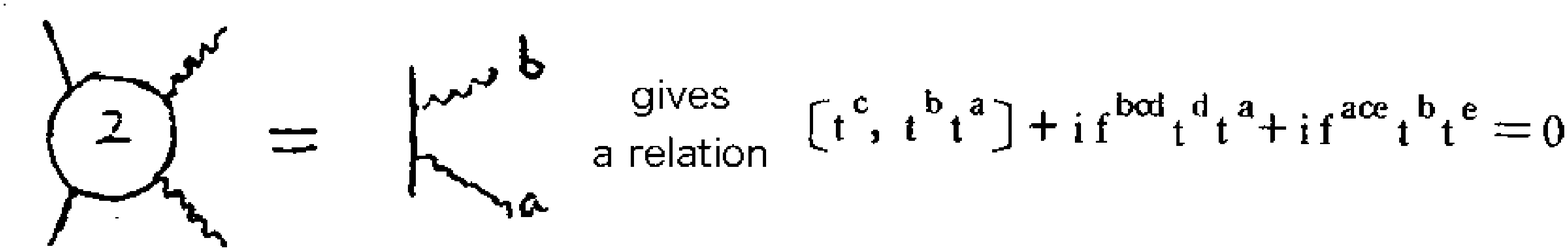}
\end{minipage}
\label{eq3.3.13}  \\
&\hspace{1.3cm} \begin{minipage}{12cm}
\centering
\includegraphics[width=8cm, clip]{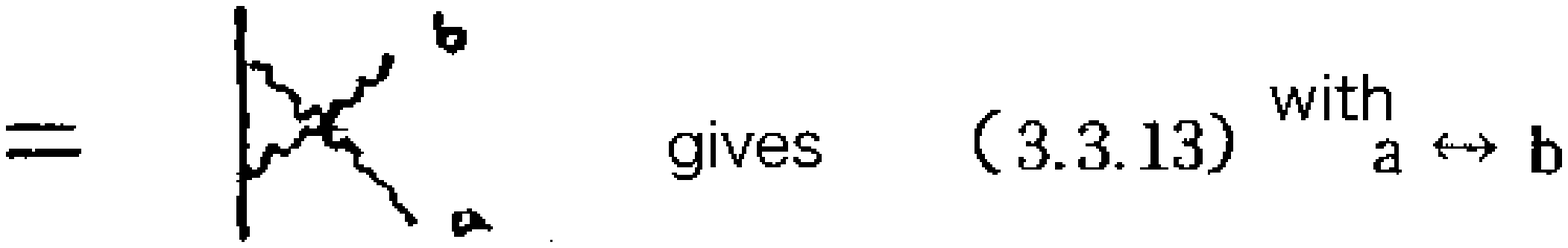}
\end{minipage}
\label{eq3.3.14}  \\
&\hspace{1.3cm} \begin{minipage}{12cm}
\centering
\includegraphics[width=8cm, clip]{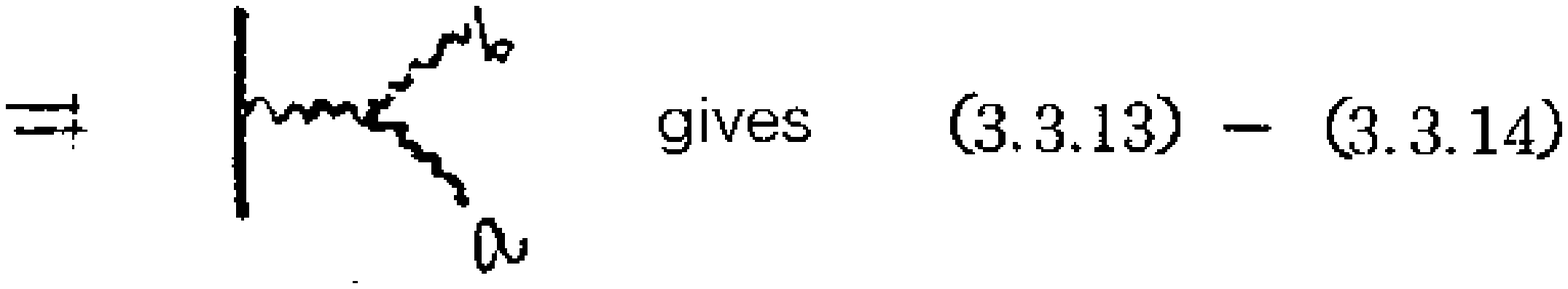}
\end{minipage}
\label{eq3.3.15}  
\end{align}
which show the identities for the color factors.

About the cancellation of infrared divergences for the graphs with corrections in external lines [Figures 1--4 in Figure 3.1.7] and that of Coulomb divergences existing in Figures 7 and 8 in Figure 3.1.7, please refer to the Reference paper I.  Here we omit the explanation.  Next, let us discuss what kinds of divergences were cancelled in this section.

First, the integral after using the factorization rules, becomes
\begin{align}
& I= \int \frac{d^N k}{(2\pi)^N} \frac{1}{k^2 + i \varepsilon} \frac{1}{\pm 2 k \cdot p +  i \varepsilon} \frac{1}{ 2 k \cdot p' +  i \varepsilon} \label{eq3.3.16}, \\
& \approx  \int \frac{d^N k}{(2\pi)^N} \frac{1}{k^2 + i \varepsilon} \frac{1}{(k \pm p)^2 -m^2 + i\varepsilon}  \frac{1}{(k + p') -m^{'2}  +  i \varepsilon} \label{eq3.3.17} \\
&    \hspace{2.5cm}   \gamma    \hspace{2.5cm}    \beta        \hspace{3cm} \alpha
\end{align}
(\ref{eq3.3.16}) and (\ref{eq3.3.17}) are equal when $k$ is soft, and $p^2=m^2$ and $p^{'2}=m^{'2}$ hold.

For $\pm$, $+$ shows the cases of $p$ going out and $p'$ coming in and of $p$ coming in and $p'$ going out, while $-$ shows the case of both $p$ and $p'$ are going out or coming in.  From (\ref{eq3.3.17}) let us see how the singularities become stronger, in three cases I) $m=m' \ne 0$, II) $m \ne 0, m'=0$, III) $m=m'=0$.  I) is the usual infrared divergences of QED, and if the divergences become stronger for II) and III), the increased part can be classified as a new divergence induced by the self-coupling of zero mass particles.  In classifying the divergences, it is helpful to compare a number of cases.  That is, by considering various cases obtained by making mass to be zero, external line be on-shell or off-shell, and so on, it becomes manifest that the divergences under consideration appears in what conditions. [For example, the ultraviolet divergences appear irrespective of these conditions such as mass is zero or not, external line is on-shell or not, and so on.  It is because the diverces are induced by the infinitely large loop momenta, which ignore mass and momentum squared.]  Now, modifying (\ref{eq3.3.17}) in terms of Feynman parameters, and integrating over $k$, we have
\begin{equation}
I= i \frac{(-1)^3}{(4\pi)^{N/2}} \Gamma(3-N/2) \int_0^1 \frac{d \alpha d\beta d\gamma \delta(1-\alpha-\beta-\gamma)}{[(\alpha p' \pm \beta p)^2 + i \varepsilon]^{3-N/2}}  \label{eq3.3.18.eps}
\end{equation}
[which is identical to (\ref{eq2.1.11}).]

Here denoting the denominator as $V$, 
\begin{equation}
V \equiv (\alpha p' \pm \beta p)^2= \alpha (\alpha+ \beta) m^{'2} + \beta (\alpha+\beta) m^2 - \alpha \beta (p' \mp p)^2, \label{eq3.3.19}
\end{equation}

Change of variables (\ref{eq2.1.13}) is performed, namely,
$$\alpha=xy, \; \beta=(1-x)y, \; \gamma=1-y, [\text{Jacobian}=y]$$
\begin{equation}
\left.
\begin{array}{ll} 
\text{Case I)}: & V_I=y^2 [m^2 - x(1-x)t ] \\
\text{Case II)}: & V_{II}=y^2(1-x)[m^2-xt] \\
\text{Case III)}: & V_{III}= y^2(1-x) x [-t]  
\end{array}
\right\},  \label{eq3.3.20}
\end{equation} 
where $t \equiv (p' \mp p )^2$

All in the cases I), II), III), divergences arise at $y=0$. 
\begin{equation}
\int_0^1 ydy \frac{1}{[y^2]^{3-N/2}} = \int_0^1 dy \; y^{N-5}= \frac{1}{N-4} ~~  [\text{Re}N >4] \label{eq3.3.21}
\end{equation}
When going to II) and III), further divergences appear near $x=1$ and $x=0$, respectively.  These divergences are characteristic in QCD, and are called ``mass singularities", if necessary to be specified:
\begin{align}
\int_0^1 dx \frac{1}{(1-x)^{3-N/2}}= \frac{1}{N/2-2}[-(1-x)^{N/2-2}]_0^1=\frac{2}{N-4}  \label{eq3.3.21} \\
\int_0^1 dx \frac{1}{(x)^{3-N/2}}= \frac{1}{N/2-2}{x^{N/2-2}}]_0^1=\frac{2}{N-4}.   \label{eq3.3.22}
\end{align}

From this
\begin{equation}
\left. 
\begin{array}{ll}
\text{Case I)}    [I_{I}]_{\text{IR}}= i \frac{-1}{(4\pi)^2} \frac{1}{N-4} \int_0^1 dx \frac{1}{m^2-x(1-x) t} \\
\text{Case II)}     [I_{II}]_{\text{IR}}= i \frac{-1}{(4\pi)^2} \frac{1}{N-4} \times \frac{2}{N-4}  \frac{1}{m^2- t} \\
\text{Case III)}     [I_{III}]_{\text{IR}}= i \frac{-1}{(4\pi)^2} \frac{1}{N-4} \times \left\{ \frac{2}{N-4}+\frac{2}{N-4} \right\}   \frac{1}{- t}
\end{array} 
\right\}  \label{eq3.3.23}
\end{equation}
Maximally, the poles of $1/(N-4)^2$ arise.

If the same calculation is performed in the momentum space, the meaning of the divergence can be clearer (\ref{eq3.3.16}).  [In the integral for $-$, the pole structure differs, but the difference between + and $-$ appears in the range of $t$, as is seen from the result (\ref{eq3.3.23}) of the parametric integration.  That is, the integral for $-$ can be obtained by an analytic continuation of the integral for +, into the region of $t$ for $-$. ]  Now, we estimate (\ref{eq3.3.16}) for +: The $k^0$-integration reads
\begin{align}
I= i \int \frac{d^{N-1} k}{4 (2\pi)^{N-2}} \frac{1}{-2|\vec{k}|}\frac{1}{2 [|\vec{k}|p^0 -\vec{k} \cdot \vec{p}]} \frac{1}{2 [|\vec{k}|p^{0'} -\vec{k} \cdot \vec{p}']} \\
=i \frac{-1}{4 (2\pi)^{N-2}}\frac{1}{2\pi} \int d^{N-1} k \frac{1}{2|\vec{k}|^3 [ p^0+ |\vec{p}| \cos \theta]  [ p^{0'}+ |\vec{p}'| \cos \theta']},  ~~~\label{eq3.3.24}
\end{align}
where $\theta$, and $\theta'$ represent angles between $\vec{k}$ and $\vec{p}$, and $\vec{k}$ and $\vec{p}'$, respectively.

Writing the denominator of the integral above by $\tilde{V}$,
\begin{equation}
\tilde{V}=2|\vec{k}|^3 [ p^0+ |\vec{p}| \cos \theta]  [ p^{0'}+ |\vec{p}'| \cos \theta'],   \label{eq3.3.25}
\end{equation}
and classifying three cases I)--III), we have
\begin{equation}
\left.
\begin{array}{ll}
\text{Case I)}: & \tilde{V}_{I}= 2|\vec{k}|^3 [ p^{0'}+ |\vec{p}'| \cos \theta']  [ p^0+ |\vec{p}| \cos \theta] , \\
\text{Case II)}: & \tilde{V}_{II}= 2|\vec{k}|^3  |\vec{p}'| [1+ \cos \theta']  [ p^0+ |\vec{p}| \cos \theta], \\
\text{Case III)}: & \tilde{V}_{III}= 2|\vec{k}|^3  |\vec{p}'| [1+ \cos \theta']  |\vec{p}|  [1+ \cos \theta].
\end{array} \right\}  \label{eq3.3.26}
\end{equation}

Here we note that $|\vec{p}| \le p^0, \;  |\vec{p}'| \le p^{'0}$ hold in general, and the equality works only for the massless case.

Comparing eq.(\ref{eq3.3.26}) with eq.(\ref{eq3.3.20}), we have understood the correspondence relations; 
\begin{equation}
\left.
\begin{array}{ll}
|\vec{k}|=\vec{0}~[\text{soft}] \leftrightarrow y=0 ~ [\alpha=\beta=0] \\
\cos \theta'=-1~[-\vec{k} \; \text{and} \; \vec{p}'\;  \text{are parallel}] \leftrightarrow x=1~[\beta=0], \\
\cos \theta=-1~[-\vec{k} \; \text{and} \; \vec{p} \;  \text{are parallel}] \leftrightarrow x=0~[\alpha=0]
\end{array} \right\}
\end{equation}
This gives the translation regulations between momentum space and parameter space.  Here we will use two cutoffs; introducing sufficiently small $\lambda'$ and $\eta$, we extract the usual infrared divergences common to I)--III) by
\begin{equation}
\int_{\lambda'} d|\vec{k}| \frac{1}{ |\vec{k}|^3}= \ln \frac{1}{\lambda'} + \text{finite quantities}  \label{eq3.3.28}
\end{equation}
and extract the new infrared divergences characteristic to QCD by 
\begin{align}
\int_{-1+\eta}^a d(\cos \theta') \frac{1}{1+ \cos \theta'}= \ln \frac{1}{\eta} + \text{finite quantities} \label{eq3.3.29} \\
\int_{-1+\eta}^b d(\cos \theta) \frac{1}{1+ \cos \theta}= \ln \frac{1}{\eta} + \text{finite quantities} \label{eq3.3.30} 
\end{align}
[ c.f. (\ref{eq3.3.21}) and (\ref{eq3.3.22})]

Accordingly, we have
\begin{equation}
\left. 
\begin{array}{ll}
\text{Case I)} \; \;     [I_{I}]_{\text{IR}}=& i \frac{-1}{(4\pi)^2} \ln \frac{1}{\lambda} \int d\Omega \frac{1}{[ p^{0'}+ |\vec{p}'| \cos \theta']  [ p^0+ |\vec{p}| \cos \theta]} \\
\text{Case II)}  \;    [I_{II}]_{\text{IR}}=& i \frac{-1}{(4\pi)^2} \ln \frac{1}{\lambda} \ln \frac{1}{\eta} \frac{1}{2(p \cdot p')},  \\
\text{Case III)}     [I_{III}]_{\text{IR}}=& i \frac{-1}{(4\pi)^2}    \ln \frac{1}{\lambda} \left\{ \ln \frac{1}{\eta} + \ln \frac{1}{\eta} \right\} \frac{1}{2(p \cdot p')}
\end{array} 
\right\}  \label{eq3.3.31}
\end{equation}
which correspond completely to (\ref{eq3.3.23}).

From the above discussions, the divergences which are shown to be cancelled in this section, are as follows:

$\circ$ Fermion-fermion scattering, $\ln \frac{1}{\lambda}$ (no other divergences appear in this case), 

$\circ$ Fermion-gauge-boson scattering, $\ln \frac{1}{\lambda}$ and $\ln \frac{1}{\lambda} \ln \frac{1}{\eta}$ (this is the highest divergences).

Here $\ln \frac{1}{\eta}$ was not discussed; [its example can be found in the radiative corrections of the external gauge bosons; see (Reference paper I)] 

Therefore, the discussion is complete for the cancellation of the highest divergences. 

%%%%%%%%%%%%%%%%%%%%
\subsection{New infrared divergences for the on-mass-shell renormalization}
   
In QED, cancellation of infrared divergences in terms of bare coupling constant implies the cancellation in terms of renormalized coupling constant [defined on the mass shell].  It is because there appear no further infrared divergences in the coefficients of the bare coupling constant expanded in renormalized coupling constant.  In other words, $Z_3$ is infrared finite. [Refer to the explicit demonstration at one loop in (\ref{eq2.2.20}) and (\ref{eq2.2.21}).]

In QCD, let us expand the bare coupling $g_B$ in terms of the renormalized coupling $g_R$.  Then, we have
\begin{equation}
g_B= Z_1Z_2^{-1}Z_3^{-1/2} g_R= g_R + A g_R^3 + \cdots,  \label{eq3.4.1}
\end{equation}
where $Z_1, Z_2, Z_3$ are renormalization constants defined on the mass shell.  They are given via the unrenormalized vertex and propatator functions as follows:
\begin{align}
&\begin{minipage}{10cm}
\centering
\includegraphics[width=8cm, clip]{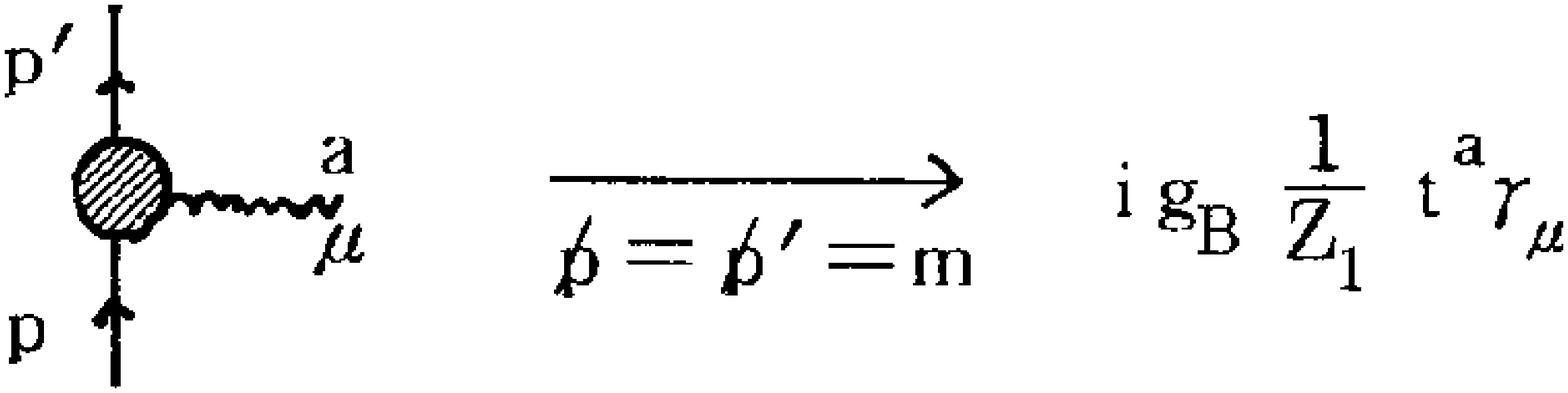}
\end{minipage}
\label{eq3.4.2}  \\
&\begin{minipage}{10cm}
\centering
\includegraphics[width=8cm, clip]{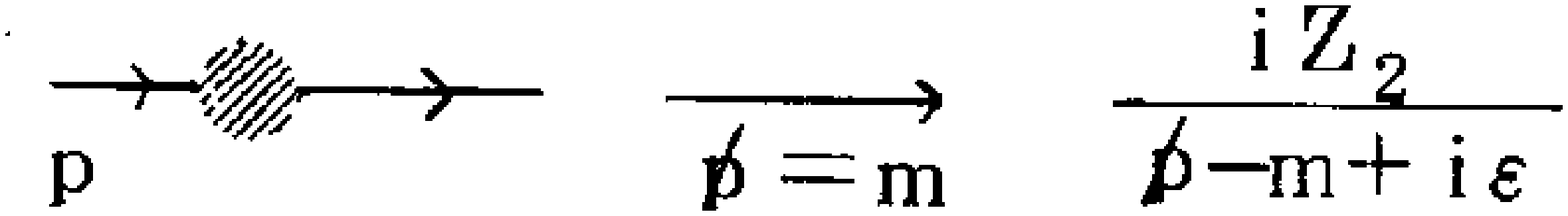}
\end{minipage}
\label{eq3.4.3}  \\
&\begin{minipage}{10cm}
\centering
\includegraphics[width=8cm, clip]{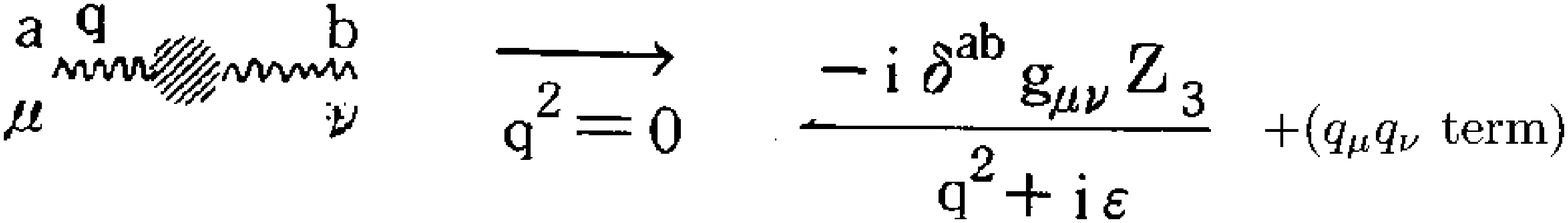}
\end{minipage}
\label{eq3.4.4}  
\end{align}
In the actual estimation shows, [referring to Appendix B of Reference paper I, in which how to extract the infrared divergences in the renormalization constants is quite explicitly written] that the infrared divergences at one loop are given by
\begin{align}
&[Z_1^{(1)}]_{\text{IR}}= \left(\frac{g_R}{4\pi}\right)^2 \frac{1}{N-4} \left[C_2(R)-\frac{1}{2} C_2(G) \right][4+2(1-\alpha)] \\
& + \left(\frac{g_R}{4\pi}\right)^2 \frac{1}{N-4} C_2(G) \left[\frac{3}{2}2(1-\alpha)\right]  \label{eq3.4.5} \\
&[Z_2^{(1)}]_{\text{IR}}= \left(\frac{g_R}{4\pi}\right)^2 \frac{1}{N-4} C_2(R)[4+2(1-\alpha)]   \label{eq3.4.6} \\
&[Z_3^{(1)}]_{\text{IR}}= \left(\frac{g_R}{4\pi}\right)^2 \frac{1}{N-4}  C_2(G) \left[\frac{10}{3}+(1-\alpha)\right],    \label{eq3.4.7}
\end{align}
where the symbols characteristic to QCD, $C_2(R)$ and $C_2(G)$ are defined by
\begin{align}
&\sum_a t^a t^a =C_2(R) 1 ~~(\text{1 is a unit matrix}),  \label{eq3.4.8} \\
&\sum_{c, d} f^{acd} f^{bcd} =C_2(G) \delta^{ab}, \label{eq3.4.9}
\end{align}
which give $C_2(R)=\frac{N^2-1}{2N}, \; C_2(G)=N$ in case of $SU(N)$ group.  [The symbols correspond to angular momentum squared in case of $SU(2)$; if taking the sum over all $a$ as $\sum_a t^a t^a$, it commutes with all $t^a$s, and is proportional to a unit matrix; its coefficient takes various value depending on the representation.]

The first and the second terms of (\ref{eq3.4.5}) correspond, respectively to the following figures:

\begin{figure}[h]
\centering
\includegraphics[width=100mm, clip]{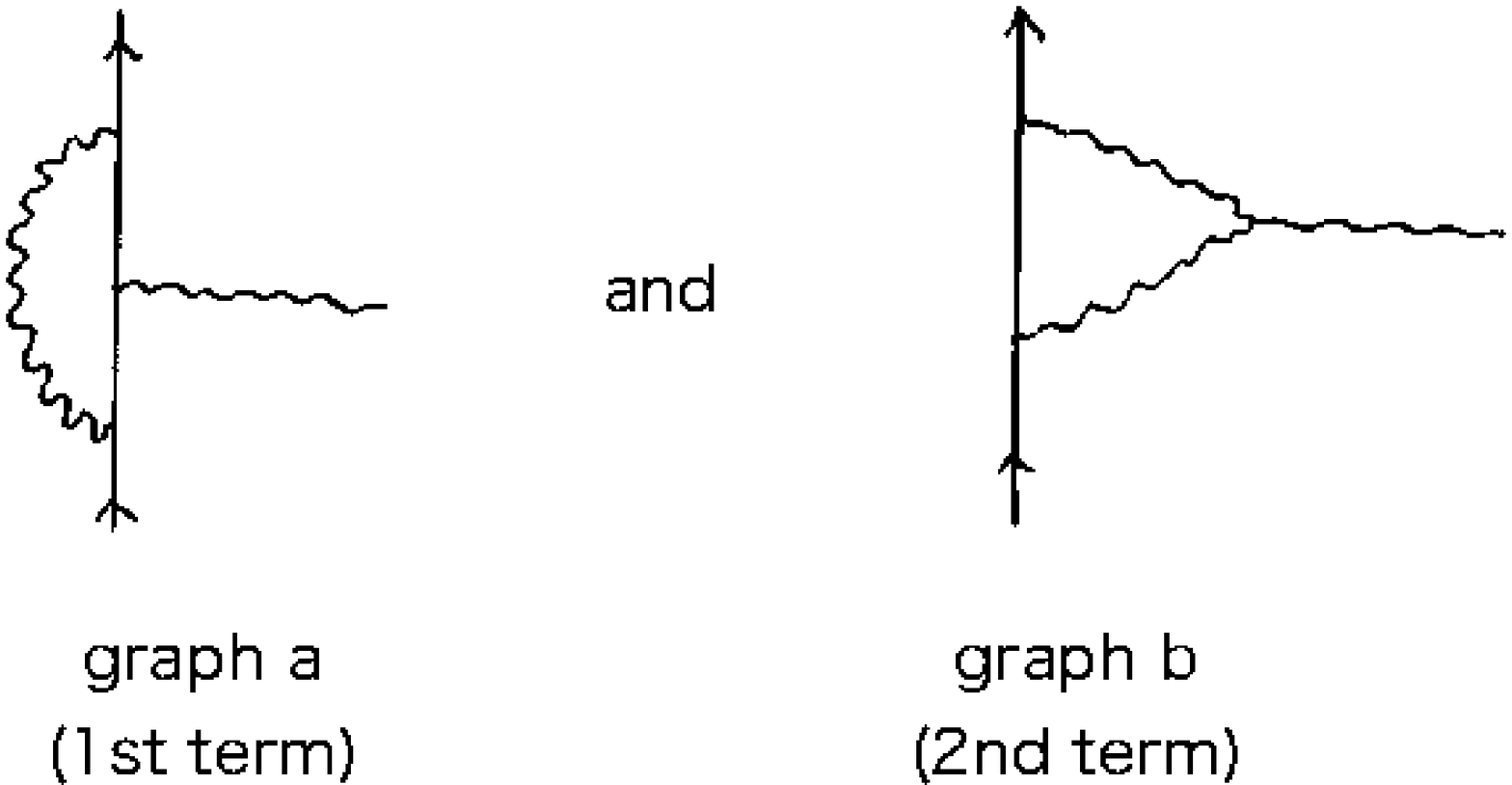}
\caption{}
\end{figure}
\[
\begin{minipage}{12cm}
\centering
\includegraphics[width=3cm, clip]{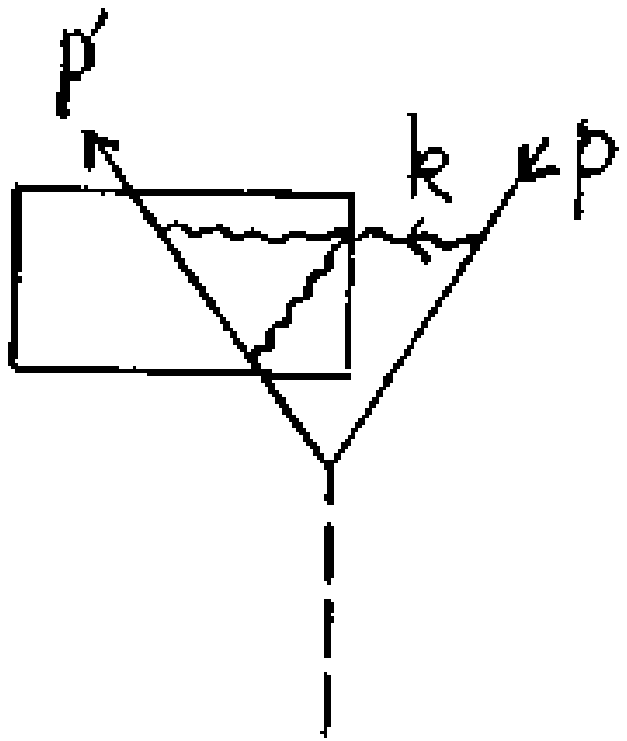}
\text{(Figure 3.4.2)}
\end{minipage}
\]
Especially, it is characteristic that the second term in (Figure b) is zero for Feynman gauge $(\alpha=1)$. 
[Writing as a reference, by jumping to a little higher level, where it is pointed out (by Kinoshita-Ukawa \cite{13}) that the two-loop graphs estimated by \underline{Feynman gauge}, give $\frac{1}{N-4}$ singularities for the non-leading infrared divergences; a simple reason of this is that the part encircled with a square in Figure 3.4.2 or (Figure b) in Figure 3.4.1, becomes zero in the limit of $k_{\mu} \to 0$; this is what we have mensioned above as a characteristic feature in QCD.]

By using (\ref{eq3.4.5})--(\ref{eq3.4.7}), the coefficient $A$, connecting the bare coupling with the renormalized coupling defined on the mass shell, can be estimated as
\begin{align}
&[A]_{\text{IR}} = [Z_1^{(1)}]_{\text{IR}} -  [Z_2^{(1)}]_{\text{IR}}-\frac{1}{2} [Z_3^{(1)}]_{\text{IR}} \\
&=\frac{1}{(4\pi)^2}\frac{1}{N-4}C_2(G)\left(-\frac{11}{3} \right) ~~~[ \text{where~Re} N > 4].  \label{eq3.4.10}
\end{align}
This does not depend on the gauge parameter.  Furthermore, it has the same form as the ultraviolet divergence in QCD \underline{without fermions}:
\begin{equation}
[A^{\text{pure Y-M}}]_{\text{UV}}=\frac{1}{(4\pi)^2}\frac{1}{4-N}C_2(G)\left(-\frac{11}{3} \right) ~~~[\text{where~ Re} N < 4].  \label{3.4.11}
\end{equation}
In other words, if the infrared divergence is regularized by introducing a mass $\lambda$ to the gauge boson, and the ultraviolet divergence is regularized by a momentum cutoff $\Lambda$, we have
\begin{align}
&[A]_\text{{IR}}=\frac{1}{(4\pi)^2} \ln \frac{1}{\lambda} C_2(G)\left(-\frac{11}{3} \right)  \label{eq3.4.12} \\
&[A^{\text{pure Y-M}}]_{\text{UV}}=\frac{1}{(4\pi)^2} \ln \Lambda C_2(G)\left(-\frac{11}{3} \right). \label{3.4.13}
\end{align} 

Therefore, this infrared divergence can be said to be controlled by the $\beta_{\text{pure Y-M}}$ function of the pure Y-M theory.  Namely, 
\begin{align}
\beta(g) \equiv \lambda \frac{\partial}{\partial \lambda} g_R =-\Lambda \frac{\partial}{\partial \Lambda} (g_R)^{\text{pure Y-M}} \equiv \beta_{\text{pure Y-M}}(g) \\
=\frac{1}{(4\pi)^2}  C_2(G)\left(-\frac{11}{3} \right) (g_R)^3 + \cdots.    \label{eq.3.4.14}
\end{align}

Now, due to this new divergece, the cancellation of infrared divergences is broken in the on-mass-shell \underline{renormalization}. 

As is well-known, the \underline{renormalization} is to store all the divergences into the redefinition of mass and charge.  
In the study of infrared divergences, the mass is renormalized on the mass shell [$\delta m$ is IR free], so that the remaining problem is the charge renormalization.
If the scattering amplitudes, $S_V$ and $S_B$ expanded in the bare coupling constant $g_B$ [see (\ref{eq3.3.1}) and (\ref{eq3.3.2})], is re-expanded in the renormalized coupling constant $g_R$, then we have
\begin{align}
S_V=& S_2 (g_B)^2 + S_4  (g_B)^4 + \cdots \notag \\
=& S_2 (g_R)^2 + [S_4 + S_2 \times 2A ]  (g_R)^4 + \cdots,~\label{eq3.4.15}  \\
S_B=& S_3 (g_B)^3 + \cdots = S_3 (g_R)^3 + \cdots,   \label{eq3.4.16}
\end{align}
where (\ref{eq3.4.1}) has been used.  In this way, the coefficients expanded in $g_R$ have no ultraviolet divergences, that implies the reormalizability. However, for the infrared divergences the cancellation in Section 3.3 implies
\begin{equation}
[S_2^* S_4 + c.c. ]_{\text{IR}} + |S_3|^2_{\text{IR}} =0,   \label{eq3.4.17}
\end{equation}
and hence the infrared divergence of the renormalized cross section becomes
\begin{equation}
[d\sigma]_{\text{IR}} = [d\sigma_V + d\sigma_B ]_{\text{IR}} = 4 \text{Re} [A]_{\text{IR}} d\sigma_0   \label{eq3.4.18}
\end{equation}
That is, the cancellation of infrared divergences violates via $[A]_{\text{IR}}$.  This result is derived in the perturbation theory at one-loop, when the on-mass-shell renormalization is adopted, that is, considering $g_R$ [the charge defined on the mass-shell] to be finite.  On the other hand, if the off-mass-shell renormalization is performed, all the Z-factors are well-defined and infrared finite, so that the cancellation of infrared divergences holds for $g(\mu)$ [charge defined off the mass-shell \cite{37}, as seen from the relation (without infrared divergences) between $g_B$ and $g(\mu)$.  In this latter case, $[A]_{\text{IR}}$ calculated in this section represents the infrared divergence which the on-mass-shell charge has.  That is, 
\begin{equation}
[g_R]_{\text{IR}} = - [A]_{\text{IR}}(g_B)^3 + \cdots 
=- [A]_{\text{IR}}[g(\mu)]^3 + \cdots,   \label{eq3.4.19}
\end{equation}
so that $g_R$ has infrared divergence, if $g(\mu)$ is considered to be finite. [Here, $[A]_{\text{IR}}$ is defined by (\ref{eq3.4.10}).]

---------------------------------------------------------------------

As a reference, we comment on the relation of infrared divergeces to quark confinement. Let us draw (\ref{eq3.4.1})-(\ref{eq3.4.4}) as figures:    
\begin{equation}
\begin{minipage}{14cm}
\centering
\includegraphics[width=12cm, clip]{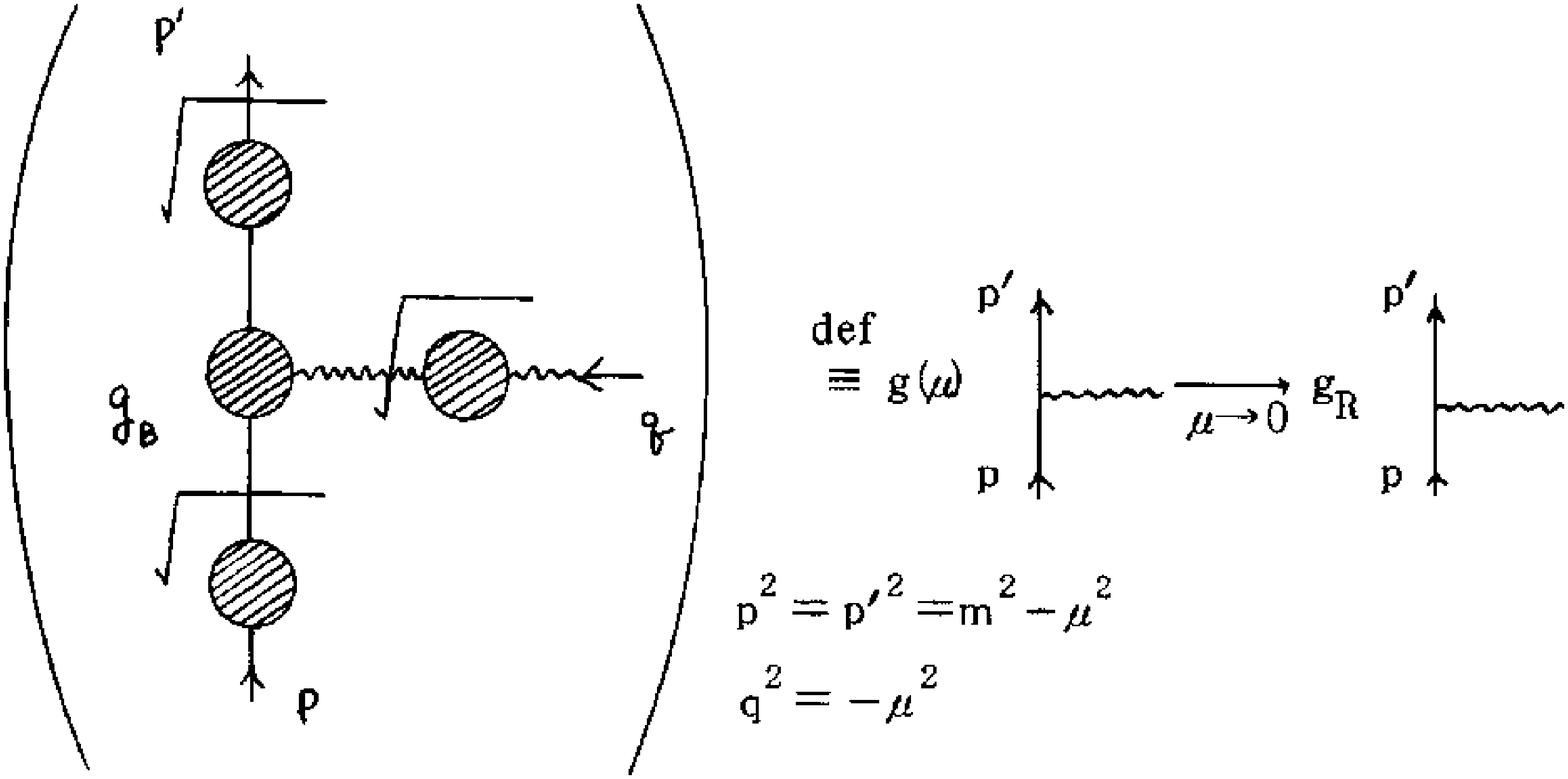}
\end{minipage}
\label{eq3.4.20}  
\end{equation}
is the definition of $g_R$.  In the above calculation $N$ dimensional method was used, but if $\mu$ [deviation from the mass shell] is used as a infrared regularization, the pole $\frac{1}{N-4}$ is replaced by $\ln \frac{1}{\mu}$.\cite{38} If these logarithmic infrared divergences are summed up properly to give $g(\mu) \propto \frac{1}{\mu}$ [on mass shell charge has a linear infrared divergence], then it can give an evidence on the linearly rising potential between quarks [inference by Miyazawa and Cornwall \cite{14}].  That is, 
\begin{equation}
\begin{minipage}{14cm}
\centering
\includegraphics[width=12cm, clip]{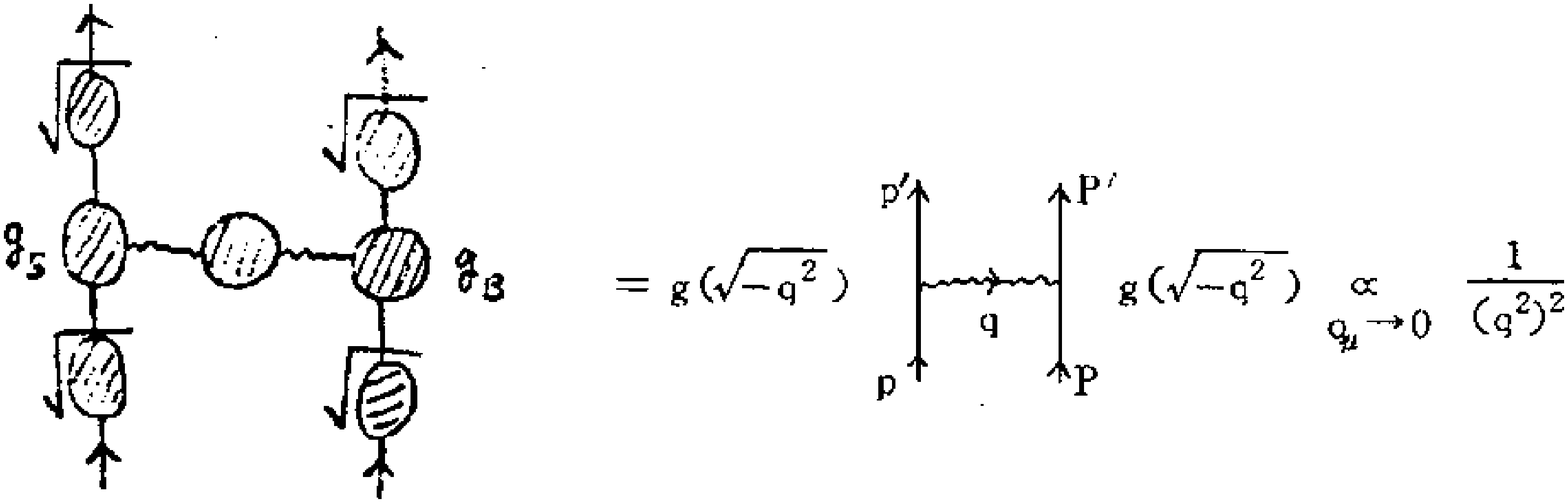}
\end{minipage}
\label{eq3.4.21} 
\end{equation}
where the deviations from the on-shell are taken to be (momentum transfer$)^2$,  
\[
\left.
\begin{array}{ll}
p^2=p^{'2}=P^2=P^{'2}=m^2-\mu^2 \\
q^2=-\mu^2 
\end{array}  \right\} 
\]
Accordingly, this inferrence may give the behavior of $1/(q^2)^2$ to the gluon propagator.  So far this can not be proved rigorously, but it gives a hint to the quark confinement.

%%%%%%%%%%%%%%%%%%%%%%%%%%%%%
% Section 4
%%%%%%%%%%%%%%%%
%%%%%%%%%%%%%%%%%%%%%%%%%%%%%%%%%%%

\section{Infrared divergence and low energy theorem in QCD (quest for the properties in all orders)}

In the previous section, a number of typical properties of infrared divergences in QCD are obtained, by the examinination at one loop level:

1) Factorization rule similar to QED [(\ref{eq3.2.2}) and (\ref{eq3.2.2})] exists. [Section 3.2] 

2) Cancellation of infrared divergences occurs in the expansion of the bare coupling constant [or of the coupling constant defined off the mass shell]. 

3) The cancellation of infrared (IR) divergences is violated in the on mass shell renormalization. The amount of the violation is equal to the UV divergence of on the mass shell charge in the off mass shell renormalization; which is controlled by the $\beta(g)$ function of the pure Yang-Mills theory without fermions.  [Section 3.4]

Now, in this section, we intend to discuss a number of properties which can be generalized to all orders in QCD.  This is based on the (Reference paper II) of the author performed in collaboration with Norio Nakagawa and Hiroaki Yamamoto \cite{15}, and on the development afterwards \cite{16}. 

\subsection{Axial gauge QCD}
First examine QCD in the axial gauge condition \cite{22}--\cite{24}, since it is convenient in the follwoing discussions. The Lagrange density is given by
\begin{equation}
\mathcal{L}= \bar{\psi} ( i \Slash{\nabla} -m ) \psi - \frac{1}{4} F^a_{\mu\nu} F^{a, \mu\nu} - \frac{1}{2\alpha}(n_{\mu} A^{a, \mu})^2 + g \bar{\psi} \gamma_{\mu} t^a \psi A^{a, \mu}.  \label{eq4.1.1}
\end{equation}
Here, the vector $n_{\mu}$ which fixes the gauge, is chosen to be time-like ($n^2 >0$), and $\alpha$ is a gauge parameter.

Then, QCD represented by (\ref{eq4.1.1}) has a notable feature of not having \underline{Fadeev-Popov ghosts}.  First we consider a simple case of $\alpha \to 0$.  The Lagrange density for the ghost fields is, following the prescription by Fadeev and Popov \cite{39},
\begin{align}
\mathcal{L}_g &\propto c_a^{\dagger} \frac{\delta(n_{\mu}A^{a, \mu})}{\delta \theta^b} c_c    \propto c_a^{\dagger} n_{\mu} [ \partial^{\mu} \delta^{ac} + g f^{abc} A_{\mu}^b ] c_c,  \label{eq4.1.2}
\end{align}
where $\theta^c(x)$ is an infinitesimal gauge transformation parameter, and $c_a$ is the Fadeev-Popov ghost fields.  If we can show the following gauge condition holds for $\alpha \to 0$, 
\begin{equation}
	n_{\mu} A^{b, \mu}=0, \label{eq4.1.3}
\end{equation}
the ghost fields do not couple to gauge fields, and are excluded from the Lagrange density.

The path integral expression of the generating functional for all the Green functions (including connected and disconnected digrams), $W[J_{\mu}, \eta, \bar{\eta}]$ (with $J_{\mu}, \eta, \bar{\eta}$ external souces of gauge field and fermions), is given by
\begin{equation}
W[J_{\mu}, \eta, \bar{\eta}] = \int [dA_{\mu}] [d\psi] [d\bar{\psi}] e^{i \int d^4 x \{ \mathcal{L}(x) + \mathcal{L}_g(x) + J^a_{\mu}A^{a, \mu} + \bar{\eta} \psi + \bar{\psi} \eta \} }.  \label{eq4.1.4}
\end{equation}
Here we take a limit of $\alpha \to 0$, then the $\alpha$ depending term becomes
\begin{equation}
e^{i \int d^4 x \frac{-1}{2\alpha} (n_{\mu} A^{a, \mu})^2} \underset{\alpha \to 0}{\propto} \prod_{x} \prod_a \delta(n_{\mu} A^{a, \mu}(x)),   \label{eq4.1.5}
\end{equation}
giving (\ref{eq4.1.4}).

To show the ghost fields do not exist in case of $\alpha \ne 0$, by introducing ghost fields in the loop diagrams following Fadeev and Popov, we show that the loop integral vanishes for the ghosts.  First, the relevant Feynman rules for ghosts, constructed from (\ref{eq4.1.3}), are
\begin{align}
& \begin{minipage}{14cm}
\centering
\includegraphics[width=6cm, clip]{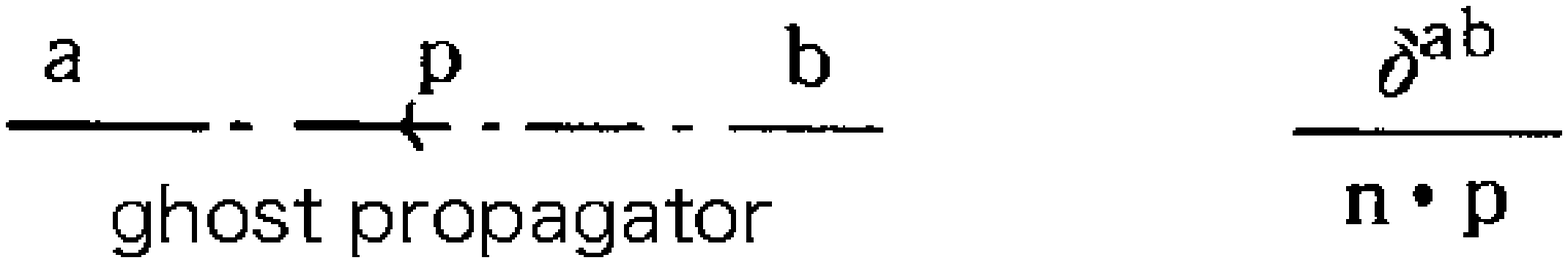}
\end{minipage}
\label{eq4.1.6} \\
&\begin{minipage}{14cm}
\centering
\includegraphics[width=7cm, clip]{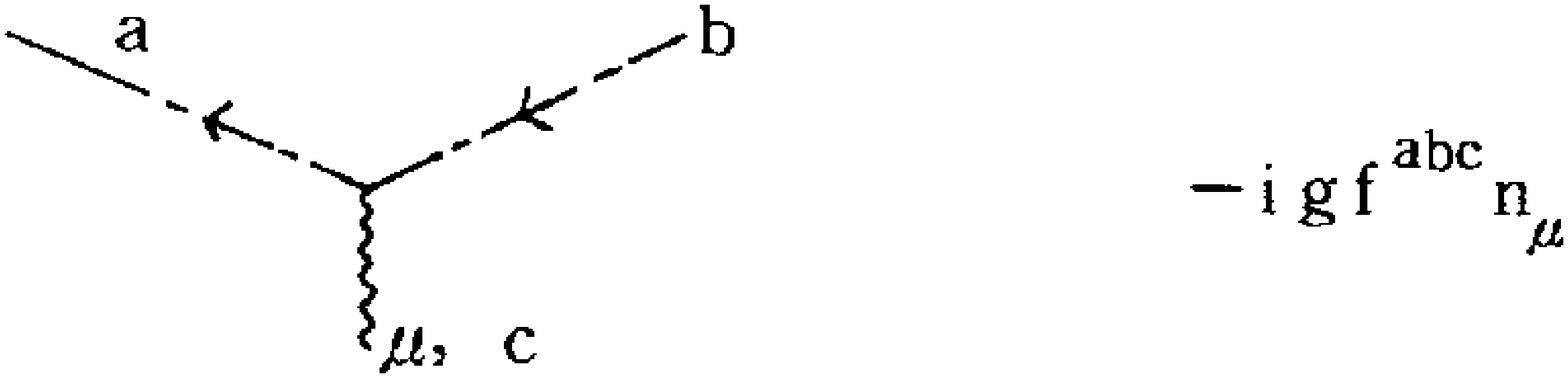}
\end{minipage}
\label{eq4.1.7}
\end{align}
With the rules, let us integrate the ghost loop.  For the ghost loop, to which gauge bosons with momena $p_1, \cdots, p_m$ come in, the numerator does not depend on the momentum.  Therefore, it is enough to consider the denominator, 

%
%\begin{wrapfigure}[6]{l}[2mm]{30mm}
%\centering
%\includegraphics[keepaspectratio, width=30mm]{figghost.eps}
%\caption{}
%\end{wrapfigure}
%
\begin{align}
&\begin{minipage}{5cm}
\includegraphics[width=3cm, clip]{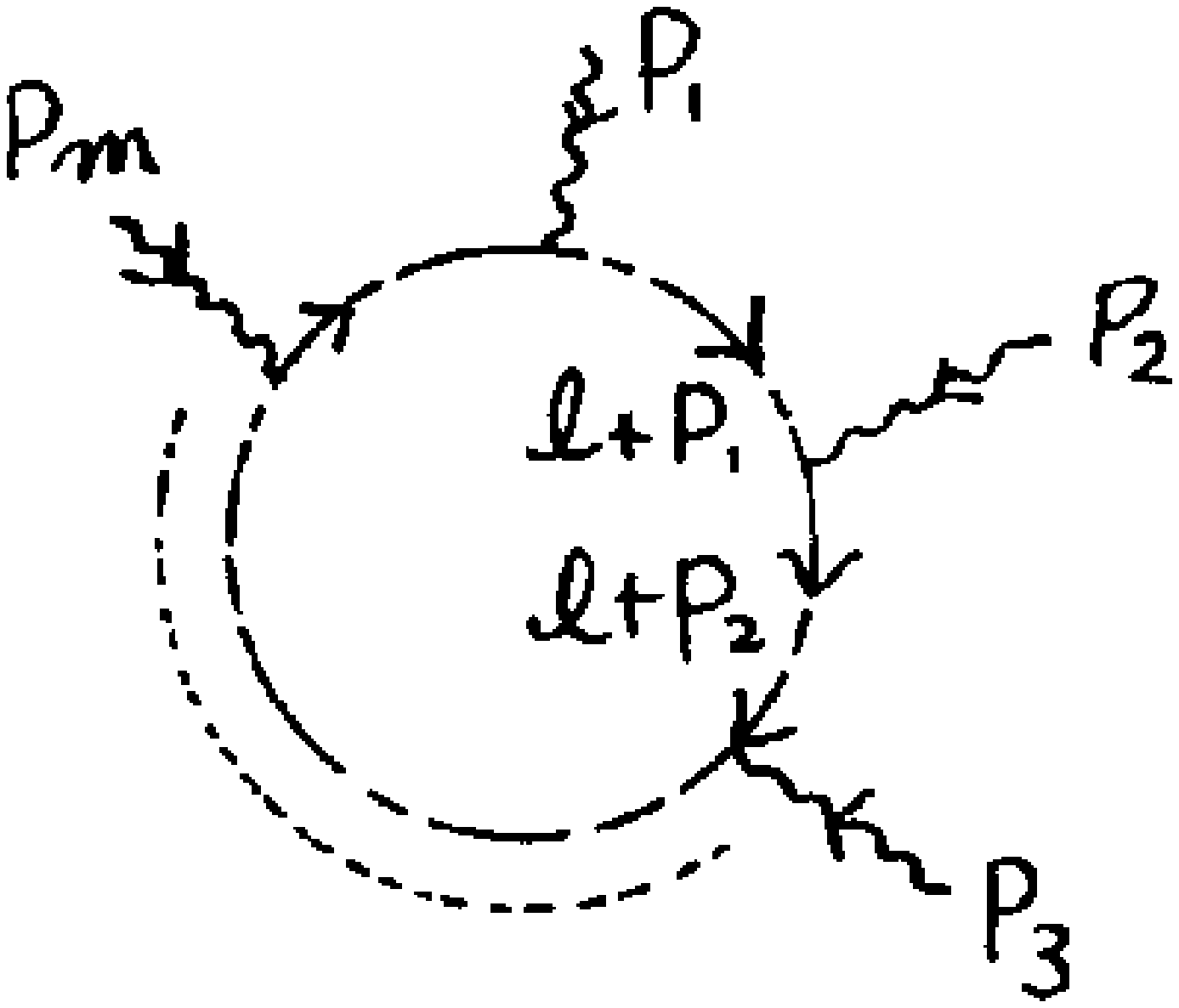}
\end{minipage} \notag \\
& \propto \int d^N l \frac{1}{n \cdot l} \frac{1}{n \cdot (l+p_1)} \frac{1}{n \cdot (l+p_1+p_2)} \cdots \frac{1}{n \cdot (l+p_1+ \cdots + p_{m-1})} \notag \\
&=\int_0^1 d\alpha_1 \cdots d\alpha_m \delta(1-\alpha_1-\cdots -\alpha_m) \int d^N l \frac{1}{[ n \cdot (l+ c(\alpha))]^m },  \label{eq4.1.8}
\end{align} 
where $c(\alpha)= \alpha_2 p_1 + \alpha_3 (p_1+p_2) + \cdots + \alpha_m (p_1 + p_2 + \cdots + p_{m-1})$.  
The $l$-integral in (\ref{eq4.1.8}) reads, after shifting $l$, 
\begin{equation}
 \int d^N l \frac{1}{[n \cdot l]^m} \propto \frac{1}{(n^2)^{m/2}} \int \frac{d^N l}{(l^2)^{m/2}}=0 \label{eq4.1.9}
\end{equation}
This result holds in case of using N dimensional method; in this method, ghost fields do not couple to gauge fields, [which is based on the study by Frenkel \cite{24}].
The other way is to show the unitarity, as will be done in Section 4.3, by combining the Feynman rules without ghosts obtained from (\ref{eq4.1.1}).  If this can be done, \underline{the theory is consistently} \underline{closed, without introducing the ghosts}.  [In the past, Feynman pointed out that Feynman rules of the Feynman gauge violate the unitarity (in gravity and gauge theory), so that the ghost fields should be introduced to recover the unitarity. \cite{40}.  On the other hand, if the unitarity is proved to hold in the axial gauge, then the ghosts are not necessary to be introduced in this gauge. ] (Refer to Section 4.3 and Reference paper II.] \\
  Based on the above consideration, we start with the follwoing Feynman rules without ghosts.  From (\ref{eq4.4.1}) we have the following rules:
\begin{align}
&\begin{minipage}{14cm}
\includegraphics[width=12cm, clip]{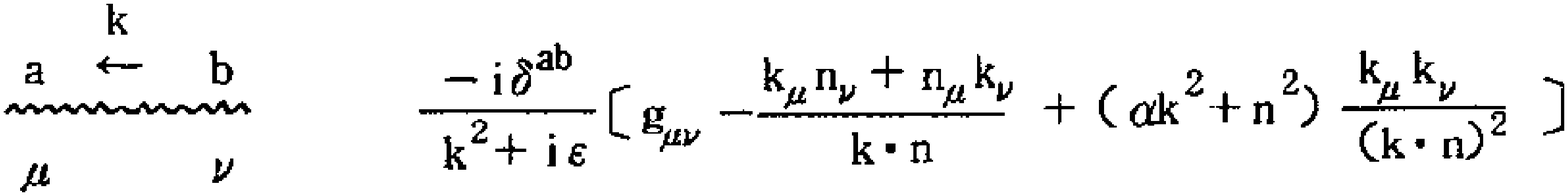}
\end{minipage}
\label{eq4.1.10} \\
&\begin{minipage}{14cm}
\includegraphics[width=5cm, clip]{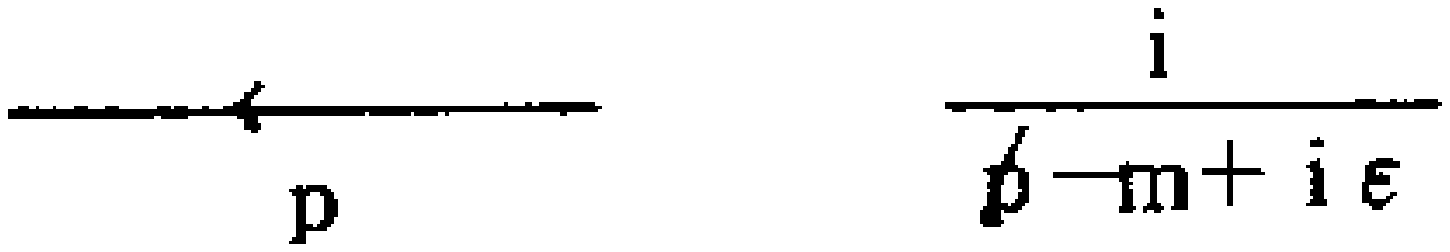}
\end{minipage}
\label{eq4.1.11} \\
&\begin{minipage}{14cm}
\includegraphics[width=12cm, clip]{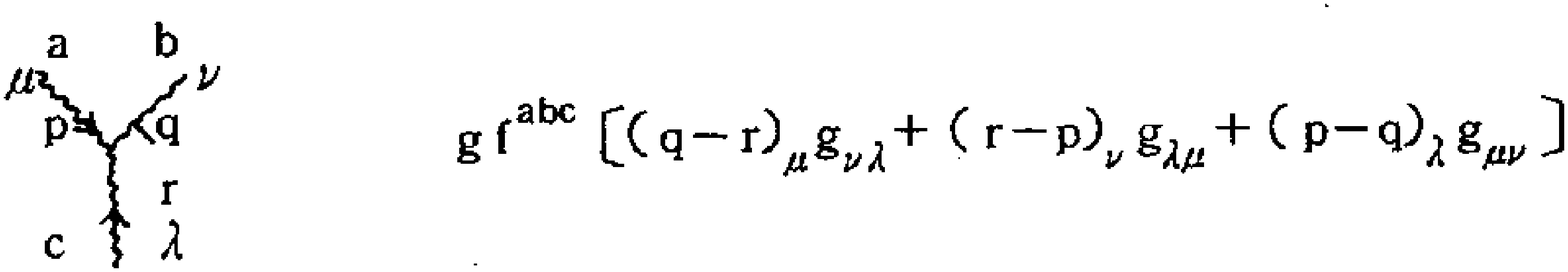}
\end{minipage}
\label{eq4.1.12} \\
&\begin{minipage}{14cm}
\includegraphics[width=12cm, clip]{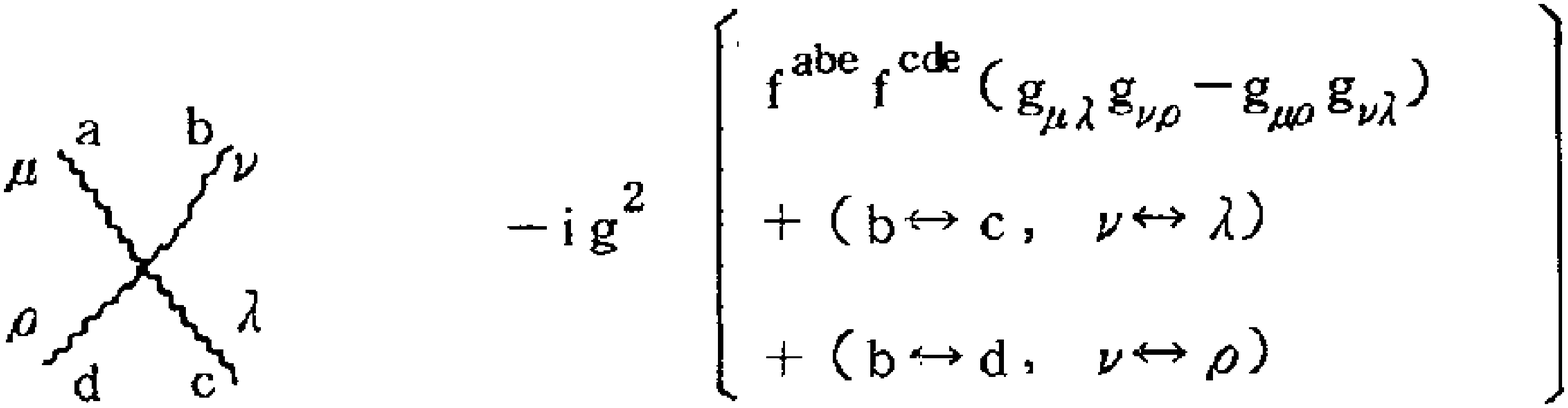}
\end{minipage}
\label{eq4.1.13} \\
&\begin{minipage}{14cm}
\includegraphics[width=5cm, clip]{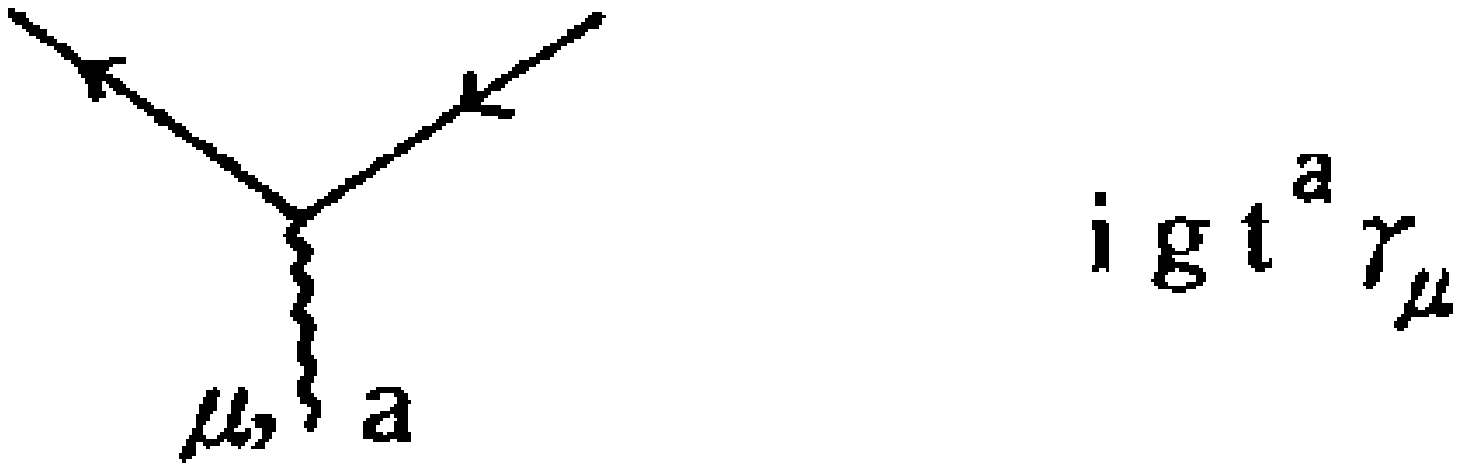}
\end{minipage}
\label{eq4.1.14}
\end{align}
The propagator of gauge boson in the axial gauge takes a complicated form (\ref{eq4.1.10}), which is, however, a very covenient form to develop the general theory.   Here the $i\varepsilon$-rule for $1/ k \cdot n$ and  $1/ (k \cdot n)^2$ becomes the following, in order for the unitarity to hold (see the general theroty in Section 4.3):
\begin{align}
&\frac{1}{k \cdot n} \to p_f \frac{1}{k \cdot n} \equiv \frac{1}{2} \left[ \frac{1}{k \cdot n + i \varepsilon} + \frac{1}{k \cdot n - i \varepsilon} \right], \label{eq4.1.15} \\
&\frac{1}{(k \cdot n)^2} \to p_f \frac{1}{(k \cdot n)^2} \equiv \frac{1}{2} \left[ \frac{1}{(k \cdot n + i \varepsilon)^2} + \frac{1}{(k \cdot n - i \varepsilon)^2} \right]. \label{eq4.1.16}
\end{align} \par
\noindent
\subsection{Diagramatical proof of Ward-Takahashi identities}
\noindent
Next, combining the Feynman rules given in (\ref{eq4.1.10})--(\ref{eq4.1.14}) given in the previouse subsection, we will derive the Ward-Takahashi identities (W-T identities).  [The identities are not directly related to the infrared divergences, but will be connected to the proof of unitarity in the next subsection, and will finally lead to the cancellation of infrared divergeces.] \\
  The proof given in the following is a remake in the axial gauge of the method which was used by 't Hooft in the Feynman gauge \cite{25}. [In the axial gauge, the proof by 't Hooft becomes extremely simple due to the absence of ghost fields.] \\
  First we introduce some notations:
\begin{align}
& \begin{minipage}{14cm}
\includegraphics[width=10cm, clip]{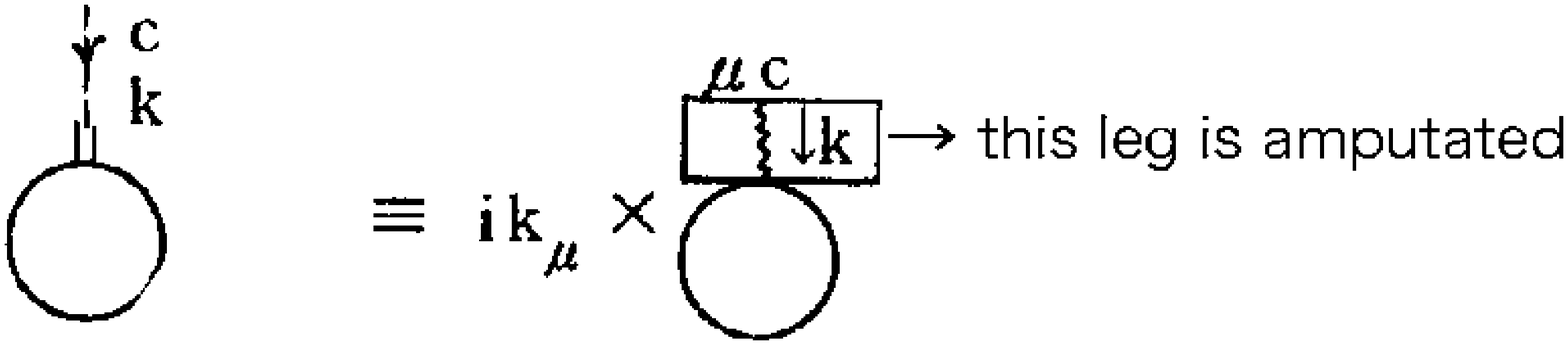}
\end{minipage}
\label{eq4.2.1} \\
&\begin{minipage}{14cm}
\includegraphics[width=7cm, clip]{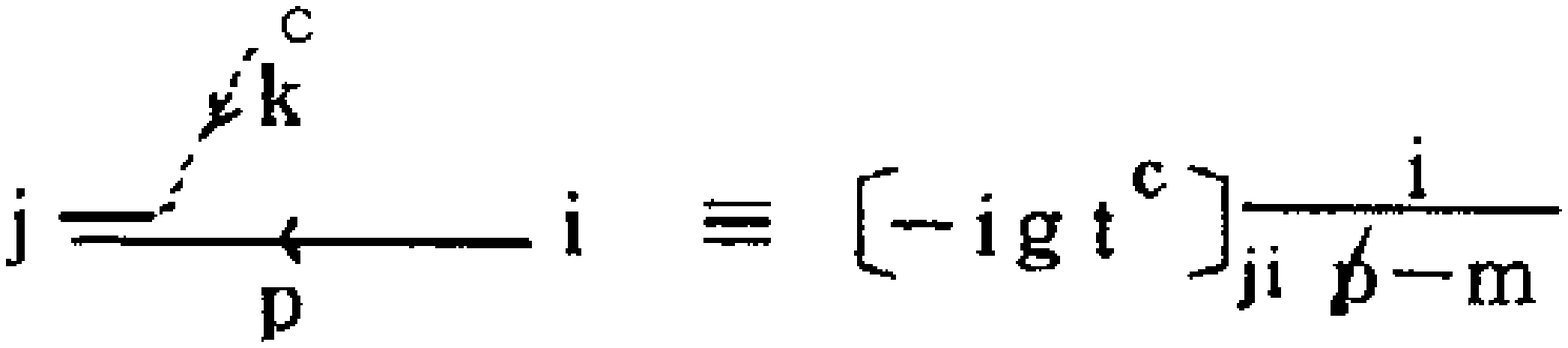}
\end{minipage}
\label{eq4.2.2} \\
& \begin{minipage}{14cm}
\includegraphics[width=7cm, clip]{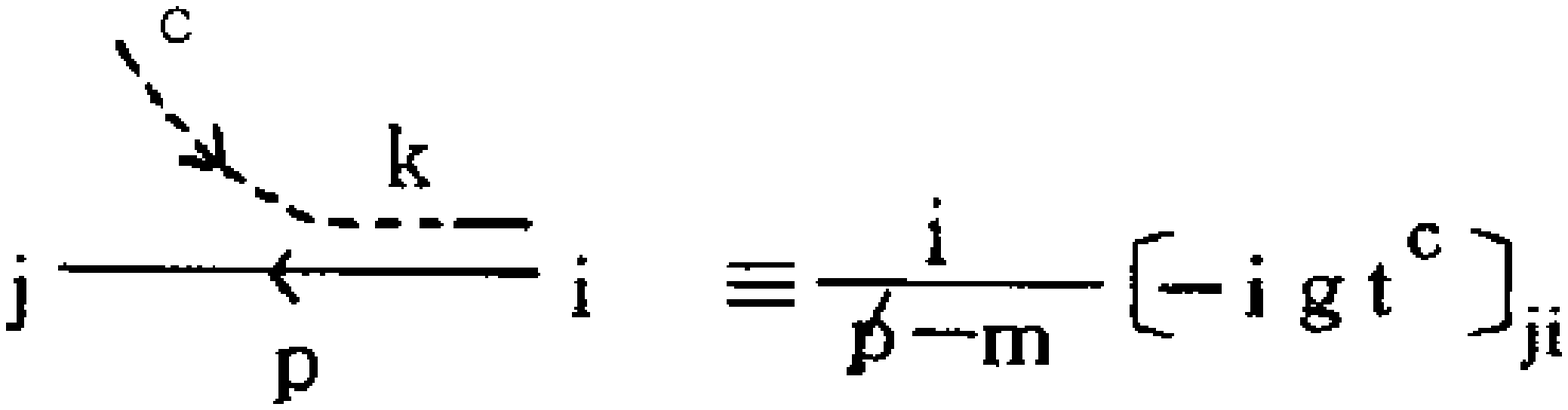}
\end{minipage}
\label{eq4.2.3} \\
& \begin{minipage}{14cm}
\includegraphics[width=7cm, clip]{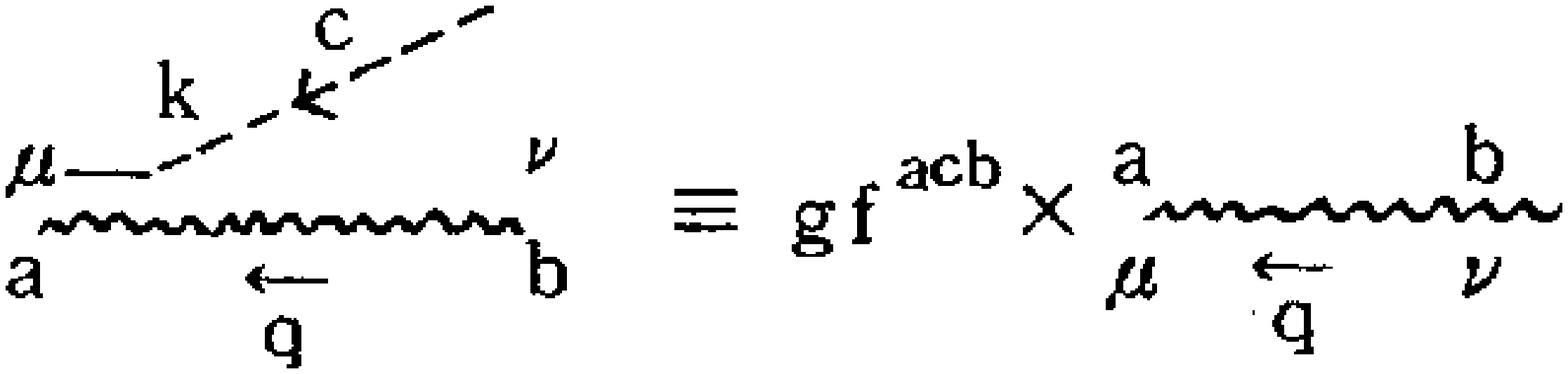}
\end{minipage}
\label{eq4.2.4}
\end{align}
Since the fermion and gauge boson have different representations, there appears a difference between (\ref{eq4.2.2}) and (\ref{eq4.2.3}), and (\ref{eq4.2.4}). [Here $f^{abc}=-i (T^c)_{ab}$, with $T^c$ the representation matrix for gauge field.]  Following 't Hooft, starting from tree graphs and combining of them, we will give a proof in all orders.  First, for the tree graphs, we have the following identities, by the explicit estimation:
\begin{align}
& \begin{minipage}{14cm}
\includegraphics[width=8cm, clip]{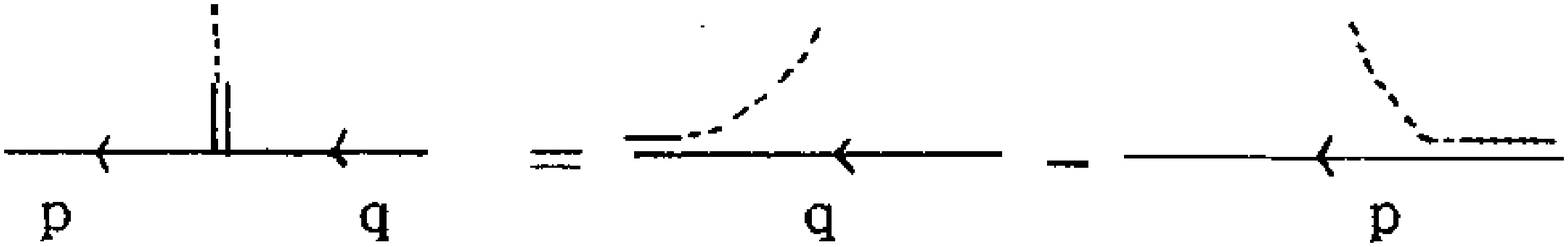}
\end{minipage}
\label{eq4.2.5} \\
& \begin{minipage}{14cm}
\includegraphics[width=8cm, clip]{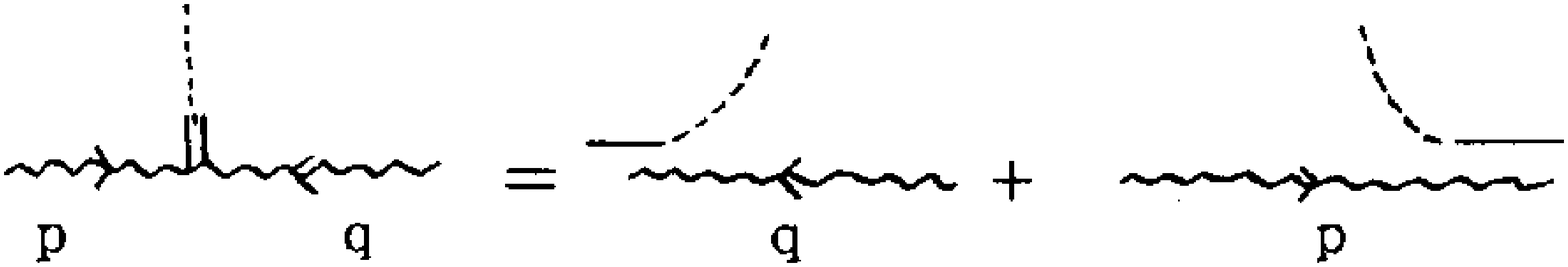}
\end{minipage}
\label{eq4.2.6} \\
&\begin{minipage}{14cm}
\includegraphics[width=9cm, clip]{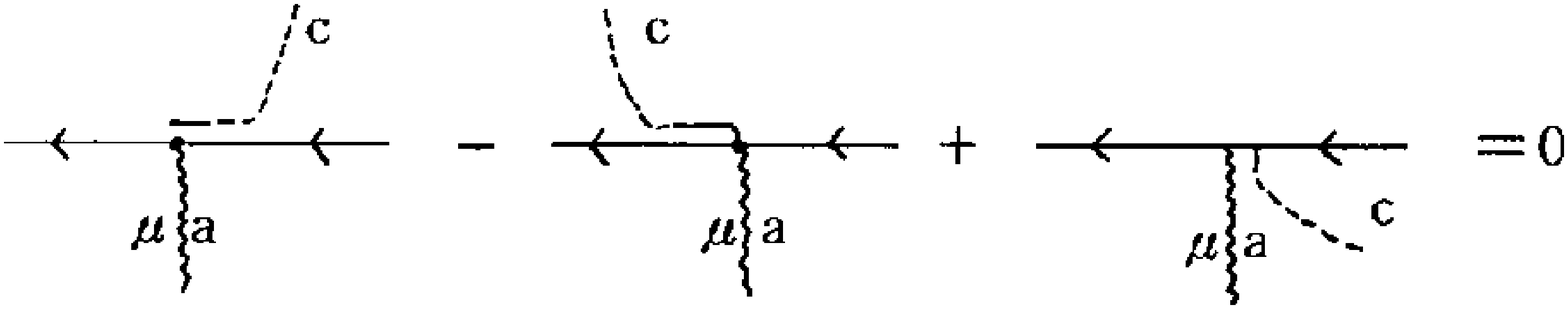}
\end{minipage}
\label{eq4.2.7} \\
& \begin{minipage}{14cm}
\includegraphics[width=8cm, clip]{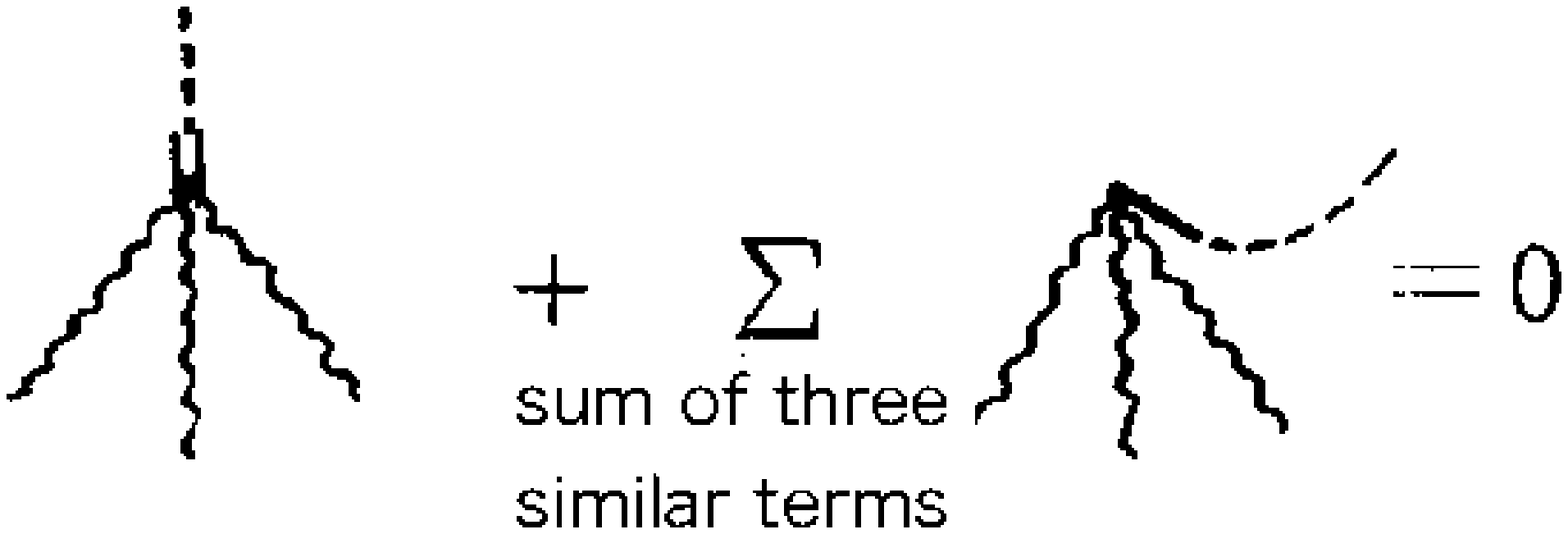}
\end{minipage}
\label{eq4.2.8} \\
& \begin{minipage}{14cm}
\includegraphics[width=5cm, clip]{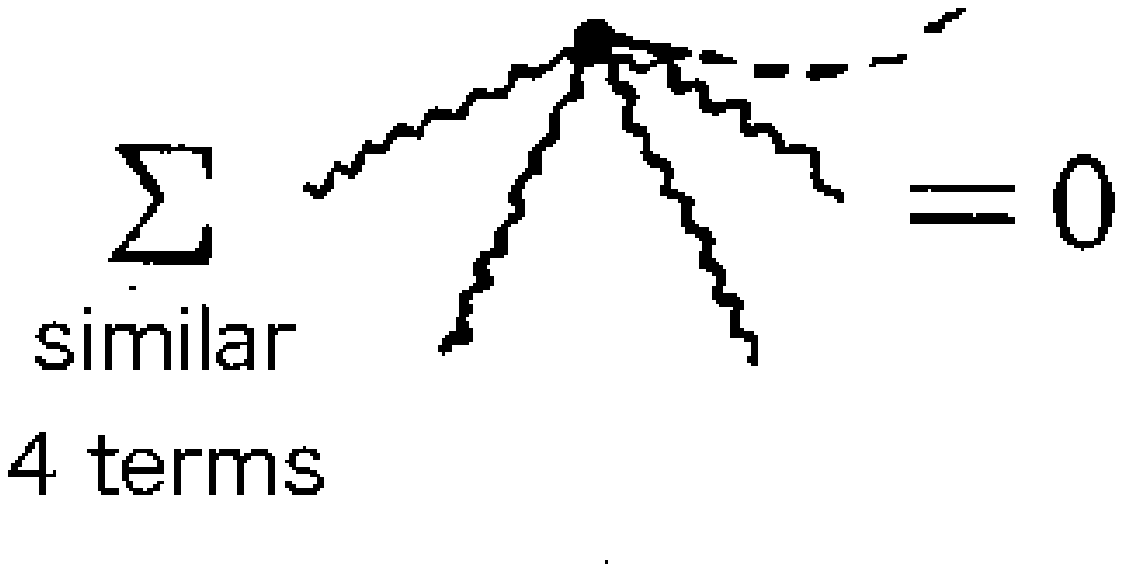}
\end{minipage}
\label{eq4.2.9}
\end{align}
Let us add a number of coments:  First, (\ref{eq4.2.5}) is a pictorial description of
\begin{equation}
\frac{i}{\Slash{p} -m } igt^c\gamma_{\mu} i (p-q)^{\mu} \frac{i}{\Slash{q} -m }= [-igt^c] \frac{i}{\Slash{q} -m }-\frac{i}{\Slash{p} -m } [-igt^c]  \label{eq4.2.10}
\end{equation}
Second, (\ref{eq4.2.6}) is characteristic identity in the axial gauge [but it should be examined whether it holds only in this gauge].  Let us compare it to the corresponding identity in the Feynman gauge (of 't Hooft) which is (\ref{eq3.2.4}) used previously. Adding the gauge boson propagators, we have
%
%\begin{wrapfigure}[4]{l}[2mm]{30mm}
%\centering
%\includegraphics[keepaspectratio, width=30mm]
%{figcontraction.eps}
%\caption{}
%\end{wrapfigure}
%
\begin{align}
& \begin{minipage}{14cm}
\includegraphics[width=5cm, clip]{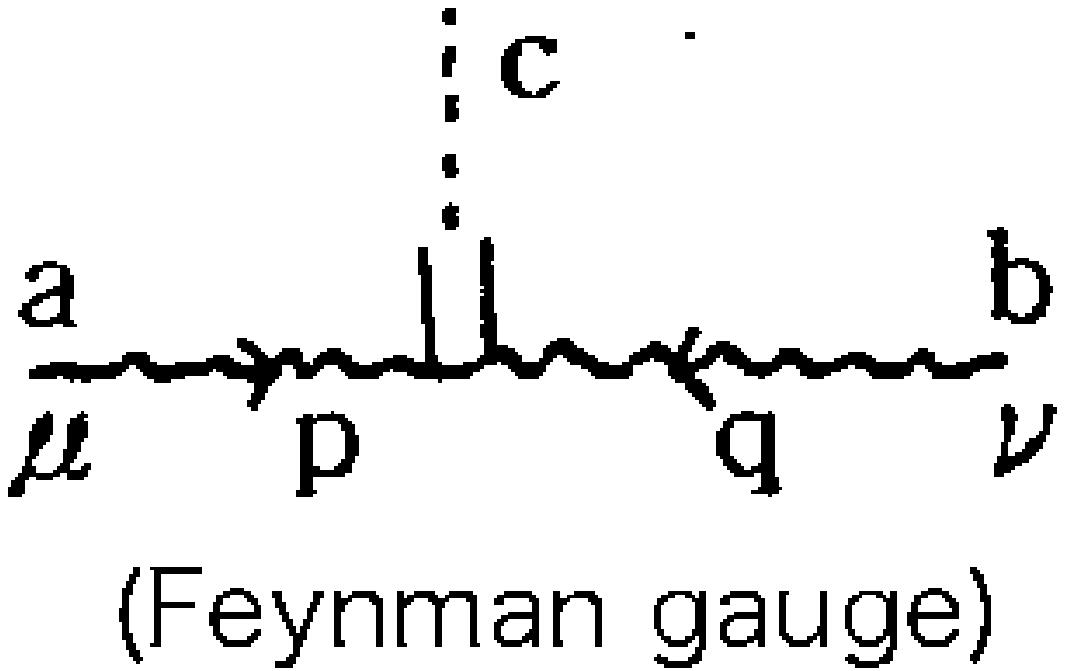} \notag
\end{minipage} \\
&= \frac{-ig_{\mu\mu'}}{p^2 + i \varepsilon} \times ig f^{cab} [(q^2 g_{\mu'\nu'}-q_{\mu'}q_{\nu'}) - (p^2 g_{\mu'\nu'}-p_{\mu'}p_{\nu'}) ] \times \frac{-ig_{\nu'\nu}}{q^2 + i \varepsilon}  \notag \\
&= g f^{acb} \frac{-ig_{\mu\nu}}{q^2 + i \varepsilon} + \frac{-ig_{\mu\nu}}{p^2 + i \varepsilon} g f^{bca} + (-ip_{\mu}) \frac{-i}{p^2 + i \varepsilon}  (g f^{acb} p_{\nu'}) \frac{-ig_{\nu'\nu}}{q^2 + i \varepsilon} \notag \\
&~~ + \frac{-ig_{\mu\mu'}}{p^2 + i \varepsilon}  (g f^{bca} q_{\mu'}) \frac{-i}{q^2 + i \varepsilon} (-i q_{\nu}) \notag \\
&\begin{minipage}{13cm}
\includegraphics[width=12cm, clip]{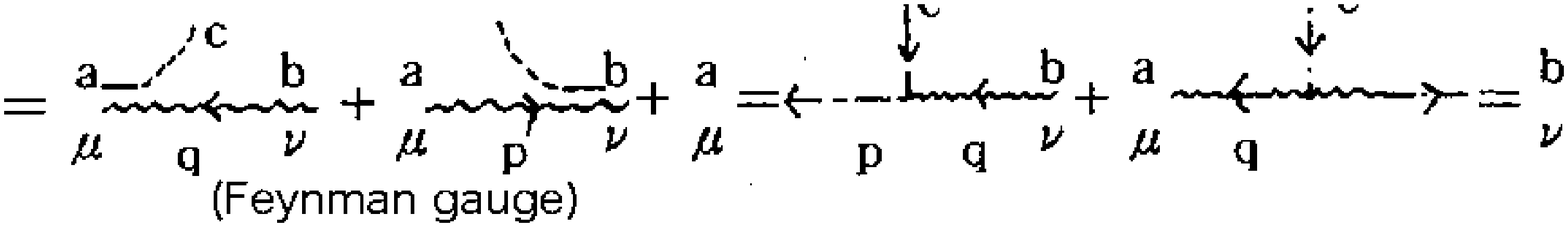} 
\end{minipage}
\label{eq4.2.11}
\end{align}
Since the figures correspond to the equations, one by one, so that we can skip the explanation, but notice that the terms corresponding to Fadeev-Popov ghosts appear in the third and forth terms in the last equation or in the last figure.  The last two terms, associated with the internal ghost loops, make the W-T identities complex.  In the axial gauge, however, such a complexity does not exist in (\ref{eq4.2.6}), so that (\ref{eq4.2.6}) completely correspons to (\ref{eq4.2.5}) for the fermions.  The proof of (\ref{eq4.2.6}) will not be given explicitly, but is easily understood  
if factorizing the gauge boson propagator as follows:
\begin{equation}
D_{\mu\nu}(p) = \frac{-i}{p^2 + i \varepsilon} \left[ g_{\mu\lambda} - \frac{p_{\mu} n_{\lambda}}{p \cdot n} \right] \left[ g_{\nu}^{\lambda} - \frac{ n^{\lambda} p_{\nu}}{p \cdot n} \right]  - i \alpha \frac{p_{\mu} p_{\nu}}{(p \cdot n)^2}.  \label{eq4.2.12}
\end{equation}
Third, (\ref{eq4.2.7}) can be understood, from (\ref{eq4.2.2})--(\ref{4.2.4}), to represent the fundamental relation of the group
\begin{equation}
t^a t^c - t^c t^a + i t^b f^{bca}=0.  \label{eq4.2.13}
\end{equation}
Forth, we omit the proof of (\ref{eq4.2.8}) and (\ref{eq4.2.9}), but they can be proved from the Feynman rules and (\ref{eq4.2.13}), by using the Jacobi identity:
\begin{equation}
f^{abe} f^{ecd} + f^{ace}f^{edb} + f^{ade} f^{ebc}=0  \label{eq4.2.14}
\end{equation}
Now, let us derive the W-T identities for an arbitrary graph, by combining W-T identities at tree level.  Substituting into the following circles \ctext{}, the sum of all the connected graphs, then we have
\begin{equation}
\begin{minipage}{14cm}
\includegraphics[width=7cm, clip]{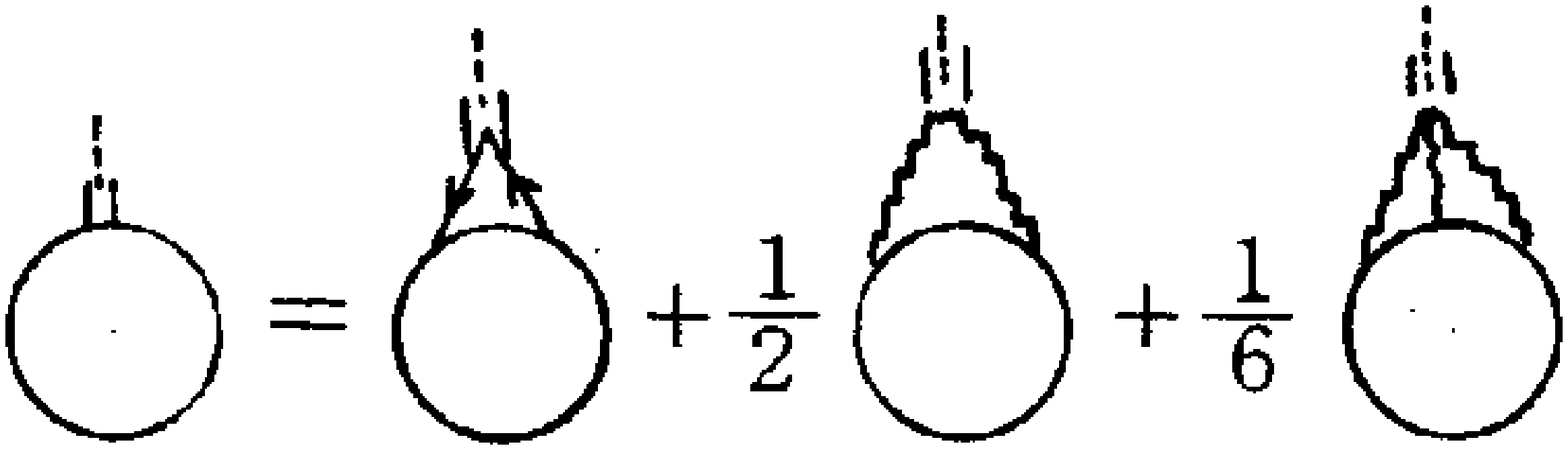}
\end{minipage}
\label{eq4.2.15}
\end{equation}
where $\frac{1}{2}, \frac{1}{6}$ are combinatorial coefficients.  Using (\ref{eq4.2.5})--(\ref{eq4.2.9}) here, we have
\begin{equation}
\begin{minipage}{14cm}
\includegraphics[width=7cm, clip]{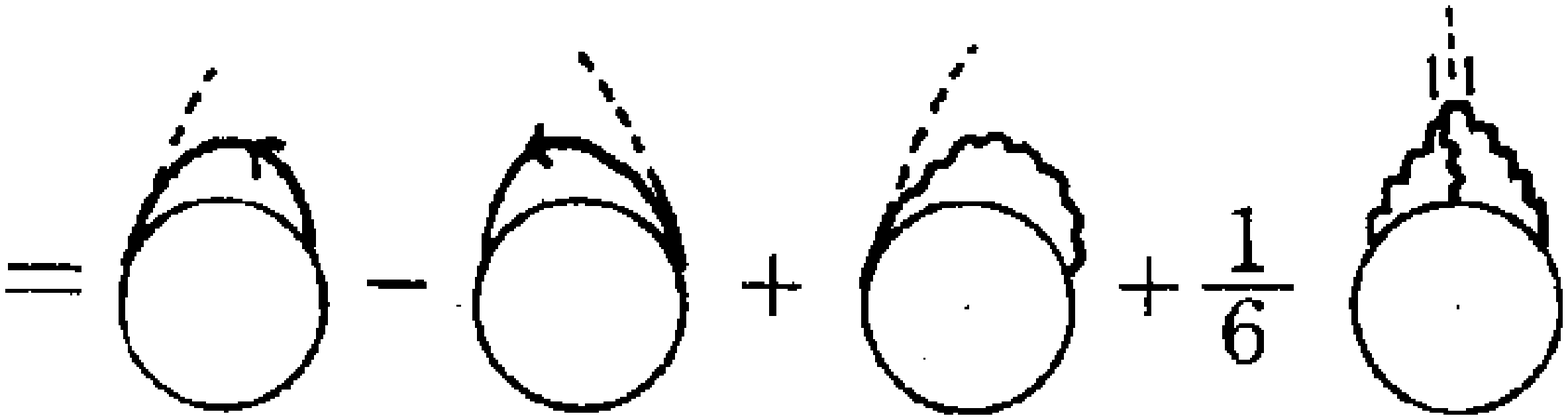}
\end{minipage}
\label{eq4.2.16}
\end{equation}
Differentiating the line to which a dotted line attachs be external or internal, and identifying the vertices to which the dotted line attach, then we have
\begin{equation}
\begin{minipage}{14cm}
\includegraphics[width=10cm, clip]{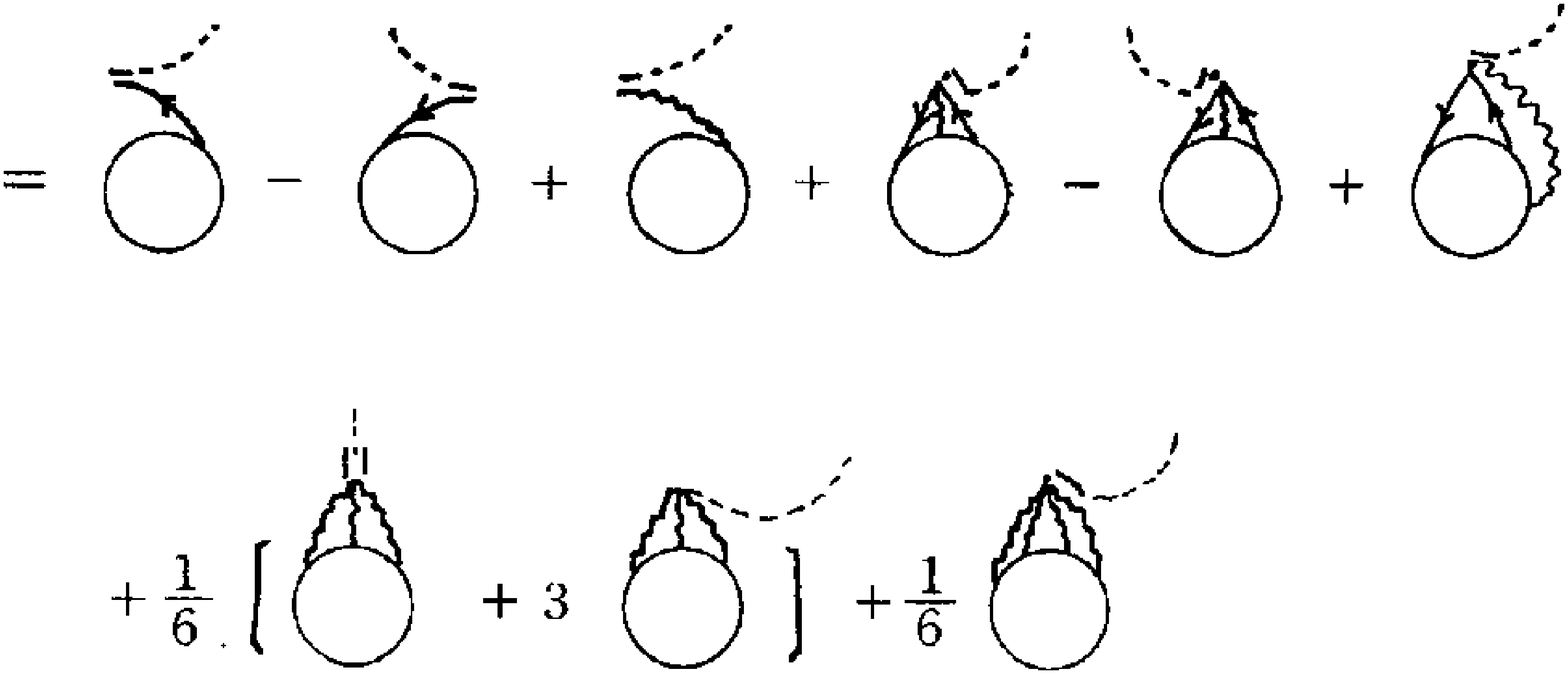}
\end{minipage}
\label{eq4.2.17}
\end{equation}
Here, we use (\ref{eq4.2.7}) in the second line of the above equation, (\ref{eq4.2.8}) in the first term of the third line, (\ref{eq4.2.9}) in the second term of the same line, then the second and third lines cancell, leading finally to
\begin{equation}
\begin{minipage}{14cm}
\includegraphics[width=10cm, clip]{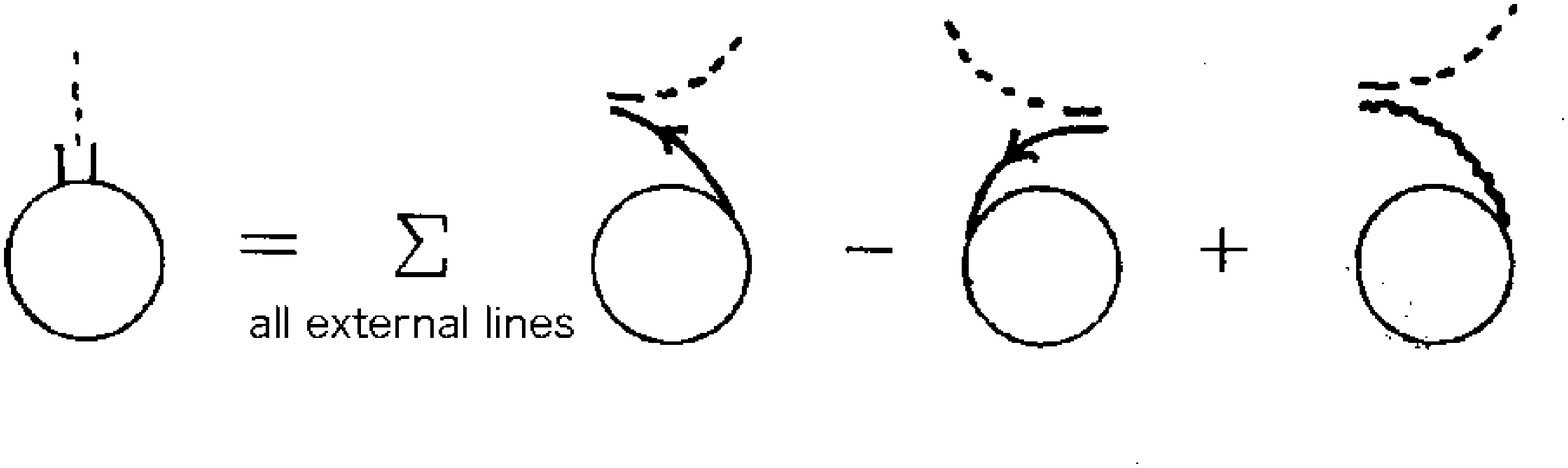}
\end{minipage}
\label{eq4.2.18}
\end{equation}
These are the \underline{general W-T identities} which we want to derive.  They are quite similar W-T identities in QED.  

------------------------------------------------------

As a reference, let us prove the W-T identities in QED, using the notations in this section.  In QED, we replace 
$t^c \to 1$ in (\ref{eq4.2.2}) and (\ref{eq4.2.3}), the identities at tree level are only 
\begin{align}
&\begin{minipage}{14cm}
\includegraphics[width=7cm, clip]{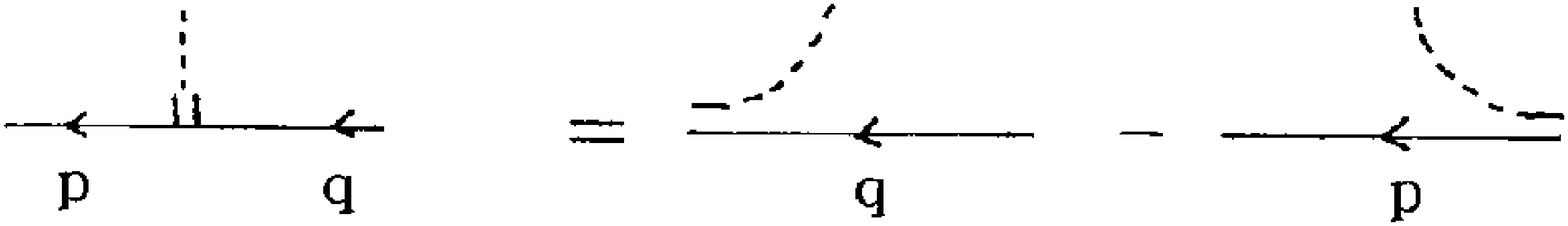}
\end{minipage}
\label{eq4.2.19}  \\
&\begin{minipage}{14cm}
\includegraphics[width=6cm, clip]{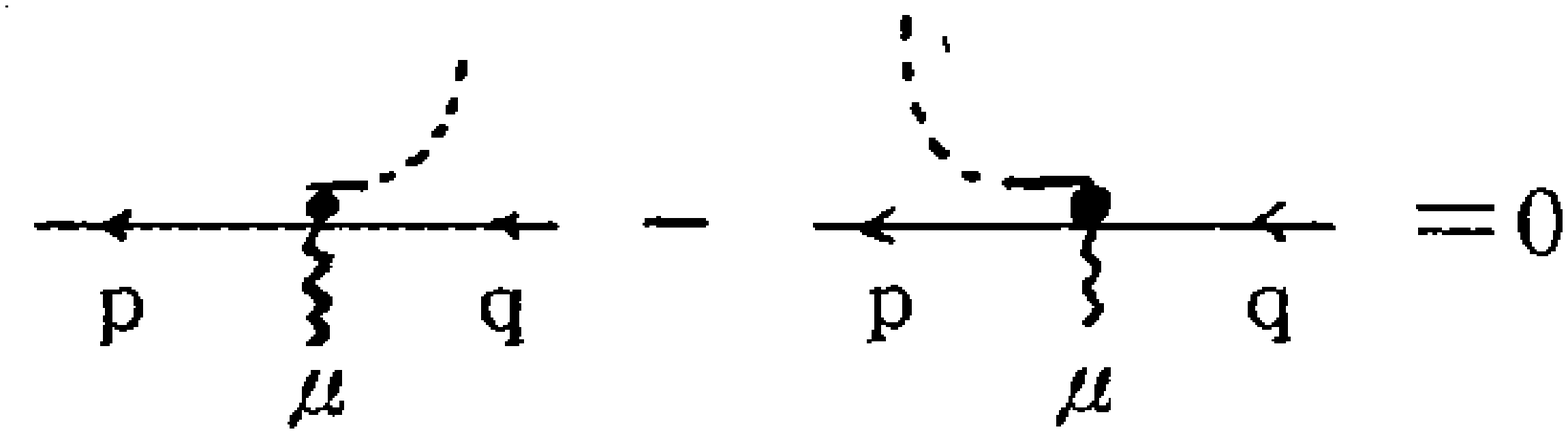}
\end{minipage}
\label{eq4.2.20}
\end{align}
Using them, for a general sum of connected graphs in the circle \ctext{}, we have
\begin{equation}
\begin{minipage}{14cm}
\includegraphics[width=7cm, clip]{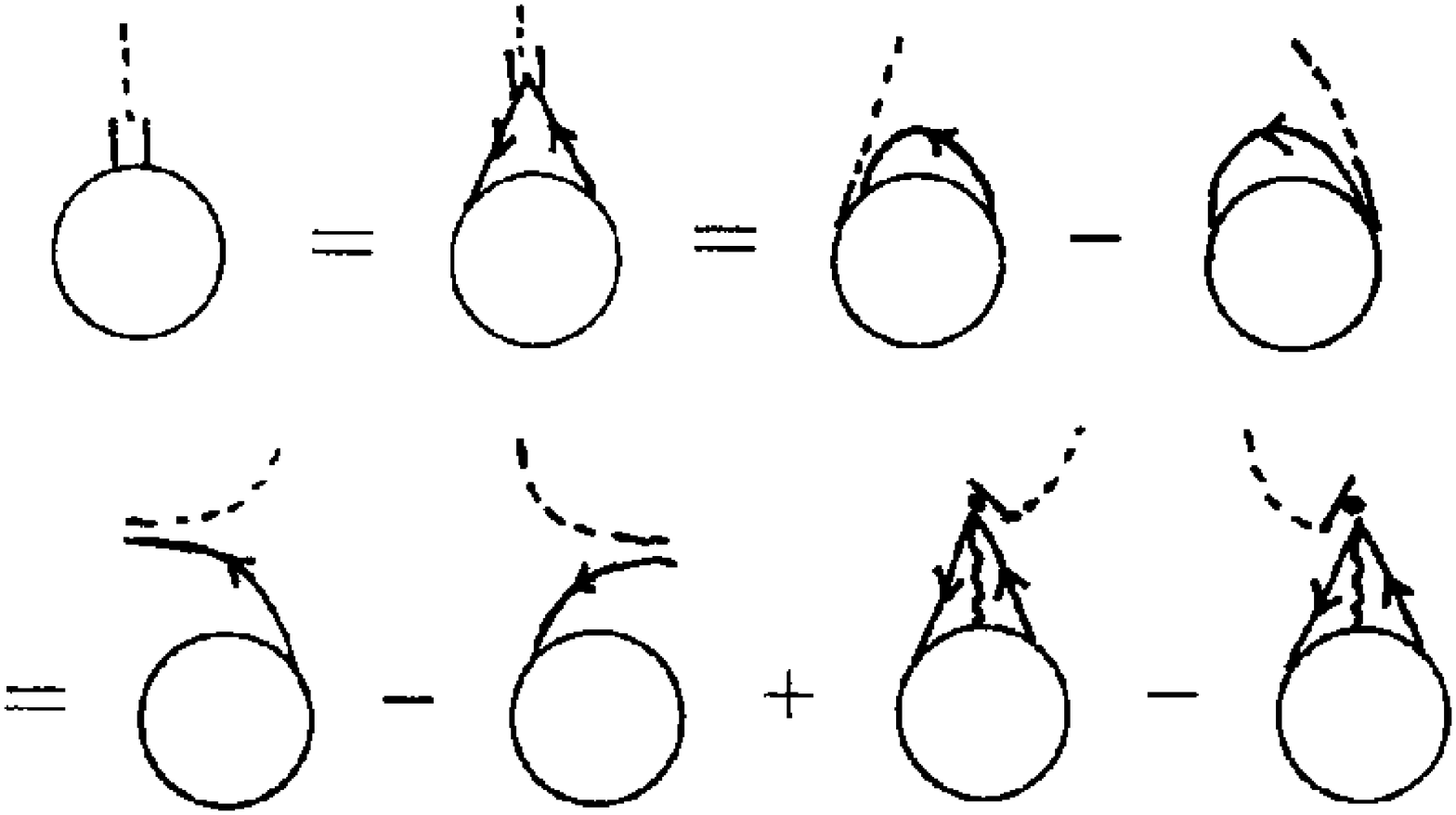}
\end{minipage}
\label{eq4.2.21}
\end{equation}
From (\ref{eq4.2.20}), we have the W-T identities in QED as follows:
\begin{equation}
\begin{minipage}{14cm}
\includegraphics[width=8cm, clip]{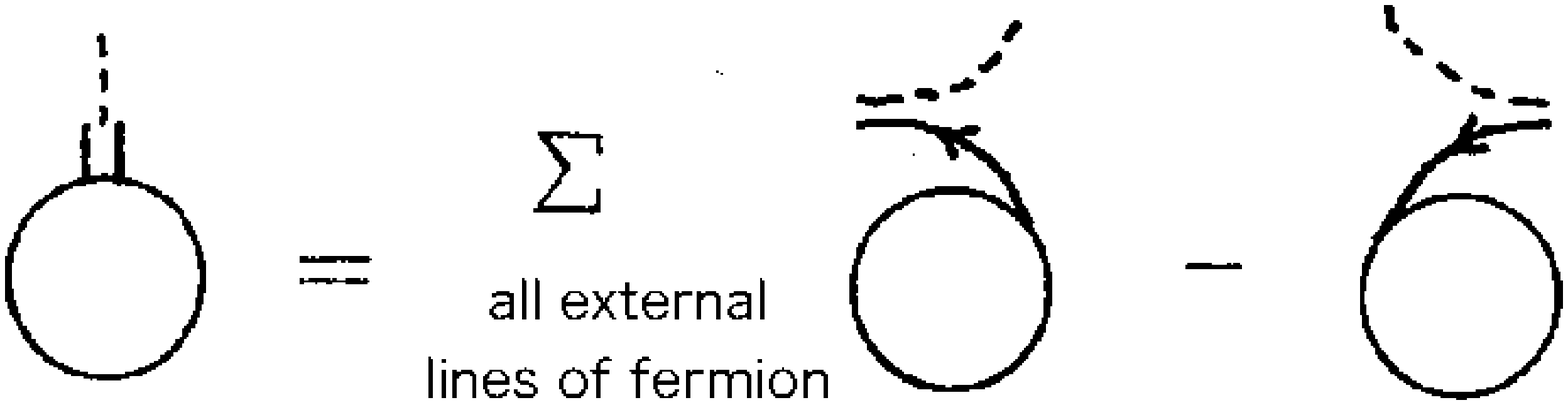}
\end{minipage}
\label{eq4.2.22}
\end{equation} \par
\noindent
\subsection{Proof of unitarity}
\noindent
Let us check the unitarity in the axial gauge QCD, using the W-T identities given in the last subsection.  The proof is a little complex, so that we restrict here to give an outline of it, but instead add the more fundamental issues which are not written in the Reference paper II. \par
  To begin with introduce $F(x_1, \cdots, x_n)$ which stands for, corresponds to a Feynman diagram, the products of propagators connecting the vertex points at $x_1, \cdots, x_n$, and of vertex functions on each vertex point.  [To obtain the scattering amplitude from the function $F$, we cut the external lines, multiply the wave functions $\frac{1}{\sqrt{Z_2}} u(p, s) e^{-i px}, \frac{1}{\sqrt{Z_2}} \bar{u}(p, s) e^{i px}, \frac{1}{\sqrt{Z_3}} \varepsilon_{\mu} e^{-i px}$ {\it e.t.c.}, and integrate over $x_1, \cdots, x_n$.] \par
  For this $F(x_1, \cdots, x_n)$, the following cutting rule holds. [Cutting rule of Nishijima and Veltman \cite{27}]
\begin{equation}
\underset{\text{all the ways to underline}}{\sum} F(x_1, \cdots, \underline{x_i}, \cdots, \underline{x_j}, \cdots, x_n)=0.  \label{eq4.3.1}
\end{equation}
Here, the meaning of underline is as follows:\\
1) The vertex function at the underlined point is the hermition conjugate of that without underlining.  If the Lagrangian is hermitian, the vertex function has a factor $i$, so the underlined vertex function becomes $(-1)$ times the original vertex function. \par
2) The propagator function $D(x_i, x_j)$ changes, according to the way of underlining, as follows:
\begin{align}
\begin{cases}
D(x_i-x_j)^{(+)} ~~\text{for the line}~~(\underline{x_i}, x_j),  \\
D(x_i-x_j)^{(-)} ~~\text{for the line}~~(x_i,\underline{x_j}),  \\
D(x_i-x_j)^{*} ~~~\text{for the line}~~(\underline{x_i},
\underline{x_j}), ~\text{and of course} \\
D(x_i-x_j) ~~~~\text{for the line}~~(x_i, x_j).
\end{cases} \notag
\end{align}
Here, $D^{(\pm)}$ are defined from the original propagator $D$, by
\begin{equation}
D(x)=\theta(x_0)D(x)^{(+)} + \theta(-x_0)D(x)^{(-)}  \label{eq4.3.2}
\end{equation}
The boundary conditions for the propagators [$i \varepsilon$-rule] must satisfy 
\begin{equation}
[D(x)^{(+)}]^*=D(x)^{(-)}  \label{eq4.3.3}
\end{equation}
It is because, (\ref{eq4.3.3}) is compulsory in the proof of cutting rule (\ref{eq4.3.1}).  Putting aside the proof of (\ref{eq4.3.1}), we will show that the $i \varepsilon$-rule for $1/(k \cdot n)$ and $1/(k \cdot n)^2$ can be obtained from (\ref{eq4.4.3}). \par
  First, examine the relation between $D(x), D(x)^{(+)}$ and $D(x)^{(-)}$, in the momentum space.  Propagators in the momentum space $D(k)$s are connected by $D(x)$s through the Fourier transformation:
\begin{align}
D(x)=& \int d^4 k \; e^{-ikx} D(k)  \notag \\
=& \theta(x_0)   \underset{\text{clockwise along the lower semi-circle}}{\int d^3 k \int dk^0 \; e^{-ikx} D(k)} + \theta(-x_0)   \underset{\text{anti-clockwise along the uppe semi-circle}}{\int d^3 k \int dk^0 \; e^{-ikx} D(k)}. \label{eq4.3.4}
\end{align}
Depending on $x_0 > 0$ or $x_0 < 0$, the contour of the integral over $k^0$ should be chosen in the lower half plane or in the upper half plane.  This decomposition by $x_0 > 0$ or $x_0 < 0$ corresponts to that in (\ref{eq4.3.2}). Consider the following situation in which $D(k)$ has a  first order pole (snigle pole) $1/(k^0 -s_+ + i \varepsilon)$, and a second order pole (double pole) $1/(k^0 -d_+ + i \varepsilon)^2$ on the lower plane, while it has a first order pole $1/(k^0 -s_- - i \varepsilon)$ and a second order pole $1/(k^0 -d_- - i \varepsilon)^2$ on the upper plane, namely, 
\begin{figure}{h}
\centering
\includegraphics[width=8cm, clip]{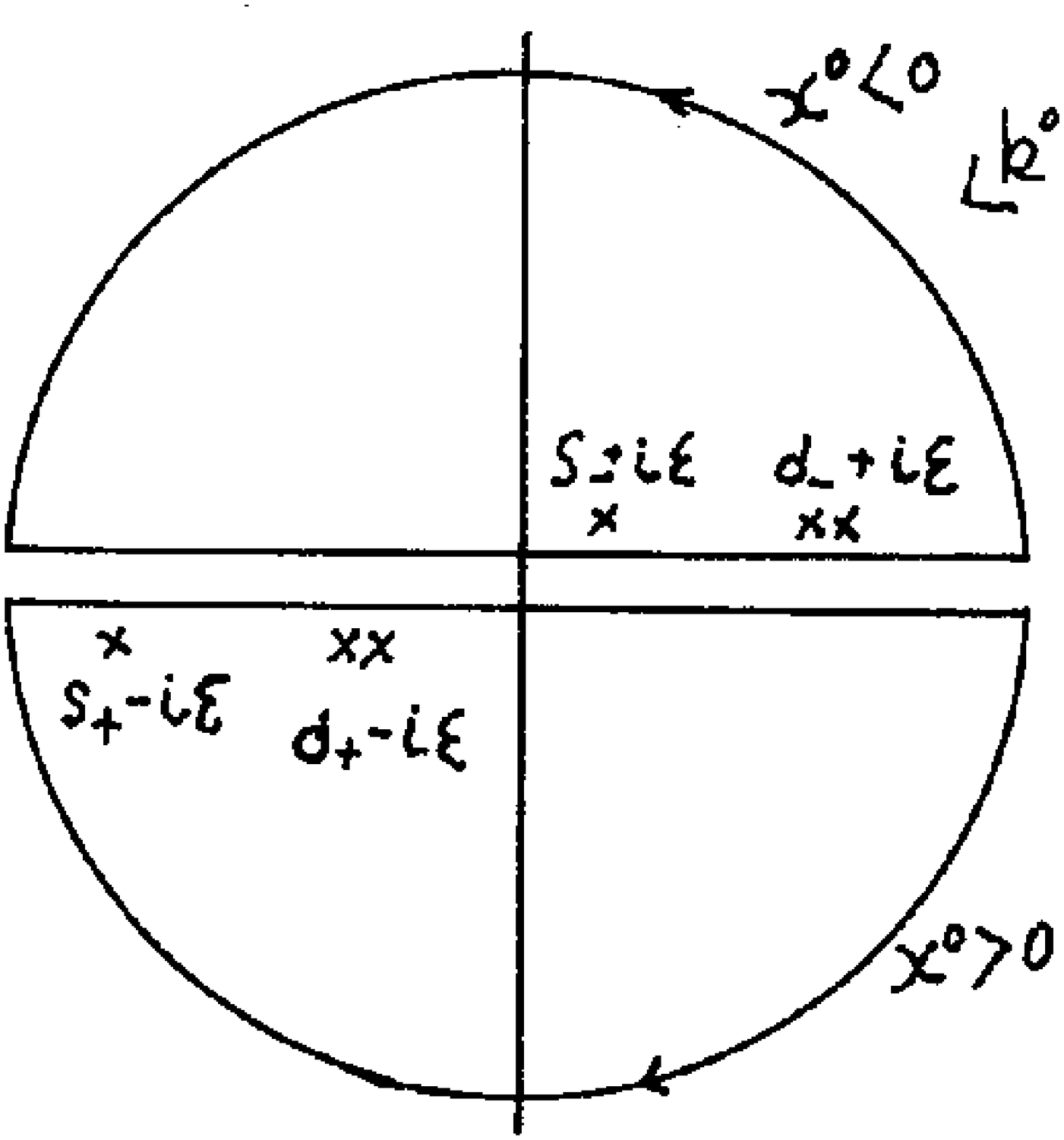}
\caption{The first order (single) pole and the second order (double) pole in the propagator}
\end{figure}
\begin{align}
D(k) = & \frac{A_+(k)}{ k^0 -s_+ + i \varepsilon } +  \frac{B_+(k)}{ (k^0 -d_+ + i \varepsilon)^2 }  \notag \\
& +  \frac{A_- (k)}{ k^0 -s_-  i \varepsilon } +  \frac{B_- (k)}{ (k^0 -d_- + i \varepsilon)^2 }   \label{eq4.3.5} 
\end{align}
Then, substituting this expression into (\ref{eq4.3.4}) yields the momentum representation $D^{(\pm)}(k)$ of $D^{(\pm)}(x)$ to be
\begin{align}
&D^{(+)}(k)= - 2\pi i \delta(k^0-s_+) A_+(k) \notag \\
&+ 2\pi i \delta' (k^0-d_+) B_+(k),   \label{eq4.3.6} \\
&D^{(-)}(k)=  2\pi i \delta(k^0-s_-) A_-(k) \notag \\
& - 2\pi i \delta' (k^0-d_-) B_-(k),   \label{eq4.3.7}
\end{align}
where $D^{(\pm)}(x) = \int d^4 k \;  e^{-ikx} D(k)^{(\pm)}$. \par
  Since the condition (\ref{eq4.3.3}) in the momentum space is
\begin{equation}
D^{(+)}(-k)^* = D^{(-)} (k),  \label{eq4.3.8}
\end{equation}
in order for (\ref{eq4.3.3}) [or (\ref{eq4.3.8})] to hold, we can find, by using (\ref{eq4.3.7}), the following equations should hold:
\begin{align}
&\delta(k^0-s_+)A_+(k)^*|_{k \to -k} = \delta(k^0-s_-)A_-(k),  \label{eq4.3.9}  \\
& \delta' (k^0-d_+)B_+(k)^* |_{k \to -k} = \delta' (k^0-d_-)B_-(k).  \label{eq4.3.10}
\end{align}
In other words, the following relations have to hold
\begin{align}
&\circ \text{Position of the first order pole}: s_+(\vec{k})|_{\vec{k} \to -\vec{k}}=-s_-(\vec{k}), \label{eq4.3.11} \\
&\circ \text{Coefficient of the first order pole}: A_+(\vec{k})^*|_{\vec{k} \to -\vec{k}}=A_-(\vec{k}), \label{eq4.3.12} \\
&\circ \text{Position of the second order pole}: d_+(\vec{k})|_{\vec{k} \to -\vec{k}}=d_-(\vec{k}), \label{eq4.3.13} \\
&\circ \text{Coefficient of the second order pole}: B_+(\vec{k})^*|_{\vec{k} \to -\vec{k}}=B_-(\vec{k}). \label{eq4.3.14}
\end{align}
As is understood from the form of gauge boson's propagator, it has a first order pole at $k^0=\pm (|k| -i \varepsilon)$, which represents the poles of the physical states, and satisfies (\ref{eq4.3.11}).  The coefficients also satisfy (\ref{eq4.3.12}).  [ The condition (\ref{eq4.3.3}) is of course not enough to exclude the other possibibily of $k^0=\pm (|k| +i \varepsilon)$, which is rejected by the causality.] \par
  Next, let us determine the $i\varepsilon$-rule for the first order pole $\frac{1}{(k \cdot n)}$ and the second order pole $\frac{1}{(k \cdot n)2}$. \par
  To begin with, for the first order pole to satisfy (\ref{eq4.3.11}), there should exist poles at $k^0=\frac{\vec{k} \cdot \vec{n}}{n^0} \pm i \varepsilon$.  The coefficients of the poles satisfy (\ref{eq4.3.12}), since we have
\begin{equation}
A_+(k)_{\mu\nu}=A_-(k)_{\mu\nu}= \frac{-i}{k^2 + i \varepsilon} \left[ (-1) 
 \frac{k_{\mu}n_{\nu}+n_{\mu}k_{\nu}}{n^0} \right] \times \frac{1}{2}.  \label{eq4.3.15}
\end{equation}
Here the last factor $\frac{1}{2}$ is added so that it may reproduce the original (\ref{eq4.1.10}), when the $i \varepsilon$-rule is forgotten.  Thus, the $\frac{1}{k \cdot n}$ is understood to be
\begin{equation}
\frac{1}{k \cdot n} \to p_f \frac{1}{k \cdot n} =\frac{1}{2} \left[\frac{1}{k \cdot n + i\varepsilon} + \frac{1}{k \cdot n - i\varepsilon} \right],  \notag
\end{equation} 
[which is (\ref{eq4.1.15}).]  In the same manner, $1/(k \cdot n)^2$ should have the second order pole at $k^0=\frac{\vec{k} \cdot \vec{n}}{n^0} \pm i \varepsilon$.  The corresponding coefficients satisfy (\ref{eq4.3.14}), provided
\begin{equation}
B_+(k)_{\mu\nu}=B_-(k)_{\mu\nu}= \frac{-i}{k^2 + i \varepsilon} \left[ (\alpha k^2 +n^2) \frac{k_{\mu}k_{\nu}}{(n^0)^2} \right] \times  \frac{1}{2}.  \label{eq4.3.16}
\end{equation}
That is, $\frac{1}{(k \cdot n)^2}$ in (\ref{eq4.1.10}) must be
\begin{equation}
\frac{1}{(k \cdot n)^2} \to p_f \frac{1}{(k \cdot n)^2} =\frac{1}{2} \left[\frac{1}{(k \cdot n + i\epsilon)^2} + \frac{1}{(k \cdot n - i\epsilon)^2} \right],  \notag
\end{equation}
[which is the previous (\ref{eq4.1.16}).]  In this way, a general theory of determining $i \varepsilon$-rule has been afforded by the unitarity. \par
  Next, the proof of the cutting rule (\ref{eq4.3.1}) [by Veltman \cite{27} ] is recapiturated. \par
  Let the vertex be $x_*$ whose time $x^0$ is the maximum among the vertices in the right-hand-side of (\ref{eq4.3.1}).  {Finally, the scattering amplitude is obtained by integrating over $x$ of (\ref{eq4.3.1}), so that $x_*$ can be one of $x_1, \cdots, x_n$ in the different integration regions.] \par
  The left-hand-side of (\ref{eq4.3.1}) is the sum of all the possible underlines, which can be a sum of two kinds of graphs with and without underlining for $x_*$, having the common part except for the vertex $x_*$.  This is depicted in Figure 4.3.2.
\begin{figure}{h}
\centering
\includegraphics[width=10cm, clip]{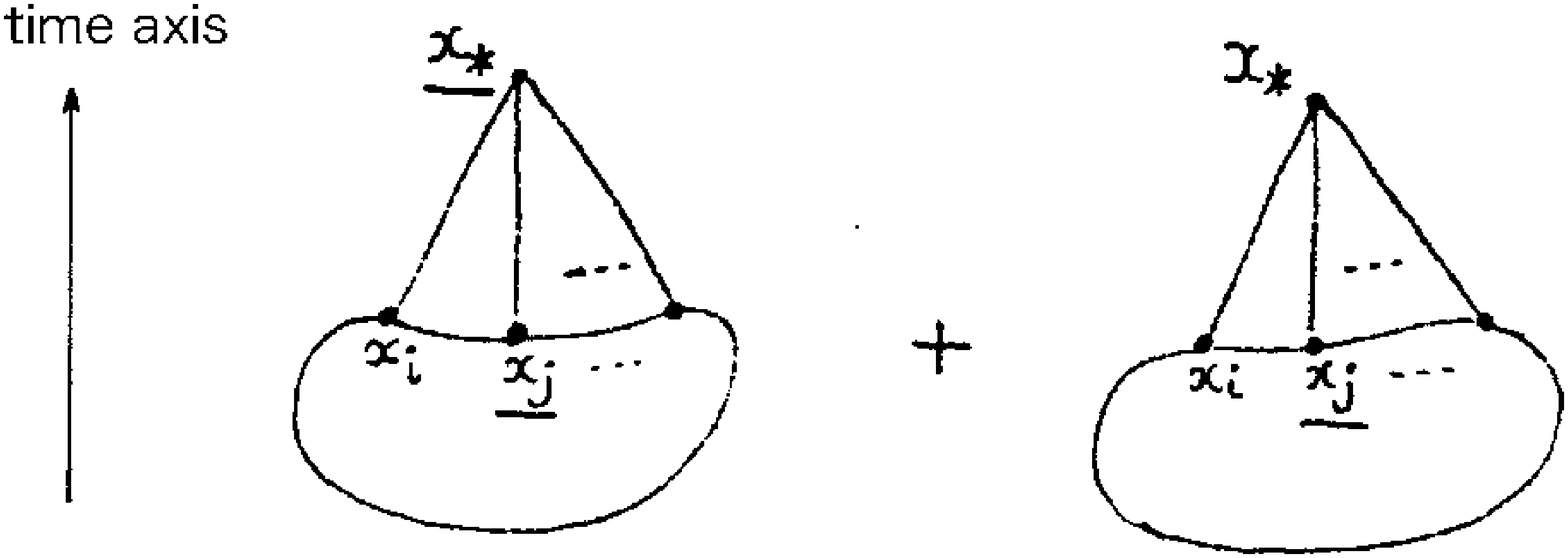}
\caption{Pictorial proof of the cutting rule}
\end{figure}
Here, the graphs inside a squashed circle are common about the underlining of the vertices.  Comparing the first term and the second term in Figure (\ref{eq4.3.2}), since $x_*$ has the maximum time $x_*^0$, the following relations hold:
\begin{align}
&\text{the first term} & & \text{the second term} \notag \\
& D(\underline{x_*}, x_i)^{(+)} &= &  D(x_*, x_i)  \label{eq4.3.17} \\
& D(\underline{x_*}, x_j)^{*} &= &  D(x_*, \underline{x_j}) , \label{eq4.3.18} 
\end{align}
and hence the propagators connecting to $x_*$ are common in the first and the second terms.  The vertex functions at $x_*$ have opposite signs with or withot the underline. [Please refer to the meaning of underline 1).] Thus, the first term and the second term in Figure (4.3.2) cancell with each other.  In this way, (\ref{eq4.3.1}) has been proved. \par
  To derive the unitarity from this cutting rule, first we make the scattering amplitude, by amputating the external lines and multiplying the proper wave functions, such as $\frac{1}{\sqrt{Z_2}}u(p, s) e^{-ipx}$, \par  
$\frac{1}{\sqrt{Z_2}}\bar{u}(p, s) e^{ipx}, \; \frac{1}{\sqrt{Z_3}}\varepsilon^{\mu} e^{-ipx}$, and integrating over $x_1, \cdots , x_n$.  Next, understand that $D(k)^{(+)}$ for the gauge boson propagator satisfies the following relation, under the $i\varepsilon$-rule fixed in the above [by substituting (\ref{eq4.3.15}) and (\ref{eq4.3.16}) into (\ref{eq4.3.6}) and (\ref{eq4.3.7}), respectively],
\begin{align}
D^{ab}_{\mu\nu}(k)^{(+)} &= D^{ab}_{\mu\nu}(-k)^{(-)} \notag \\
&= - 2\pi \delta^{ab} \left\{ \theta(k^0) \delta(k^2) \left[g_{\mu\nu} - \frac{k_{\mu}n_{\nu}+ n_{\mu}k_{\nu} }{k \cdot n} + (\alpha k^2 + n^2) \frac{k_{\mu}k_{\nu}}{(k \cdot n)^2} \right] \right.  \notag \\
& ~~~~\left. -\frac{1}{2} \delta(k \cdot n) \frac{1}{k^2} (k_{\mu}n_{\nu}+ n_{\mu}k_{\nu}) - \frac{1}{2} \delta'(k \cdot n) \frac{1}{k^2}  (\alpha k^2 + n^2)(k_{\mu}k_{\nu}) \right\},  \label{eq4.3.19}
\end{align}
[where $n^0 >0$.] \par
  Then, the conditions for the cutting rule to give the unitarity for the scattering amplitude yield the following two: \\
1) The terms of $\delta(k \cdot n)$ and $\delta'(k \cdot n)$ cancell in the cutting rule when applied to the scattering amplitude, [since these terms do not correspond to the physical emission of particles.] \\
2) Furthermore, the term of $\theta(k^0) \delta(k^2)$ becomes a sum of the product of physical polarization vectors.  That is, in the cutting rule of the scattering amplitude, the following substitution is allowed to reproduce the emission process of the physical particles, 
\begin{equation}
D^{ab}_{\mu\nu}(k)^{(+)} \to + 2 \pi \theta(k^0) \delta(k^2) \sum_{\text{pol.}} \varepsilon_{\mu}(k)^* \varepsilon_{\nu}(k)  \label{eq4.3.20}
\end{equation}
These two conditions are required to hold.  The proof of the conditons is a little complecated, so that we leave it to p.12--p.15 of the Reference paper II.  Then, the proof of the unitarity has finished.  Here, let us supplement the more fundamental things on the \underline{physical polarization vectors in the axial gauge}.\\
  It is known [by Kummer {\it et al.} \cite{23} and Frenkel \cite{24}] that the physical polarization vectors in the axial gauge QCD satisfy 
\begin{equation}
n^{\mu} \varepsilon^{a(\lambda)}_{\mu} = k^{\mu} \varepsilon^{a(\lambda)}_{\mu} ~~(\lambda=1, 2)  \label{eq4.3.21}
\end{equation}
Let us make this fact more familiar, by showing how to obtain the physical state condition, starting from the Feynman graphs. \par
  To begin with, the following things should hold, that is, \\
1) Physical state is invariant under the time development. For this to hold, physical state $\varepsilon_{\mu}^a(k)^{\text{phys.}}$ is the eigenvector of the dressed propatator $\tilde{D}^{ab}_{\mu\nu}(k)$, near the mass shell:
\begin{equation}
\tilde{D}^{ab}_{\mu\nu}(k) \varepsilon^{b, \nu}(k)^{\text{phys.}} \underset{k^2=0}{\approx} \Lambda(k) \varepsilon^{a}_{\nu}(k)^{\text{phys.}}.   \label{eq4.3.22}
\end{equation}
2) Propagator of connecting physical state by physical state is equal to the free propagator up to a factor.  That is, the following should hold,
\begin{equation}
\varepsilon^{a, \mu}(k)^{\text{phys.}} \tilde{D}^{ab}_{\mu\nu}(k) \varepsilon^{b, \nu}(k)^{\text{phys.}} \underset{k^2=0}{\approx} \Lambda(k) \frac{i Z_3}{k^2 + i \varepsilon}.  \label{eq4.3.23}
\end{equation}
Since the physical external filed is $\frac{1}{\sqrt{Z_3}}\varepsilon^a_{\mu}$, the propagator between the real physical states is equal to the free propagator. \par
  Let us derive (\ref{eq4.3.21}) from the physical state conditions 1) and 2). \\
  First we construct the dressed propagator $\tilde{D}^{ab}_{\mu\nu}(k)$.  With the proper self-energy tensor $\Pi^{ab}_{\mu\nu}$ and bare propagator $D^{ab}_{\mu\nu}(k)$, we have
\begin{equation}
\tilde{D}^{ab}_{\mu\nu}(k) = D^{ab}_{\mu\nu}(k) + \tilde{D}^{ac}_{\mu\lambda}(k) \Pi^{cd}_{\lambda\rho}(k)  D^{db}_{\rho\nu}(k)  \label{eq4.3.24}
\end{equation}
From this we have easilly
\begin{equation}
[\tilde{D}^{ab}_{\mu\nu}(k)]^{-1} = [D^{ab}_{\mu\nu}(k)]^{-1} - \frac{1}{i}   \Pi^{ab}_{\mu\nu}(k)   \label{eq4.3.25}
\end{equation}
[In the complicated gauge theories, it is better to use (\ref{eq4.3.25}).] \par
  Now, using the Ward-Takahashi identity [(\ref{eq4.2.11})] for $\Pi^{ab}_{\mu\nu}(k)$, we have
\begin{equation}
k^{\mu} \Pi^{ab}_{\mu\nu}(k)=0  \label{eq4.3.26}
\end{equation}
[It is because the color tensor for $(a, b)$ is not other than $\delta_{ab}$, so that it vanished as $f^{bca}\delta_{ab}=0$, after multipied by $f^{abc}$ in the W-T identity.] \par 
  In the axial gauge QCD, there exists one more vector $n_{\mu}$ which fixes the gauge, in additon to the incoming momentum vector $k_{\mu}$, so that we have to decompose $\Pi^{ab}_{\mu\nu}(k)$ into two tensors, using these two vectors, so that it satisfies (\ref{eq4.3.26}): 
\begin{equation}
\Pi^{ab}_{\mu\nu}= \delta^{ab} \left\{ \Pi^{(1)} [k^2 g_{\mu\nu} -k_{\mu} k_{\nu} ] + \Pi^{(2)} \frac{ [k_{\mu}(k\cdot n)-k^2 n_{\mu} ] [k_{\nu}(k\cdot n)-k^2 n_{\nu} ]}{(k \cdot n)^2} \right\}   \label{eq4.3.27}
\end{equation} 
[This is exactly same problem as in the deep inelastic scattering of electron and proton, in which two form factors $W_1$ and $W_2$ exist and are expressed by the two momenta of proton $n_{\mu}$ and of virtual photon $k_{\mu}$.]  Using (\ref{eq4.1.10}) and (\ref{eq4.3.27}) in (\ref{eq4.3.25}), we have obtained
\begin{align}
\tilde{D}^{ab}_{\mu\nu}(k)=& \frac{-i}{k^2 + i \varepsilon} \frac{1}{1+ \Pi^{(1)}}  \notag \\
& \times \left\{ \left[ g_{\mu\lambda} - \frac{k_{\mu}n_{\lambda}}{k\cdot n} \right] \left[ g_{\nu}^{\lambda} - \frac{n^{\lambda}k_{\nu}}{k\cdot n} \right] + \frac{\Pi^{(2)}}{ [1- k^2 n^2 /(k\cdot n)^2] \Pi^{(2)} - [ 1+ \Pi^{(1)} ] }  \right. \notag \\
&\left. ~~~\times \frac{k^2}{(k \cdot n)^4} [ (k\cdot n) n_{\mu} - n^2 k_{\mu} ]  [ (k\cdot n) n_{\nu} - n^2 k_{\nu} ] \right\}  \notag \\
& +  \frac{-i}{k^2 + i \varepsilon} \alpha k^2 \frac{k_{\mu} k_{\nu}}{(k \cdot n)^2}.   \label{eq4.3.28}
\end{align}
Using this, let us make the physical states $\varepsilon_{\mu}^{a(\text{phys.})}$ so as to meet the physical state conditons 1) and 2). \par
  First write down all the eigen-values and eigen-vectors for $\tilde{D}_{\mu\nu}(k)$ [up to $\delta^{ab}$].  It is clear that $\varepsilon_{\mu}(k)^{\lambda} ~[\lambda=1, 2]$ are the eigen-vectors, and the remaining eigen-vectors  $\varepsilon_{\mu}(k)^{\lambda} ~[\lambda=0,3]$ can be obtained as linear combinations of $(n_{\nu}, k_{\nu})$, as follows: 
\begin{align}
&\varepsilon_{\nu}^{(0)} = N^{(0)} [ n_{\nu} + c_+ k_{\nu} ]  \label{eq4.3.29} \\
&\varepsilon_{\nu}^{(3)} = N^{(3)} [ n_{\nu} + c_- k_{\nu} ],  \label{eq4.3.30}
\end{align}
where $c_{\pm}$ satisfy the following second order equation:
\begin{align}
&c^2 \delta + c(\delta n^2 + \alpha k^2) /(k \cdot n) + \alpha=0 ,  \label{4.3.31} \\
\text{and the solutions are} & \notag \\
& c_{\pm} = \frac{1}{2\delta} [ - (\delta n^2 + \alpha k^2) /(k \cdot n) \pm \sqrt{ ( \delta n^2 + \alpha k^2)^2/(k \cdot n)^2 - 4 \alpha \delta}  \label{eq4.3.32}
\end{align}
Here, the $\delta$ is defined by
\begin{align}
&\delta \equiv 1/ 1+\Pi^{(1)} -(1-\gamma)\Pi^{(2)}  \label{eq4.3.33} \\
&\gamma \equiv k^2 n^2 /(k \cdot n)^2 \label{eq4.3.34}
\end{align}
The normalization constants $N^{(0)},~ N^{(3)}$ are fixed so as to satisfy the following normalization condition and the completeness conditon,
\begin{align}
& g^{\mu\nu} \varepsilon_{\mu}^{(i)} \varepsilon_{\nu}^{(j)} = g^{ij}  \label{eq4.3.35} \\
&g_{ij} \varepsilon_{\mu}^{(i)} \varepsilon_{\nu}^{(j)} = g_{\mu\nu}  \label{eq4.3.36}
\end{align}
The eigen-values for these eigen-vectors are
\begin{align}
& \Lambda^{(0)} = \frac{-i}{k^2 + i \varepsilon} \frac{k^2}{(k\cdot n)^2}  \times (-1) c_+ \delta,  \label{eq4.3.37} \\
& \Lambda^{(1)} = \Lambda^{(2)} = \frac{-i}{k^2 + i \varepsilon} \frac{1}{1+ \Pi^{(1)}},  \label{eq4.3.38} \\
&\Lambda^{(3)} = \frac{-i}{k^2 + i \varepsilon} \frac{k^2}{(k\cdot n)^2}  \times (-1) c_- \delta,  \label{eq4.3.39}
\end{align}
where the pole $k^2=0$ is cancelled for the unphsysical polarizations $\Lambda^{(0)}$ and  $\Lambda^{(3)}$.  The physical state conditon  2) of having the pole at $k^2=0$ is satisfied for $\varepsilon_{\nu}^{(1)}$ and $\varepsilon_{\nu}^{(2)}$, so that these two are physical state polarizations.  This is a way to construct the physical states. 
[In case of the gauge parameter $\alpha \to 0$, the discussion is simplified to 
\begin{equation}
c_+(\alpha=0) =0, ~ c_-(\alpha=0) = - \frac{n^2}{k \cdot n} \label{eq4.3.40}
\end{equation}
In this case our physical states reproduce those by Kummer {\it et al.} \cite{23}. \\
  Here, we note that if $\varepsilon_{\mu}$ is an eigen-vector of $\tilde{D}_{\mu\nu}$, then $n^{\mu} \cdot \varepsilon_{\mu}=0 $ can be derived from $k^{\mu} \cdot \varepsilon_{\mu}=0$; indeed from
\begin{equation}
\Lambda (n^{\mu} \cdot \varepsilon_{\mu}) = n^{\mu} \tilde{D}_{\mu\nu} \varepsilon^{\nu} = -i \alpha \frac{1}{k \cdot n} (k^{\nu} \cdot \varepsilon_{\nu})=0  \label{eq4.3.41}
\end{equation}
we have $n^{\mu} \cdot \varepsilon_{\mu}=0.$ \\
  The unitarity proved in this way, can be used to show the cancellation of infrared divergences in the \underline{total cross sections} \cite{41}. 
For example, let us consider the infrared divergences included in the total cross section of $\sigma(e^+ e^- \to$ hadrons).  It is related to the photon's self-energy $\Pi(q^2)$ by unitarity, and is proved to have no infrared divergences for $q^2 \ne 0$ \cite{41}, which shows that at least the total cross section has no infrared divergences.  There is, however, no general theory exists about the cancellation of infrared divergences, when the mmomenta of the final quarks are fixed.  [ Appelquist {\it et al.} have checked explicitly at two loops for this electron-positron scattering process.  The content of Section 3 is the check of this cancellation of infrared divergeces for the other processes.] 
\subsection{Generalization of F. E. Low's low energy theorem to QCD (Part 1)}
Using the axial gauge QCD studied in Section 4.1--Section 4.3, we will prove the F. E. Low"s low energy theorem (abbreviated  as \underline{Low's theorem} in the following) at all orders of the perturbation.  [There exists in the past the erxplicit check of it at one loop in the high energy limit \cite{20}, but the proof at all orders is a new performance.] 

The process which we are going to consider is the emission of one or two soft gauge bosons in the fermion scattering process by a colorless external field.  In case of the emission of two soft gauge bosons, the proof is restricted for the two to have the same color indices.  [In the reference paper II, only the single soft gauge boson emission process was studied.]  Let us give the Low's theorem in figures. 

That is, they are given in the axial gauge as follows:
\begin{align}
&\begin{minipage}{14cm}
\includegraphics[width=12cm, clip]{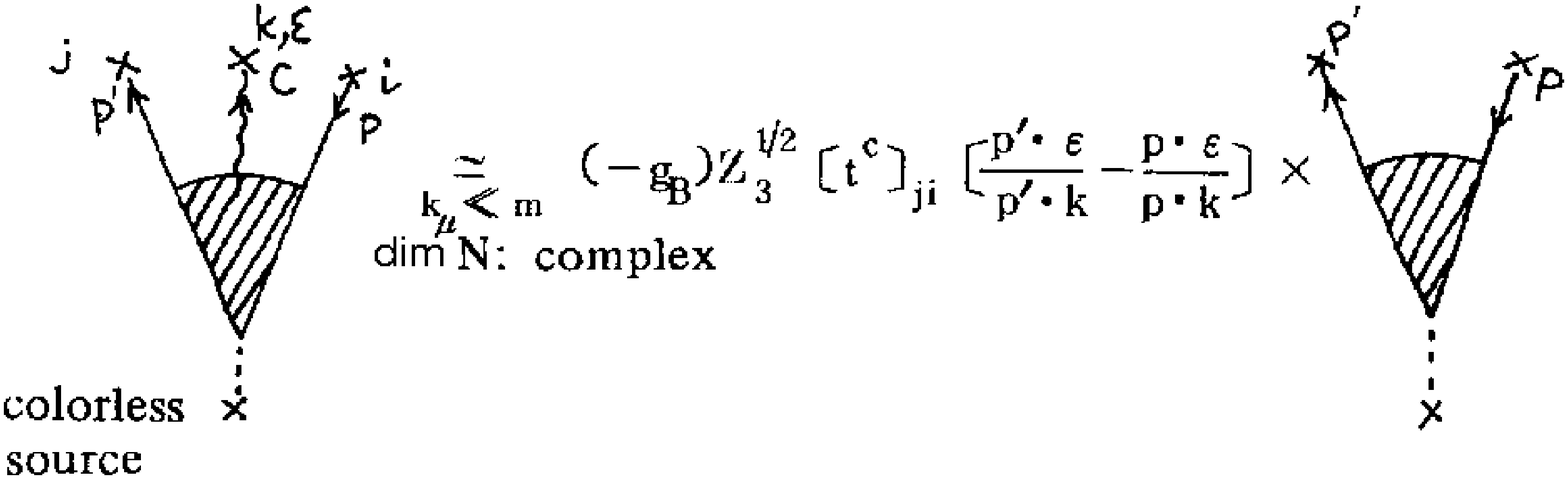}
\end{minipage}
\label{eq4.4.1}  \\
&\begin{minipage}{14cm}
\includegraphics[width=12cm, clip]{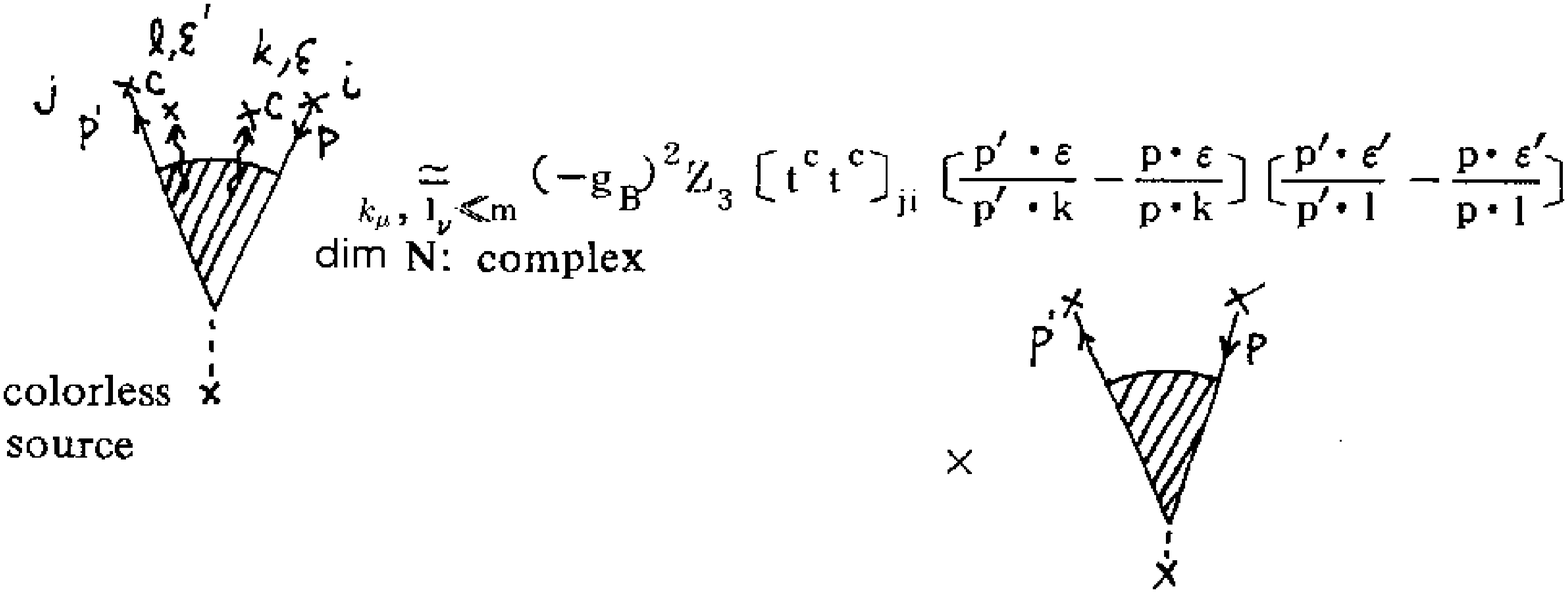}
\end{minipage}
\label{eq4.4.2}
\end{align}
[The proof of (\ref{eq4.4.2}) will be given in Section 4.5.]

The meaning of the notations is clear, but $\{p, p', k, l \}$ are 4-momenta, $\{i, j, c\}$ are color indices, and $\{\varepsilon, \varepsilon' \}$ are polarization vectors.  (\ref{eq4.4.1}) and (\ref{eq4.4.2}) are \underline{unrenormalized relations}, being expanded in the bare coupling constant $g_B$.  The both hand sides of these equations have infrared divergences, since the external lines are on the mass shell [and the transverse wave for gauge bosons], which are assumed to be regularized by the \underline{N-dimensional method}. 

That is, these are relations which hold in the limit of $k_{\mu}$ (and $l_{\mu}$) be soft, after relularized by N-dimensional method [or performing the analytical continuation of N to complex].  [In case of the regularization of infrared divergence by introducing a mass $\lambda$ to gauge bosons, the limit of (\ref{eq4.4.1}) and (\ref{eq4.4.2}) to hold is $|k_{\mu}| \ll \lambda \ll m$. The N dimensoinal method is more convenient, since the ultraviolet divergences are regularized at the same time.]  $Z_3$ represents the corrections for the external line of gauge boson, being the renormalization constant defined on the mass shell, [please refer to (\ref{eq4.3.23}) and (\ref{eq4.3.38}).]

For the later convenience, we will write a formula of (\ref{eq4.4.2}) after \underline{amputated the external line} of the gauge boson. That is, 
\begin{equation}
\begin{minipage}{14cm}
\includegraphics[width=12cm, clip]{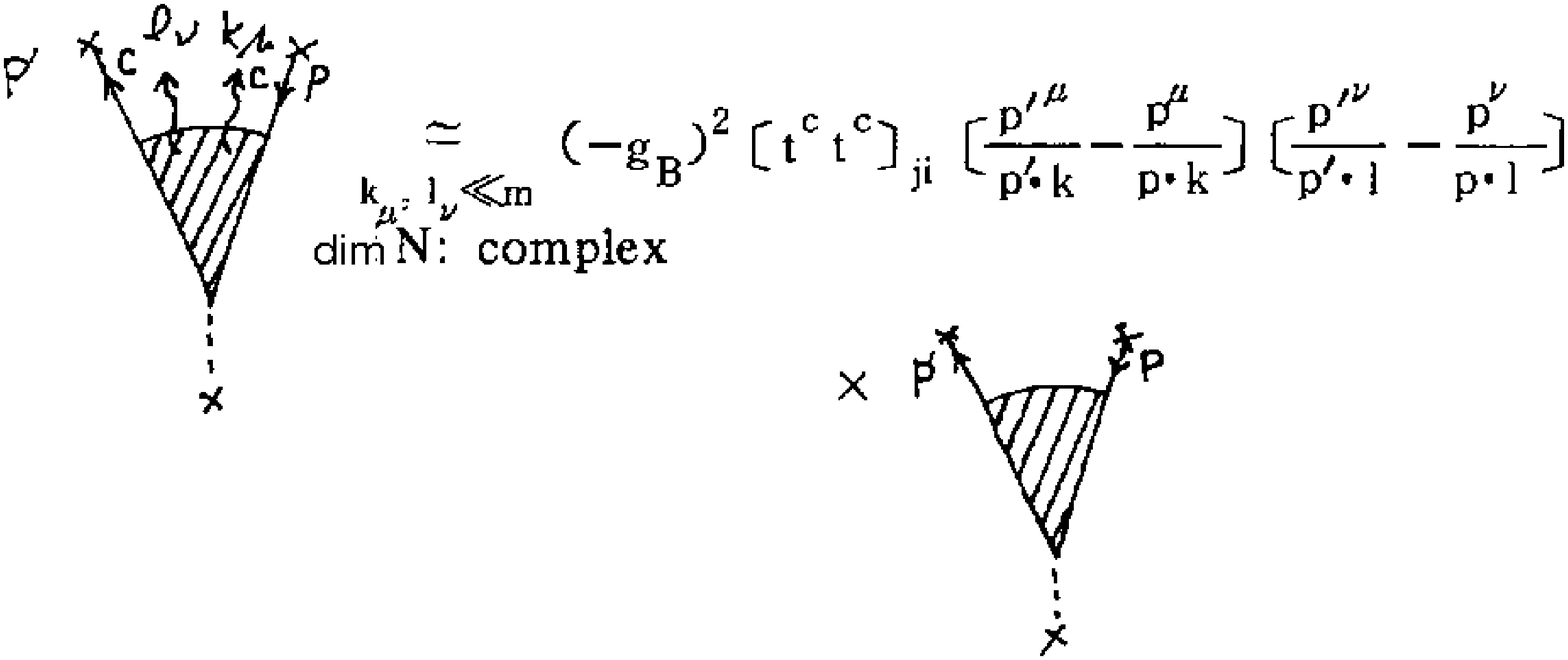}
\end{minipage}
\label{eq4.4.3}
\end{equation}
[This will be proved in Section 4.5.]

Before beginning the proof, let us remind of the relations (\ref{eq2.3.1}), (\ref{eq2.3.7}), (\ref{eq2.2.17}) in QED discussed in Section 2.  Note that the above mentioned (\ref{eq4.4.1})--(\ref{eq4.4.3}) are the generalization of these relations to QCD.  It is however, the Low's theorem in QED [(\ref{eq2.3.1}), (\ref{eq2.3.7}), (\ref{eq2.2.17})] can be proved, as was explained in details in Section 2, by using the Eikonal identities (\ref{eq2.2.14}), (\ref{eq2.2.15}), but in QCD, these Eikonal identities can be applied \underline{directly}.  The Eikonal identities hold when summing up all the possible soft photon emissions from the external electron lines, but the same strategy does not work in QCD, in which two soft gauge bosons emitting from the external quark line have different color indices under the exchange of the two, for the non-Abelian group. In QCD, the soft gauge boson can be also emitted from the inner lines of the gauge boson.  Therefore, we have to sum over all the possible soft gauge boson emissions [not only from the external fermion lines, but also from the inner gauge boson lines].

It is easily understood that a soft photon emission from all the possible places in QED can be realized by differentiating the 
independent external momenta for fermions, as well as the independent loop-momenta of the inner fremion loops.

As an example, let us consider the electron scattering by an external field in Section 2.  The process with an additional photon emission with zero momentum is given by
\begin{equation}
\begin{minipage}{14cm}
\includegraphics[width=12cm, clip]{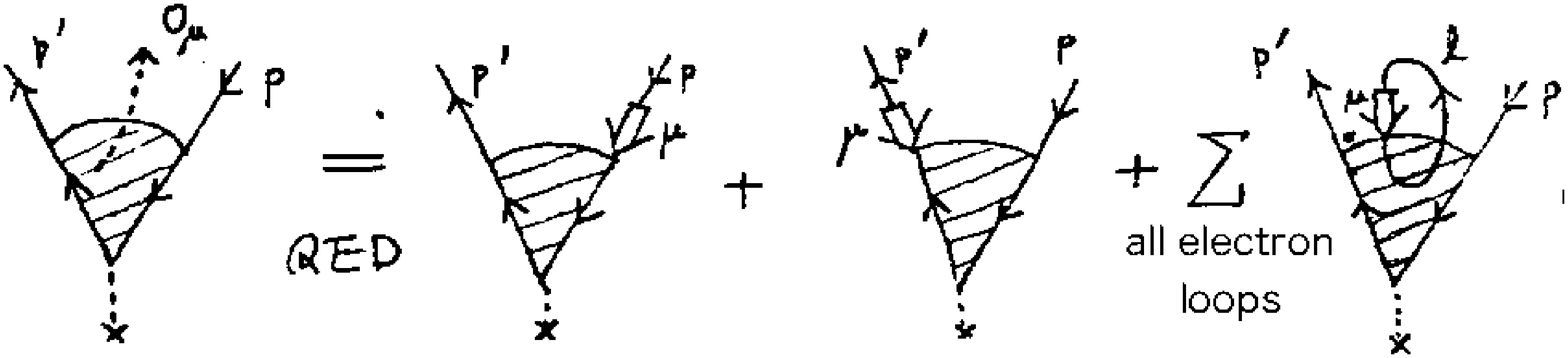}
\end{minipage}
\label{eq4.4.4}
\end{equation}
Here, the arrow symbol $\Rightarrow$ in QED is defined by
\begin{equation}
\begin{minipage}{14cm}
\includegraphics[width=10cm, clip]{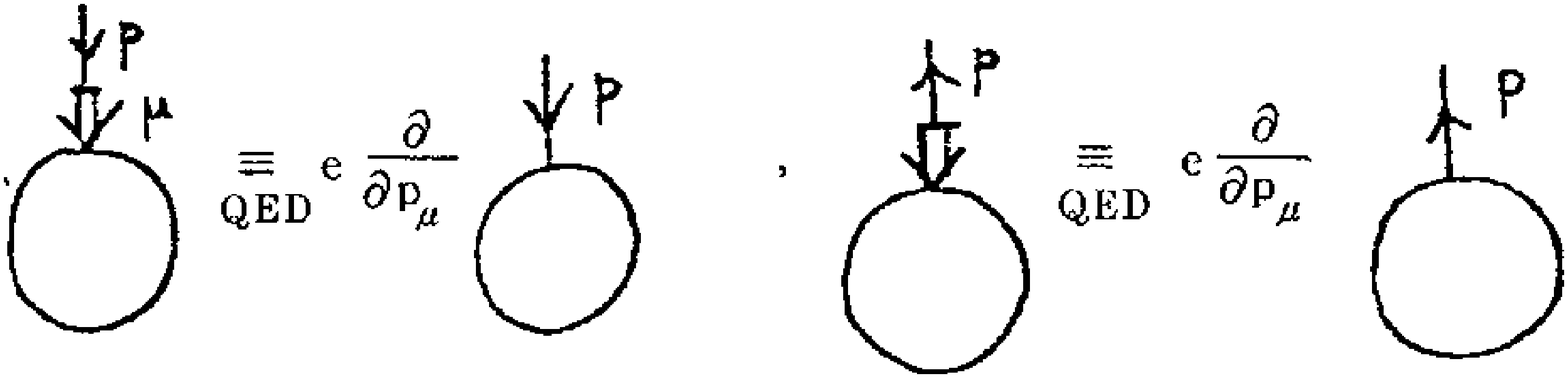}
\end{minipage}
\label{eq4.4.5}
\end{equation}
[Remind of (\ref{eq2.2.9}) and (\ref{eq2.2.10a}) or  (\ref{eq2.2.10b}) in Section 2]   The momentum $p$ is assumed to flow along the electron line and going out to the external source, while the momentum $p'$ enters from the source and flow out to the emitted electron line.  
  Then, the first term represents the emission of soft photons from the orbit of electron on the right, the second term does from the orbit of the electron on the left, and the third term does the emission of photon from the inner loops of electrons.

In QCD we need to have the \underline{differential operation with color index} in order to generate a gauge boson emission with zero momentum and color index $c$.  That is, corresponding to (\ref{eq4.4.5}), we have to introduce the following arrow symbol $\Rightarrow$ in QCD.  Namely,
\begin{align}
&\begin{minipage}{14cm}
\includegraphics[width=6cm, clip]{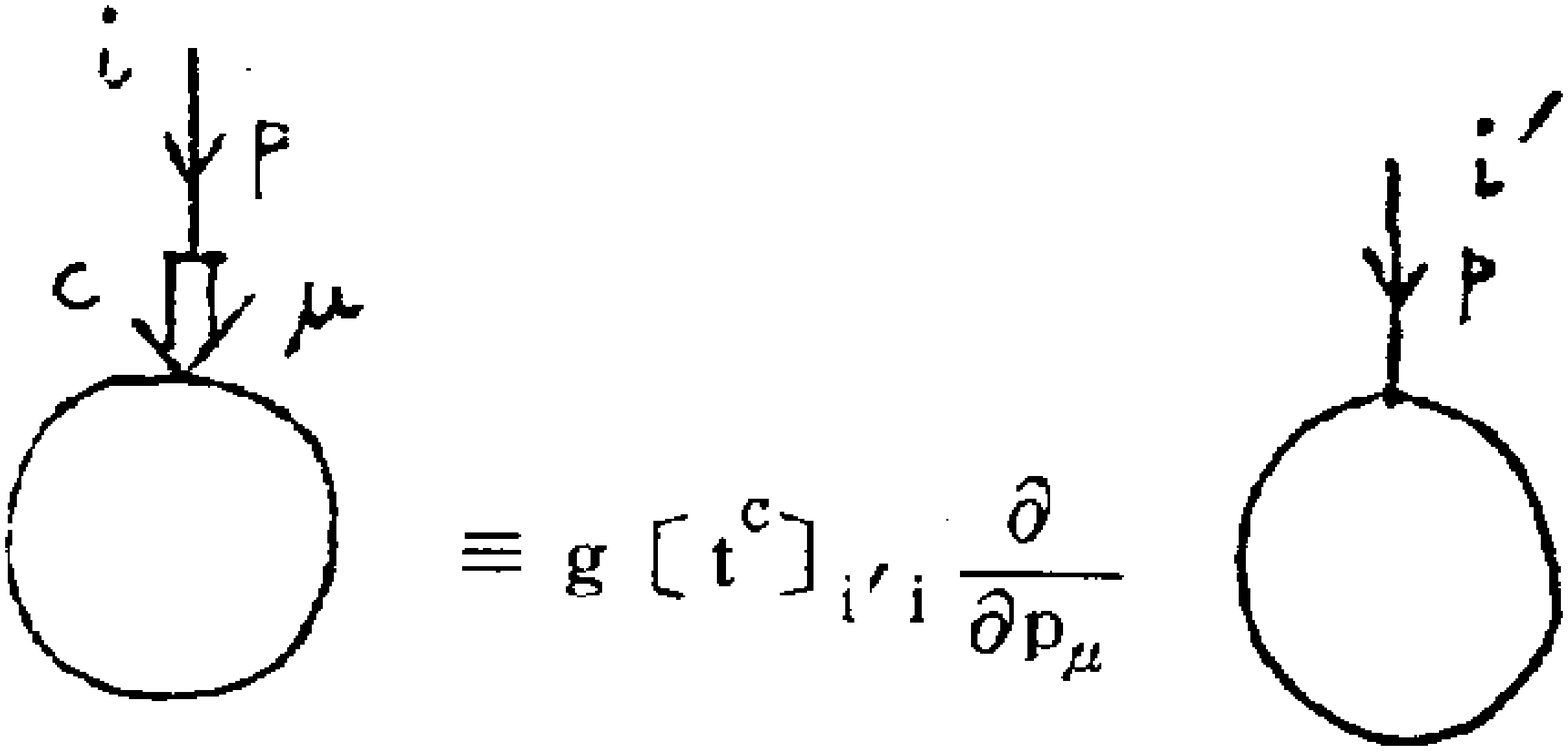}
\end{minipage}
\label{eq4.4.6} \\
&\begin{minipage}{14cm}
\includegraphics[width=6cm, clip]{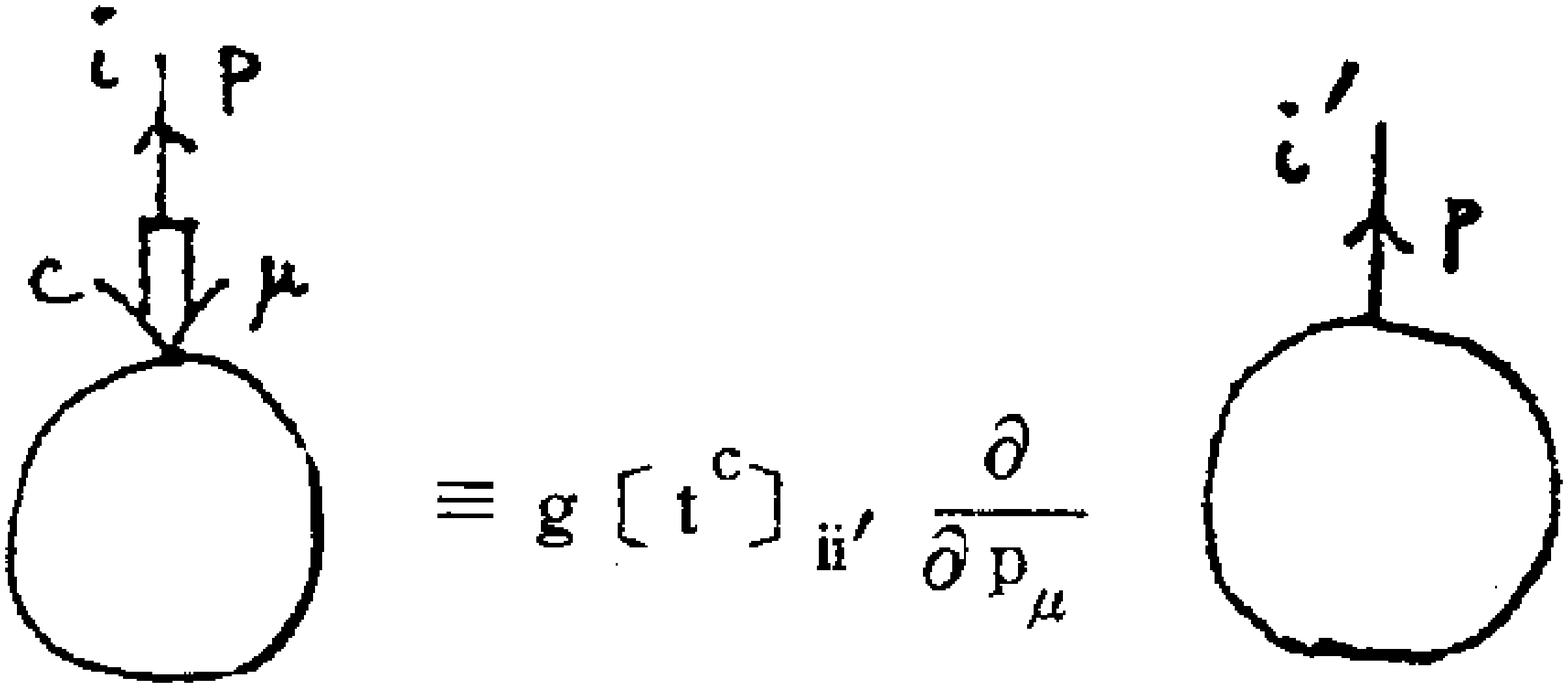}
\end{minipage}
\label{eq4.4.7} \\
&\begin{minipage}{14cm}
\includegraphics[width=6cm, clip]{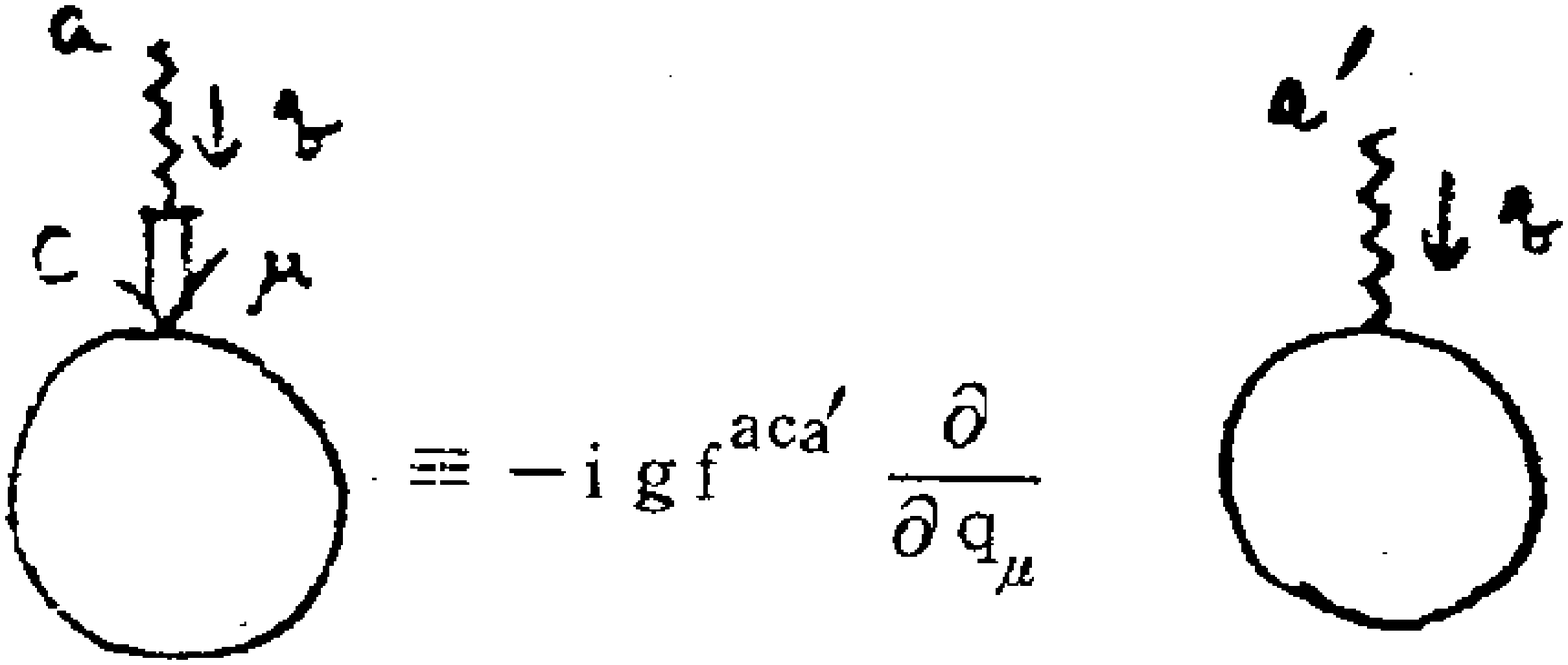}
\end{minipage}
\label{eq4.4.8}
\end{align}
The direction of the arrow symbol points along the part to differentiate.  Now, using this arrow symbol in the following, let us generate the emission of gauge boson with zero momentum and a color $c$.  First, for the simplest graphs, the following relations hold \cite{42}:
\begin{align}
&\begin{minipage}{14cm}
\includegraphics[width=10cm, clip]{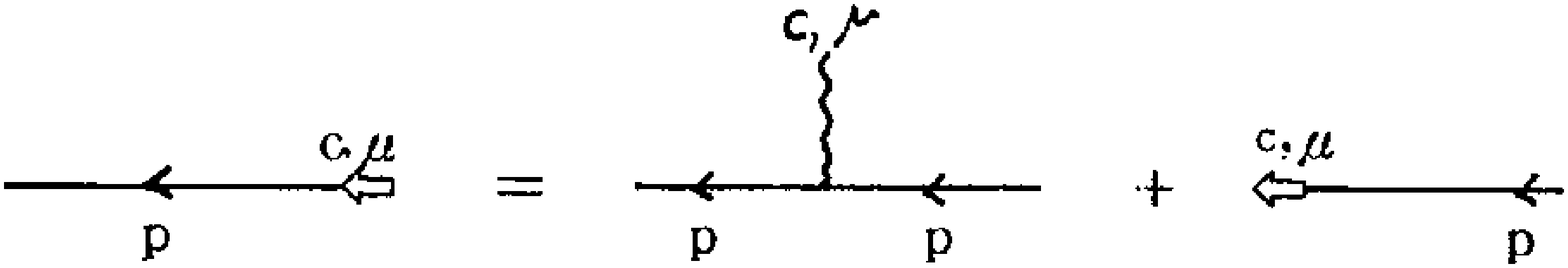}
\end{minipage}
\label{eq4.4.9} \\
&\begin{minipage}{14cm}
\includegraphics[width=10cm, clip]{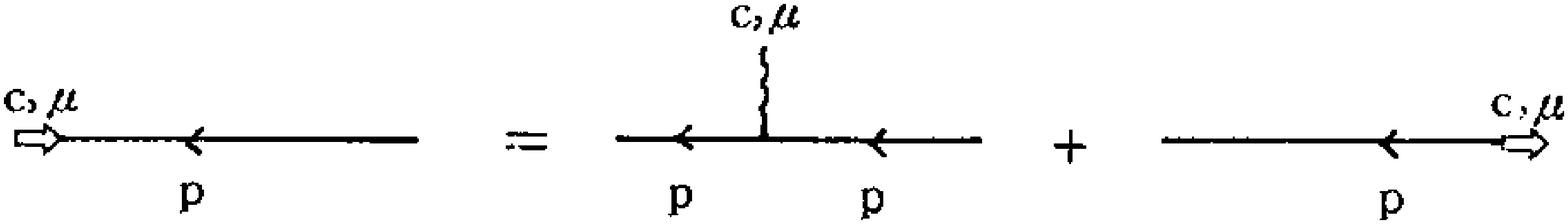}
\end{minipage}
\label{eq4.4.10} \\
&\begin{minipage}{14cm}
\includegraphics[width=10cm, clip]{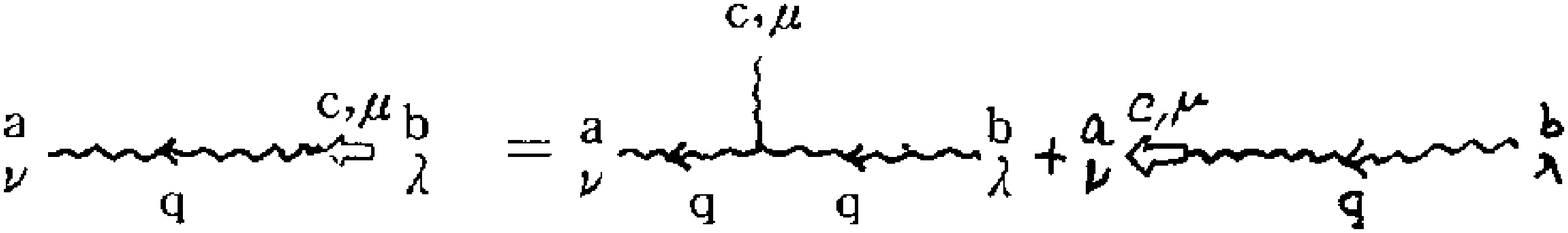}
\end{minipage}
\label{eq4.4.11} \\
&\begin{minipage}{14cm}
\includegraphics[width=10cm, clip]{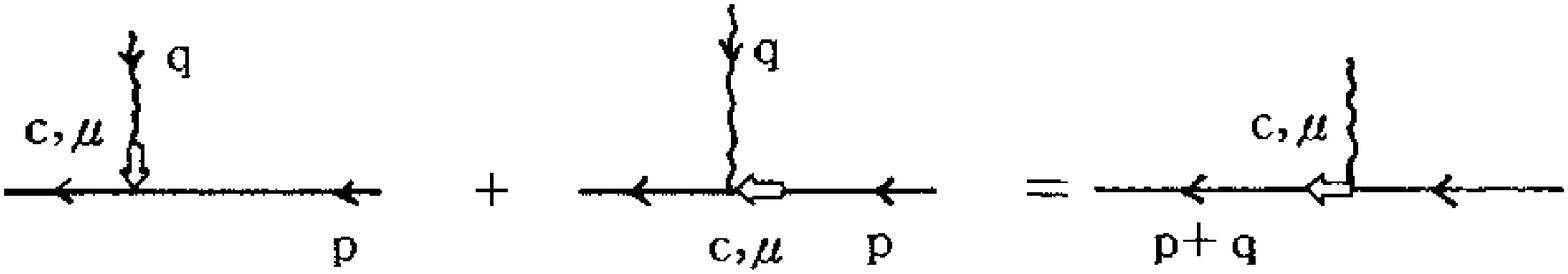}
\end{minipage}
\notag \\
&\begin{minipage}{14cm}
\includegraphics[width=10cm, clip]{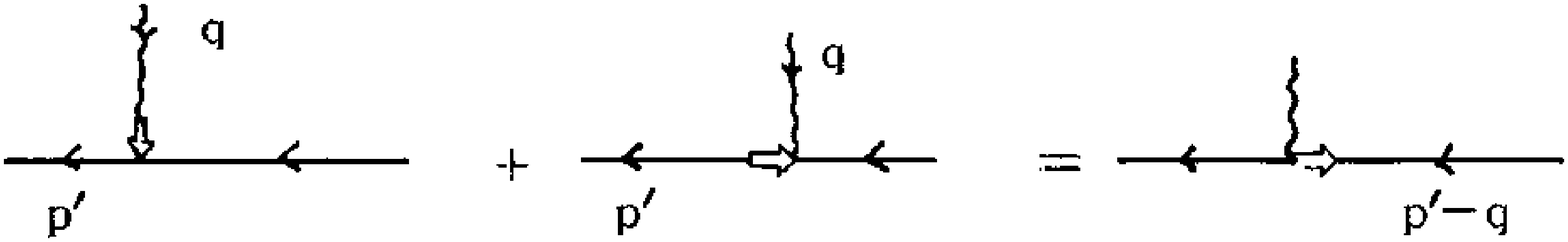}
\end{minipage}
\notag \\
&\begin{minipage}{14cm}
\includegraphics[width=10cm, clip]{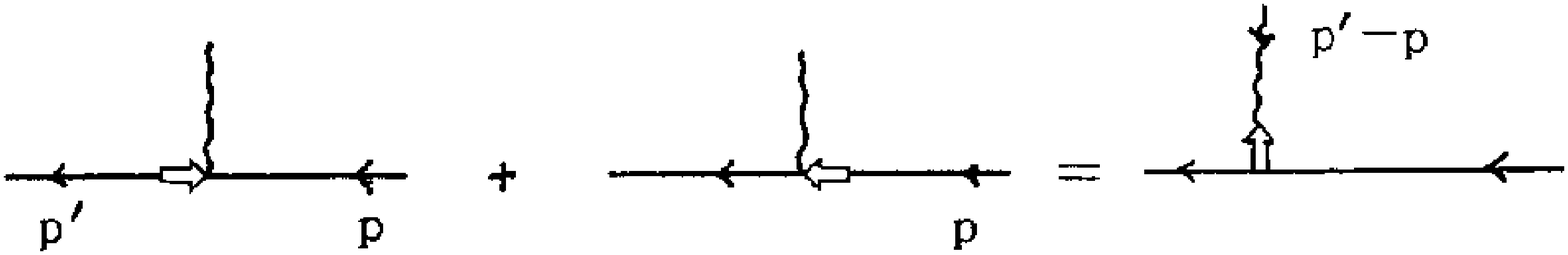}
\end{minipage}
\label{eq4.4.12} \\
&\begin{minipage}{14cm}
\includegraphics[width=10cm, clip]{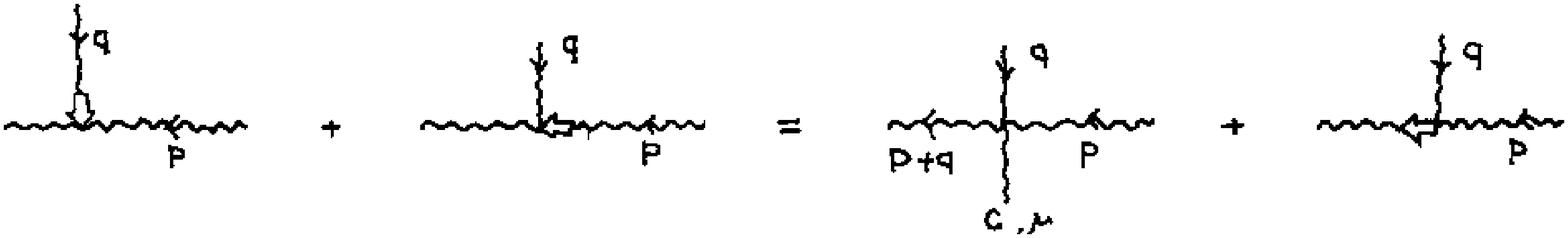}
\end{minipage}
\label{eq4.4.13} \\
&\begin{minipage}{14cm}
\includegraphics[width=10cm, clip]{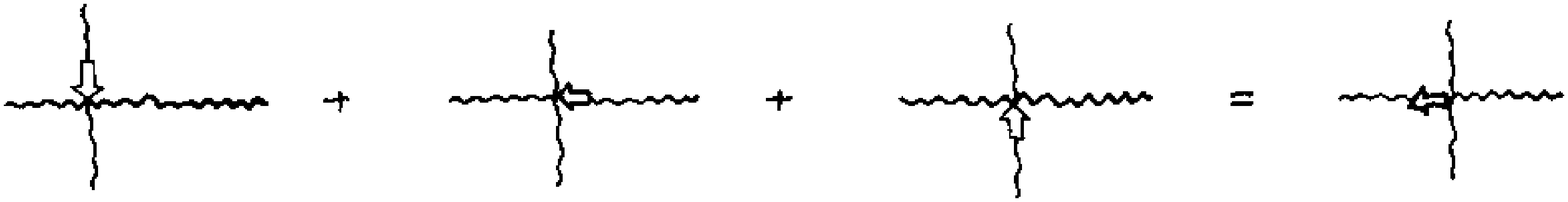}
\end{minipage}
\label{eq4.4.14}
\end{align}
These formulae correspond to (\ref{eq4.2.5})--(\ref{eq4.2.9}) in Section 2, and can be chcked directly by using the Feynman rules (\ref{eq4.1.10})--(\ref{eq4.1.14}), the fundamental relation of the group $[t^a, t^b]=i f^{abc} t^c$, and Jacobi identities.  
The characteristic of the axial gauge is (\ref{eq4.4.11}), which becomes complicated in the other gauge, by including additional terms.  [ Among (\ref{eq4.4.9})--(\ref{eq4.4.14}), only (\ref{eq4.4.11}) depends on the gauge condition, and the others are common in any gauge. It should be examined, however, if there are no gauge conditions other than the axial gauge which satisfies (\ref{eq4.1.11}).]

With the use of these lowest order identities (\ref{eq4.4.6})--(\ref{eq4.4.14}), the following identities can be derived for any tree graph, by acting the arrow operations (\ref{eq4.4.6})--(\ref{eq4.4.8}) to one of the external lines:
\begin{equation}
\begin{minipage}{14cm}
\includegraphics[width=10cm, clip]{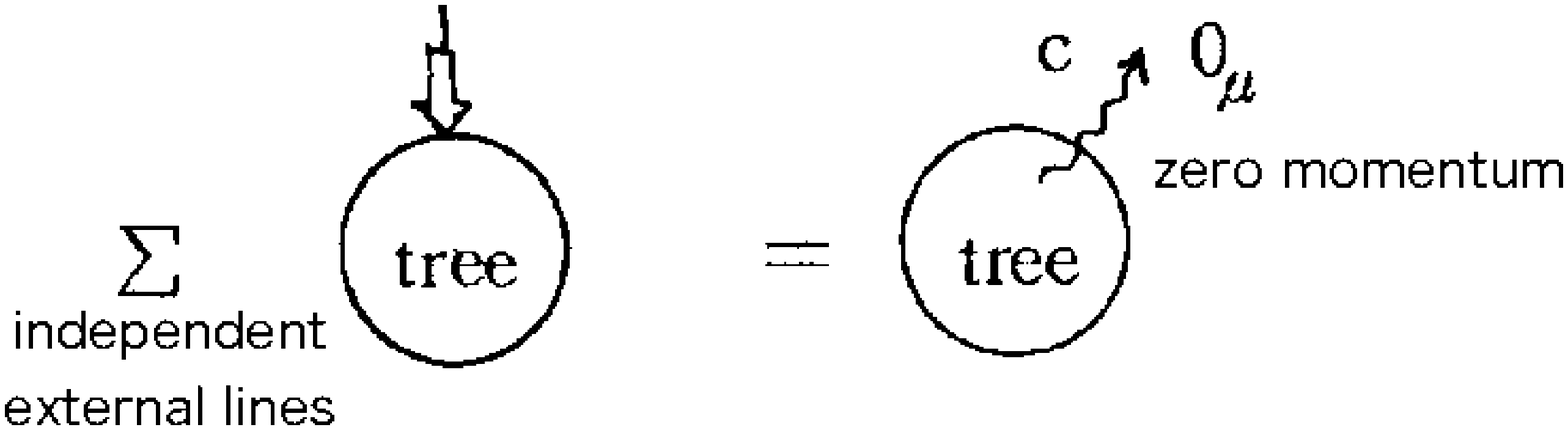}
\end{minipage}
\label{eq4.4.15}
\end{equation}
The proof of it is simple.\footnote{Footnote added (2022): The left-hand-side of (\ref{eq4.4.15}) with the arrow symbol $\Rightarrow$ gives QED-like soft photon emissions from the external lines, while the right-hand-side affords all the possible soft gluon emissins.}  First, understand the characteristics of the relations (\ref{eq4.4.9})--(\ref{eq4.4.14}).  They are summarized to 

1)When the arrow symbol passes through the propagator, the gauge boson with zero momentum and color $c$ is emitted from the propagator.   [(\ref{eq4.4.9})--(\ref{eq4.4.11})] 

2) As for the vertex, the arrow symbols approach along the flows of independent momenta, join and go out to the line with a not-independent momentum.  Only when a gauge boson emission is possible from the vertex, the graph with the emission of a gauge boson with zero momentum and color $c$, remains additionally. [(\ref{eq4.4.12})--(\ref{eq4.4.14})]

First we fix the flow of independent momenta, then there remain all the possible gauge boson emission graphs, and the arrows are joined and gone out to the unique external line with non-independent momentum.  This is the proof of (\ref{eq4.4.15}).

Next, we generalize (\ref{eq4.4.15}) to the more general graphs with loops.  First, write down the result, and prove it afterwards.  The general result can be depicted as follows \cite{43}:
\begin{equation}
\begin{minipage}{14cm}
\includegraphics[width=10cm, clip]{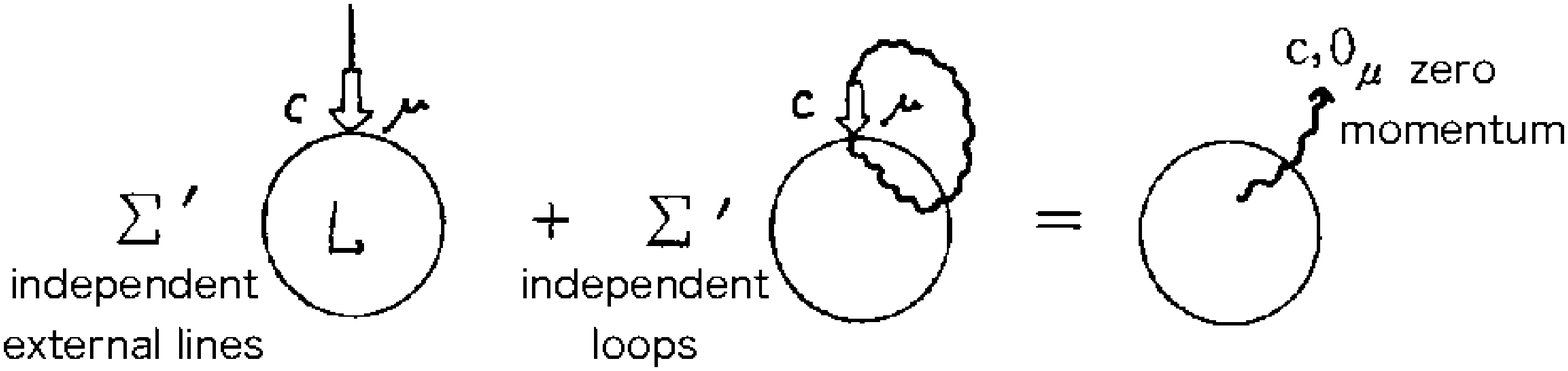}
\end{minipage}
\label{eq4.4.16}
\end{equation}
The prove is done by induction with respect to the number of loops.  
First, for a tree graph (the number of loops $L=0$), (\ref{eq4.4.16}) is equal to (\ref{eq4.4.15}) and is correct. So, assuming (\ref{eq4.4.16}) holds for the graphs with loops less than $L$, consider a graph with  $L+1$ loops.  The left hand side of (\ref{eq4.4.16}) for the graph with $L+1$ loops, can be deformed as follows:
\begin{equation}
\begin{minipage}{14cm}
\includegraphics[width=12cm, clip]{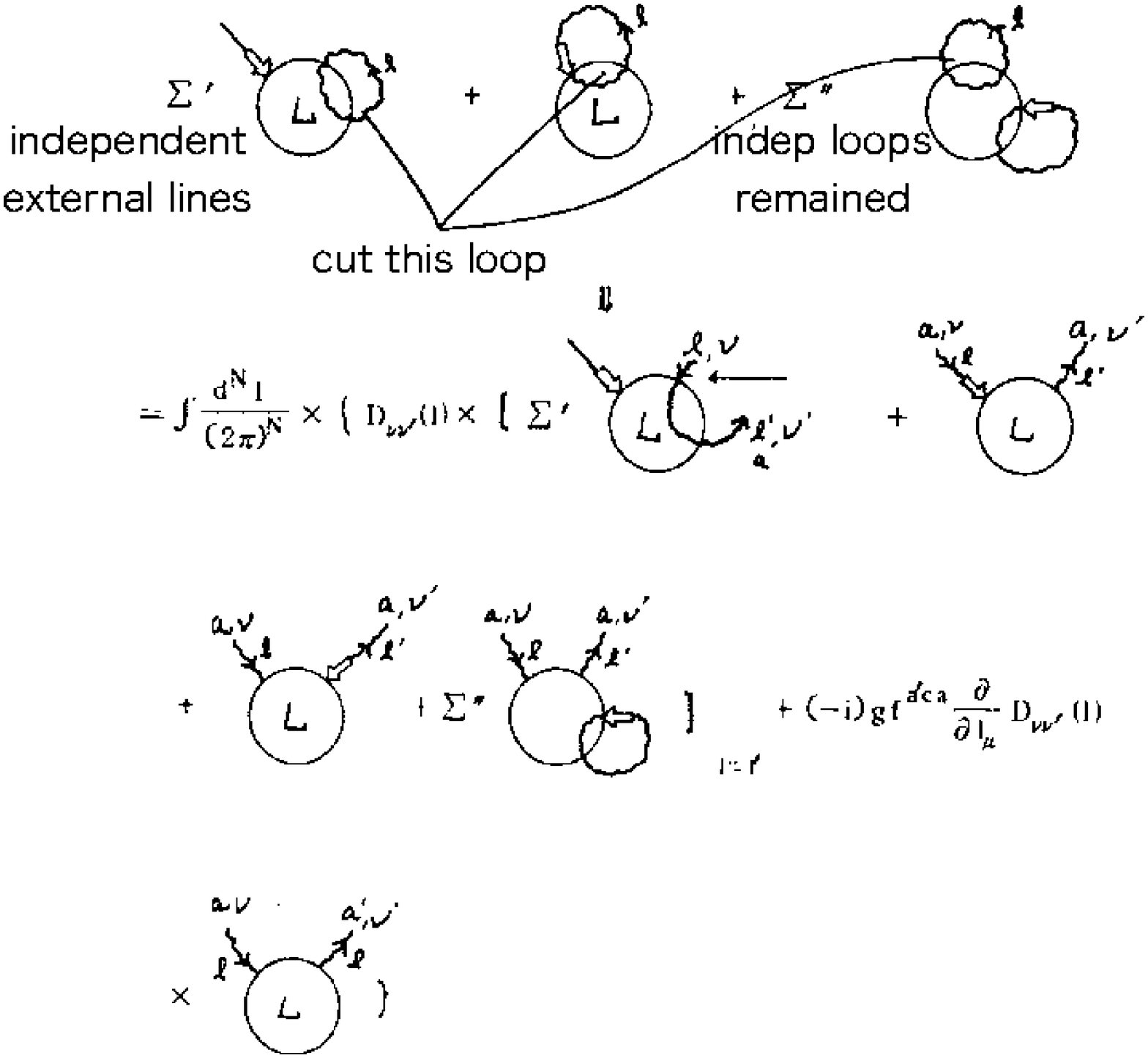}
\end{minipage}
\label{eq4.4.17}
\end{equation}
Here, we have used the following identity:
\begin{equation}
\begin{minipage}{14cm}
\includegraphics[width=10cm, clip]{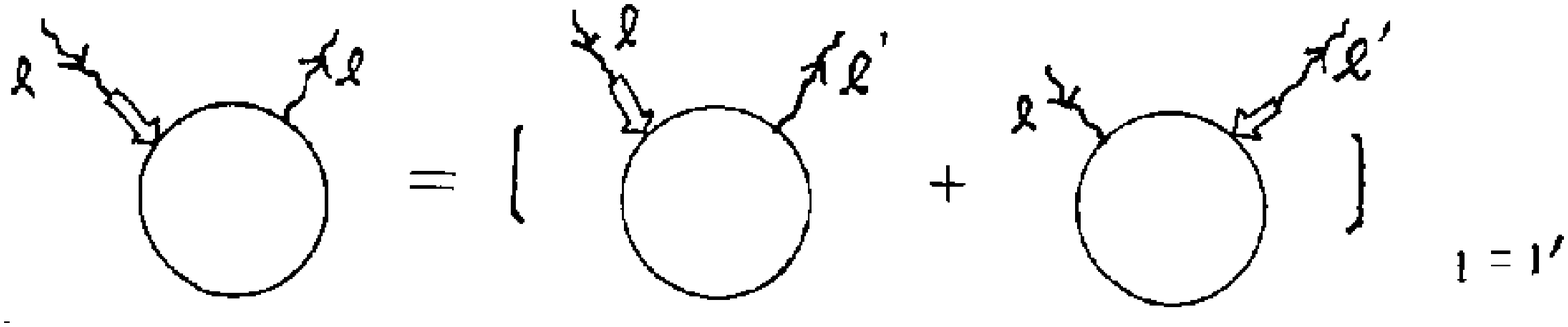}
\end{minipage}
\label{eq4.4.18}
\end{equation}
which is a pictorial representation of the following equation,
\begin{equation}
\frac{\partial}{\partial l_{\mu}} F(l, l)= \left( \frac{\partial}{\partial l_{\mu}} + \frac{\partial}{\partial l'_{\mu}} \right) F(l, l') | _{l=l'}.    \label{4.4.19}
\end{equation}
The external lines of gauge bosons with momenta $l, \; l'$ in the right hand side of (\ref{eq4.4.17}) are amputated. So, the last term in the right hand side of (\ref{eq4.4.17}) represents the differentiation of the amputated propagator.  Now, using the assumption of the induction in (\ref{eq4.4.17}), we understand
\begin{equation}
\begin{minipage}{14cm}
\includegraphics[width=10cm, clip]{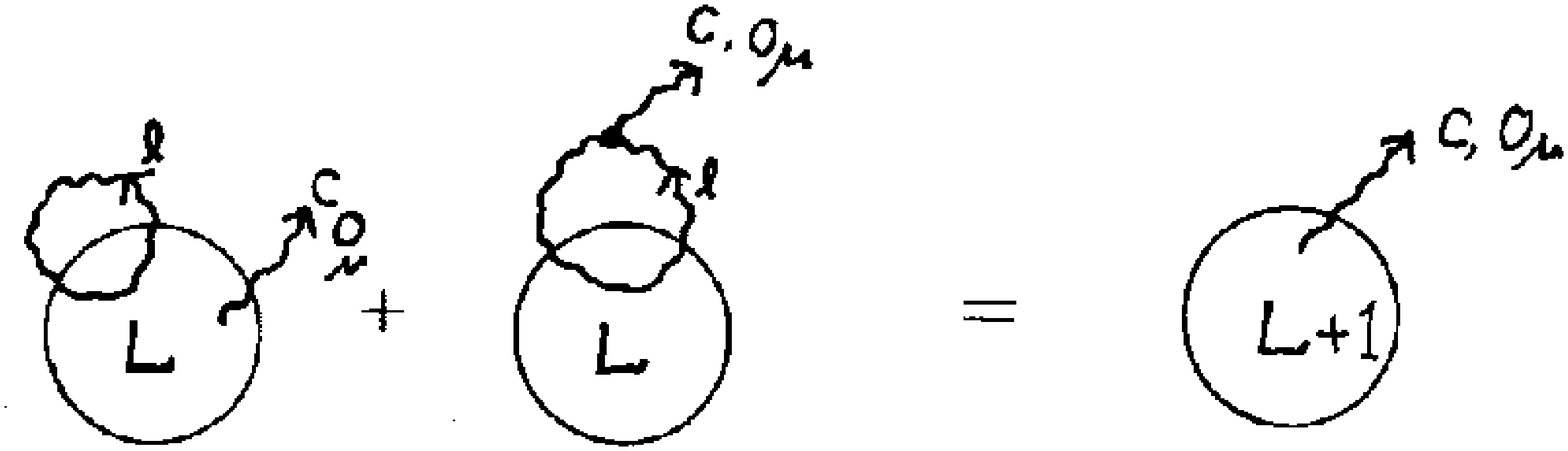}
\end{minipage}
\label{eq4.4.20}
\end{equation}
which shows that the (\ref{eq4.4.16}) is valid for the case with $(L+1)$ loops, and the proof of (\ref{eq4.4.16}) for the graph with any loop has been completed.  [In the above proof we add gauge boson loops, but we can add fermion loops in the same way.]

Comparing this (\ref{eq4.4.16}) with (\ref{eq4.4.1}) in QED, we can understand that the contribution from the independent loops in QCD,
\begin{equation}
\begin{minipage}{14cm}
\includegraphics[width=6cm, clip]{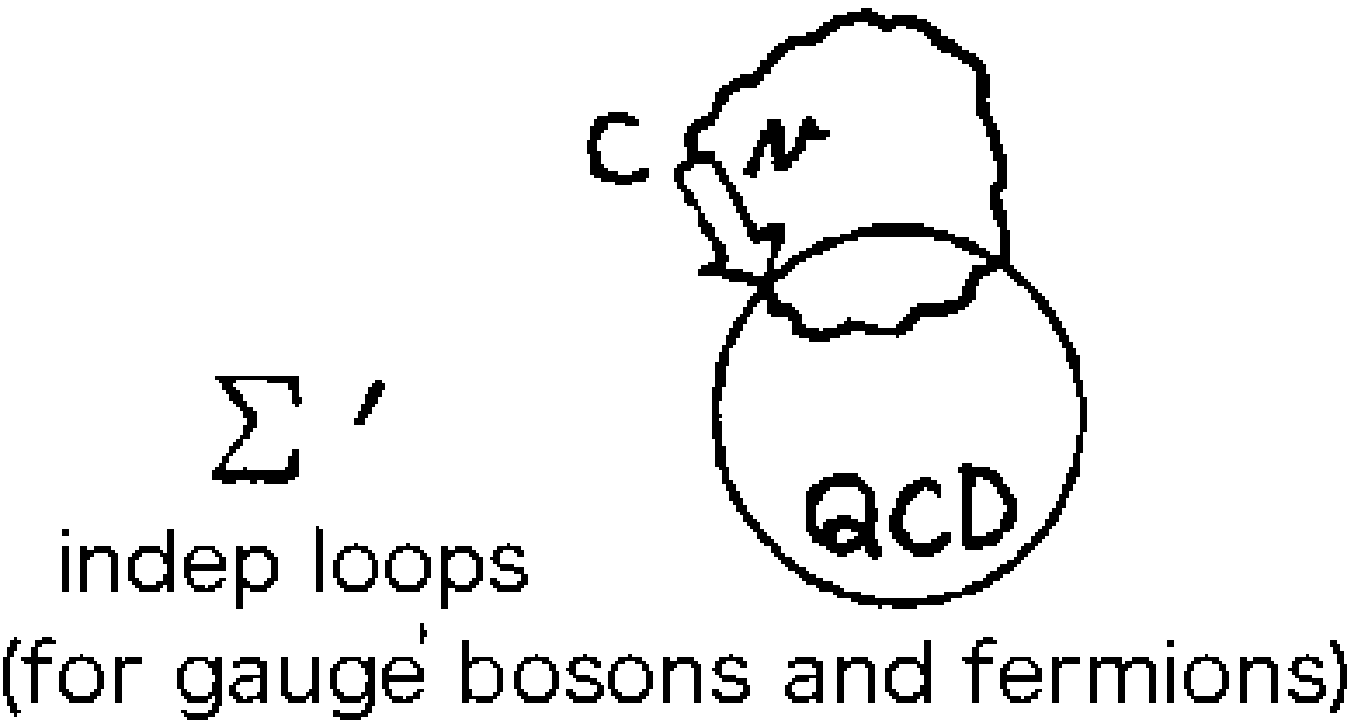}
\end{minipage}
\label{eq4.4.21}
\end{equation} 
corresponds just to the zero momentum photon emission from the fermion loops in QED,
\begin{equation}
\begin{minipage}{14cm}
\includegraphics[width=6cm, clip]{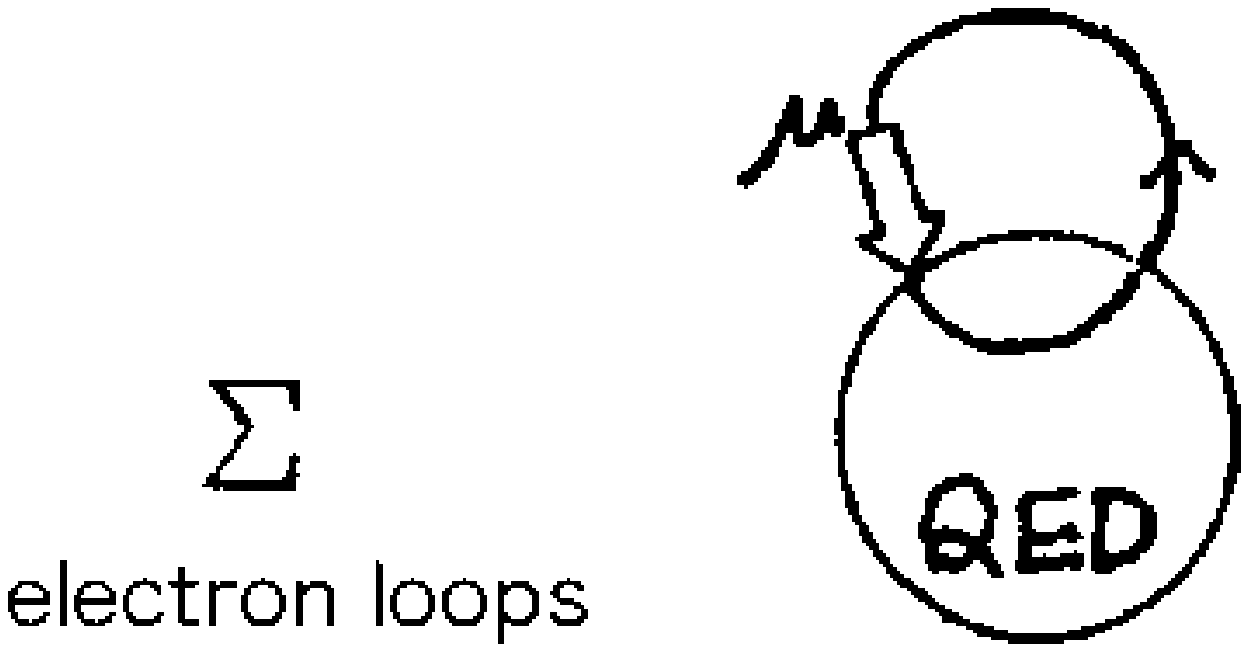}
\end{minipage}
\label{eq4.4.22}
\end{equation} 
Furthermore, (\ref{eq4.4.22}) becomes zero due to a surface integral of the loop integral, as was discussed in Section 2, [please refer to (\ref{eq2.2.8}) and its proof by (\ref{eq2.2.11})--(\ref{eq2.2.13})].  In QCD if (\ref{eq4.4.21}) is regularized by N dimensional method, it becomes zero, since  (\ref{eq2.2.11})--(\ref{eq2.2.13}) hold in the same way.  Namely, we have
\begin{equation}
\begin{minipage}{14cm}
\includegraphics[width=8cm, clip]{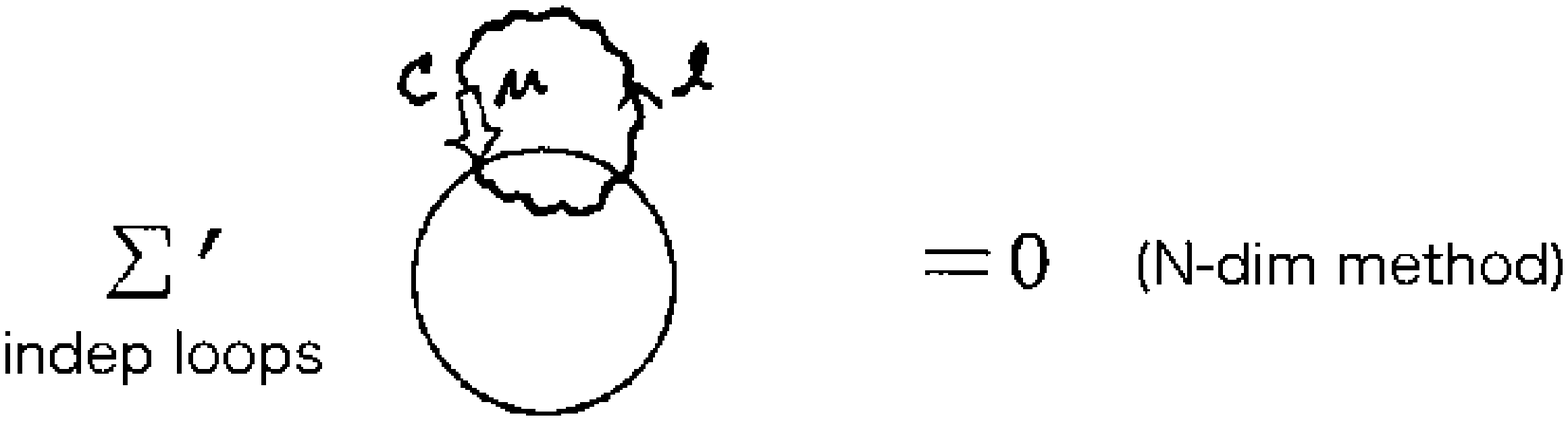}
\end{minipage}
\label{eq4.4.23}
\end{equation}
This (\ref{eq4.4.23}) has the important meaning.

As a simple example, let us consider a two loop graph,
\begin{equation}
\begin{minipage}{14cm}
\includegraphics[width=3cm, clip]{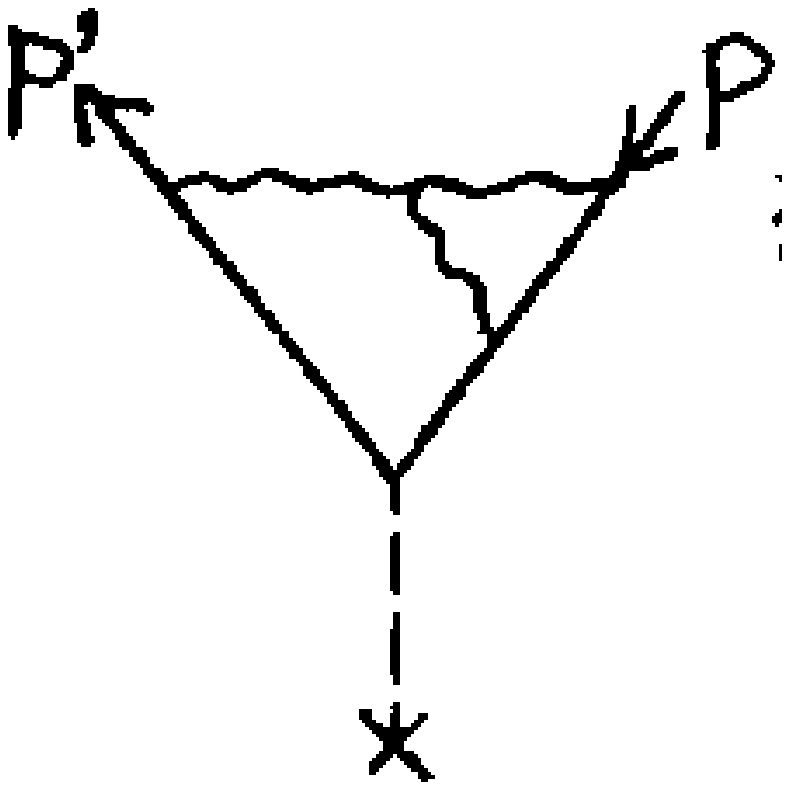}
\end{minipage}
\notag
\end{equation} 
and take the following limit, 
\begin{equation}
\begin{minipage}{14cm}
\includegraphics[width=12cm, clip]{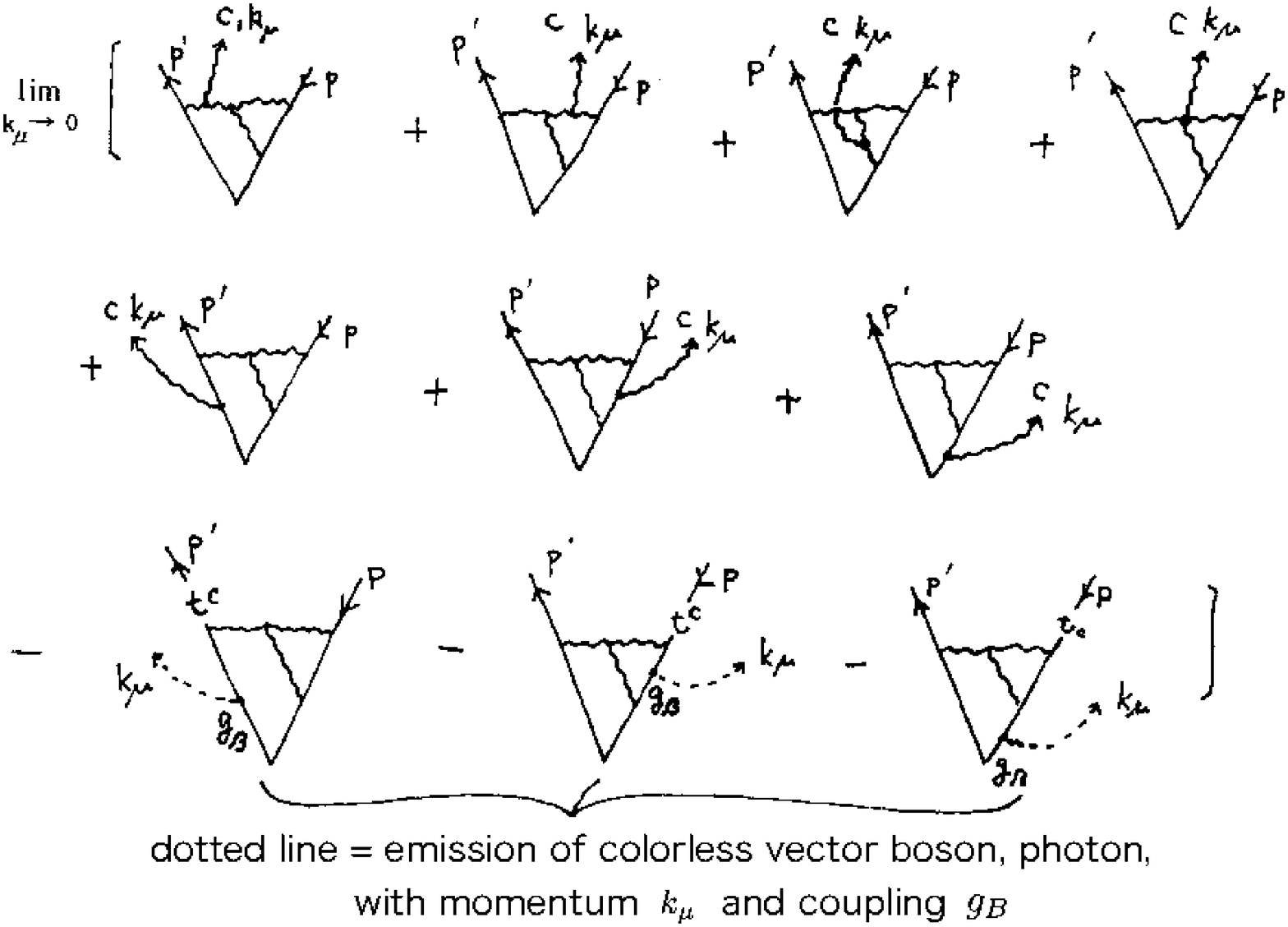}
\end{minipage}
\label{eq4.4.24}
\end{equation}

Here, since the loop integrals in each graph are regularized by the N dimensional method, the limit $k_{\mu} \to 0$ can be taken continuously, \underline{as far as $N$ is complex}.  

If the divergences, $\frac{1}{|k|}, \ln |k|$, {\it e.t.c.}, in the limit of $k_{\mu} \to 0$ in four dimensions, appear in the N- dimensional method as poles at $N=5, N=4$, {\it e.t.c.} in the complex $N$ plane, after taking the limit $k_{\mu} \to 0$ and the integration.  So, using (\ref{eq4.4.23}), (\ref{eq4.4.24}) implies, when the limit $k_{\mu} \to 0$ is taken while keeping $N$ be complex, 
\begin{equation}
\begin{minipage}{14cm}
\includegraphics[width=14cm, clip]{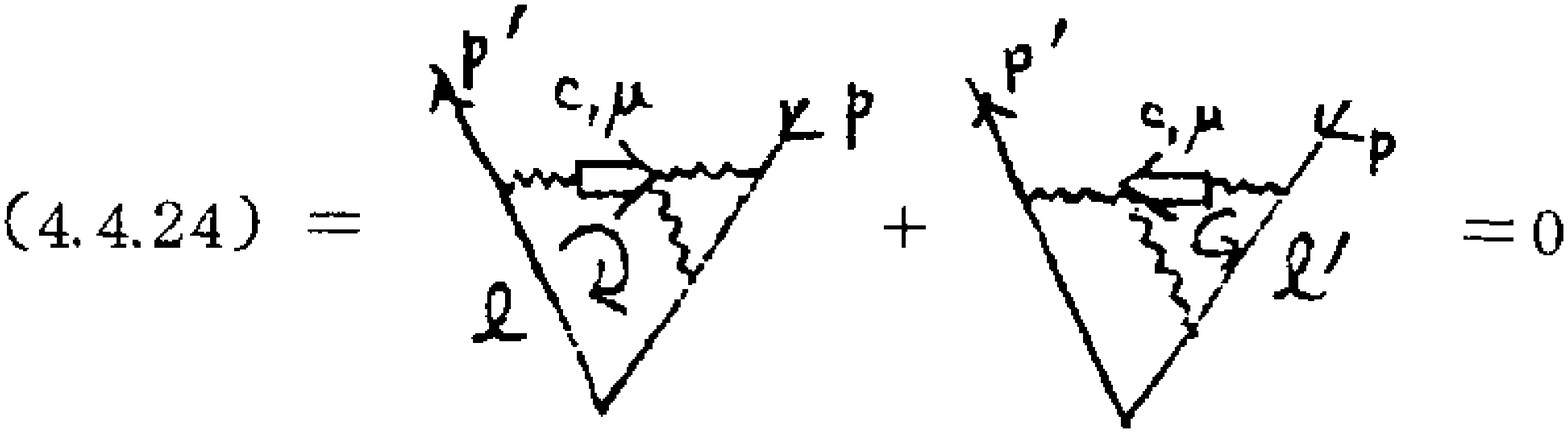}
\end{minipage}
\label{eq4.4.25}
\end{equation}
which has no poles for $N$; namely, the singularities for $|k|$, $\frac{1}{|k|}, \ln |k|$, {\it e.t.c.} in (\ref{eq4.4.24}), which may appear in the calculation in four dimensions, disappear in the N-dimensional method.  Therefore, in the limit of $|k_{\mu}| \ll m$,  it is possible to replace the following soft gauge boson emission process by the soft photon emission processes. 
\begin{equation}
\begin{minipage}{14cm}
\includegraphics[width=14cm, clip]{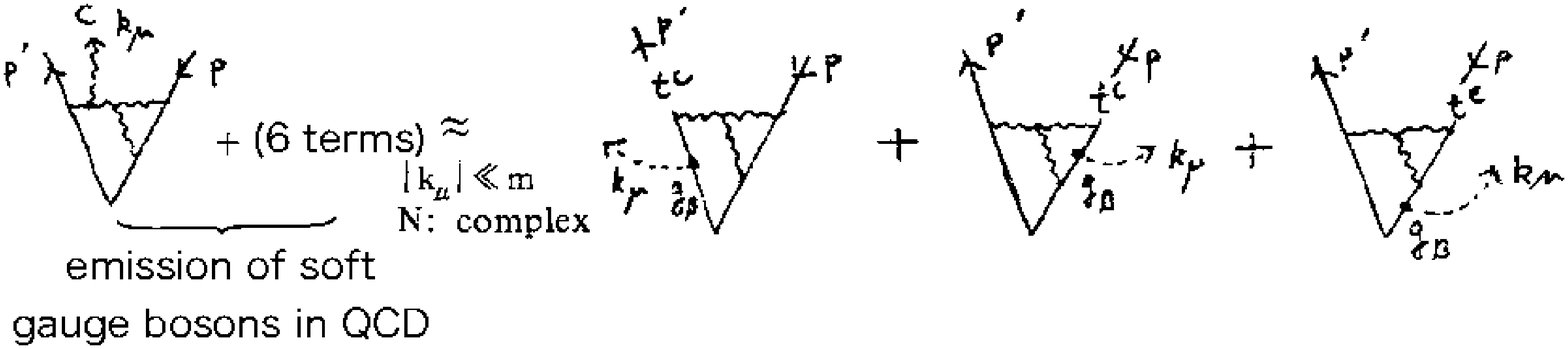}
\end{minipage}
\label{eq4.4.26}
\end{equation}
 [The dotted lines in the right hand side represent the colorless photon emission with the bare coupling $g_B$.]  

As is known from this example, (\ref{eq4.4.23}) represents the \underline{cancellation mechanism} among the soft gauge boson emission graphs, with which the emission of soft gauge boson in QCD can be reduced to soft photon emission in QED, up to a color factor $[t^c]_{ji}$ [ $i, j$ are color indices of fermions coming in and out]. (\ref{eq4.4.26}) is one example.  In the following, we will consider this mechanism in the genaral case.

To begin with, let us prove the equality of the two renoamalizatin constats $Z_1$ and $Z_2$ defined on the mass shell; the former $Z_1$ is the renormalization constant for the fermion-fermion-gauge boson vertex, and the latter $Z_2$ is the wave function renormalization constant for the fermion.
This is depicted as
\begin{equation}
\begin{minipage}{14cm}
\includegraphics[width=10cm, clip]{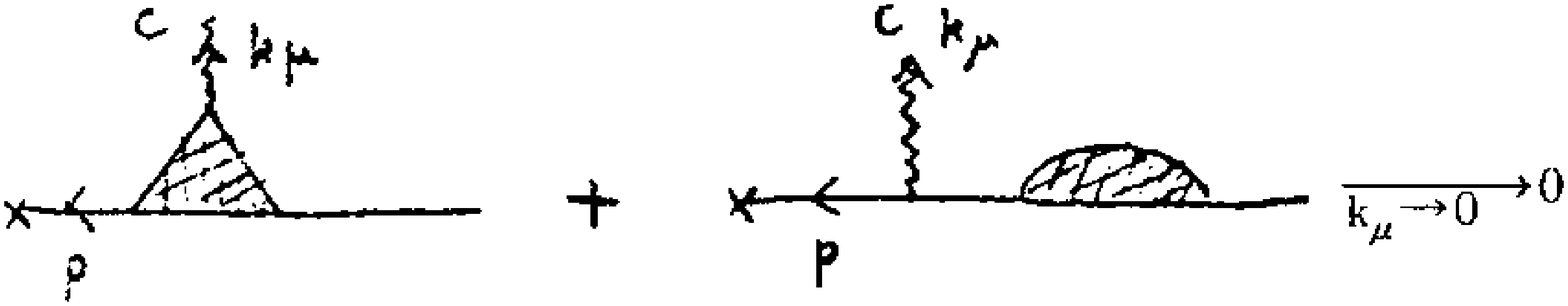}
\end{minipage}
\label{eq4.4.27}
\end{equation}
The first term represents the proper vertex without the lowest graph, and the second term does the proper self-energy.  Using (\ref{eq4.4.16}) and (\ref{eq4.4.23}), this first term can be expressed in terms of the proper self-energy $\Sigma(p)$, 
\begin{equation}
\bar{u}(p) g t^c \frac{1}{i} \frac{\partial}{\partial p_{\mu}} \times \cdots,         \label{eq4.4.28}
\end{equation}
and the second term can be
\begin{equation}
\bar{u}(p) g\gamma_{\mu} t^c \frac{1}{\Slash{p} - m} \times \cdots,         \label{eq4.4.29}
\end{equation}
The mass $m$ in $\Sigma(p)$ is renormalized on the mass shell, so that $\Sigma(\Slash{p}=m)=0$ holds, and the following expansion is possible,
\begin{equation}
\Sigma(p)= ( \Slash{p}-m) \Sigma_1(p) +  ( \Slash{p}-m)^2 \Sigma_2(p) + \cdots
\end{equation}
Substitution of this expansion into (\ref{eq4.4.29}) and (\ref{eq4.4.29}), and comparison of the two, give the proof of (\ref{eq4.4.27}).  Indeed, if (\ref{eq4.4.27}) is expressed in terms of $Z_1, \;  Z_2$, we have
\begin{equation}
0= \bar{u}(p) ig\gamma_{\mu} t^c \left[ \frac{1}{Z_1} -1 \right] \times \cdots +  \bar{u}(p) ig\gamma_{\mu} t^c \left[ 1- \frac{1}{Z_2} \right] \times \cdots  \label{eq4.4.31}
\end{equation}
which gives $Z_1=Z_2$.  Using this, the following replacement is allowed, for the soft bauge boson emission from the external lines:
\begin{align}
&\begin{minipage}{14cm}
\includegraphics[width=14cm, clip]{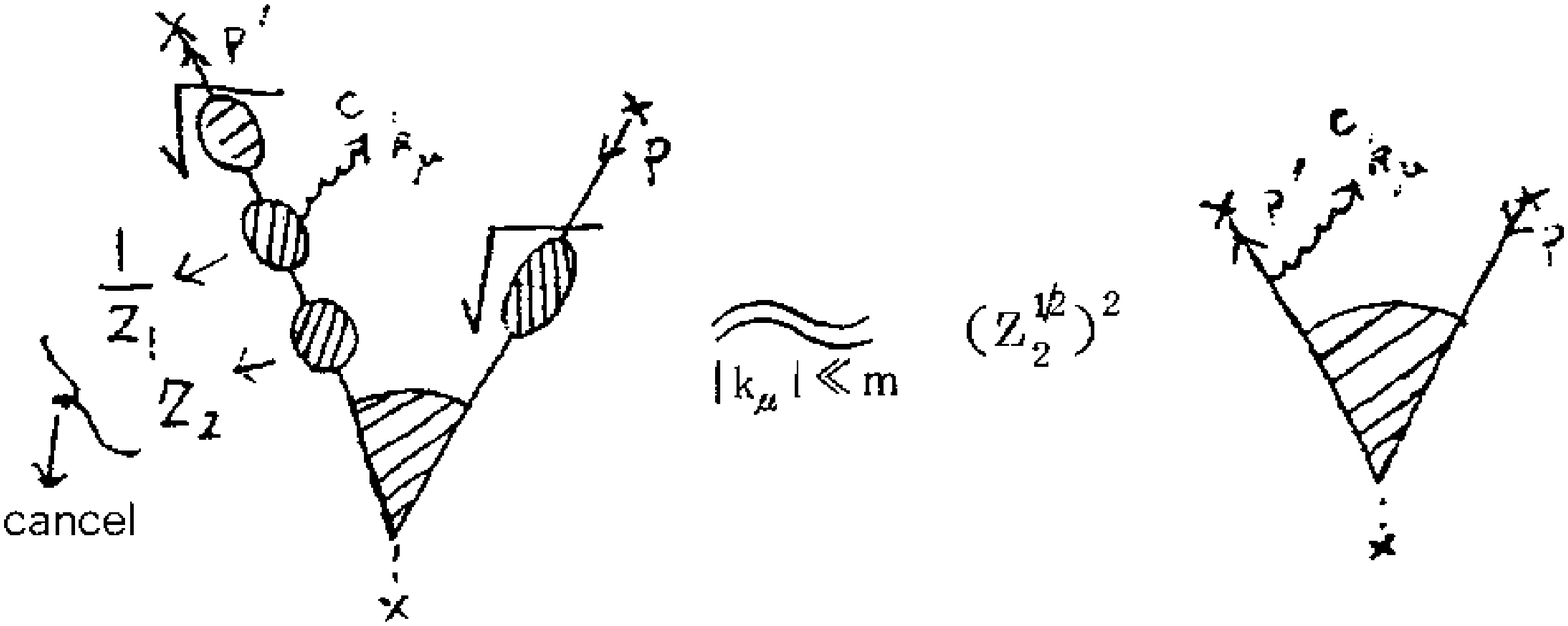}
\end{minipage}
\label{eq4.4.32}  \\
&\begin{minipage}{14cm}
\includegraphics[width=14cm, clip]{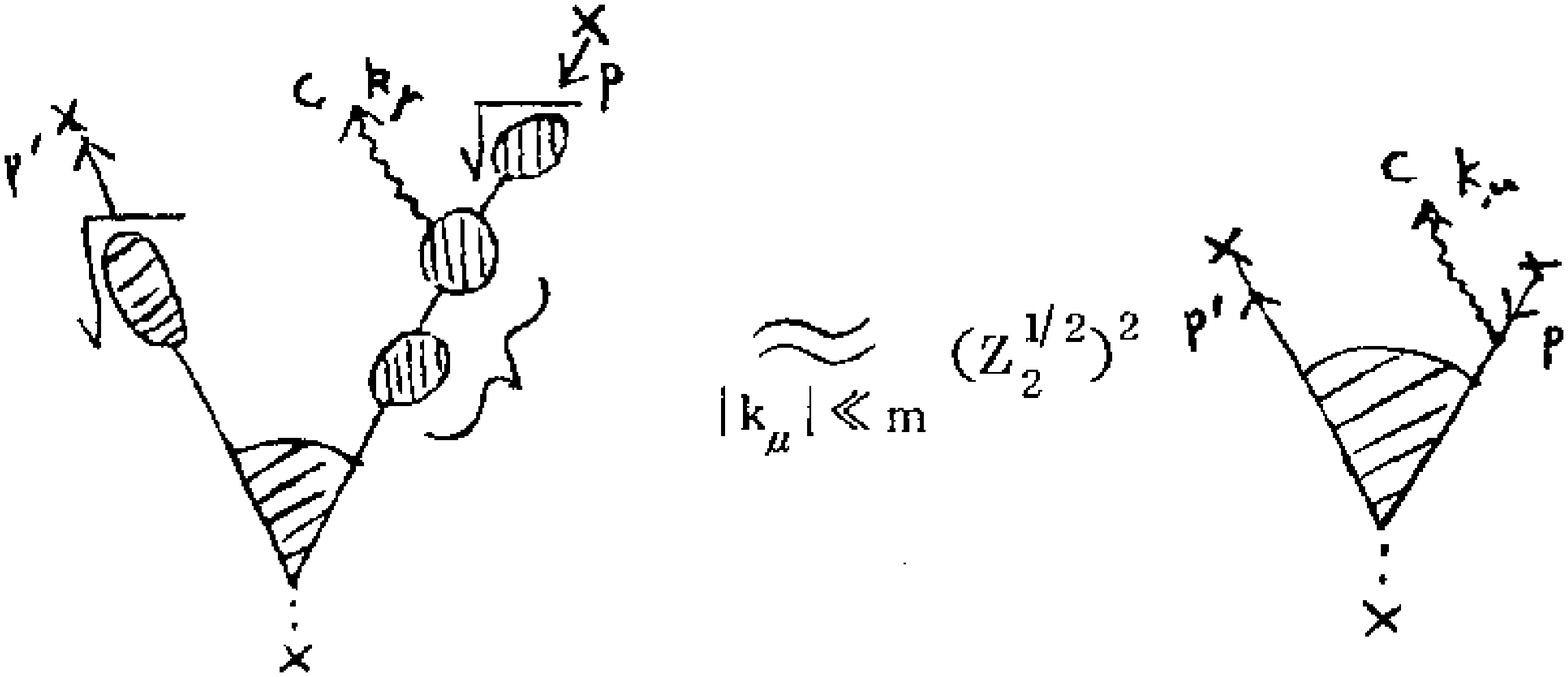}
\end{minipage}
\label{eq4.4.33}
\end{align}
[The $\sqrt{\cdots}$ in the corrections of the external lines implies the multiplication of $\sqrt{Z_2}$ to the external line.]

Next, for the soft gauge boson emission from the inside of a graph, the generaalization of the aforementioned (\ref{eq4.4.24}) holds.   

That is, the following limit of $k_{\mu} \to 0$ can be taken \underline{continuously in the N-dimensional} \underline{regularization}, 
\begin{equation}
\begin{minipage}{14cm}
\includegraphics[width=12cm, clip]{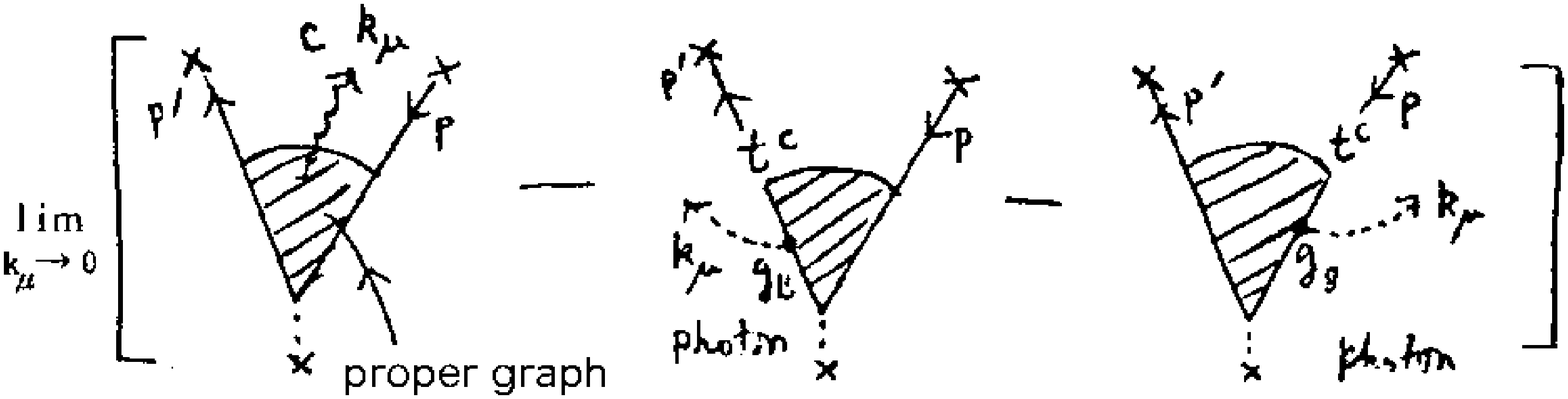}
\end{minipage}
\label{eq4.4.34}
\end{equation}
we obtain the following relation vanishes,
\begin{equation}
\begin{minipage}{14cm}
\includegraphics[width=12cm, clip]{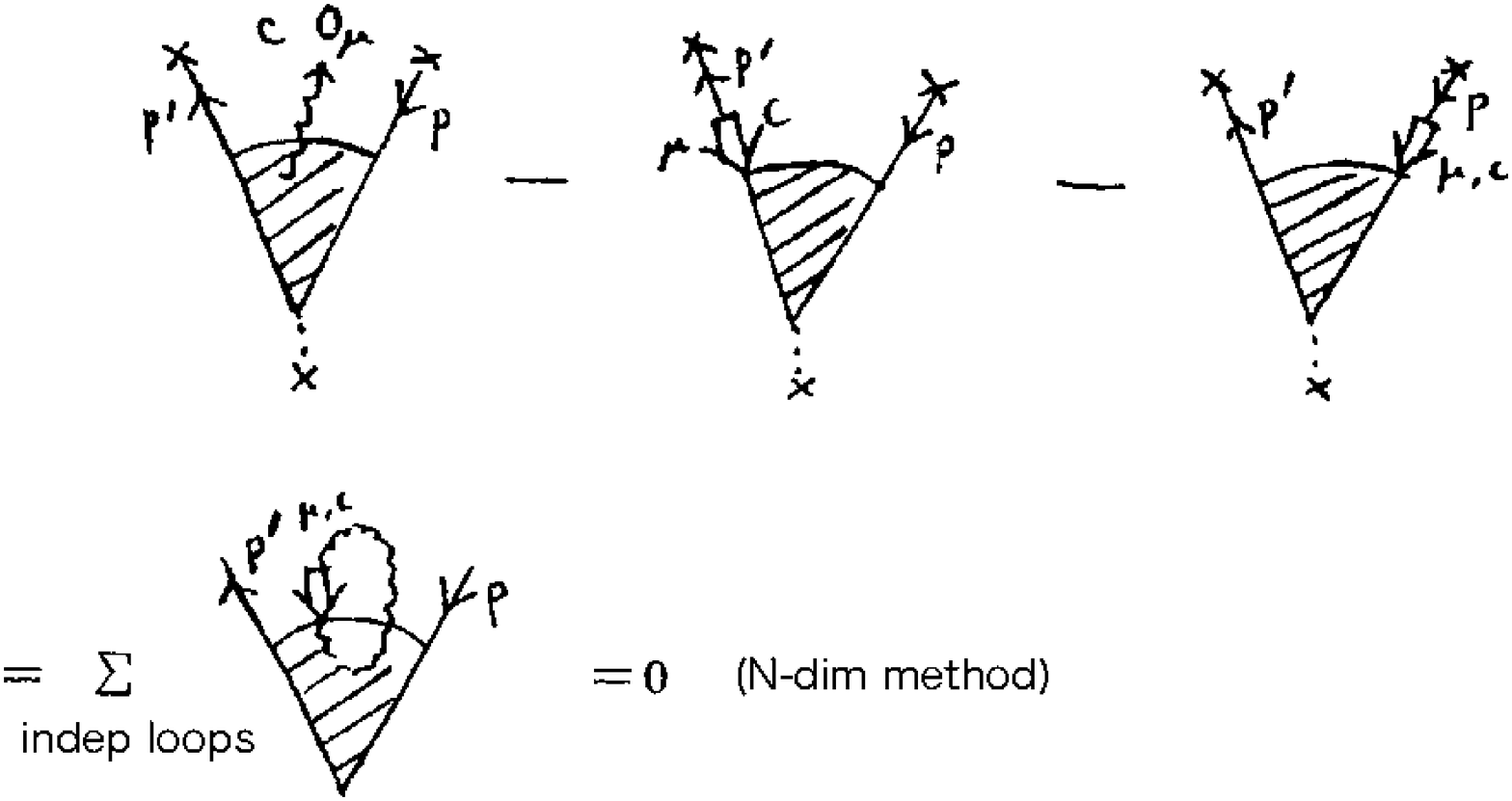}
\end{minipage}
\label{eq4.4.35}
\end{equation}
Therefore, in the limit of $|k_{\mu}| \ll m $, the first term of (\ref{eq4.4.34}) can be replaced by the soft photon emission graphs of the second and the third terms.  That is,
\begin{equation}
\begin{minipage}{14cm}
\includegraphics[width=12cm, clip]{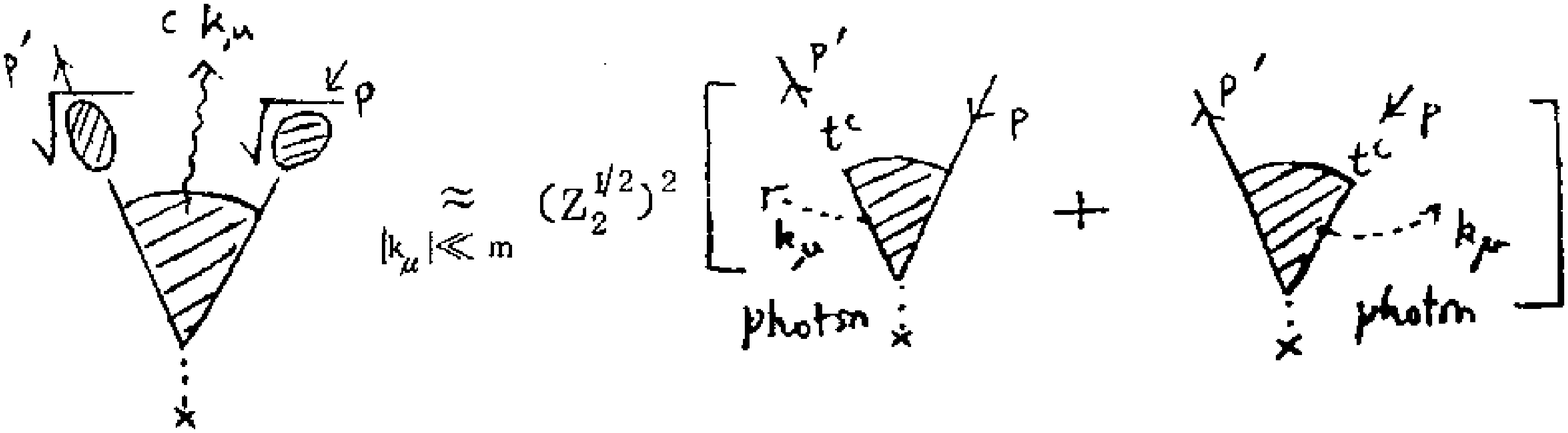}
\end{minipage}
\label{eq4.4.36}
\end{equation}
Here, the corrections for the external lines are additionally added to (\ref{eq4.4.35}).  [Frequently appeared notation of dotted arrow, of course, represents the emission of photon without color index, which interacts to the external lines of fermions with the bare coupling $g_B$.]

Just to be sure, let us discuss about the limit of (\ref{eq4.4.34}).
If we take a limit $k_{\mu} \to 0$ in $N=4$, the divergences, such as $\frac{1}{|k|}, (\ln |k|)^m$, {\it e.t.c.}, appear in each term in (\ref{eq4.4.3}).
However, before the calculation is carried out, if $N$ is analytically continued to a complex value, then the limit $k_{\mu} \to 0$ can be taken continuously, but as the aftereffects various poles at $N=5, N=4$ {\it e.t.c.} arise, which correspond to the divergences $\frac{1}{|k|}, (\ln |k|)^m$, {\it e.t.c.} in the calculation for $N=4$.  
Therefore, the fact that the limit of (\ref{eq4.4.34}) vanishes, due to the cancellation mechanism of (\ref{eq4.4.23}), without having poles in the N-dimensional method implies that the divergences $\frac{1}{|k|}, (\ln |k|)^m$, {\it e.t.c.} cancell in (\ref{eq4.4.34}) in the usual calculation for $N=4$. [The property mentioned above holds always in the N-dimensional method.  For example, when some graph gives an ultraviolet divergence $\ln \Lambda$ in a calculation by introducing an ultraviolet cutoff $\Lambda$ in $N=4$ dimensions,  the same calculation performed in the N-dimensional method induces that \underline{the limit of $\Lambda \to \infty$ can be taken continuously}, but a pole at $N=4$ appears.] 

Now, combining (\ref{eq4.4.32}), (\ref{eq4.4.33}), and (\ref{eq4.4.36}), 
we obtain 
\begin{equation}
\begin{minipage}{14cm}
\includegraphics[width=12cm, clip]{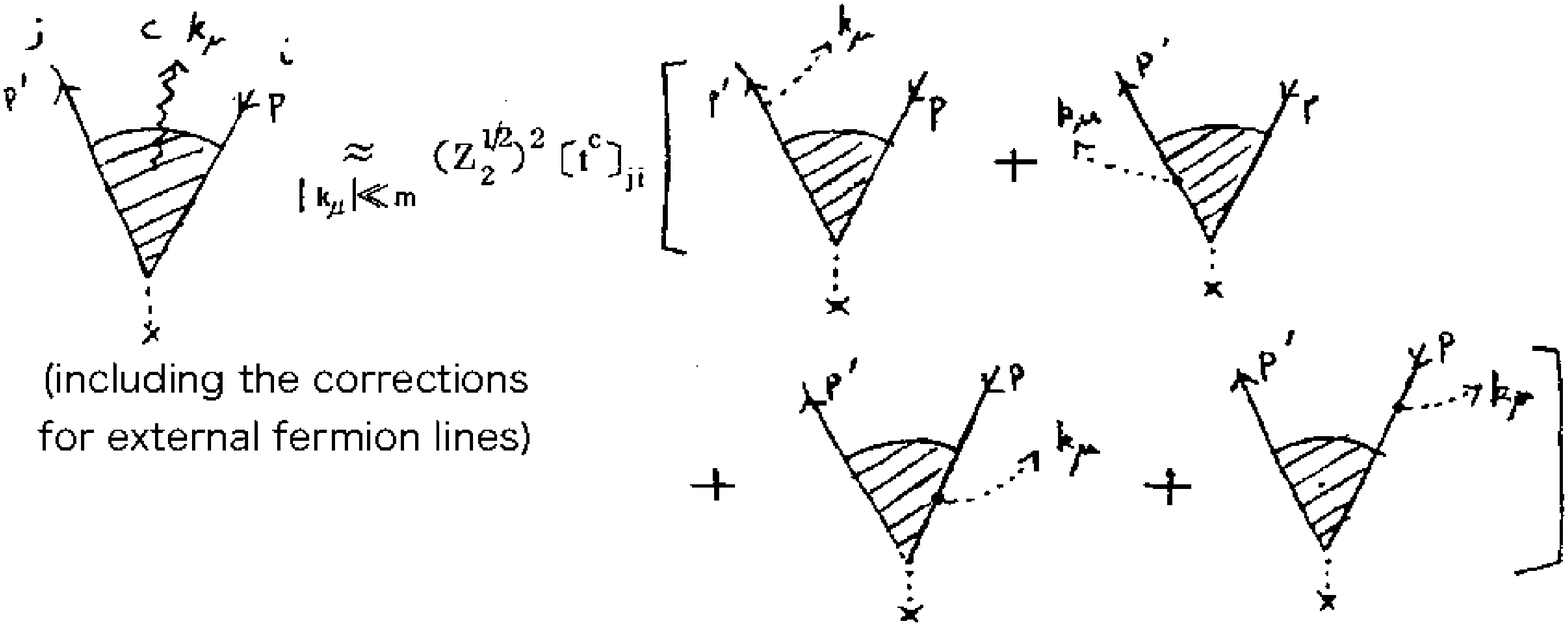}
\end{minipage}
\label{eq4.4.37}
\end{equation}
in which the soft gauge boson emission with color has been reduced to the soft photon emission witout color.  In the right hand side, the soft photon emission processes from the left and right fermion orbits, are summed up, so that the Eikonal identities in Section 2  [(\ref{eq2.2.14}) and (\ref{eq2.2.15})] can be applied, and as the result, the following identity is obtained:
\begin{equation}
\begin{minipage}{14cm}
\includegraphics[width=12cm, clip]{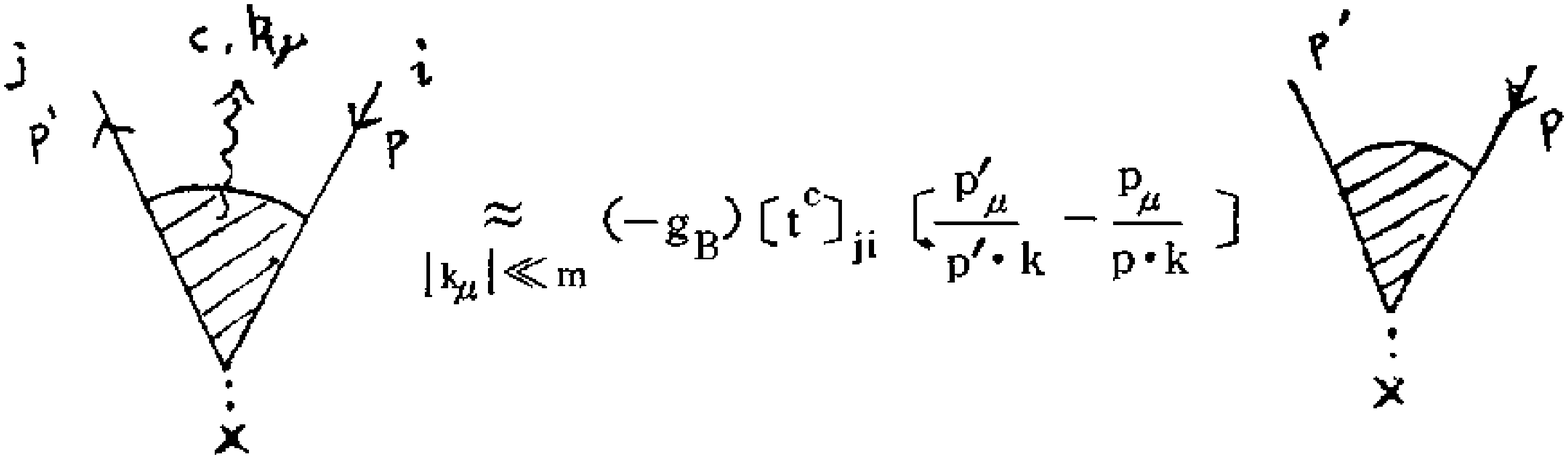}
\end{minipage}
\label{eq4.4.38}
\end{equation}
[In the both hand sides, the corrections of the external fermion lines are taken into account.] Adding furthermore the corrections of the external gauge boson lines, and multiplying the physical polarization vectors, we obtain (\ref{eq4.4.1}), one of the Low's theorem in QCD.

Next, we considr the renormalization of this unrenormalized Low's theorem (\ref{eq4.4.1}).  The mass $m$ is renormalized on the mass shell, and so examine the renormalization of the coupling constant.  (]ref{eq4.4.1}) is expanded in the bare coupling constants $g_B$.  
To renormalize (]ref{eq4.4.}), it is enough to re-expand it in the coupling constants defined under various conditions.  The coupling constant defined on the mass shell $g_R$ and the coupling constant $g(\mu)$ defined off the mass shell, are connected to the bare coupling constant $g_B$ by the following relations:
\begin{align}
g_B = & g_R Z_3^{-1/2}   \label{eq4.4.39} \\
       = & g(\mu) Z_1(\mu) Z_2(\mu)^{-1} Z_3(\mu)^{-1/2}   \label{eq4.4.40}
\end{align}
There is a relation of $Z_1=Z_2$ [(\ref{eq4.4.27}) or (\ref{eq4.4.31})] between the renormalization constants defined on the mass shell.  Off the mass shell renormaliation constants are defined as follows:
\begin{align}
&\begin{minipage}{14cm}
\includegraphics[width=12cm, clip]{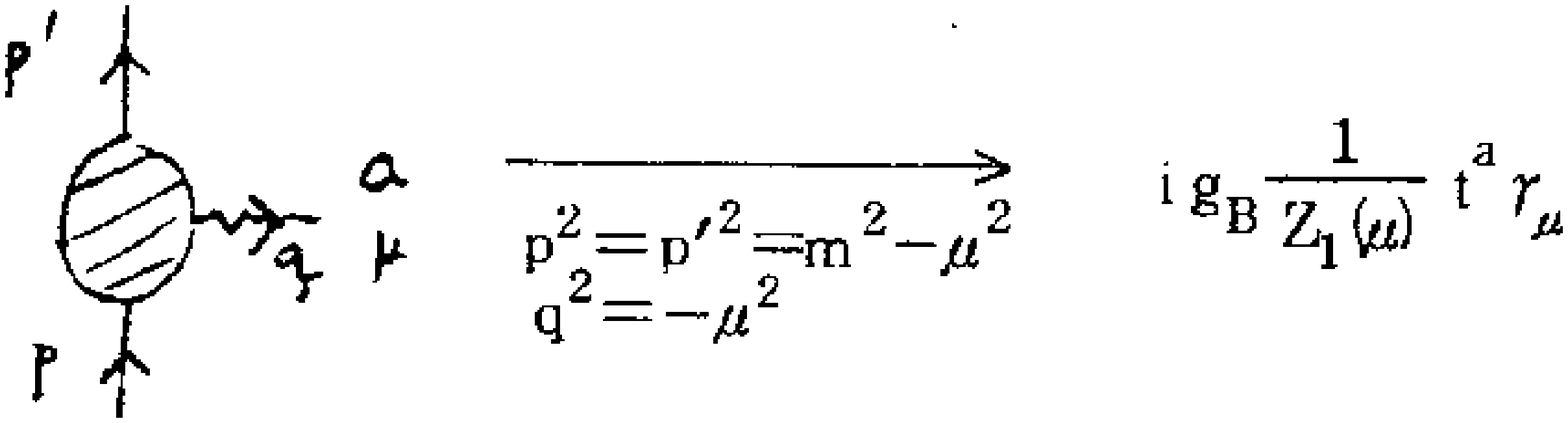}
\end{minipage}
\label{eq4.4.41}  \\
&\begin{minipage}{14cm}
\includegraphics[width=12cm, clip]{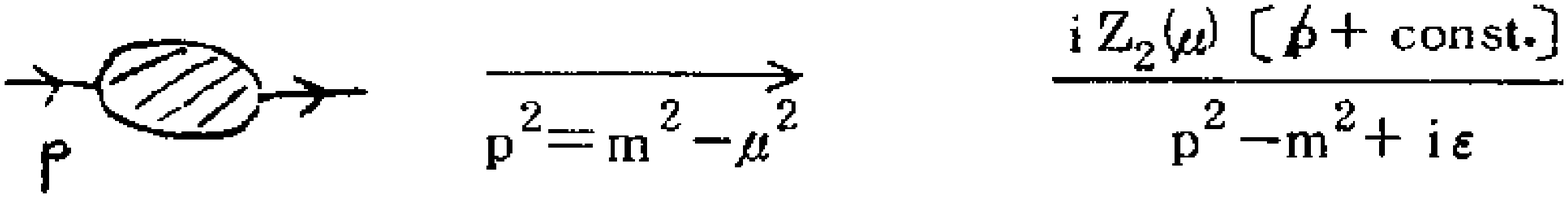}
\end{minipage}
\label{eq4.4.42}  \\
&\begin{minipage}{14cm}
\includegraphics[width=12cm, clip]{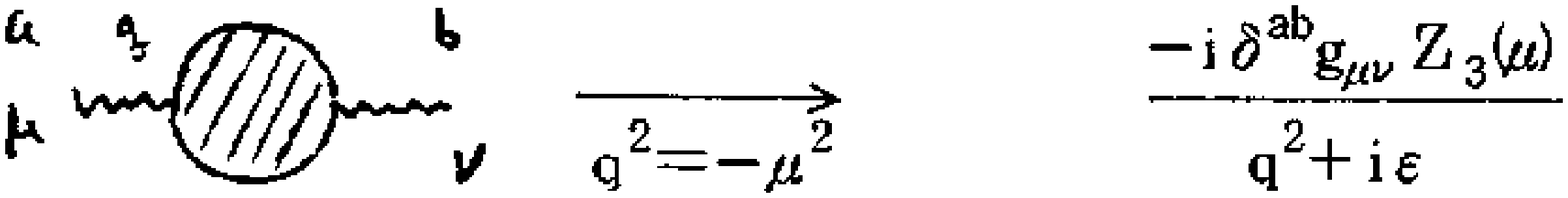}
\end{minipage} 
\label{eq4.4.43} \\
&\hspace{6cm} + [\text{terms} \propto q_{\mu}q_{\nu}, q_{\mu}n_{\nu}+n_{\mu}q_{\nu}, n_{\mu}n_{\nu}].   \notag
\end{align}
[Various choices for the off-mass-shell coupling constant (\ref{eq4.4.40}) are possible.  We heve to find a good choice so that the coupling constant may not depend on the gauge parameter $\alpha$.]  The gauge parameter is renormalized as
\begin{equation}
\alpha_B=Z_3 \alpha= Z_3(\mu) \alpha(\mu)  \label{eq4.4.44}
\end{equation}
which is the same as in the covariant gauge.  [Please refer to (\ref{eq4.3.28}).]   What we are discussing is the scattering amplitudes for which they do not depend on $\alpha_B$ and also not on the gauge fixing vector $n_{\mu}$, when they are expanded in $g_B$ \cite{23}, \cite{24}; so we can ignore the $\alpha_B$ dependence in the following.  {As was pointed out in the above, an additional gauge dependence may appear by the introduction of the off-mass-shell coupling $g(\mu)$.]
The $Z_1(\mu), Z_2(\mu, Z_3(\mu)$ in (\ref{eq4.4.39}) and (\ref{eq4.4.40}) are defined off the mass shell, so that thery are free from the infrared divergences, while $Z_3$ defined on the mass shell includes infrared divergences.  This is a special property in QCD, not observed in QED.  The explicit calculation \cite{29} [by Frenkel and Meuldermans] gives, for  
\begin{align}
g_BZ_3^{1/2}=& g_R (\text{on the mass shell charge}) \notag \\
= & \sum_{n=0}^{\infty} \; c^{(n)}(\mu) [ g(\mu) ]^{2n+1}   \label{eq4.4.45}
\end{align}
the following infrared divergence exists,
\begin{equation}
 [c^{(1)}(\mu)]_{\text{IR}} = \frac{1}{4 \pi^2} c_2(G) \frac{1}{(N-4)^2}  \label{eq4.4.46}
\end{equation}
That is, the coefficient of expanding the on-mass-shell coupling $g_R$ in terms of the off-mass-shell coupling $g(\mu)$, includes the infrared divergence \cite{46}.  Therefore, the Low's theorem (\ref{eq4.4.1}) takes a different form for the on-mass-shell renormalization [taking $g_R$ be finite], or for the off-mass-shell renormalization [taking $g(\mu)$ be finite].  To explain this, let us express (\ref{eq4.4.1}) formally as follows:
\begin{equation}
\sum_{n=0}^{\infty} (g_B)^{2n+1} A^{(n)} = - g_B Z_3^{1/2} \left( \frac{p' \cdot \varepsilon}{ p' \cdot k}-\frac{p \cdot \varepsilon}{ p \cdot k} \right) \sum_{n=0}^{\infty} (g_B)^{2n} B^{(n)}   \label{eq4.4.47}
\end{equation}
Then, expand it in $g(\mu)$ by using (\ref{eq4.4.45}). [Let us include the case of on mass shell renormalization for $\mu=0$ in the following.]

First, introduce the renormalized amplitudes $A^{(n)}(\mu)$ with soft photon emission, and the renormalized amplitudes $B^{(n)}(\mu)$ without soft photon emission, both of which are connected to the corresponding unrenormalized ones $A^{(n)}$ and $B^{(n)}$, as follows;
\begin{align}
&\sum_{n=0}^{\infty} (g_B)^{2n+1} A^{(n)}=\sum_{n=0}^{\infty} [g(\mu)]^{2n+1} A^{(n)}(\mu)  \label{eq4.4.48} \\
&\sum_{n=0}^{\infty} (g_B)^{2n} B^{(n)}=\sum_{n=0}^{\infty} [g(\mu)]^{2n} B^{(n)}(\mu)  \label{eq4.4.49}
\end{align}
Substituting these into (\ref{eq4.4.17}), we have the following renormalized Low's theorem (\ref{eq4.4.1}):
\begin{equation}
A^{(n)}(\mu) = - \left( \frac{p' \cdot \varepsilon}{ p' \cdot k}-\frac{p \cdot \varepsilon}{ p \cdot k} \right) \sum_{k=0}^{n} c^{(k)}(\mu) B^{(n-k)}(\mu)   \label{eq4.4.50}
\end{equation}

If we take $\mu=0$, the renormalization becomes on the mass shell, for which 
\begin{equation}
A^{(n)}(0) = - \left( \frac{p' \cdot \varepsilon}{ p' \cdot k}-\frac{p \cdot \varepsilon}{ p \cdot k} \right)  B^{(n)}(0)   \label{eq4.4.51}
\end{equation}

In summary, Low's theorem in QCD coincides with that of QED as in (\ref{eq4.4.51}), up to the color factor $[t^c]_{ji}$.  Applying the off mass shell renormalization,  however, the Low's theorem (\ref{eq4.4.5}0) [for $\mu \ne 0$], deviates from the QED type of (\ref{eq4.4.51}) by the dependence on $c^{(n)}(\mu)$. The deviations $c^{(n)}(\mu)$ are the expanson coefficients of the on mass shell coupling $g_R$ in the off-mass-shell coupling $g(\mu)$, and include the infrared divergences characteristic in QCD.  This is the end of the proof of  Low's theorem in QCD in case of single soft gauge boson emission. 

The proof of Low's theorem (\ref{eq4.4.2}) and (\ref{eq4.4.3}) in case of emitting two soft gauge bosons, can be done with a little generalization of the proof of (\ref{eq4.4.1}), which will be given in the next Subsection.

\subsection{Generalization of F. E. Low's low energy theorem in QCD (Part 2)--[Application to differential equation which controlls infrared divergence]}

In this subsection,a lillte generalizing the proof of the Low's theorem of (\ref{eq4.4.1}) given in the previous subsection for a single soft gauge boson emission, we will prove the Low's theorem of (\ref{eq4.4.2}) and (\ref{eq4.4.3}), wihch describes the emission of two soft gauge bosons, [with a restriction that two gauge bosons have the same colors].  Furthermore, the theorem is applied to derive the renormalization group like diffferential equation which controles the infrared divergences in QCD.  [This is the generalization of Section 2.2 in QED to QCD.  (\ref{eq4.4.3}) in QCD corresponds to (\ref{eq2.2.7}) in QED.]

To begin with, by generalizing the essential formula (\ref{eq4.4.16}) in the proof of (\ref{eq4.4.1}) [where in the second term, the cancellation mechanism (\ref{eq4.4.23}) works], we will prove 
\begin{equation}
\begin{minipage}{14cm}
\includegraphics[width=12cm, clip]{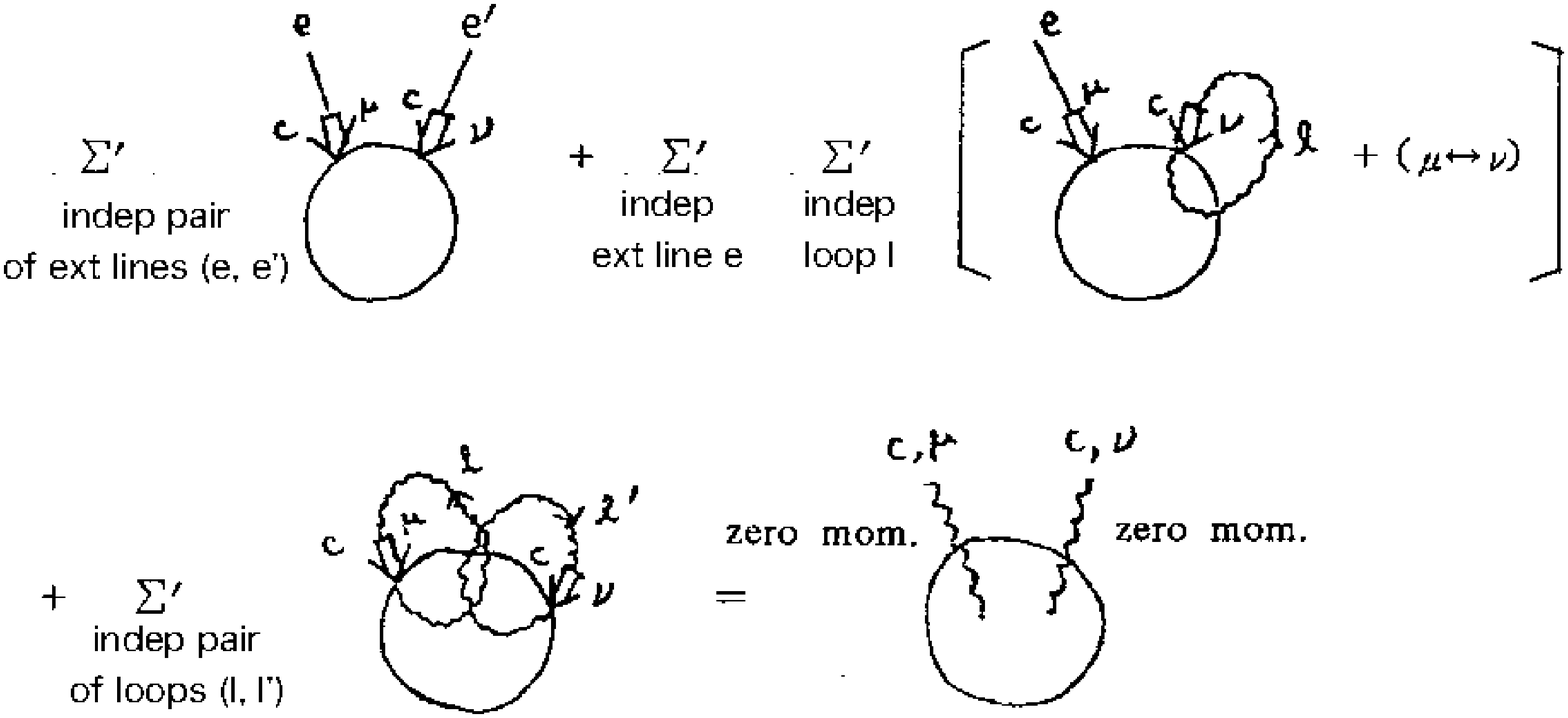}
\end{minipage}
\label{eq4.5.1}
\end{equation}
[Here, the sum over $(e, e')$ and $(l, l')$ includes the duplication, such as $(1, 2) + (1, 1) + (2, 1) + (2, 2)$.]

Since (\ref{eq4.4.16}) has been already proved, to obtain (\ref{eq4.5.1}) we have to show
\begin{equation}
\begin{minipage}{14cm}
\includegraphics[width=12cm, clip]{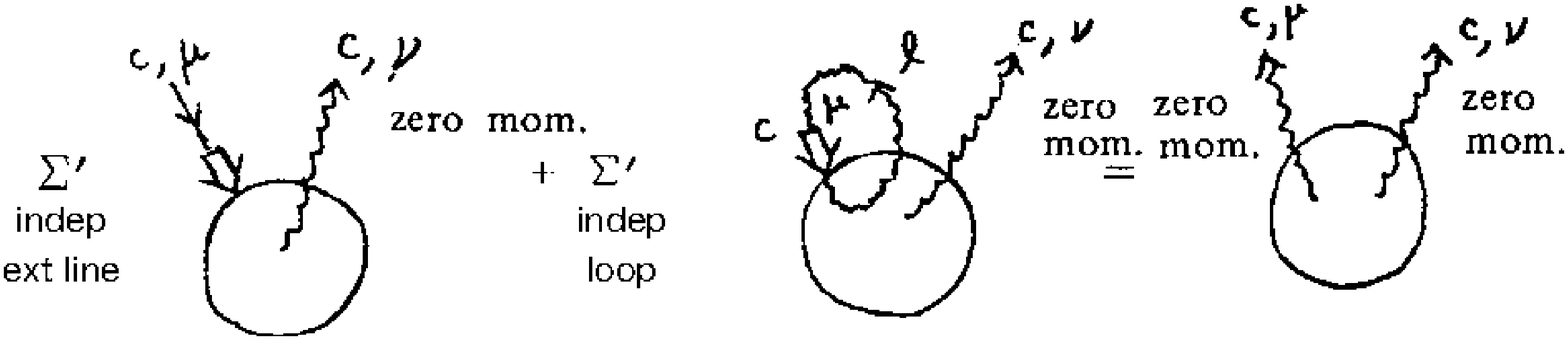}
\end{minipage}
\label{eq4.5.2}
\end{equation}
That is, we repeat the proof in Subsection 4.4, starting from the following graph,
\[
\begin{minipage}{12cm}
\centering
\includegraphics[width=3cm, clip]{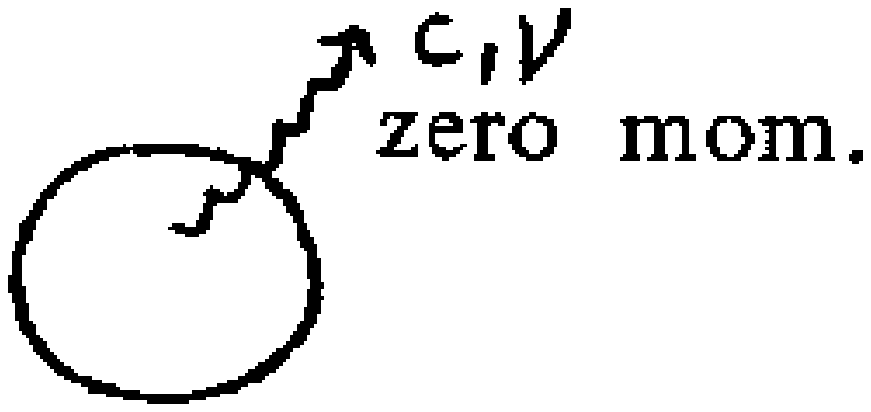}
\end{minipage}
\]
depicting a gauge boson emission with color $c$, Lorentz index $\mu$ and zero momentum.

First, for the simplest graphs, in additon to (\ref{eq4.4.9})--(\ref{eq4.4.14}), the following identities are needed. That is, we require additionally the following four identities: 
\begin{align}
&\begin{minipage}{14cm}
\includegraphics[width=14cm, clip]{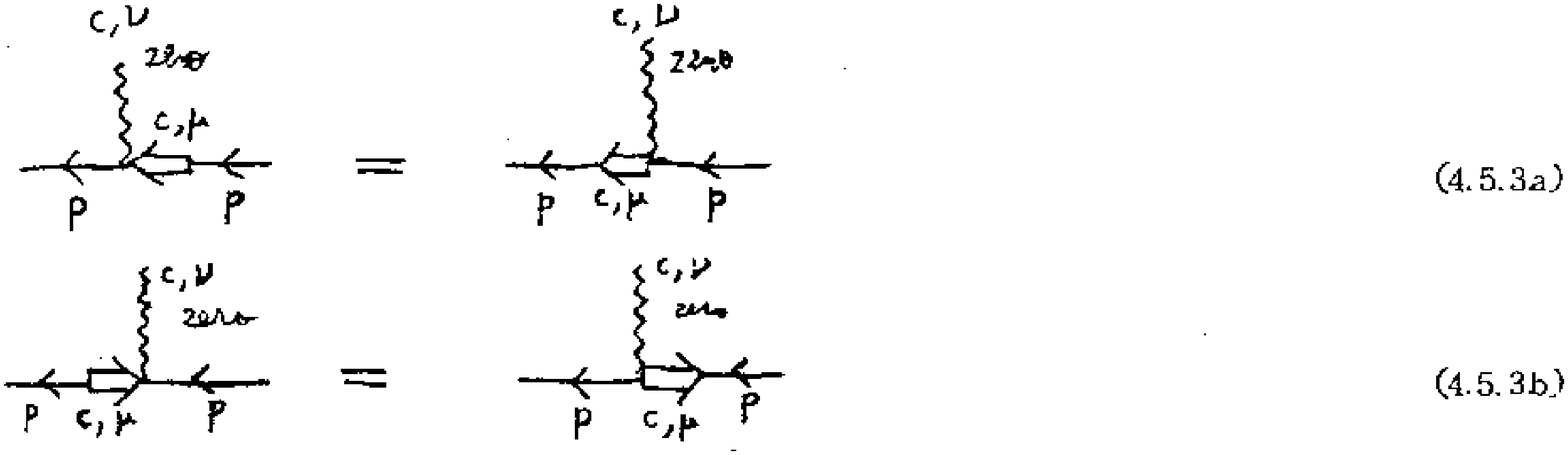}
\end{minipage}
\label{eq4.5.3} \\
&\begin{minipage}{14cm}
\includegraphics[width=12cm, clip]{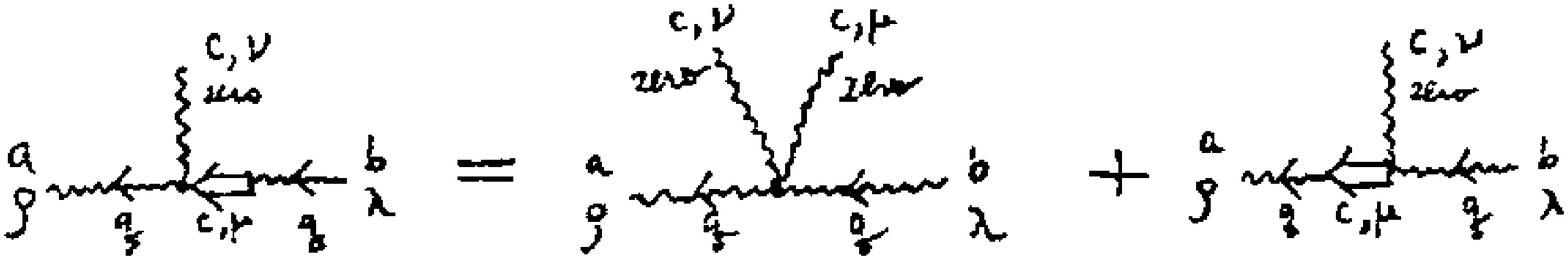}
\end{minipage}
\label{eq4.5.4} \\
&\begin{minipage}{14cm}
\includegraphics[width=12cm, clip]{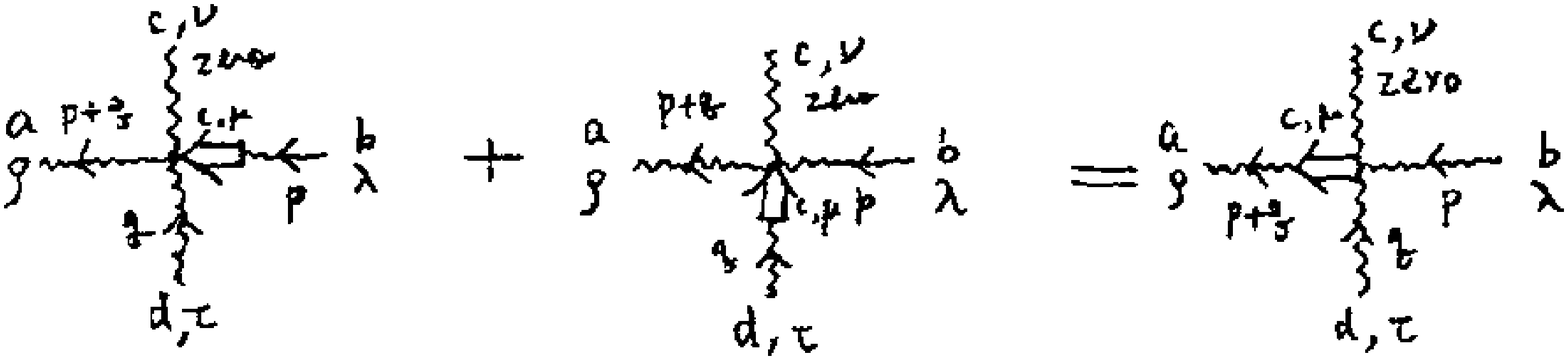}
\end{minipage}
\label{eq4.5.5} 
\end{align}
These identities can be proved, similar to the previous (\ref{eq4.4.9})--(\ref{eq4.4.14}), by using the Feynman rules, the fundamental relation of the group $[t^a, t^b]=if^{abc} t^c$, and the Jacobi identity.  
These identities do not hold, unless the color of the two emitting gauge bosons are equal.  It should be noted that [for example, suppose the two color indices $c$ and $d$ are different with $c \ne d$ in (4.5.3a), then there appears an additional term proportional to $[t^c, t^d]$, and the (4.5.3a) is broken].

In the same manner as in the previous subsection, using (\ref{eq4.4.9})--(\ref{eq4.4.14}) and (\ref{eq4.5.3})--(\ref{eq4.5.5}), we can derive
\begin{equation}
\begin{minipage}{14cm}
\includegraphics[width=10cm, clip]{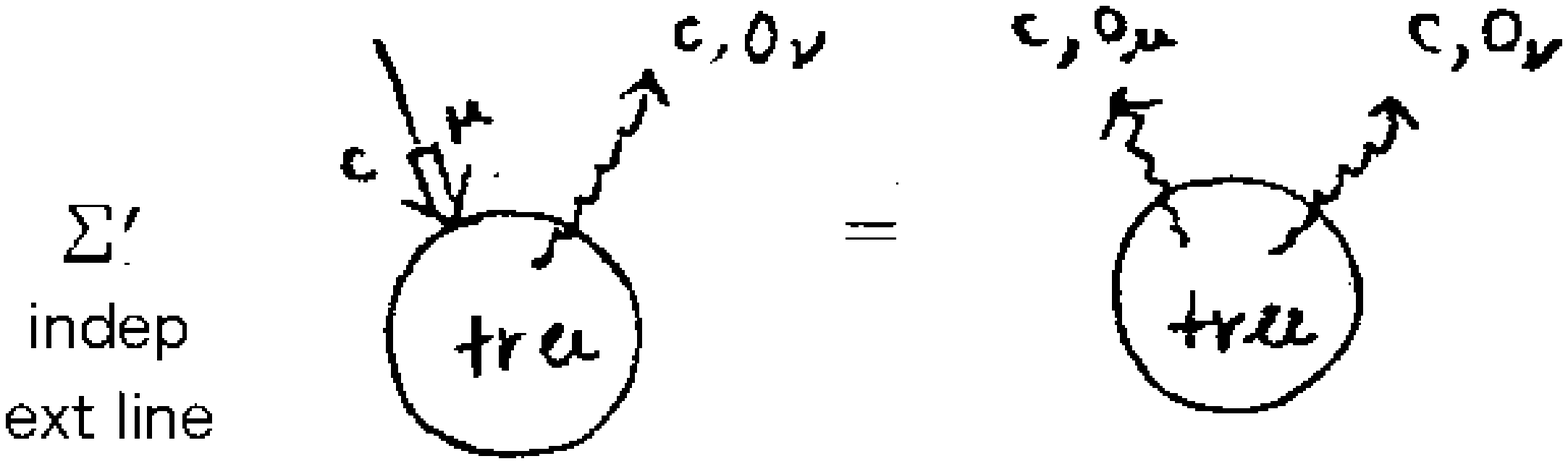}
\end{minipage}
\label{eq4.5.6}
\end{equation}
[This is a generalization of (\ref{eq4.4.15}).  Please remind of the proof just after (\ref{eq4.4.15}).  The characteristics 2) mentioned there holds also for the vertex, which can emitts an additonal gauge boson with color $c$, Lorentz index $\nu$, and the zero momentum.] 
Next, to generalize (\ref{eq4.5.6}) to (\ref{eq4.5.2}), we use the induction with respect to the number of loops; in this case it is only to repeat the method used in Subsection 4.4 from (\ref{eq4.4.17}) to (\ref{eq4.4.20}), starting from the following graph,
\[
\begin{minipage}{12cm}
\centering
\includegraphics[width=3cm, clip]{figsoftemission.eps}
\end{minipage}
\]
[Therefore, we can skip to write the proof here.]  Then, the proof of (\ref{eq4.5.3}) has been finished.  Here, we have used the fact that the second term in (\ref{eq4.5.2}) has no surface term in the N-dimensional method, which is a generalization of (\ref{eq4.4.23}) in Subsection 4.4.  This is depicted in this case as  
\begin{equation}
\begin{minipage}{14cm}
\includegraphics[width=12cm, clip]{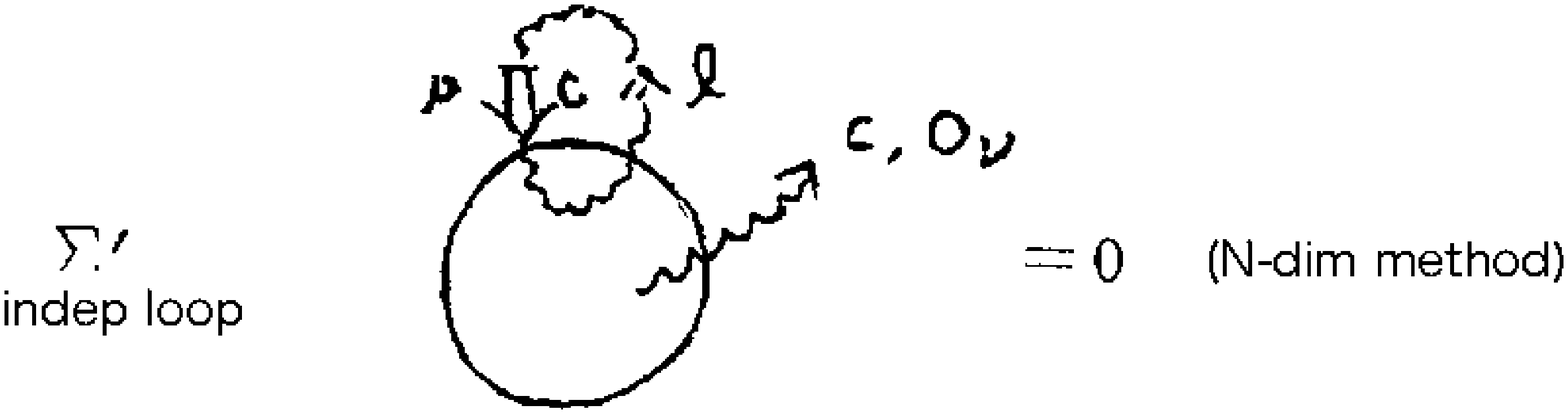}
\end{minipage}
\label{eq4.5.7}
\end{equation}
[Please refer to (\ref{eq2.2.8}) and (\ref{eq2.2.11})--(\ref{eq2.2.13}).]
Accordingly, using the cancellation mechanism by the surface term (\ref{eq4.4.23}), (\ref{eq4.5.7}), we have the following formula from (\ref{eq4.5.2}), 
\begin{equation}
\begin{minipage}{14cm}
\includegraphics[width=12cm, clip]{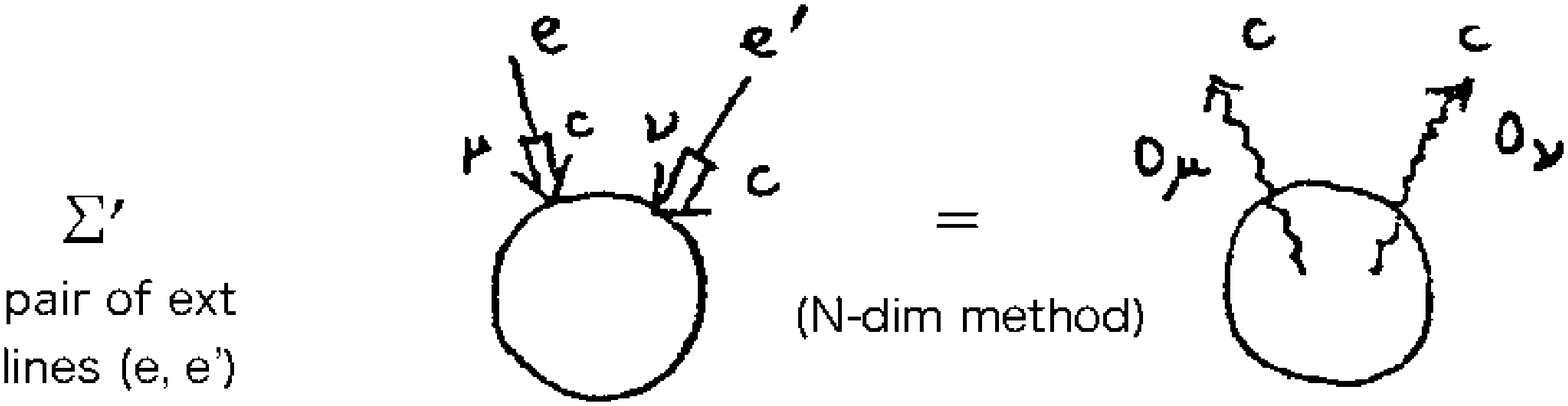}
\end{minipage}
\label{eq4.5.8}
\end{equation}

Let us apply this to our process of the fermion scattering by the colorless external source. 
That is, we consider the difference of two processes in the soft momentum limit of two gauge bosons; one process is the emission of two soft gauge bosons, having the same color $c$ with momenta $k_{\mu}$ and $l_{\nu}$, while the other process is the emission of colorless two soft photons with the same momenta.  Pictorially, it is
\begin{align}
&\begin{minipage}{14cm}
\includegraphics[width=12cm, clip]{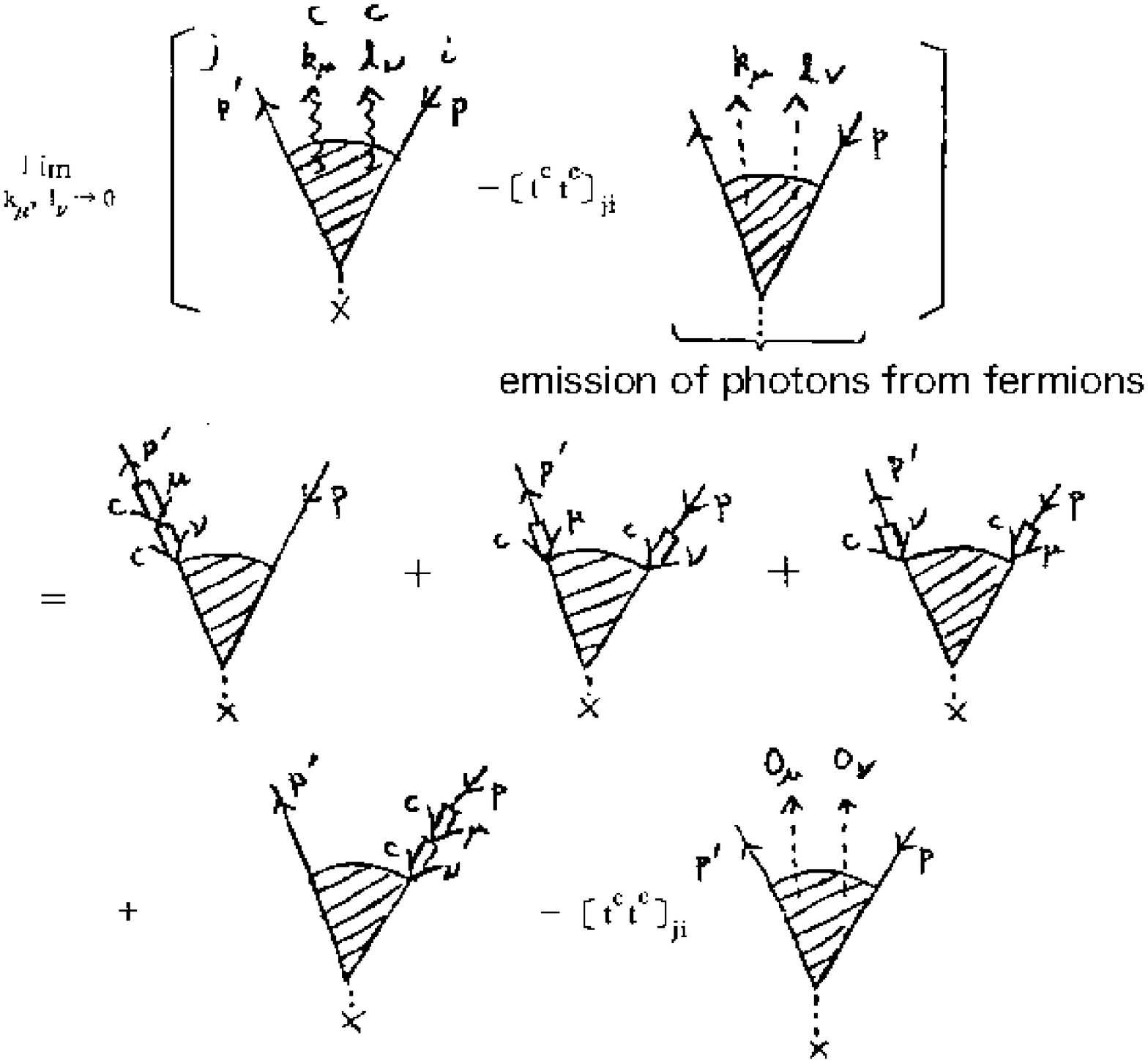}
\end{minipage}
\notag \\
&\begin{minipage}{14cm}
\includegraphics[width=12cm, clip]{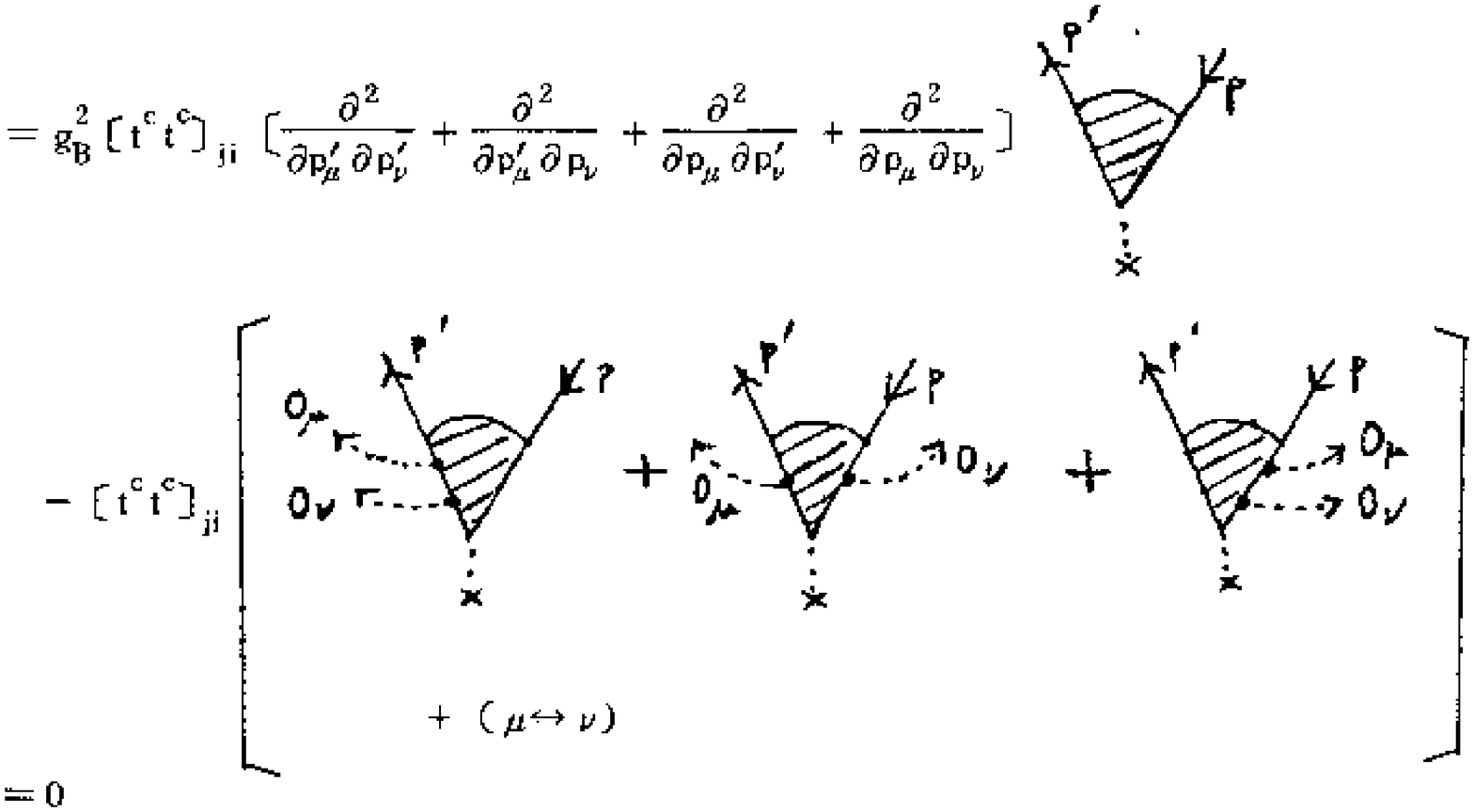}
\end{minipage}
\label{eq4.5.9}
\end{align}
Here, (\ref{eq4.5.8}) has been used in the deformation of the first line.  The meaning of the limit $k_{\mu}, l_{\mu} \to 0$ is to take the limit under the regularization by the N-dimensional method, as was discussed in the last subsection.  In (\ref{eq4.5.9}), two cases of soft gauge boson emissions, one from the external lines and the other from the internal lines, are not specified. For example, it includes the following two soft gauge bosons' emission, which implies
\begin{equation}
\begin{minipage}{14cm}
\includegraphics[width=10cm, clip]{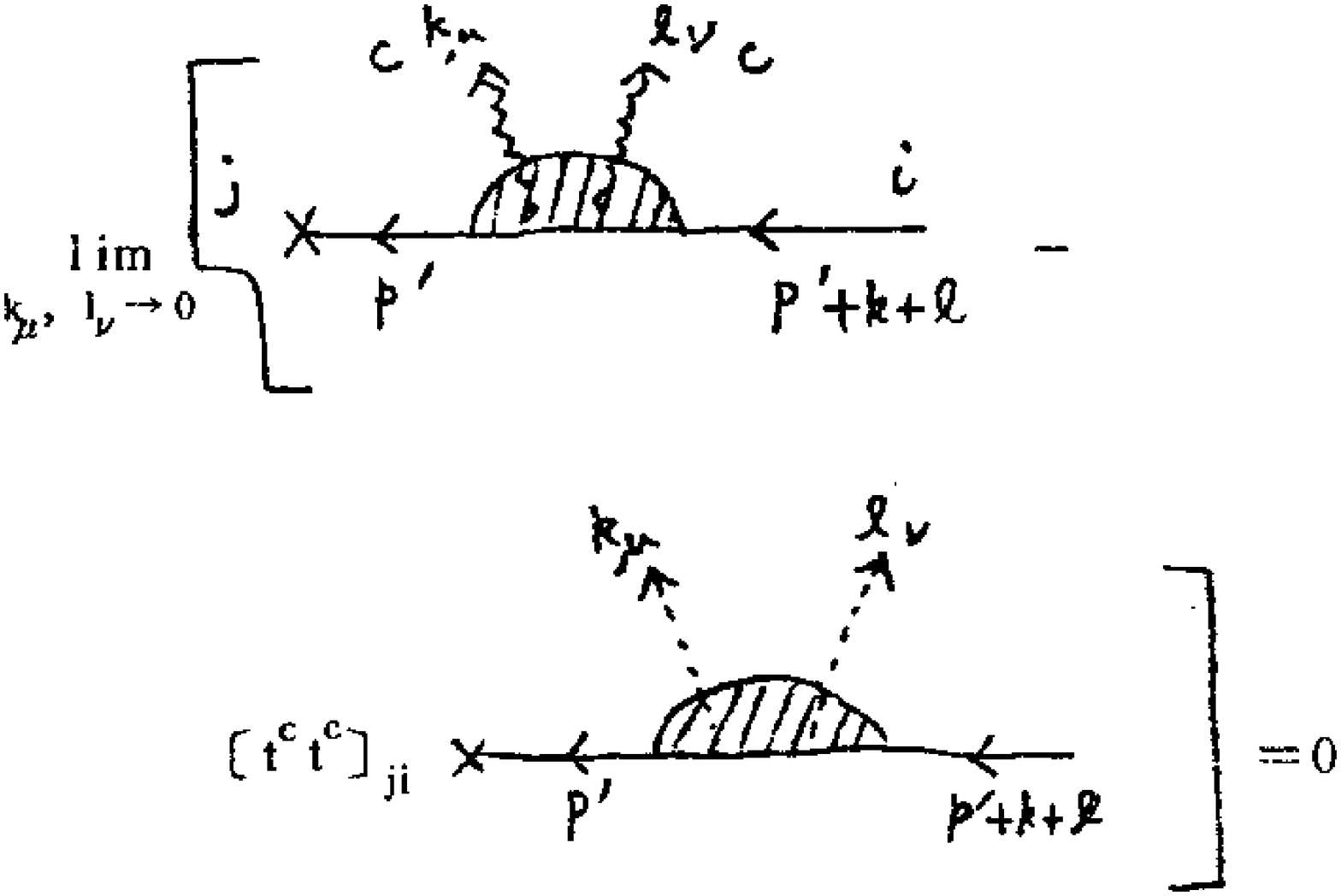}
\end{minipage}
\label{eq4.5.10}
\end{equation}
[We should examine the Thomson limit in QCD, using this formula.  A difficulty appears, however, that a soft gauge boson can be easilly resolved into two or three soft gauge bosons via the self-interaction between the soft gauge bosons. Therefore, it is difficult to restrict the number of soft gauge bosons to be one in the initial or in the final state.]

By this (\ref{eq4.5.9}), soft gauge boson emission in QCD can be reduced to the soft photon emission as follows:
\begin{equation}
\begin{minipage}{14cm}
\includegraphics[width=12cm, clip]{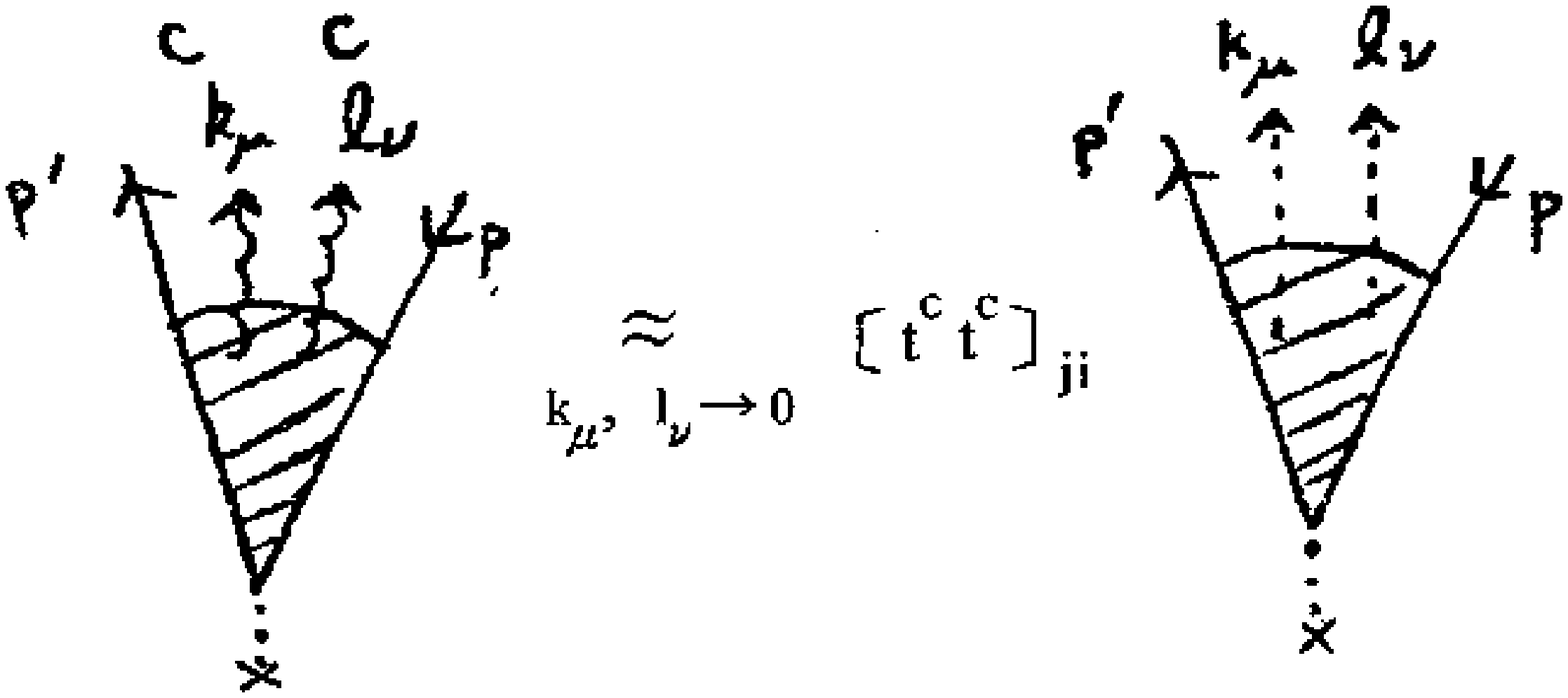}
\end{minipage}
\label{eq4.5.11}
\end{equation}
Apply the Eikonal identities (\ref{eq2.2.14}), (\ref{eq2.2.15}) to the right hand side of this equation, we obtain 
\begin{equation}
\begin{minipage}{14cm}
\includegraphics[width=12cm, clip]{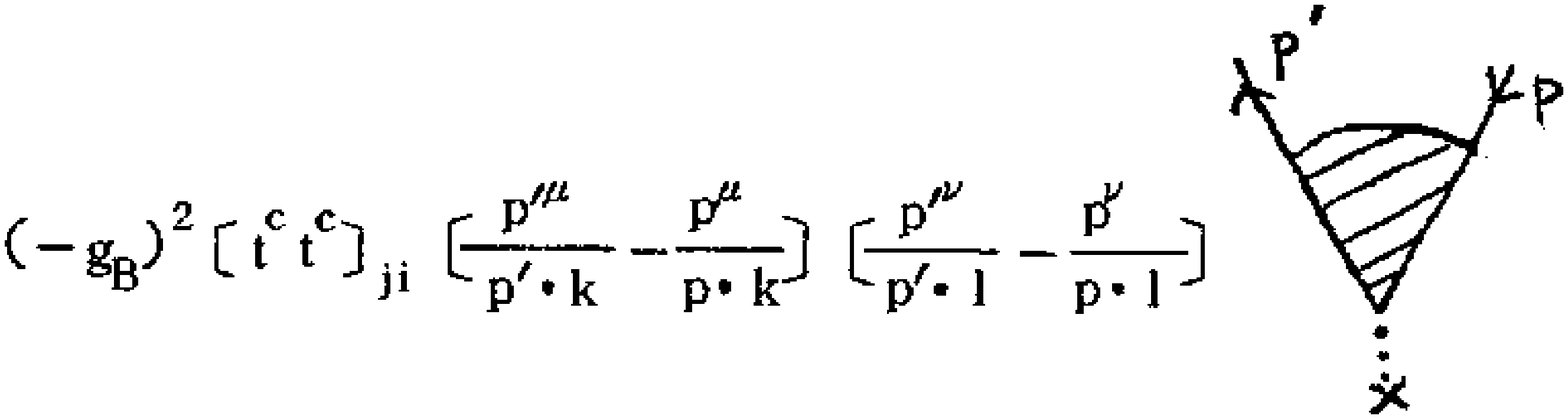}
\end{minipage}
\label{eq4.5.12}
\end{equation}
[when using (\ref{eq2.2.14}), (\ref{eq2.2.15}), the particle with momenta $l_1, l_2, \cdots, l_n$, should be considered as a gauge boson with a color].  In this way (\ref{eq4.4.3}) is shown to hold in case of two soft gauge bosons' emission.  Including the external corrections to the external gauge bosons, and multiplying physical polarization vectors, it becomes (\ref{eq4.4.2}). \cite{47}

Here, we will notice by comparison that (\ref{eq4.4.3}) obtaiend here and (\ref{eq2.2.7}) or (\ref{eq2.2.17}) in Subsection 2.2 have the same form up to the color factor $[t^ct^c]_{ji}$. In Section 2.2, this (\ref{eq2.2.7}) plays an important role, which contributes to derive the differential equation (\ref{eq2.2.3}) [or (\ref{eq2.2.23})] controling the heighest infrared divergences in all orders; this differential equation is easily solved to give (\ref{eq2.2.1}) [or (\ref{eq2.2.24})].  Therefore, using (\ref{eq4.4.3}) in QCD, corresponding to (\ref{eq2.2.7}) in QED, it is easily expected to be able to derive the differential equation which controlls the highest inftared divergences in QCD.  [This derivation was given by Cornwall and Tiktopoulos \cite{20}, but the proof of Low's theorem in (\ref{eq4.4.3}) was not given at that time.  On the other hand, Kinoshita and Ukawa \cite{48} develpped a general theory to estimate the highest logarithmic infrared divergences, and derived the above differentail equation. Their derivation is, however, too complicated to understood easilly.  The merit of the following derivation of the differential equation is its simpleness.  The new point here is that the Low's theorem (\ref{eq4.4.3}) has been proved in all orders. Why the proof becomes successful largely owes the usage of the axial gauge.] 

Now, let us repeat the derivation of the differential equaion in QCD, as was done in QED in Subsection 2.2.  There, we have introduced the gauge boson mass $\lambda$ as the infrared regularization.  Then, applying $\lambda \frac{\partial}{\partial \lambda}$ to the fermionic scattering amplitude $S$ by the colorless external source, [even if it has the same notation in Subsection 2.2, but it is here the amplitude in QCD], we obtain, in the similar way to (\ref{eq2.2.6}), the following
\begin{equation}
\lambda \frac{\partial}{\partial \lambda} S = \frac{1}{2} \int \frac{d^N k}{(2\pi)^N} \sum_c K^{\mu\nu}_{cc} (p, p' ; k, -k) \lambda \frac{\partial}{\partial \lambda} \tilde{D}_{\mu\nu} (k ; \lambda).       \label{eq4.5.13}
\end{equation}
[Here, $N$ was used as the reularization of the ultraviolet divergences.]  This $K^{\mu\nu}_{cc} (p, p' ; k, -k)$ in QCD corresponds to Figure (\ref{eq4.5.1}) in QED.  

\[
\begin{minipage}{12cm}
\centering
\includegraphics[width=3cm, clip]{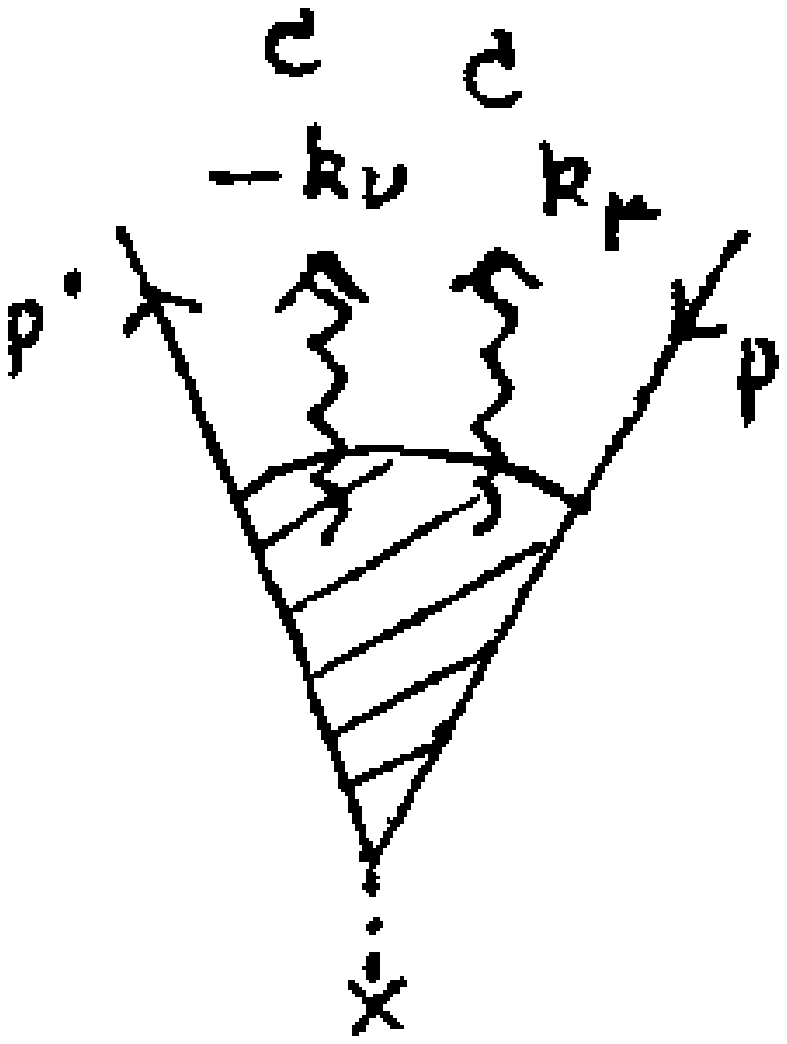}
\text{(Figure 4.5.1)}
\end{minipage}
\]
Two leggs of gauge bosons are connected by the dressed propagator, and have the same color $c$.  In the following we proceed by referring to Subsection 2.2 in QED.  To examine the infrared divergences of (\ref{eq4.5.13}), we have to know the behavior of $K^{\mu\nu}_{cc} (p, p' ; k, -k)$ in the limit of $k_{\mu} \to 0$ and the behavior of $\lambda \frac{\partial}{\partial \lambda} \tilde{D}_{\mu\nu} (k ; \lambda)$.  First, for the $K^{\mu\nu}_{cc} (p, p' ; k, -k)$, we can use (\ref{eq4.4.3}), the Low's theorem for two gauge bosons' emission with the same color, proved in this subsection.  In the present notation it reads
\begin{align}
&  \sum_{c} K^{\mu\nu}_{cc} (p, p' ; k, l)  \underset{k_{\mu}, l_{\nu} \to 0}{\approx}  + (g_B)^2 \sum_{c} [t^c t^c]  \notag \\
& \times \left[ \frac{p^{'\mu}}{p' \cdot k} - \frac{p^{\mu}}{p \cdot k} \right] \left[ \frac{p^{'\mu}}{p' \cdot l} - \frac{p^{\mu}}{p \cdot l} \right] S   \label{eq4.5.14}
\end{align}
Here, the color index $c$ is summed over, so $\sum_c t^c t^c=C_2(R) \bm{1}$ holds, [$\bm{1}$ is a unit matrix].  [Please refer to (\ref{eq3.4.8}) and the subsequent explanation.]  This corresponds to (\ref{eq2.2.7}).  When using (\ref{eq4.5.14}) in (\ref{eq4.4.13}), it is necessary to have an attention.  It is because in the proof of (\ref{eq4.5.14}) in this subsection, we used the N-dimensional method for both ultraviolet and infrared regularizations.  In deriving the differential equation to control the infrared divergences, we will keep the N-dimensional method for ultraviolet regularization, but the different regularization method will be used for the infrared divergences; introduction of non-zero mass $\lambda$ to the gauge boson for the infrared regularization.  Therefore, it is better to try again the proof of (\ref{eq4.5.14}) [or Low's theorem (\ref{eq4.4.13})].  However, the difference appears only in (\ref{eq4.4.11}) among the relations (\ref{eq4.4.9})--(\ref{eq4.4.14}) for simplest graphs; for which the $O(\lambda^2)$ breaking appears in its right hand side.  Namely, (\ref{eq4.4.11}) is replaced by the following expression:
\begin{align}
&\begin{minipage}{14cm}
\includegraphics[width=12cm, clip]{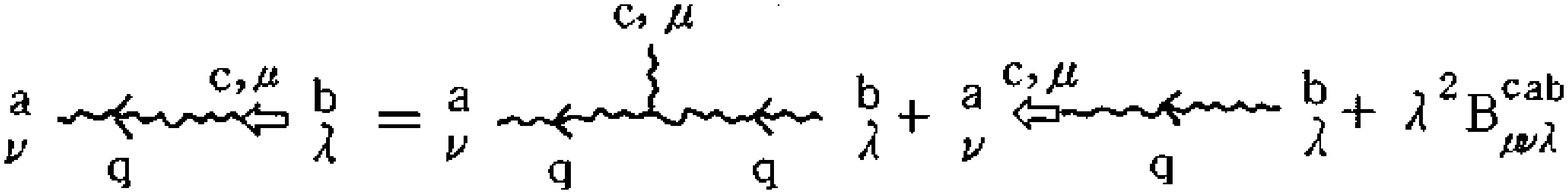}
\end{minipage}
\label{eq4.5.15} \\
& \text{where} ~ B^{cab}_{\mu\nu\lambda}= g f^{cab} \frac{1}{(q^2 - \lambda^2 + i \varepsilon)^2} \frac{\partial}{\partial q_{\mu}} K_{\nu\lambda}(q),  \label{eq4.5.16}  \\
& K_{\nu\lambda}(q) \equiv g_{\nu\lambda} - \frac{q_{\nu}n_{\lambda} + n_{\nu}q_{\lambda} }{q \cdot n} +(\alpha q^2 + n^2 ) \frac{q_{\nu}q_{\lambda}}{(q \cdot n)^2} ~[\text{numerator of the propagator}]   \label{eq4.5.17}
\end{align}
Accordingly, each step of proof and the final result of Low's theorem have always $O(\lambda^2)$ breakings. These breqakings do not finally remain after taking the limit of the cutoff $\lambda \to 0$, so that we can ignore these breakings.   This proof of Low's theorem in this subsection corresponds to 
 (\ref{eq2.2.8})--(\ref{eq2.2.17}) in Section 2.2.

Now, we can use (\ref{eq4.5.14}) in (\ref{eq4.5.13}), and obtain the following differential equation, 
\begin{align}
\lambda \frac{\partial}{\partial \lambda} \ln [S]_{\text{IR}} =& -\frac{1}{2} g_B^2 C_2(R) \int_0^{\delta} \frac{d^N k}{(2\pi)^N}  \left[ \frac{p^{'\mu}}{p' \cdot k} - \frac{p^{\mu}}{p \cdot k} \right] \left[ \frac{p^{'\nu}}{p' \cdot k} - \frac{p^{\nu}}{p \cdot k} \right] \notag \\
& \times \lambda \frac{\partial}{\partial \lambda} \tilde{D}_{\mu\nu} (k ; \lambda).       \label{eq4.5.18}
\end{align}
[Here, the part of $k_{\mu} \to 0$ contributes to the highest infrared singularities.]

This differential equation can be solved easily, and we have
\begin{equation}
[S]_{\text{IR}}= e^{-\frac{1}{2} g_B^2 C_2(R) \int_0^{\delta} \frac{d^N k}{(2\pi)^N}  \left[ \frac{p^{'\mu}}{p' \cdot k} - \frac{p^{\mu}}{p \cdot k} \right] \left[ \frac{p^{'\nu}}{p' \cdot k} - \frac{p^{\nu}}{p \cdot k} \right] 
\times \tilde{D}_{\mu\nu} (k ; \lambda) }  \times S^{(0)},    \label{eq4.5.19}
\end{equation}
where $S^{(0)}= \bar{u}(p') \Slash{J}_{\text{colorless}} u(p)$.  There is the arbitrariness in the initial condition.  [(\ref{eq4.5.19}) implies that only the highest divergences are equal in both hand sides and the arbitrariness for the finite parts is allowed.]

In summary, the highest infrared divergences included in the fermion scattering amplitude by a colorless source is determined, as in (\ref{eq4.5.18}), by the behavior of the \underline{dressed gauge boson} \underline{propagator}, $\tilde{D}_{\mu\nu} (k ; \lambda)$.  [Of cource this result holds under the axial gauge condition.]  

Comparing with QED, the results are formally equal up to the group factor $C_2(R)$ [$\frac{N^2-1}{2N}$ for $SU(N)$].  The essential difference exists, however, in $\tilde{D}_{\mu\nu} (k ; \lambda) $.  That is, in QED $Z_3$ is infrared finite [as in (\ref{eq2.2.20}) and (\ref{eq2.2.21})], while in QCD the infrared divergence exists in $Z_3$. As was shown by Frenkel and Meuldermans \cite{29} in the one loop calculation, the infrared divergence included in $\Pi^{(1)}$ near the mass shell, is
\begin{equation}
\Pi^{(1)}(k^2=0; N)_{\text{IR}} = - \frac{g^2}{2 \pi^2} c_2(G) \frac{1}{(N-4)^2},   \label{eq4.5.20}
\end{equation}
while $\Pi^{(2)}(k^2=0; N)_{\text{IR}}$ has no double pole at $N=4$, so that it can be ignored, in comparison with $\Pi^{(1)}$.  In our regularization adopted here, this gives
\begin{equation}
\Pi^{(1)}(k^2=0; \lambda)_{\text{IR}} = - \frac{g^2}{2 \pi^2} c_2(G) \left(\ln \frac{m}{\lambda}\right)^2  \label{eq4.5.21}
\end{equation}
[Please refer to the definition of $\Pi^{(1)}, \; \Pi^{(2)}$ in (\ref{eq4.3.27}).]  Thus, there is the infrared divergence in $Z_3=1/1+ \Pi^{(1)} $ (on-shell), not appearing in QED.  Of course, if we perform the renormalization on the mass shell, we have 
\begin{align}
g_B^2 \tilde{D}_{\mu\nu}(k ; \lambda) =& (g_B Z_3^{1/2})^2 \frac{\tilde{D}_{\mu\nu}(k ; \lambda)}{Z_3} \notag \\
=& (g_R)^2 \tilde{D}_{\mu\nu}(k ; \lambda)_{R},  \label{eq4.5.22}
\end{align}
so that divergence of $Z_3$ in $\tilde{D}_{\mu\nu}(k ; \lambda)$ is renormalized into $g_R$. If we perform the off-mass-shell renormalization [as in (\ref{eq4.4.40})--(\ref{eq4.4.43})], however, we have 
\begin{equation}
g_B^2 \tilde{D}_{\mu\nu}(k ; \lambda) = g(\mu)^2 \tilde{D}_{\mu\nu}(k ; \lambda, \mu) \left[\frac{Z_1(\mu)}{Z_2(\mu)} \right]^2,  \label{eq4.5.23}
\end{equation}
so that the infrared divergence remains in $\tilde{D}_{\mu\nu}(k ; \lambda, \mu)$, not being renormalized into the off-shell coupling $g(\mu)$. [Here, $\tilde{D}_{\mu\nu}(k ; \lambda, \mu)$  is the dressed gauge boson propagator, renormalizd at a renormalization point $\mu$.]  Because, for $\mu \ne 0$, $Z_1(\mu), Z_2(\mu), Z_3(\mu)$ are infrared finite.   Here again, the characteristic feature of QCD, that the infrared divergence appears in the charge renormalization, plays the important role.  In short, the problem is reduced to investigate the behavior at $k_{\mu} \sim 0$ of the \underline{dressed gauge boson propagator, $\tilde{D}_{\mu\nu}(k ; \lambda, \mu)$}.  This is the same kind of problem as that of investigating how does $g(\mu)$ behave for $\mu \to 0$ [with referring (\ref{eq3.4.20}) and (\ref{eq3.4.21})], so that this will be the important target from now on.

\section*{Acknowledgements}

At the end of the thesis, the author gives his sincere thanks to Dr. Ukawa for introducing the author to this field, and to Mr. Nakagawa and Mr. Yamamoto for their collaboration in Reference paper II.  He is grateful to thank Dr. Shimizu, Dr. Higashijima, Dr. Sato, Dr. Arafune, Dr. Fukugita, Mr. Nishimura (A.), Dr. Nakazawa, Dr. Okamura, and Dr. Okabayasi for their valuable discussions and advices as well as encouragements to him, since his beginning to work in this field.  He gives his gratitude to Professor Yamaguchi and the members of the particle physics group at University of Tokyo for their warmly watching the author since his entrance to the university.

%%%%%%%%%%%%%%%%%%%%%%%%%%%%%%%%%
%
%\addcontentsline{toc}{section}%
%

\section{Appendix: Application of M. Sato's microfunction
to the cancellation problem of infrared divergences}
%
%%%%%%%%%%%%%%%%%%%%%%%%%%%%%%%%%%%

This Appendix includes the latter half of the author's report, which was submitted to Prof. Mikio Sato in August (1976), to get the credit of the graduate course in physics at U. of Tokyo.  The title of the course was ``Microlocal Analysis and Quantum Field Theory" \cite{Sato}.

The first half of the report (Section 1 and 2) was an introductory explanation of the cancellation of infrared divergences in QED; $d\sigma_V + d\sigma_B=$ infrared finite, where $d\sigma_V$ is the cross section with the virtual corrections by photon of fermion scattering by an external source, while $d\sigma_B$ is the cross section associated with the emission of an additonal photons (``B" means Bremmsstrahlung), and hence this part can be omitted, except for important sentences and equations.

\subsection{A claim stated in Section 1 of the report}
The singular spectrum (S.S.) of a micro function gives the place where the function becomes not microanalytic.  Since the infrared divergence appearing in the scattering amplitude belongs to the S.S. of a Feynman integral (which is a micro function), the analysis by S.S. is especially effective, when the divergent part and its cancellation are dicussed without referring to the finite part.

\subsection{Definition of $J_a$ and $J_b$ given in Section 2 of the report}
The cancellation of infrared divergences can be understood by the Cutkosky rule.  For example, consider the forward amplitude, $I$, of the scattering of an electron by an external source, 
%%%%%%%%%%%%%%%
%
\begin{equation}
\begin{minipage}{12cm}
\centering
\includegraphics[width=10cm,clip]{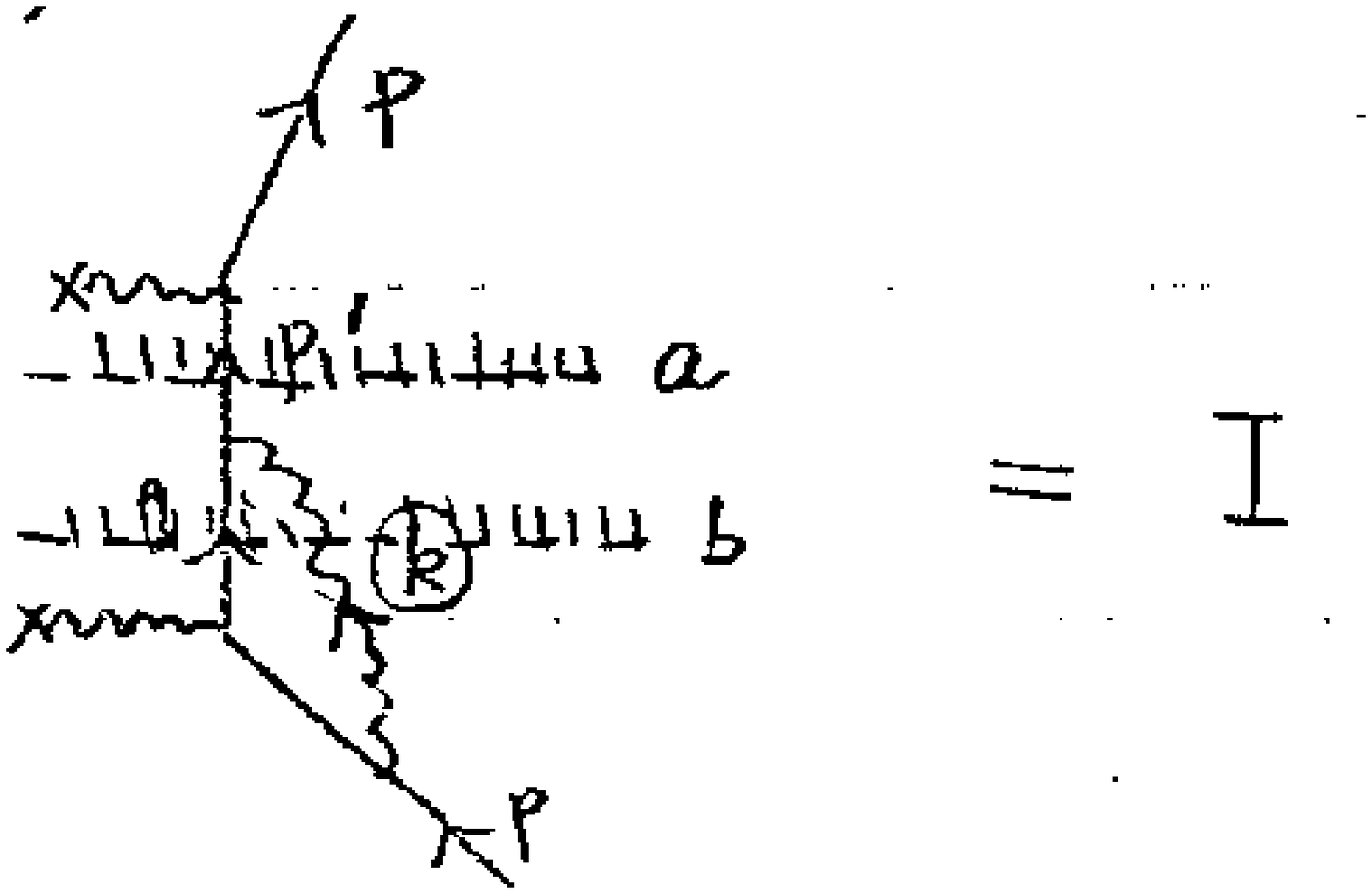} \label{eq.a.1}
\end{minipage},
\end{equation}
%
%%%%%%%%%%%%%%%%%%%%%%%
where the two diffferent Cutkosky cuts at $a$ and $b$ are taken. The numerator is common to both cases, so that we consider $J$, which is $I$ up to the numerator:
\begin{eqnarray}
J=\int dp_0' \; dq_0 \; dk_0 \; \frac{1}{p^{'2}-m^2 + i 0} \cdot \frac{1}{q^{2}-m^2 + i 0} \cdot \frac{1}{k^{2} + i 0} \; \delta(p'-q-k). \label{eq.a.2}
\end{eqnarray}
Application of Cutkosky cuts at $a$ and $b$ gives,
\begin{eqnarray}
&&J_a=\int dp_0' \; dq_0 \; dk_0 \; (-2\pi i) \theta(p'_0) \delta(p^{'2}-m^2) \cdot \frac{1}{q^{2}-m^2 + i 0} \cdot \frac{1}{k^{2} + i 0} \nonumber \\
&&~~~~~~~\hspace{2cm} \times \delta(p'-q-k),  \label{eq.a.3} \\
&&J_b=\int dp_0' \; dq_0 \; dk_0 \; \frac{1}{p^{'2}-m^2 + i 0} \cdot (-2\pi i) \theta(q_0) \delta(q^{2}-m^2 )  \cdot (-2\pi i) \theta(k_0) \delta(k^{2}) \nonumber \\
&&~~~~~~~\hspace{2cm}\times \delta(p'-q-k). \label{eq.a.4}
\end{eqnarray}
If the sum of $J_a$ and $J_a$ reduces the degree of singularity by one, then the infrared divergences are cancelled between $a$ (virtual process) and $b$ (emission process of photon).

\subsection{Section 3-- Application of S.S. of micro function to infrared divergences}
We will estimate the singular spectra (S.S.) of Eq.(\ref{eq.a.3}) and Eq.(\ref{eq.a.4}).  Before that we summarize the formulae for S.S \cite{Sato}:\footnote{(Footnote added in 2022) The singular spectrum (S.S.) gives the ``singularity structure" of a microfunction, to which Feynman amplitudes belong.  Feynman amplitude is given by a product of propagators with $i\epsilon(=i0)$ rule, which means the singularities are avoided by shifting the pole positions a little in the imaginary direction.  Thus, the microfunction is defined as a boundary value on the real axis of an analytic function given in the complex plane.  To manifest the shift in the complex direction of the pole, the differential (cotangent vector) giving the pole shift in the imaginary direction, is specified additionally in the definition of S.S..  

An example of the differential can be seen for the propagator as 
\begin{eqnarray} 
\sqrt{-1} \frac{\alpha}{2} d(p^2-m^2)=\sqrt{-1} \left(\alpha \sqrt{\vec{p}^2+m^2}\right) \left\{-d(p^0+\sqrt{\vec{p}^2+m^2})+d(p^0-\sqrt{\vec{p}^2+m^2})\right\}, \label{eq.a.5}
\end{eqnarray}
which shows the directions of the shifts for the positive and negative energy poles are opposite.}
\begin{eqnarray}
&&\widehat{S.S.} \left( \frac{1}{p^{2}-m^2 + i 0}\right)=\left\{ \left(p, \sqrt{-1} \frac{\alpha}{2} d(p^2-m^2) \right) \Big| \; \alpha \ge 0,\;  \alpha(p^2-m^2)=0 \right\}, \label{eq.a.6} \\
&&\widehat{S.S.}  \left(\theta(p_0) \delta(p^{2}-m^2)\right)=\left\{ \left(p, \sqrt{-1} \frac{\alpha}{2} d(p^2-m^2) \right) \Big| \;  \alpha \in \mathbb{R},\;  p^2-m^2=0,\;  p_0 >0 \right\}. ~~~~~~~~~\label{eq.a.7}
\end{eqnarray}

The product and integral formulae are given as follows \cite{Sato}:
\begin{eqnarray}
&&\widehat{S.S.} \left(f(x) g(x)\right)=\left\{ (x, \sqrt{-1} \xi dx ) |\;  \xi=\xi_1+\xi_2~ s.t. \right. \nonumber \\
&&\hspace{3.5cm} \left. (x, \sqrt{-1} \xi_1 dx )=\widehat{S.S.} f(x), ~ (x,\sqrt{-1} \xi_2 dx )=\widehat{S.S.}g(x)\right\}, \label{eq.a.8} \\
&&\widehat{S.S.} \left(\int f(x, y) dy \right)=\left\{ (x, \sqrt{-1} \xi dx ) \Big| ~\exists y ~ s.t.~ (x, y; \sqrt{-1} \xi dx ) \in \widehat{S.S.} f(x, y)\right\}.~~~~~ \label{eq.a.9}
\end{eqnarray}

\subsubsection{$\widehat{S.S.}(J_a)$}
If we choose the coefficients of cotangent vectors in the directions of $p^{'2}-m^2$, $q^2-m^2$, $k^2$, and $p'-q-k$, be $\frac{\alpha}{2}$, $\frac{\beta}{2}$, $\frac{\gamma}{2}$, and $y^{\mu}$, respectively, then we have
\begin{eqnarray}
\widehat{S.S.} (J_a)=\left\{ \left(\vec{p'}, \vec{q}, \vec{k}; \; \sqrt{-1} [ \vec{u}d\vec{p'}+\vec{v}d\vec{q}+\vec{w}d\vec{k}] \right) \right\}, \label{eq.a.10}
\end{eqnarray}
where there exist $\alpha \in \mathbb{R}$, $\beta \ge 0$, $\gamma \ge 0$, $y^{\mu} \in \mathbb{R}, \; p^{'0}, \; q^0, \; k^0$, such that the following condition holds,
\begin{eqnarray}
\vec{u}d\vec{p'}+\vec{v}d\vec{q}+\vec{w}d\vec{k}=\frac{\alpha}{2}d(p^{'2}-m^2) + \frac{\beta}{2}d(q^{2}-m^2) +\frac{\gamma}{2}d(k^2) + \vec{y} d(\vec{p'}-\vec{q}-\vec{k}), \label{eq.a.11}
\end{eqnarray}
which is satisfied, when the following relations are satisfied,
\begin{eqnarray}
&& \vec{u}=\alpha \vec{p'} + \vec{y}, \; \vec{v}=\beta \vec{q}-\vec{y}, \; \vec{w}= \gamma \vec{k} -\vec{y}, \label{eq.a.12}\\
&& 0=\alpha p^{'0}+y^0, \; 0=\beta q^{0}-y^0, \; 0=\gamma k^{0}-y^0, \label{eq.a.13}\\
&&p^{'0}>0, \; p^{'2}-m^2=0;  \; \beta(q^2-m^2)=0, \; \gamma(k^2)=0, \label{eq.a.14}\\
&&(p'-q-k)^{\mu}=0. \label{eq.a.15}
\end{eqnarray}

Let us examine the conditions Eq.(\ref{eq.a.12})-Eq.(\ref{eq.a.15}).  

Eq.(\ref{eq.a.13}) appears due to the integrations over $p^{'0}, \; q^0$, and~$k^{0}$.  Eqs.(\ref{eq.a.14})(\ref{eq.a.15}) give the on shell condition, or means the singularity arises when the particle flowing the propagator becomes on the mass shell.  The kinematics shows that three momenta, $p', q, k$, can not be on the mass shell at the same time.  Therefore, the following two cases remain,
\begin{eqnarray}
(Case 1): \beta=0, \; k^2=0, ~~(Case2) : q^2-m^2=0, \; \gamma=0.  \label{eq.a.16}
\end{eqnarray}

First we consider (Case2); from Eq.(\ref{eq.a.13}), $y^0=0$ and $\alpha p^{'0}=\beta q^0=0$.  Since on the mass shell conditions tell $p^{'0}, \; q^0 \ne 0$, we have
\begin{eqnarray}
\alpha=\beta=\gamma=y^0=0. \label{eq.a.17}
\end{eqnarray}
This implies the corresponding cotangent vector (the direction to avoid the singularity) vanishes,
\begin{eqnarray}
\vec{u}d\vec{p'}+\vec{v}d\vec{q}+\vec{w}d\vec{k}=\vec{0}, \label{eq.a.18}
\end{eqnarray}
so that $J_a$ is \underline{microanalytic} in this case.

Thus we consider (Case1); from Eq.(\ref{eq.a.13}), $y^0=0$ and since $p^{'0} \ne0$, we have 
 \begin{eqnarray}
 \alpha=0, \; k^{0}=0,  \label{eq.a.19}
 \end{eqnarray}
 where $\gamma=0$ holds as before.
 
 Now we have
 \begin{eqnarray}
 \alpha=\beta=y^0=0; \; |\vec{k}|=0, \; \vec{u}=\vec{v}=\vec{0}, \; \vec{w}=\gamma \vec{k}. \label{eq.a.20}
 \end{eqnarray}

Of course, if $|\vec{k}|=0$, then we have $\vec{w}=\vec{0}$, and $J_a$ becomes microanalytic.  This happens because we have not do the regularization.  In reality we have to introduce the regularization (photon mass) and consider
\begin{eqnarray}
|\vec{k}| \sim \lambda, \; \vec{w}=\gamma \vec{k}. \label{eq.a.21}
\end{eqnarray}

Summarizing the above consideration, we have arrived at 
\begin{eqnarray}
\widehat{S.S.}(J_a)=\left\{ \left(\vec{p'}, \vec{q}, \vec{k}; \sqrt{-1} \gamma (\vec{k} d\vec{k})\right) \big| \;  \gamma >0, \; |\vec{k}| \sim \lambda, \; k^2=0, \; \vec{p'}=\vec{q}+\vec{k} \right\}. \label{eq.a.22}
\end{eqnarray}

Taking into account the supplementary condition of
\begin{eqnarray}
\alpha=\beta=y^0=0, \label{eq.a.23}
\end{eqnarray}
the result of $\widehat{S.S.} (J_a)$ becomes quite reasonable from the physical view point. 

We will examine this in the following:
$S.S (J_a)$ arises from the region of $\alpha=\beta=0, \; \gamma \ne0, \;  |\vec{k}| \sim \lambda, \; k^2=0$.  If we consider $\alpha, \; \beta, \; \gamma$ be ``resistance" in the circuit analogy, $\alpha=\beta=0$ implies that the large electric currents $p', \; q$ flow along the corresponding paths, while $\gamma \ne0$ implies the small current $|\vec{k}| \sim \lambda$ flows along the path, which gives the region of infrared divergence in the limit of soft photon.  That is,
\begin{eqnarray}
\widehat{S.S} (J_a) \Leftrightarrow \mathrm{Infrared~divergence~region~of}~ J_a. \label{eq.a.24}
\end{eqnarray}

\subsubsection{$\widehat{S.S.}(J_b)$}
In the same manner, $\widehat{S.S} (J_b)$ can be obtained.  Namely we have
\begin{eqnarray}
\widehat{S.S.} (J_b)=\left\{ \left(\vec{p'}, \vec{q}, \vec{k}; \; \sqrt{-1} [ \vec{u}d\vec{p'}+\vec{v}d\vec{q}+\vec{w}d\vec{k}] \right) \right\}, \label{eq.a.25}
\end{eqnarray}
where there exist $\alpha \ge 0$, $(\beta, \gamma, y^0)  \in \mathbb{R}$, \; $p^{'0}, \; q^0, \; k^0$, such that the following conditions hold,
\begin{eqnarray}
&& \vec{u}=\alpha \vec{p'} + \vec{y}, \; \vec{v}=\beta \vec{q}-\vec{y}, \; \vec{w}= \gamma \vec{k} -\vec{y}, \label{eq.a.26}\\
&& 0=\alpha p^{'0}+y^0, \; 0=\beta q^{0}-y^0, \; 0=\gamma k^{0}-y^0, \label{eq.a.27}\\
&&\alpha (p^{'2}-m^2)=0;  \; q^0 >0, \; q^2-m^2=0; \; k^0>0, \; k^2=0, \label{eq.a.28}\\
&&(p'-q-k)^{\mu}=0. \label{eq.a.29}
\end{eqnarray}

Let us examine these conditions. From Eq.(\ref{eq.a.28}) we have $p^{'2} \ne m^2$, so $\alpha=0$, which implies $y^0=0$ and $\beta=0$ and $\gamma k^0=0$.

In the same manner as before, if $\gamma=0$, $J_b$ is microanalytic, so that we consider $\gamma \ne 0$.  Since $k^0 \sim \lambda$ and $k^2=0$, we have $|\vec{k}| \sim \lambda$.  Summarizing this, we have $\alpha=\beta=\gamma=0$, yielding 
\begin{eqnarray}
\widehat{S.S.}(J_b)=\left\{ \left(\vec{p'}, \vec{q}, \vec{k}; \sqrt{-1} \gamma (\vec{k} d\vec{k})\right) \big| \;  \gamma \ne 0, \; |\vec{k}| \sim \lambda, \; k^2=0, \; \vec{p'}=\vec{q}+\vec{k} \right\}. \label{eq.a.30}
\end{eqnarray}
This result implies $p'$ and $q$ are hard and $k$ is soft, giving the infrared divergence of $J_a$.  Schematically, we have
\begin{eqnarray}
\widehat{S.S} (J_b) \Leftrightarrow \mathrm{Infrared~divergence~region~of}~ J_b. \label{eq.a.31}
\end{eqnarray}

It is important to note that $\widehat{S.S.}(J_a)$ and $\widehat{S.S.}(J_b)$ overlap with each other, so that the sum of these two can reduce \underline{the degree of infrared divergence}.

\subsubsection{Sum of $J_a$ and $J_b$}
%%%
Instead of making a sum of $J_a$ and $J_b$, consider the forward scattering amplitude $J$ without cut at $a$ or $b$. Tnen, we have
\begin{eqnarray}
\widehat{S.S.} (J)=\left\{ \left(\vec{p'}, \vec{q}, \vec{k}; \; \sqrt{-1} [ \vec{u}d\vec{p'}+\vec{v}d\vec{q}+\vec{w}d\vec{k}] \right) \right\}, \label{eq.a.32}
\end{eqnarray}
where there exist $(\alpha, \beta, \gamma) \ge 0, \; y^0 \in \mathbb{R}; \;   p^{'0}, \; q^0, \; k^0$, such that the following conditions hold,
\begin{eqnarray}
&& \vec{u}=\alpha \vec{p'} + \vec{y}, \; \vec{v}=\beta \vec{q}-\vec{y}, \; \vec{w}= \gamma \vec{k} -\vec{y}, \label{eq.a.33}\\
&& 0=\alpha p^{'0}+y^0, \; 0=\beta q^{0}-y^0, \; 0=\gamma k^{0}-y^0, \label{eq.a.34}\\
&&\alpha (p^{'2}-m^2)=0;  \; \beta(q^2-m^2)=0; \; \gamma k^2=0, \label{eq.a.35}\\
&&(p'-q-k)^{\mu}=0. \label{eq.a.36}
\end{eqnarray}

Let us examine these conditions. We can consider four cases.

(Case1): $\alpha=\beta=\gamma=0$ gives the zero cotangent vector, so that this case can be dropped. 

(Case2): $p'$ is on the mass shell, $p^{'2}-m^2=0$, and others are off shell, $q^2-m^2 \ne0$, $k^2 \ne 0$.  This case also gives the zero cotangent vector and is dropped.

(Case3): $q$ is on the mass shell, but others are off shell, $p^{'2}-m^2 \ne 0$, and $k^2 \ne 0$. This is also irrelevant.

(Case4): $k$ is on the mass shell $k^2=0$, and the others are off shell, $p^{'2}-m^2 \ne 0$ and $q^{2}-m^2 \ne 0$.  This implies $\alpha, \beta =0$, $y^0=0, \; \gamma k^0=0$.  Choose $\gamma\ne0$ and $k^0 \sim \lambda$, we have $|\vec{k}| \sim \lambda$.  This can give the infrared divergence.

(Case5): $p^{'2}=m^2, \; q^{2}=m^2$, but $k^0\ne 0$.  This implies $k^0 \ne0, \; \gamma=0, \; y^0=0, \; \alpha=\beta=0$, so that this is an irrelevant case.

(Case6): The conditions $p^{'2}=m^2, \; k^2=0$, but $q^{2}\ne m^2$ implies $\beta=0, \; y^0=0, \; \alpha=0$, and $\gamma k^0=0$, so that $\gamma\ne0$ and $k^0 \sim \lambda$, and $|\vec{k}| \sim \lambda$.  This is a relevant case.

(Case7): We have $q^2=m^2, \; k^2=0$, and $p^{'2}\ne m^2$.  These are the same conditions to (Case6).

Since three vectors can not be on the mass shell at the same time, and hence we have exhausted all the cases.  Now the conditions lead to
\begin{eqnarray}
&&\alpha=\beta=y^0=0, \; \gamma >0, \label{eq.q.37} \\
&&k^2=0, \; k^0\sim |\vec{k}|\sim \lambda, \label{eq.q.38}
\end{eqnarray}
and (Case4) $p'$ and $q$ are both off the mass shell, and (Case6) or (Case7) One of $p'$ and $q$ is on the mass shell, remain as the possible cases to generate infrared divergences. (Both $p'$ and $q$ are off the mass shell is prohibited.) Thus, we have
\begin{eqnarray}
\widehat{S.S.}(J)=\left\{ \left(\vec{p'}, \vec{q}, \vec{k}; \sqrt{-1} \gamma (\vec{k} d\vec{k})\right) \big| \;  \gamma > 0, \; k^2=0, \; |\vec{k}| \sim \lambda, \;  \vec{p'}=\vec{q}+\vec{k} \right\}. \label{eq.a.39}
\end{eqnarray}

Schematically, 
\begin{eqnarray}
\widehat{S.S} (J) \Leftrightarrow \mathrm{Candidate~of~infrared~contribution~to}~ J. \label{eq.a.40}
\end{eqnarray}

By the Cutkosky rule, we have 
\begin{eqnarray}
J + J^* \propto  J_a+J_b, \label{eq.a.41}
\end{eqnarray}
and know that the infrared singularities cancel between $J_a$ and $J_b$, and that $J$ has no infrared singularities. 

The microlocal analysis clarifies the position of singularities, but we need to know furthermore ``Order of Singularities" at the singular position.

Physically, we know the ``Ord." of singularities by power counting with respect to the photon momentum $k$, that is, 
\begin{eqnarray}
&&Ord.[J_a] =Ord.[J_b]= -2, \label{eq.a.42} \\
&&\mathrm{and}~~~Ord.[J]=-1. \label{eq.a.43}
\end{eqnarray}
These equations show the appearance of infrared divergences and the cancellation of them.  

Next, we will give an ``primitive way of power connoting".  First we introduce the following criterions:

(C1) Assign -2 to the photon propagator, 

(C2) Assign -1 to the internal fermion propagator which connects to the photon on the mass shell and to the outgoing electron on the mass shell,

(C3) Assign +1 to the integration by photon momentum.

With these criterions we have
\begin{eqnarray}
&&Ord.[J_a]=Ord.[J_b]=-2-1+1=-2, \label{eq.a.44} \\
&&Ord.[J_a+J_b] \le Ord.[J] = -2+1=-1, \label{eq.a.45}
\end{eqnarray}
where the Cutkosky rule is taken into account.  This reproduces the result obtained by the explicit integration of the amplitudes.

We hope to have the technique in the micro functions to estimate the Order of  Singularities (Ord.).

(Note added in 2022) Ten years later, a similar work to this Appendix appeared \cite{Kawai}. 

%%%%%%%%%%%%

-----------------------------------------------------

\section*{Acknowledgements and epilogue in 2022}
The author deeply thanks Shiro Komata who kindly offered to translate a good part of this thesis into English, written originally in Japanese in 1978.  Without his help, the English translation would not have been achieved. 

He also thanks So Katagiri, Yoshiki Matsuoka, and Kimiko Yamashita for reading and refining of the draft.

\subsection*{Epilogue (2022)}

Even now, the interplay between the infrared (IR) behavior and the ultraviolet (UV) behavior in some theories (especially in quantum gravity), seems not to be fully understood. During the past 45 years, Wilson's lattice gauge theory \cite{Wilson} was so successful in doing the strong coupling expansion, which gives not only hadron masses but also scattering amplitudes consistent to the experimental results.  In this lattice gauge theory, a linearly rising potential between quarks is given, 
$$V_{q\bar{q}}(k) \propto \frac{g(|k|)^2}{|k|^2} \propto \frac{1}{|k|^4} + \cdots, ~\text{or}~V_{q\bar{q}}(r) \propto r + \cdots,$$
which was a mere guess in this thesis. Also the evolution equation, similar to the differential equation derived in this thesis, was developed by Dokshitzer, Gribov-Lipatov, and Altarelli-Parisi (DGLAP) \cite{DGLAP}, describing the evolution of the parton distribution functions, consistently with experiments in QCD. 

This thesis claims the importance of Ward-Takahashi identities in
the cancellation of infrared as well as of mass singularities to occur, while guaranteeing the unitarity. For the cancellation to occur \underline{the gauge invariant set of graphs} is essential. 

To represent the set, however, Wilson operators, such as W[Path=open] and \\
W[Path=closed] for non-Abelian gauge theory and W[Path=closed] for gravity, seem not completely satisfactory. The more suitable definition of variables surely exists, which is made up of massless quarks, gluons and gravitons, moving almost in
parallel with each other. Such variables include the soft momentum cutoff
$\lambda'$ and the angle cutoff $\eta$ from the parallel direction, or the fictitious mass $\lambda$ introduced to quarks, gluons and gravitons.
The new variables, if denoted as 
$$X[\text{open}; \lambda] ~ \text{and} ~ X[\text{closed}; \lambda],$$
are expected to follow the string dynamics, with a tension given by the cutoff $\lambda$,
$$\alpha' \propto \frac{1}{\lambda^2}.$$ 
The theories are IR finite from the beginning, due to the cancellation of IR divergences for the gauge invariant sets, studied in this thesis, where the Ward-Takahashi identities were fully used.

Furthermore, if the modular symmetry ($\tau \to -1/\tau$) is imposed on the partition function, then the UV finiteness can also be achieved. Here,
$\tau$ is an evolution parameter in the partition function. The parameter is the ratio per unit length $L$ of the object, of a time difference at finite temperature $T$, {\it i.e.} 
$$\tau=\frac{c\Delta t}{L} +i\frac{\hbar c}{k_B TL}.$$
It is quite plausible that the partition function for an infrared free set or a linearly extended object can be consistently given when the modular invariance is satisfied.

Recently, Ward-Takahashi identities associated with the Virasoro-like symmetry, arising in the soft graviton limit of gravity, has been understood by A. Strominger and others \cite{Strominger}, as the soft graviton theorem.   Some 60 years ago, the symmetry was known by Bondi, van der Burg, Metzner and Sacks (BMS) \cite{BMS}, and the soft graviton theorem was also known in \cite{2}, but the interrelaton between them is clarified only recently. This new topics called ``Celestial Holography" uses the similar terminology as in this thesis, and seems to pursuit the similar issue mentioned in the above. 

Frankly speaking, many researchers of the infrared slavery at that time, were surely aimimg at the issue mentioned above. A reason why they (including the author) would not finish the issue is the appearance of the numerically calculable lattice gauge theories.  History repeats itself.

This motivates the author for translating his old thesis recently. He hopes this English translation contributes to the study of the issue.

%%%%%%%%%%%%%%%%%%%%%%%%%%
%%%%%%%%%%%%%%%%%%%
% References
%%%%%%%%%%%%%%%%%%%%%%%%%%

%%%%%%%%%%%%%%%%%%%%%%%%%%%


\begin{thebibliography}{99}
\bibitem{30}
Among the divergences produced by the parameter integral, it is the ultraviolet divergence (UV) that is independent on the momentum configuration
of the external line,
and on the mass of the particle on the internal line.
This is because the external momentum configurations
and the mass of the internal particles, are irrelevant, when the internal loop momentum becomes infinitely large in the UV limit.
\bibitem{31}
Other than the ultraviolet divergence mentioned in \cite{30},
another divergence occurs, when the two conditions hold simultaneously; one is
the external line being on the mass shell, and the other is
a particular particle being massless.  This divergence is called
infrared divergence (IR), an example of which is the divergence
occurring at $\alpha=\beta=0$.  By the circuit theory,
Feynman parameter acts as a resistance against the current (momentum flow).
Therefore $\alpha=\beta=0$, $\gamma=1$ corresponds to the
momentum configuration, in which $p'+k$ and $p+k$
associated with $\alpha$ and $\beta$, respectively, being hard,
and the momentum $k$ with $\gamma$ being soft.
\bibitem{32}
This symbol will work as a powerful tool for QCD in Section 4.
\bibitem{33}
However, in electron-positron pair creation and pair annihilation processes,
two poles at $k^0=\frac{\bm{p'}\cdot \bm{k}}{p'^0}$ and
$\frac{\bm{p}\cdot \bm{k}}{p^0}$ are split into the upper and lower
half planes.  Thus one of the two poles contributes to infrared
divergence.  This contribution acts as pure imaginary for
(\ref{eq2.3.8}) and produces a pure imaginary phase $e^{i\infty}$
in the scattering amplitude.  This is called the ``Coulomb phase''.
In the cross section the divergence vanishes since
$|e^{i\infty}| = 1$.  For the details, see reference \cite{2}
(in particular, the Appendix of S.\ Weinberg's paper).
\bibitem{36}
The Feynman rule is shown in, for instance, Reference paper I.
\bibitem{38}
From the experience from QED, $\frac{1}{N-4} \leftrightarrow \frac{1}{4}\ln \frac{1}{\mu^2}$ \ref{eq2.1.18}, but the coefficients in the relation are not clear.
\bibitem{42}
It is noted that varous choices are possible for the independent momenta for the vertex.
\bibitem{43}
It is necessary to pay attention to the position where the arrow symbol is attached.  First, fix the flow of independent external momenta and independent internal loop momenta.  Then, the momentum flowing a given internal line is a linearly indepemdent sum of these external momenta and internal loop momenta.  
The arrow symbol associated with a given independent momentum is attached to the lines along which the momentum flow.  It is recommended to make some examples for the better understanding.
\bibitem{44}
It is noted that, since the external field does not have a color, the first and the second terms in (\ref{eq4.4.36}) have the same color factor $[t^c]_{ji}$.
\bibitem{45}
It is not clear in the off-mass-shell renormalization, however, whether a relation $Z_1(\mu)=Z_2(\mu)$ holds or not; which depends on how $\mu$ is chosen.
\bibitem{46}
The $-[A]_{\text{IR}}$ in Subsection 3.4 corresponds to (\ref{eq4.4.46}) of the covariant gauge.  However, there is no proper explanation of why $1/(N-4)$ in the covariant gauge becomes $1/(N-4)^2$ in the axial gauge.  Therefore,  it is necessary to recheck the calculation of this $[c^{(1)}(\mu)]_{\text{IR}}$.
\bibitem{47}
The application of off-mass-shell or on-mass-shell renormalization for (\ref{eq4.4.2}) is the same as the renormalization of (\ref{eq4.4.1}) in Subsection 4.4.

Express formally (\ref{eq4.4.3}) as
\begin{equation}
\sum_{n=0}^{\infty} (g_B)^{2n+2} D^{(n)}= (g_B Z_3^{1/2})^2 \left( \frac{p' \cdot \varepsilon}{p' \cdot k} -  \frac{p \cdot \varepsilon}{p \cdot k} \right) \left( \frac{p' \cdot \varepsilon}{p' \cdot l} -  \frac{p \cdot \varepsilon}{p \cdot l} \right) \sum_{n=0}^{\infty} (g_B)^{2n} B^{(n)},  \notag
\end{equation}
and define the renormalized amplitudes of $D^{(n)}$ as $D^{(n)}(\mu)$, 
\begin{equation}
\sum_{n=0}^{\infty} D^{(n)} = \sum_{n=0}^{\infty}[g(\mu)]^{2n+2} D^{(n)}(\mu), \notag
\end{equation}
then the pertuabative expansion of the renormalized (\ref{eq4.4.3}) with $\mu \ne 0$, is given for the off-mass-shell renoramalization by
\begin{equation}
 D^{(n)(\mu)}= \left( \frac{p' \cdot \varepsilon}{p' \cdot k} -  \frac{p \cdot \varepsilon}{p \cdot k} \right) \left( \frac{p' \cdot \varepsilon}{p' \cdot l} -  \frac{p \cdot \varepsilon}{p \cdot l} \right) \sum_{i, j \ge 0}^{\infty}  c^{(i)}(\mu) c^{(j)} (\mu) B^{(n-i-j)},  \notag
\end{equation}
while for the on-mass-shell renormalization with $\mu=0$, by
\begin{equation}
 D^{(n)(0)}= \left( \frac{p' \cdot \varepsilon}{p' \cdot k} -  \frac{p \cdot \varepsilon}{p \cdot k} \right) \left( \frac{p' \cdot \varepsilon}{p' \cdot l} -  \frac{p \cdot \varepsilon}{p \cdot l} \right)  B^{(n)}(0).  \notag
\end{equation}
\end{thebibliography}
\end{document}